\documentclass[11pt,twoside,bibliography=totoc]{report} 
\usepackage{pifont}
\usepackage{natbib}
\usepackage{graphicx}
\usepackage{amssymb}
\usepackage{epstopdf}
\usepackage{amsmath}
\usepackage{subfigure}
\usepackage{lscape}
\usepackage{latexsym}
\usepackage{amsmath}
\usepackage{yhmath}
\usepackage{bm}
\usepackage{epsfig}
\usepackage{fancyhdr}
\usepackage{sectsty,lscape,url}
\usepackage[hmargin=1in,vmargin=1in]{geometry}

\usepackage{enumitem}
\usepackage[utf8]{inputenc}
\usepackage[T1]{fontenc}

\usepackage{datetime}

\newcommand{\mychapter}[2]{
    \setcounter{chapter}{#1}
    \setcounter{section}{0}
    \chapter*{#2}
    \addcontentsline{toc}{chapter}{#2}
}

\begin{document}

\author{Ilia Mindlin}
\title{Water Waves: Nonlinear Theory \\  *** \\ {\it {-- Monograph --}} }
\date{2022/02/01}
\maketitle

\thispagestyle{empty}

{\it{Mindlin, Ilia Michailovich. Water Waves: Nonlinear Theory, 2022. \\

\bigskip

This book presents a novel mathematical nonlinear theory of surface gravity waves in deep water. 
It performs analytical analysis of the classical nonlinear equations of fluid dynamics under less restrictive assumptions than those applied by existing theories. In particular, the new theory ensures uniqueness of the solution without the need to employ the so-called radiation condition, 
and its solutions are such 
that the liquid always remains at rest at infinity, and the energy supplied to the water by a source of disturbances is finite at all times - all that in contrast with conventional approaches that operate in terms of spatially-infinite harmonic waves. The new theory accounts for the non-linearity of the problem, and yields solutions valid at all times, from zero (the time of setting initial conditions) to infinity. The author describes previously unknown patterns in wave evolution, and confirms those patterns with the experimental results by other researchers. \\

The book is intended for graduate students and researchers specializing in hydrodynamics, from wind-generated waves to tsunamis.}}


\fancyhead{}
\tableofcontents

\fancypagestyle{main}{
\fancyhf{}
\fancyhead[RO]{\textsl{\rightmark}}
\fancyhead[LE]{\textsl{\leftmark}}
\fancyfoot[CO,CE]{\thepage}
\renewcommand{\headrulewidth}{0pt}
\renewcommand{\footrulewidth}{0pt}
}

\fancypagestyle{intro}{
\fancyhf{}
\fancyfoot[CO,CE]{\thepage}
\renewcommand{\headrulewidth}{0pt}
\renewcommand{\footrulewidth}{0pt}
}

\def\pd#1#2{\frac{\partial #1}{\partial #2}}
\def\hpd#1#2{\frac{\hat \partial #1}{\partial #2}}

\pagestyle{intro}
%
%
%
%
%
%
%
%
%
%
%
%
%
\mychapter{0}{Introduction}   

This monograph presents a mathematical nonlinear theory of  surface gravity waves in deep water, which continues the monograph ``Integro-differential Equations in Dynamics of a Heavy Layered Liquid'' by the author published in Russia in 1996. 
\vspace{3mm}

A typical flow geometry considered in this book is that of a heavy liquid occupying a half-space below a lighter liquid separated by the liquid-liquid interface. The two liquids are supposed to be non-viscous, immiscible, and homogeneous with constant densities. The system, initially at rest with a horizontal surface, is set moving 
 either by an initial displacement of its surface, or by initial disturbance to the particle velocities due to, say, a solid body moving below the interface or  by variable pressure acting on the free surface of the liquid
 (the concept of ``free surface'' is used when upper of the two liquids separated by the interface is a fictitious liquid of zero density). 
The ensuing motion is  irrotational. 
The position of the interface is not known a priori 
and is to be found while solving the differential or integro-differential
(depending on the formulation of the problem) equations
of fluid dynamics. In the book, all problems are formulated in terms of
integro-differential equations supplemented by physically reasonable boundary and initial conditions. 
\vspace{3mm}

The theory presented in this monograph is referred to as nonlinear theory of wave packets  (NTWP), to distinguish it from 
 linear wave theory (LWT) and weakly nonlinear wave theory (WNWT) 
commonly applied to the same class of problems.
The three  theories stem from the classical equations of fluid dynamics, which are: 
\begin{itemize}
\item
the Laplace equation for the velocity potential;
\item 
 the nonlinear kinematic condition on the fluid surface which means that a liquid particle in the fluid surface can have no velocity relative to the surface in the direction of the normal;
\item
 the nonlinear condition of the pressure continuity across the fluid surface;
\item
conditions at infinity;
and
\item 
 initial conditions. 
\end{itemize}
When the problem includes a solid body moving inside the liquid, the above equation list  also includes the kinematic condition on the surface of the solid, as well as the dynamic equation(s) for the solid. 
The nonlinear conditions on the interface present a barrier for finding the solution to the problem analytically, unless some simplifying assumptions are applied.
\vspace{3mm}

The  three  theories are quite different in technique of simplifying the classical equations and mathematical tools employed for finding solutions. The resulting solutions behave differently in time, and are being applied to different kinds of specific problems. 
 The differences between the  Packet Theory and LWT (we concentrate primarily on LWT here, since, as a matter of fact, mathematical formulation of WNWT deviates quite far from the original governing equations of the problem) follow:
\begin{enumerate}
\item
The cardinal difference is  
how conditions at infinity along the interface are prescribed in the theories. 

In the theory of wave packets, the solution of the equations is obtained under boundary conditions  which assure that the liquid remains at rest at infinity, and the energy supplied to the water by a source of disturbances is finite at any time.
  
In  the linear theory, the fluid velocity is required to be bounded at infinity along the interface. Under the condition of boundedness, the problem has a set of steady state solutions in the form of harmonic waves periodic in time and along the free surface. To produce  waves periodic along the free surface infinite in extent, infinite energy must be supplied to the liquid.   The physics here is somewhat mildly violated for the convenience of expressing the solution with the Fourier integral. 
\item
The boundary condition on the free surface imposed in NTWP ensure the uniqueness of the solution of initial-value problem. 

In the LWT, solution of an initial-value problem obtained as a Fourier integral is a linear combination of waves traveling to $\pm \infty$ in opposite directions.  
That is why in the LWT, radiation condition is prescribed  which means that only waves outgoing  from disturbed body of water represent  physically reasonable solutions. The waves coming from infinity are  ignored.

The inconsistency of the LWT which represents spatially and temporarily finite motion with a set of infinite harmonics had been noted so long ago that it might had been forgotten by now. Citing Stoker, ``the difficulty arises because the problem of determining simple harmonic motions is an unnatural problem in mechanics... The steady state problem is unnatural - in the author's  view, at least - because a hypothesis is made about the motion that holds for all time, while Newtonian mechanics is basically concerned with  the prediction - in a unique way, furthermore - of the motion of a mechanical system from given initial condition...

One should in principle rather formulate and solve an initial value problem by assuming the medium to be originally at rest everywhere outside a sufficiently large sphere, ...''  (Stoker, Water Waves, p.176)
\item
 In the theory of wave packets, it is assumed that the waves start to
propagate away from an initially disturbed body of liquid. 
It is also assumed that the characteristic horizontal dimension $D$ of the   body is much larger than the vertical displacement $d$ of the fluid surface in the wave originating area. 
Equations of the  linear theory are  obtained from classical equations assuming that amplitude $a$ of a wave is small compared to the wave length $\lambda$.  Small parameters $d/D<<1$ and $a/\lambda<<1$,
respectively, are used in the theories to simplify the classical equations. 
\item
In NTWP, equations are obtained with nonlinear boundary conditions on the evolving interface unknown in advance. The shape of the interface is thought in a parametric form which is equivalent to a nonlinear transformation of the governing equations.

In LWT, the conditions on the interface are linearized and shifted from the interface to the horizontal plane - the equilibrium position of the interface. Thus the vertical departure of the free surface becomes an explicit function of the horizontal coordinate (and time).
\item
In NTWP, the nonlinear transformation maps the trace of the liquid-liquid interface in a vertical plane onto a half of a circle. The mapping allows us to find a countable (with respect to spacial argument) set of solutions (specific wave packets). Any wave group is a nonlinear combination of the wave packets.
 
The LWT solution is assembled from a continuum of harmonic waves. 
\item
Nonlinear Theory of Wave Packets (NTWP) accounts for the non-linearity and provides the solution valid at all times, from zero (the time of setting initial conditions) to infinity.
\end{enumerate}

The theory of weakly nonlinear waves (WNWT) in deep water 
  is based on the concept that the frequency, amplitude, and wave number of a weakly nonlinear wave train are slowly varying functions of time. On this assumption, the classical equations were reduced to equations governing the evolution of the frequency, amplitude, and wave number. These equations in turn were reduced to nonlinear Schredinger equation for complex-valued wave envelope. 
Asymptotic behavior of solutions of the Schredinger equation was analyzed by Zakharov and Shabat (1972). Provided that the initial free surface displacement decays sufficiently rapidly with the distance from its peak, it was found, in particular, that an initial wave group with an arbitrary envelope eventually disintegrates into a number of final packets. Each final packet has a specific envelope which travels at a constant velocity, keeps its shape, and retains its shape and velocity after interaction with other final packets. Disintegration of a wave group into final packets traveling at different velocities is also the dispersion. But the final packets don't disperse. 
\vspace{3mm}

WNWT is not bridged to the LWT, but rather the two theories represent two unconnected patterns of wave behavior. Moreover, WNWT predictions contradict the observed behavior of water waves: even long waves in mid-ocean (wave length of order of 100-150 km for the ocean depth of 4-5 km) are dispersive, as shown by their records. 
\bigskip

 {\bf{Book content overview}}\\
 
The book presents a novel mathematical theory of nonlinear surface gravity waves in deep water. It starts with elaborating the mathematical tools in Chapters 1 and 2.
\vspace{3mm}
 
In Chapters 3 - 5,  equations of nonlinear initial-boundary value problem on deep-water gravity waves of finite amplitude are solved approximately (up to small terms of higher order) assuming that the waves on the interface are generated by an initial disturbance to the liquid and the horizontal dimensions of the initially disturbed body of the liquid  are much larger than the magnitude of the interface displacement.

A set of dispersive wave packets is found with one-to-one correspondence between the packets and positive integers, say, packet numbers, such that any initial  surface displacement gradually disintegrates into a number (limited or unlimited, depending on initial conditions) of the wave packets. The greater the packet number, the shorter the wavelength of the packet's carrier wave component, the slower the packet travels, the slower the envelope of the packet is widening; evolution of any packet is not influenced by evolution of another packet. The speed of any packet is less than the speed of packet number 1 which, in turn, depends on typical horizontal length of the wave origin. 

The analytical solution and original ``law of similarity'' for the wave packets presented in Chapter 5 are used to test the theory of packets 
against experiments performed in a water tank (waves on deep water, in terms of linear theory) by Feir (1967) and  Yuen, H.C. and Lake, B.M. (1982). 
The agreement between theoretical predictions based on NTWP and the  experimental  measurements in a water tank is surprisingly good. 
\vspace{3mm}  
   
Chapter 6 focuses on applying Packet Theory for estimating wave origin and the traveling waves parameters in a particular scenario of wave generation. The analytical solution and the ``law of similarity'' for the wave packets  are used  to estimate an assemblage of dimensional parameters of the wave origin and waves radiated from the origin.

It is assumed that initially the water surface has not yet been displaced from its mean level, but the velocity field has already become different from zero. This means that the motion of a body of water is triggered by a sudden change in the velocity field.
The scenario resembles a tsunami generation event. Some characteristics of the latter are in line with the assumptions of the theory of nonlinear wave packets, such as energy supplied to the water by the earthquake being finite, and the horizontal extent of the wave origin (tens to hundreds kilometers) immensely exceeding the departure of the water surface from the mean sea level (order of meters).

Considering the resulting wave as a specific wave packet and applying the ``law of similarity'', we develop a procedure for using water surface measurements at a location distant from the wave originating area for estimating duration of the wave origin formation, size of the origin, water elevation in the origin, energy supplied to the water by the motion trigger,  distribution of wave heights in the wave packet, speed of the waves. Relation between the speed of the wave of maximum height and its length is obtained. 

In section 8 of Chapter 6, we consider a situation when wave records at one or several DART{\footnote{Deep-ocean Assessment and Reporting of Tsunamis - bottom-pressure recorders paired with transmitting surface buoys used for monitoring oceans for tsunamis}} buoys are already known, but the records at the next buoys are not obtained yet.  Starting from the first record, a line of forecasts of the amplitude and arrival time of the wave of maximum height at the next buoys is produced corresponding to the timeline of the DART  records. 
 
 We exercise our  parameter estimation procedure with DART records of waves triggered by the earthquakes of November 2006 and January 2007 near Kuril Islands and of august 2007 near central Peru. 
 


Though the analytical solution to the problem and  the ``law of similarity'' was obtained for an idealized bottom-less basin, the above estimates agree with observations of long waves in the Pacific as well as with some results of numerical simulations of the Kuril tsunamis.
\vspace{3mm}

In Chapter 7, the problem on gravity waves 
 on the free surface of a  liquid initially at rest is solved analytically in cases where the external pressure force of limited power is distributed over a large area in the free surface but is otherwise arbitrary  
(it is also supposed that the pressure acts on the free surface of the water for infinite time interval)
 According to the solution,  the crests of the forced waves move faster than the troughs, the horizontal distance between a crest and the following  trough increases, and a chain of waves with ``overhanging''  develops when the packets  run away from the pressure zone. 
It is proved that at any fixed moment of time, the wave packet in two dimensions dies out like inverse square  of the distance from the pressure zone.
By passing to the limit in the solution as time goes to infinity, the form of the nonlinear steady-state waves is obtained. 
  At a distance from the pressure zone, the steady-state wave
  seems as infinite chain of Kelvin-Helmholtz billows. 
The chain is unstable. 
   
  The averaged energy absorbed by the liquid per unit time (energy absorption functional) is obtained from the solution. 
 In general, the waves transport energy toward infinity.
But a discrete set of specific frequencies is found at which the energy functional is equal to zero. This means that the steady-state wave does not transport energy toward infinity. Only standing waves have this property. 

The frequency spectrum and the form of the forced standing waves are found explicitly. The waves  have a finite number of nodes in the free surface infinite in extent. The amplitude of the standing waves dies out like inverse square  of the distance from its peak.
\vspace{3mm}
 
In Chapter 8, a problem on gravity waves excited in two layered liquid by oscillating solid cylinder is solved analytically assuming that the ratio of the cylinder radius to the distance to the interface is small. 
In the available literature, this problem is considered at linearized conditions on the interface, that is, in Cartesian rectangular coordinates. But in the Cartesian coordinates, one can't linearize a surface of a cylinder, so different simplifications are made.
For example, in \cite[]{tyv}, the cylinder is modeled as a dublet,  and solution to the problem on free surface waves generated by a cylinder initially at rest is sought in the form of power series of time; only first three terms of the series are obtained. 

On the contrary, in this book,  the solution is obtained with  no-flow condition on the  true surface of cylinder (as well as with nonlinear conditions on the liquid-liquid interface) and is valid at all times, from zero (the time of setting initial conditions) to infinity. According to the solution, a chain of waves with ``overhanging''  develops when the  wave  gets away from the oscillating cylinder.
Letting the time go to infinity, the steady state waves are found. 


\pagestyle{main}
\setcounter{chapter}{0}

\chapter{Mathematical Theory of a Vortex Sheet in an Ideal Liquid}

{\small{This chapter introduces the analytical approach to solving the initial value problem being developed in the book. This approach is based on expressing the fully-nonlinear classical  equations for flows with a free surface in a particular system of curvilinear coordinates. 
The resulting solution takes a form which does not involve the Fourier integral nor the so-called radiation condition, but instead expresses the motion in terms of functions vanishing at infinity along the free surface - as different from spatially-infinite sines and cosines. }}

\section{Generalized velocity potential }

Motion of two non-viscous incompressible liquids separated by a liquid-liquid interface is considered below.
\vspace{3mm}

The interface is infinite in horizontal directions. The half-space above  the interface is occupied by a homogeneous liquid of density $\gamma_1$, while a  homogeneous liquid of density $\gamma_2$ fills the half-space below the interface. 
The two-layered liquid is stably stratified, i.e., $\gamma_1 <\gamma_2 $. 
Diffusion across the interface and surface tension are neglected.
\vspace{3mm}

Let initially (at $t=0$) the flow inside each of the liquids is irrotational.
When the liquids are acted upon by no external force other than gravity, 
the flow inside each of the liquids  remains irrotational at $t>0$ .
It does not mean that the flow of the two layered liquid is irrotational: the circulation in a closed
circuit intersecting the interface is nonzero, and, consequently, the velocity potential does not exist. 
But it is possible to construct the generalized velocity potential $\Phi (M)$ such that 
fluid velocity ${\bf q}={\bf grad}\,\Phi$ inside  each of the liquids,  and both kinematic and dynamic conditions are satisfied on the interface. With all that, the potential $\Phi (M)$ and ${\bf grad}\,\Phi$ jump across the interface $S$.
\vspace{3mm}

A fluid surface is referred to as a vortex sheet, if there is a jump
in tangential velocity at the surface. Two nonlinear conditions are required to be satisfied on a vortex sheet:

i) kinematic condition which means that a liquid particle in the sheet can
have no velocity relative to the sheet in the direction of the normal;

ii) dynamic condition of the pressure continuity across the sheet.\\
The generalized velocity potential constructed in this section ensures that 
the nonlinear conditions on a vortex sheet evolving with time are satisfied automatically. 
\vspace{3mm}

\begin{figure}
\centering
	\resizebox{0.7\textwidth}{!}
			{\includegraphics{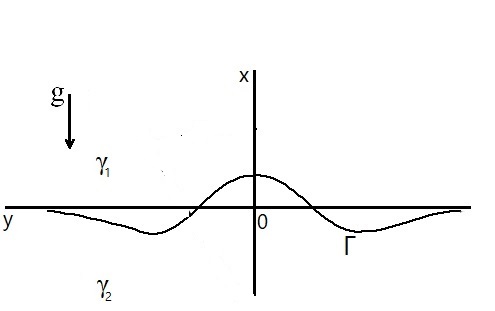}}
	\caption{Coordinate systems and sketch of the liquid-liquid interface}
	\label{c1_qu-.1-.10}
\end{figure} 

From now on we consider the flow in two dimensions. 
We think of the liquid as being contained between two vertical planes
(parallel to the $(x,y)$-plane) at a distance apart. 
Let the curve $\,\,\Gamma\,\,$ (in figure 1) be the trace of the interface in the  plane $\,\,(x,y)\,\,$,  the vertical $x$-axis be oriented upward and the $y$-axis be horizontal; 
 $\,x=f<0,\,$ $y=0$ be the coordinates of the pole $\,O_1\,$ of the polar coordinate system in the $\,(x,y)\,$ plane,
$\,\theta\,$ be the polar angle measured from the positive $x$-axis
in the counterclockwise direction, $\,t\,$ be the time, 
 $\, {\bf q}$ denote the velocity of a liquid particle relative to
the  "Earth-fixed"  coordinate system $\,(x,y)\,$.    
The equilibrium position of the interface is horizontal plane $x=0$. 
\vspace{3mm}
 
Generalized velocity potential $\Phi$ of the flow induced by the vortex sheet is sought in the form of  doublet distribution over the  sheet:
\begin{equation}
\Phi(Q)=\frac {1}{2\pi}\int\limits_{\Gamma}g(Q_1)\pd {}{n}
\left(\ln\frac {1}{r}\right)dl,
  \label{c1_1.1}                             
\end{equation}   
where  $dl$ is the infinitesimal element of the curve $\Gamma$.	 
Let $Q_1$ denote a point on the interface  $S$, $Q$ is a point outside $S$,  $r=|Q_1Q|$ is the distance between the points, $g(Q_1)$ is the density of the doublets distributed over the surface $S$. 
In the outside of the interface $S$ the Laplace's equation is satisfied
provided that the doublet density $g(Q_1)$ is a continuous function.
The interface $S$  is considered to be two-sided with
unit normals ${\bf n_{+}}$ and ${\bf n_{-}}=-{\bf n_{+}}$ directed
in the positive  and negative  sides of the interface, respectively.
\vspace{3mm}

For the one-sided limits we introduce the notations
$$
\Phi_{+}(Q_1)=\lim_{Q\to {Q_1}^{+}}\Phi(Q),\,\,\,\,\,\,\,
\Phi_{-}(Q_1)=\lim_{Q\to {Q_0}^{-}}\Phi(Q).  
        $$
where $Q\to {Q_1}^{+}$ ($Q\to {Q_1}^{-}$)  means that the point $Q$ approaches the point $Q_1$ from the positive (negative) side of the
interface.
The potentials have the following properties \cite[]{courant}  
\begin{equation}
\Phi_{+}(Q_1)- \Phi_{-}(Q_1)=g(Q_1)\,\,
\text{in traditional notations} , 
		$$
$$
\Phi_{+}(Q_1)- \Phi_{-}(Q_1)=-g(Q_1)\,\,
\text{in notations of Chapter 3}
 \label{c1_ 1.2}
 \end{equation}                               
\begin{equation}
2\Phi (Q_1)=\Phi_{+}(Q_1)+ \Phi_{-}(Q_1),\,\,\,\,\,\,\pd {\Phi}{n_{+}}+\pd {\Phi}{n_{-}}=0,     
 \label{c1_1.3}
 \end{equation}        
where $\pd{\Phi}{n_{+}} $
($\pd{\Phi}{n_{-}} $) denotes the derivative at the point 
$Q_1$ in the direction of the positive (negative) normal. 
\vspace{3mm}

The function $\Phi (Q)$ is a generalized velocity potential, if the normal component of the fluid velocity and the pressure change continuously 
across the interface. Equation  \eqref{c1_1.3}   shows that
the normal-velocity condition on the interface is satisfied automatically.
All parameters, variables, and equations have already been made
non-dimensional by the use of some reference quantities: length $L_*$, density $\gamma_*$, pressure $P_*$, and dimensional unit of time, $T_*$, defined by the relation $gT_*^2=L_*$, where $g$ is acceleration of free fall. The non-dimensional acceleration of free fall is equal to unity, so the force function for the weight per unit mass is  $-x$: ${\bf F}={\bf grad}(-x)$. 
Inside each of the two liquids  Bernoulli's equation hold \cite[]{miln}: 
\begin{equation}
\frac{P}{\gamma}+\frac{1}{2} q^2+x+\pd{\Phi}{t}=c(t).       
\label{c1_1.4}                                    
\end{equation}  

 \section {Vortex sheet in curvilinear coordinates}
 
Assuming that  the equation of a vortex sheet is  $V(x,y,t)=0$, we define curvilinear coordinates $u,\,v$ in the $x,y$-plane by the relations 
 \begin{equation}
 v=V(x,y,t),\,\,\,\,\,\,\,u=U(x,y,t),      
 \label{c1_2.1}
 \end{equation}                           
where functions $V(x,y,t),$ $U(x,y,t)$ and their first partial derivatives
are assumed to be continuous.  
The equation of the vortex sheet becomes $v=0$.  
\vspace{3mm}

The kinematic condition  may be written as
 \begin{equation}
 \left[\pd{V}{t}+\pd{V}{x} q_x+\pd{V}{y} q_y\right]_{V=0}=0,           
 \label{c1_2.2}
 \end{equation}                          
where $(q_x, q_y)$  are the components of the fluid velocity,
 ${\bf q}$, in the directions of the Cartesian axes.
\vspace{3mm}

The component of the fluid velocity normal to the vortex sheet is
$$
q_n= \frac{1}{N}\left[\pd{V}{x} q_x+\pd{V}{y} q_y
\right]_{V=0}=-\frac{1}{N}\pd{V}{t},
	$$
$$
N^2=\left(\pd{V}{x} \right)^2+\left(\pd{V}{y} \right)^2;
	$$ 
 the subscript $V=0$ shows that the expressions in the square brackets  are evaluated on the sheet. The normal velocity component is continuous on the sheet.
\vspace{3mm}

Notation $\hat G(u,v,t)$ is used for any function $G(x,y,t)$ expressed 
in curvilinear coordinates $u,\,v,\,t,\,$, i.e. 
$G(x,y,t)=\hat G[U(x,y,t),V(x,y,t),t]$.
The difference between the notations  $\pd {\hat G}{t}$ and 
$\hpd{G}{t}$ (as well as between  $\pd {\hat G}{x}$ and 
$\hpd{G}{x}$ ) should be noted:
the former means the partial derivative of $\hat G(u,v,t)$    
with respect to $t$, while the latter means that the partial derivative
$ \pd{G(x,y,t)}{t}$ is expressed in curvilinear
coordinates $u,v,t$, i.e.
 $$
 \hpd {G}{t}=\left. \pd {G}{t}\right|_{x=x(u,v,t),y=y(u,v,t)},\,\,\,\,\,\,
\hpd {G}{x}=\left. \pd {G}{x}\right|_{x=x(u,v,t),y=y(u,v,t)}.
	$$
	 The following notations are in use below
 \begin{equation}
 D=\pd{U}{x} \pd{V}{y}-\pd{U}{y} \pd{V}{x} \ne 0,\,\,\,\,\,\,
D_{11}=\left(\pd {U}{x}\right)^2+\left(\pd {U}{y}\right)^2,                                                          
 \label{c1_2.3}
 \end{equation}               
$$
D_{22}=\left(\pd {V}{x}\right)^2+\left(\pd {V}{y}\right)^2,\,\,\,\,\,\,D_{12}=\pd{U}{x} \pd{V}{x}+\pd{U}{y} \pd{V}{y}.	
	$$
	
	For one-sided limits of any function $a(u,v)$ the following notations are used:
 $$
 a_+=\lim_{v\to +0}a(u,v),\,\,\,\,\,\,\,
 a_-=\lim_{v\to -0}a(u,v),
        $$
$v\to +0$ ($v\to -0$) means that $v$ approaches $0$ from  the positive side   $v>0$ (negative side $v<0$) of the sheet.
\vspace{3mm}

Unit vectors in the normal and tangent directions to the level line $V={\rm const}$ in the $x,y$-plane are determined as
  \begin{equation}
 \vec n=\frac {1}{\sqrt { D_{22}}}\,\left(\hpd {V}{x},\,\,\,\pd {V}{y}\right),\,\,\,\,\,\,\,\,
\vec \tau=\frac {1}{\sqrt {D_{22}}}\left(-\pd {V}{y},\,\,\,\pd {V}{x}\right) .                                 
 \label{c1_2.4}
 \end{equation}     
 At a point on the level line  $V={\rm const}$ derivatives of function $f(x\,y)$ in the normal and tangential directions to the line equal  
 respectively
\begin{equation}
\hpd{f}{n}=\frac {1}{\sqrt { D_{22}}}\,\left(\hat D_{12}\,\hpd{f}{u}+\hat D_{22}\,\hpd{f}{v}\right),\,\,\,\,\,\, 
\hpd{f}{\tau}=-\frac {\hat D}{\sqrt {\hat D_{22}}}\,\hpd{f}{u}
\label{c1_2.5}                           
 \end{equation}     
In curvilinear coordinates the velocity components in the normal and tangent directions to the level line  $V={\rm const}$  are 
\begin{equation}
\hat q_n=\frac {1}{\sqrt { D_{22}}}\,\left(\hat D_{12}\,\hpd{\Phi}{u}+\hat D_{22}\,\hpd{\Phi}{v}\right), \,\,\,\,\,\,
\hat q_{\tau}=-\frac{\hat D}{\sqrt{\hat D_{22}}}\,
\pd{\hat \Phi}{u}.
 \label{c1_2.6}                            
 \end{equation}
	Using \eqref{c1_2.6} we write the condition of continuity of the normal velocity as 		
\begin{equation}
\left(\pd {\hat \Phi}{v_{+}}-\pd {\hat \Phi}{v_{-}}\right)\hat D_{22}=-
\left(\pd {\hat \Phi}{u_{+}}-\pd {\hat \Phi}{u_{-}}\right)\hat D_{12}.
 \label{c1_2.7}                                               
 \end{equation}
 The jump in the tangential velocity across the vortex sheet  is
\begin{equation}
(\hat q_{\tau})_+-(\hat q_{\tau })_-=-
\frac{\hat D}{\hat D_{22}}\left(\pd{\hat\Phi}{u}_+-
\pd{\hat\Phi}{u}_-\right)             
\label{c1_2.8}
\end{equation}
   In curvilinear coordinates the kinematic condition \eqref{c1_2.5}  governing evolution of the vortex sheet takes the form 
\begin{equation}
 \left[\hpd {V}{t}+\hat D_{12}\,\pd{\hat\Phi}{u}+
\hat D_{22}\,\pd{\hat\Phi}{v} \right]_{v=0}=0
\label{c1_2.9}                                                
\end{equation}             
     
The doublet  density $g(Q)$ is required 
to assure the pressure  continuity condition $P_+=P_-$. 
Formally, we can write two Bernoulli's equations \eqref{c1_1.4}
 (for the regions $v>0$ and $v<0$, respectively) 
$$
\frac{P(x,y,t)}{\gamma}+\frac12 q^2+x+\pd{\Phi}{t}=c(t),      	
	$$
	or, in curvilinear coordinates,
$$
\hat P(u,v,t)=-\gamma\left[\pd {\hat \Phi}{t}+
\pd {\hat \Phi}{u}\,\hpd {U}{t}+\pd {\hat \Phi}{v}\,\hpd {V}{t}\right.+
\left.\frac {1}{2}\hat q^2+\hat x(u,v,t)\right]                                                                                                   
        $$        
Using   condition of incompressibility $d\gamma/dt=0$, and equations 
\eqref{c1_2.7}, \eqref{c1_2.8},  \eqref{c1_2.9}, we obtain from the pressure continuity condition 
$$
\hat P_{+}=\hat P_{-}
	$$
the following governing equation for the doublet density
\begin{equation}
\left(\pd{g}{t}+\pd{g}{u}\hpd{U}{t}-
\frac{\hat D_{12}}{\hat D_{22}}\hpd{V}{t}\pd{g}{u}+
\frac{1}{2}\frac{\hat D^2}{\hat D_{22}}\pd{g}{u}
\left(\pd{\hat\Phi}{u}_++\pd{\hat\Phi}{u}_-\right)\right)_{v=0}+
\label{c1_2.10}                                    
\end{equation}
$$
\left(1-\frac{\gamma_2}{\gamma_1}  \right)
\left(\frac{1}{2}\,\hat q^2_-+\hat x+\hpd{\Phi}{t}_-\right)=0 
	$$				
where $\gamma_+$ 	and $\gamma_-$ are the densities of the liquids. 
Equation \eqref{c1_2.10} ensure  the pressure continuity on the evolving interface.
\vspace{3mm}

In curvilinear coordinates $u,\,v$ velocity potential \eqref{c1_2.1} 
takes the form
\begin{equation}
\hat \Phi (u,v,t)=-\frac {1}{4\pi}\int\limits_{\Gamma}
\frac {g(u_1,t)}{\hat D}
\left[ \hat D_{12}\pd {}{u_1}\ln {\hat R}+
\hat D_{22}\pd {}{v_1}\ln{\hat R}\right]_{v_1=0}du_1,                
\label{c1_2.11}                                
\end{equation}                                
$$
\hat R=(\hat x(u,v,t)-\hat x(u_1,v_1,t))^2+(\hat y^2(u,v,t)-
 \hat y^2(u_1,v_1,t))^2
        $$

In   \eqref{c1_2.11},    the derivatives 
$\partial \ln {\hat R}/ \partial u_1 $,  $\partial \ln{\hat R}/ \partial v_1$ 
 depend on five variables 
$u$, $v$, $u_1$, $v_1$, $t$ and are  evaluated at $v_1=0$; 
all other functions depend on three variables 
$ u_1$, $v_1$, $t$ only and are evaluated at $v_1=0$. 
Thus, the function $\Phi (P)$ \eqref{c1_2.1} 	is the generalized velocity potential in the region including the liquid-liquid interface $S$, if the function $g(u,t)$ satisfies the integrodifferential equation \eqref{c1_2.11}. 
\vspace{3mm}

The double-layer potential  \eqref{c1_2.1} 
is constant at the surface $S$, if the doublet density is constant, so the generalized velocity potential $\Phi$ and the density $g$ are determined by the equations of the liquid motion only up to an arbitrary addend  
specified on the condition that at infinity the pressure (inside each of the homogeneous liquids) depends linearly on the vertical coordinate $x$ as $|y|\rightarrow +\infty$.

\subsection{ Derivation of pressure continuity equation across the vortex sheet}

It follows from  Bernoulli's equation  \eqref{c1_1.4} that 
$$
P_+=\gamma_1\left( \frac{1}{2}\,q_+^2+x+\pd{\Phi_+}{t}  \right)
	$$
$$
P_--=\gamma_2\left( \frac{1}{2}\,q_-^2+x+\pd{\Phi_-}{t} \right)
	$$
	The condition of the pressure continuity reads 
$$
P_+-P_- \left(1-\frac{\gamma_2}{\gamma_1} \right)P_-=0
	$$ 
	or
$$
\pd{\Phi_+}{t}-\pd{\Phi_-}{t}+ \frac{1}{2}\,(q_+^2-q_-^2)+
 \left(1-\frac{\gamma_2}{\gamma_1} \right)
\left(\frac{1}{2}\,q_-^2+x+\pd{\Phi_-}{t}  \right)
	$$	

	At $v>0$ and $v<0$
$$
\pd{\Phi}{t}=\pd{\hat\Phi}{t}+\pd{\hat\Phi}{u}\pd{U}{t}+
\pd{\hat\Phi}{v}\pd{V}{t}
	$$
	so
$$
\pd{\Phi_+}{t}-\pd{\Phi_-}{t}=
\pd{\hat\Phi_+}{t}-\pd{\hat\Phi_-}{t}+
\left(\pd{\hat\Phi_+}{u}-\pd{\hat\Phi_-}{u}\right)\hpd{U}{t}+
\left(\pd{\hat\Phi_+}{v}-\pd{\hat\Phi_-}{v}\right)\hpd{V}{t}
	$$
	From equations \eqref{c1_2.2}, \eqref{c1_2.3}, \eqref{c1_2.7}, and \eqref{c1_2.8}  we obtain 
$$
\pd{\Phi_+}{t}-\pd{\Phi_-}{t}=\pd{g}{t}+\pd{g}{u}\hpd{U}{t}-
\frac{\hat D_{12}}{\hat D_{22}}\hpd{V}{t}\pd{g}{u}
	$$
$$
 q_+^2- q_-^2=\hat q_{\tau}^2-\hat q_{\tau}^2=
\frac{\hat D^2}{\hat D_{22}}\pd{g}{u}\left(\pd{\hat\Phi}{u}_++\pd{\hat\Phi}{u}_-\right)
	$$
The last two equalities lead to \eqref{c1_2.10}.  

%
%
%

%
%
%
%
%


\chapter{Integral Operators Connected With Potentials}

{\small{This chapter introduces and discusses properties of integral operators involved in building the solution of the transformed governing equations in the form of infinite series.}}

\section{Definition of some linear operators connected with logarithmic potential}                               

To solve the integrodifferential equations of vortex and wave dynamics, we will use filtering properties of some integral operators connected with potentials.
Let ${\varphi_k(x)},\,\,\,\,\,k=0,\,1,\,2,\,\dots$ be a complete set of
functions defined on the interval $a\le x\le b$ and $L(f)$ be a linear
operator defined in particular on the set ${\varphi_k(x)}$. To establish
the filtering properties of the operator with respect to the set
${\varphi_k(x)}$ means to find the "responce" $f_k(x)=L(\varphi_k(x))$  
of the operator to the "input" $\varphi_k(x)$ and determine the
coeffitients $c_{kj}$ of the expansion 
$$
f_k(x)=\sum_{j=1}^{+\infty}c_{kj}\varphi_j(x). 
        $$
Define $T(f(\theta))$ and  $N(f(\theta))$ to be the linear integral operators 
\begin{equation}
T(f(\theta))=\frac {1}{4\pi}\int\limits_0^{2\pi}f(\theta _1)\ln [s^2+{s_1}^{2}-
2ss_1\cos (\theta _1-\beta)]d\theta_1,    
\label{c2_1.1}
\end{equation}

\begin{equation}
N(f(\theta))=\frac {|s^2-{s_1}^2|}{2\pi}
\int\limits_0^{2\pi}\frac{f(\theta _1)d\theta_1}
{s^2+{s_1}^{2}-2ss_1\cos (\theta _1-\beta) }.  
\label{c2_1.2}                            
\end{equation}
where $s>0,$ $\,s_1>0,$ $\,\beta$ are parameters.
\vspace{3mm}

Operators $G(f)$ and  $H(f)$  are defined by
\begin{equation}
G(f;a,b,c)=\frac {1}{4\pi}\int\limits_{\Gamma}f(\theta_1)\ln Wd\theta_1,        
\label{c2_1.3}                          
\end{equation}
\begin{equation}
H(f;a,b,c)=(ab-c^2)^{1/2}\cdot \frac {1}{2\pi}\int\limits_{\Gamma}
\frac {f(\theta_1)}{ W}d\theta_1,    
\label{c2_1.4}                            
\end{equation}
$$
W=a\cos^2\theta _1+b\sin ^2\theta _1-2c\sin \theta _1\cdot \cos\theta_1.
        $$
The integrals on the right-hand sides are taken over the domain 
$\Gamma$: either $-\pi/2<\theta _1<\pi/2$ or 
  $\pi/2<\theta _1<3\pi/2$. Parameters $a,$ $\,b,$ $\,c\,$ do not depend on $\theta _1$ and satisfy
the conditions $a>0,$ $\,ab-c^2\ge 0,$ so $W\ge 0$.
  
\section{Properties of the operators}  
        
\subsection{ Properties of operator $T$.} 
\begin{equation}
T(c_1\cos (n\theta)+c_2\sin (n\theta))=
\lambda _n(c_1\cos (n\beta)+c_2\sin (n\beta)),     
\label{c2_2.1}                              
\end{equation}
where the eigenvalues $\lambda_n$ of the operator $T$ are given by
\begin{equation}
\lambda _0=\ln s,\,\,\,\,\,\,n\ge1\,\,\,\,\,\,2n\lambda_n=-(s_1/s)^n
\,\,\,\,\,\,\hbox{if}\,\,\,\,\,\,s_1<s;    
\label{c2_2.2}                              
\end{equation}       
$$
\lambda _0=\ln s_1,\,\,\,\,\,\,n\ge1\,\,\,\,\,\,2n\lambda_n=-(s/s_1)^n
\,\,\,\,\,\,\hbox{if}\,\,\,\,\,\,s<s_1\,\,\,\,\,\,(n=1,2,3,\dots),  
        $$
and the eigenfunctions of the operator $T$ corresponding to the eigenvalue
$\lambda _n$ are $\cos (n\theta)$ and $\sin (n\theta)$ as well as any linear combination of these two functions.
\vspace{3mm}

To prove formulas \eqref{c2_2.1} and \eqref{c2_2.2}, we write
$$
T[c_1\cos (n\theta)+c_2\sin (n\theta)]=
        $$
$$
[c_1\cos (n\beta)+c_2\sin (n\beta)]m_n+
[-c_1\sin (n\beta)+c_2\cos (n\beta)]l_n,
        $$
$$
m_n=\frac {1}{4\pi}\int\limits_0^{2\pi}\ln [s^2+{s_1}^{2}-2ss_1\cos \alpha]
\cos (n\alpha)d\alpha,
        $$
$$
l_n=\frac {1}{4\pi}\int\limits_0^{2\pi}\ln [s^2+{s_1}^{2}-2ss_1\cos \alpha]\sin (n\alpha)d\alpha=0,
        $$
and, consequently, $\lambda _n=m_n$. Integration by parts gives (for 
$n\ge1$) 
$$
2n\lambda _n=-I_n,\,\,\,\,\,\,
I_n=\frac {ss_1}{\pi}\int\limits_0^{2\pi}
\frac{\sin \alpha\cdot\sin (n\alpha)d\alpha}
{s^2+{s_1}^{2}-2ss_1\cos \alpha }. 
        $$
Let
$$
\varphi _n(x)=\frac {x}{\pi}\int\limits_0^{2\pi}\frac{\sin \alpha\cdot
\sin (n\alpha)d\alpha}{1+x^2-2x\cos \alpha }\,\,\,\,\,\,
(0\le x<1).
        $$
Then $I_n=\varphi _n(s/s_1) $ for $s<s_1$, and 
$I_n=\varphi _n(s_1/s) $ for $s>s_1$.
\vspace{3mm}

Using the notations
$x_1=e^{i\alpha}$, $\,x_2=e^{-i\alpha} $ $\,(i^2=-1)$, decomposing the rational function into sum of simpler terms, and applying the formula for the sum of geometric progression, we obtain the expansion
$$
\frac {1}{1+x^2-2x\cos \alpha}=\frac {1}{i\cdot 2\sin\alpha}
\left[\frac {1}{x_2}\frac {1}{1-x/x_2}-\frac {1}{x_1-x}\right]=
        $$
$$
\frac {1}{\sin \alpha}\sum _{k=0}^{+\infty}x^k\sin [(k+1)\alpha].
        $$
Inserting the expansion into the integrand and integrating term by term we get $\varphi _n=x^n$.  
This complete the proof of formulas \eqref{c2_2.1} and \eqref{c2_2.2} for $n \ge 1$.
\vspace{3mm}

For $n=0$ we have 
$$
\lambda _0=\frac {1}{2}\ln(s^2+{s_1}^2)+\frac {1}{4\pi}u(x),\,\,\,\,\,\,
u(x)=\int\limits_0^{2\pi}\ln (1-x\cos\alpha )d\alpha,
        $$
$$
x=\frac {2ss_1}{s^2+{s_1}^2}<1,\,\,\,\,\,\,u(0)=0,\,\,\,\,\,\,
u'(x)=\frac {2\pi}{x}-\frac {2\pi}{x(1-x^2)^{1/2}}.
        $$
The solution of the initial-value problem is 
$u(x)=2\pi \ln[1+(1-x^2)^{1/2}]-2\pi \ln 2$, which leads to  
\eqref{c2_2.2} for $\lambda _0$.

\subsection{ Properties of operator $N$.}  

It is easily verified that 
\begin{equation}
N(f)=s\frac {\partial T}{\partial s}-s_1\frac {\partial T}{\partial s_1}\,
\,\,\,\,\hbox{for}\,\,\,\,s_1<s.
\label{c2_2.3}                             
\end{equation}

From \eqref{c2_2.1} - \eqref{c2_2.2} it follows that 
\begin{equation}
N(c_1\cos (n\theta)+c_2\sin (n\theta))=
\mu_n(c_1\cos (n\beta)+c_2\sin (n\beta)),  
\label{c2_2.4}                                 
\end{equation}
$$
\mu _0=1,\,\,\,\,\,\,n\ge1\,\,\,\,\,\,\mu _n=-2n\lambda _n,
        $$
where $\lambda _n$ is the eigenvalue of the operator $T$ 
\eqref{c2_1.1}.

\subsection{ Properties of operators $G(f)$  and  $H(f)$}  

We define $s\,$ and $\,s_1$ by 
$$
s^2+{s_1}^2=(a+b)/2,\,\,\,\,\,\,-2ss_1\cos (2\beta)=(a-b)/2,\,\,\,\,\,\,
2ss_1\sin (2\beta)=c,
        $$
and obtain 
$$
4s^2=a+b+2\sqrt {ab-c^2},\,\,\,\,\,\,4{s_1}^2=a+b-2\sqrt {ab-c^2},
        $$
\begin{equation}
4ss_1=\sqrt {(a-b)^2+4c^2},               
\label{c2_2.5}                             
\end{equation}
$$
\cos (2\beta)=\frac {b-a}{\sqrt {(a-b)^2+4c^2}},\,\,\,\,\,\,
\sin (2\beta)=\frac {2c}{\sqrt {(a-b)^2+4c^2}}.
        $$
Now it follows from \eqref{c2_1.1} - \eqref{c2_1.4} that
\begin{equation}
2G(c_1\cos (2n\theta)+c_2\sin (2n\theta))=
\lambda _n(c_1\cos (2n\beta)+c_2\sin (2n\beta)),       
\label{c2_2.6}                           
\end{equation}
 \begin{equation}
2H(c_1\cos (2n\theta)+c_2\sin (2n\theta))=
\mu_n(c_1\cos (2n\beta)+c_2\sin (2n\beta)),       
\label{c2_2.7}                              
\end{equation}      
where factors $\lambda _n,$ $\mu _n$ are given by \eqref{c2_2.2} and \eqref{c2_2.4}. 
\vspace{3mm}

Of particular interest for our purpose 
are the following specific cases of operators \eqref{c2_1.3} and 
\eqref{c2_1.4}: 
\begin{equation}
\hat G(f)=G(f;\hat a,\hat b,\hat c),\,\,\,\,\,\,
\hat H(f)=H(f;\hat a,\hat b,\hat c),    
\label{c2_2.9}                              
\end{equation}
$$
\hbox{where}\,\,\,\,\hat a=(\sigma -\sigma _1)^2+(\sigma -h)^2\tan^2\theta,
\,\,\,\,\,\,\hat b=(\sigma _1-h)^2,
        $$
$$
\hat c=(\sigma-h)(\sigma _1-h)\tan\theta;
        $$
and
\begin{equation}
\tilde G(f)=G(f;\tilde a,\tilde b,\tilde c),\,\,\,\,\,\,
\tilde H(f)=H(f;\tilde a,\tilde b,\tilde c),    
\label{c2_2.10}                           
\end{equation}
where
$$
\tilde a=r^2-2r\cos\theta \cdot (\sigma _1-h)+(\sigma -h)^2,\,\,\,\,\,\,
\tilde b=(\sigma _1-h)^2,\,\,\,\,\,\,
\tilde c=r\sin \theta\cdot (\sigma _1-h).
        $$
For these operators from \eqref{c2_2.6} - \eqref{c2_2.7} we get 
\begin{equation}
2\hat G(c_1\cos (2n\theta)+c_2\sin (2n\theta))=
\hat \lambda _n(c_1\cos (2n\hat \beta)+c_2\sin (2n\hat \beta)),
\label{c2_2.11}                           
\end{equation}     

\begin{equation}
2\hat H(c_1\cos (2n\theta)+c_2\sin (2n\theta))=
\hat \mu_n(c_1\cos (2n\hat\beta)+c_2\sin (2n\hat\beta)),      
 \label{c2_2.12}                           
\end{equation}                   
$$
\cos (2\hat \beta)=\frac {2\alpha\cos^2\theta -1}{z},\,\,\,\,\,\,
\sin (2\hat \beta)=\frac {\alpha \sin (2\theta)}{z},
        $$
$$
\alpha=\frac {\sigma _1-h}{\sigma -h},\,\,\,\,\,\,
z=\sqrt {1-4\alpha(1-\alpha)\cos^2\theta},
        $$
$$
\hat \mu_n=\frac {1}{z^n}\,\,\,\,\hbox{for}\,\,\,\, \alpha (1-\alpha)<0;
\,\,\,\,\,\,\hat \mu_n=z^n\,\,\,\,\hbox{for}\,\,\,\, \alpha (1-\alpha)>0;
        $$

$$
\hbox{for}\,\,\,\,n\ge 1\,\,\,\, 2n\hat\lambda _n=-\mu _n;\,\,\,\,\,\,
\hat \lambda _0=\ln \left | \frac {\sigma-h}{2\cos \theta}\right |\,\,\,\,
\hbox{for}\,\,\,\,\alpha (1-\alpha)>0;
        $$

$$
\hat \lambda _0=\ln \left |\frac {\sigma-h}{2\cos \theta}\right |+\ln z\,\,\,\,
\hbox{for}\,\,\,\,\alpha (1-\alpha)<0;
        $$

\begin{equation}
2\tilde G(c_1\cos (2n\theta)+c_2\sin (2n\theta))=
\tilde \lambda _n(c_1\cos (n\tilde \beta)+c_2\sin (n\tilde \beta)),
\label{c2_2.13}                          
\end{equation}                             

\begin{equation}
2\tilde H(c_1\cos (2n\theta)+c_2\sin (2n\theta))=
\tilde \mu_n(c_1\cos (n\tilde\beta)+c_2\sin (n\tilde\beta)),
\label{c2_2.14}                        
\end{equation}              
$$
\cos (\tilde \beta)=\frac {\cos\theta -\delta}{z},\,\,\,\,\,\,
\sin (\tilde \beta)=\frac { \sin (\theta)}{z},\,\,\,\,\,\,
2\delta =\frac {r}{\sigma _1-h},
        $$
$$
z=\sqrt {1-2\delta \cos \theta +\delta ^2};
        $$
$$
\tilde \mu _n=\left(\frac {\delta}{z}\right)^n\,\,\,\,\hbox{for}\,\,\,\,
1-2\delta \cos \theta >0;\,\,\,\,\,\,
\tilde \mu _n=\left(\frac {z}{\delta}\right)^n\,\,\,\,\hbox{for}\,\,\,\,
1-2\delta \cos \theta <0;
        $$
$$
\hbox{for}\,\,\,\,n\ge 1\,\,\,\,\,2n\tilde \lambda _n=-\tilde \mu _n;
\,\,\,\,\,\,\tilde\lambda _0=\ln\frac {r}{2}\,\,\,\,\,\hbox{for}\,\,\,\,
1-2\delta \cos \theta <0;
        $$
$$
\tilde\lambda _0=\ln z+\ln(|\sigma _1-h|)\,\,\,\,\,\hbox{for}\,\,\,\,
1-2\delta \cos \theta >0.
        $$      
When $|\delta |<1$, the following expansions hold for the right-hand side
of \eqref{c2_2.13}:
\begin{equation}
\hbox{for}\,\,\,\,n\ge 1\,\,\,\,\,\,\tilde \mu _n\cos (n\tilde \beta)=
\sum_{j=0}^{+\infty}\delta ^{n+j}b_{n,n+j}\cos ((n+j)\theta),
\label{c2_2.14}                        
\end{equation}         
$$
\tilde \mu _n\sin (n\tilde \beta)=
\sum_{j=0}^{+\infty}\delta ^{n+j}b_{n,n+j}\cos ((n+j)\theta), 
        $$
where
$$
b_{n,k}=0\,\,\,\,\hbox{for}\,\,\,\,k<n; \,\,\,\,
b_{n,k}=\frac {(k-1)!}{(n-1)!(k-n)!}=
C_{k-1}^{n-1}\,\,\,\,\hbox{for}\,\,\,\,k\ge n.
        $$
The expansions converge uniformly in the region $|\delta|\le |\delta_0|<1$.
For $\tilde \lambda _n\cos (n\tilde \beta)$ and
$\tilde \lambda _n\sin (n\tilde \beta)$ similar expansions can be derived from \eqref{c2_2.14} 
since $\tilde \mu _n=-2n\tilde \lambda _n$.

The following expansion  also holds:
\begin{equation}
\tilde \lambda _0=\ln (|\sigma _1-h|)-
\sum_{j=0}^{+\infty}\frac {1}{j}\delta ^j\cos (j\theta).   
\label{c2_2.15}                      
\end{equation}

To prove formulas \eqref{c2_2.13} and \eqref{c2_2.14}, we write (for $n\ge 1,$  $|\delta|<1$ )
$$
\tilde \mu _n\cos (n\tilde \beta)=\frac {1}{2}\delta ^n
\frac {(W-\delta)^n+(W^{-1}-\delta )^n}{(1-(W+W^{-1})\delta+\delta ^2)^n}
        $$
where $W=\exp (i\theta)$ $(i^2=-1)$.  The denominator can be factored
to give
$$
\tilde \mu _n\cos (n\tilde \beta)=\frac {1}{2}\delta ^n
[(W-\delta)^{-n}+(W^{-1}-\delta)^{-n}].
        $$
Hence
$$
\frac {1}{\pi}\int\limits_0^{2\pi}
\tilde \mu _n\cos (n\tilde \beta)e^{ik\theta}d\theta=
-\frac {i\delta ^n}{2\pi}\int\limits_{|W|=1}
[(W-\delta)^{-n}+(W^{-1}-\delta)^{-n}]W^{k-1}dW.
        $$
Applying the theorem of residues we obtain \eqref{c2_2.14}.
\vspace{3mm}

Let 
$$
I_k(\delta)=\frac {1}{2\pi}\int\limits_0^{2\pi}
\ln (1-2\delta \cos \theta +\delta ^2)\cos (k\theta)d\theta,
        $$
then $I_k(0)=0$,  
$$
\frac {d}{d\delta}I_k=\frac {1}{\pi}\int\limits_0^{2\pi}
\frac {(\delta-\cos \theta)\cos (k\theta)}
{1-2\delta \cos \theta +\delta ^2}\,d\theta=-\delta ^{k-1}\,\,\,\,
(\hbox{for}\,\,\,\,k\ge 1),
        $$
and $I_k=-\delta ^k/k,$ $\,I_0=0$. The last integral can be evaluated
with the use of the residues. Formula \eqref{c2_2.15} has been proved.
\vspace{3mm}

Similar results are obtained for right-hand sides of \eqref{c2_2.10} and \eqref{c2_2.11}. Defining $W\,$ and $\delta$ by $W=\exp (i\cdot 2\theta)\,$ and  $\delta=(1-\alpha)/\alpha$, we write 
$$
\hat \mu _n\cos (2n\hat \beta)=\frac {1}{\alpha ^n}\frac {1}{2}
\frac {(W-\delta)^n+(W^{-1}-\delta)^n}{(1-(W+W^{-1})\delta+\delta ^2)^n},
        $$
and, for $-1<\delta <0$, obtain 
$$
\hat \mu _n\cos (2n\hat \beta)=\frac {1}{\alpha ^n}
\sum_{j=0}^{+\infty}\delta ^jb_{n,n+j}\cos (2(n+j)\theta).
        $$
For $0<\delta<1$  $\,(\alpha (1-\alpha)>0)$ we find 
$$
\hat \mu _n\cos (2n\hat \beta)=
\sum_{k=0}^{n}C_n^k\alpha ^k(\alpha -1)^{n-k}\cos (2k\theta).
        $$
The legitimacy of the formulas 
$$
(\cos \theta)^n\cos (n\theta)=\sum_{k=0}^n\mu_{kn}\cos (2k\theta),
        $$
$$
(\cos \theta)^n\sin (n\theta)=\sum_{k=0}^n\mu_{kn}\sin (2k\theta),
        $$
$$
\mu_{kn}=\frac {1}{2^n}\frac {n!}{k!(n-k)!}=\frac {1}{2^n}C_n^k
        $$
can be easily established with the use of the Euler's relation \linebreak  
$\exp (i\theta)=\cos \theta +i\sin \theta.$
 
\section{Properties of a series involving Laguerre polynomials} 
\subsection{  Laguerre polynomials}    
 
  The Laguerre polynomials of degree $k$,
$\,\,L_k^{(\alpha)}(u)\,\,$, are defined by the recursion relation \cite[]{abra}
$$
L_0^{(\alpha)}=1,\,\,\,L_1^{(\alpha)}(u)=\alpha+1-u,\,\,\,\,
        $$
$$
k\ge 1\,\,\,\,\,\,
(k+1)L_{k+1}^{(\alpha)}(u)=(2k+1+\alpha -u)L_k^{(\alpha)}(u)-
(k+\alpha)L_{k-1}^{(\alpha)}(u), 
	$$   
where $\alpha$ is parameter.
\vspace{3mm}

In this book, the polynomials $L_k^{(1)}(u)$ corresponding to  
$\alpha=1$ are used. The recursion relations among these polynomials read  
\begin{equation}
L_0^{(1)}(u)=1,\,\,\,L_1^{(1)}(u)=2-u,\,\,\,\,
\label{c2_3.1}                              
\end{equation}           
$$
k\ge 1\,\,\,\,\,\,
(k+1)L_{k+1}^{(1)}(u)-(2k+2-u)L_k^{(1)}(u)+(k+1)L_{k-1}^{(1)}(u)=0.
        $$
Laguerre polynomials form  a set of functions orthogonal on the
semiaxis $u\ge 0$. The orthogonality integral for Laguerre polynomials is 
\begin{equation}
\int\limits_0^{+\infty}u^{\alpha}e^{-u}L_k^{(\alpha)}(u)L_n^{(\alpha)}(u)
\,du=C_{kn}(\alpha),          
\label{c2_3.2}
\end{equation}        
$$
C_{nn}(0)=1,\,\,\,\,\,\,C_{nn}(1)=n+1,\,\,\,\,\,\,
C_{kn}(\alpha)=0\,\,\,\,\hbox{for}\,\,\,\,k\ne n.
        $$
Generating function for the Laguerre polynomials 
 $L_k^{(1)}(u)$
and the power series expansion of the function are given by 
\begin{equation}
(1-w)^{-2}\exp\left(\frac {sw}{1-w}\right)=
\sum_{n=0}^{+\infty}L_{n}^{(1)}(s)w^n,\,\,\,\,\,\, |w|<1.     \label{c2_3.3}                  
\end{equation} 
        
Asymptotic behavior of Laguerre polynomials is described by the following formula \cite[]{suetin}:
\begin{equation}
 x^{2+\alpha}e^{-x^2}L_k^{(\alpha)}(x^2)=\frac{1}{\sqrt \pi}
k^{\frac{\alpha}{2}-\frac{1}{4}}
 e^{-\frac{x^2}{2}}l_k,           
\label{c2_3.4}                  
\end{equation}        
$$
 l_k=x^\frac{3}{2}\cos\left( 2x\sqrt{N}-\beta\pi \right)+r_k,
        $$
$$
    r_k=O\left(\frac{x^\frac{1}{2}}{\sqrt k}\right)+
    O\left(\frac{x^\frac{9}{2}}{\sqrt k}\right)
    +O\left(\frac{x^7}{k^\frac{3}{4}}\right)+h_k,\,\,\,\,\,x\ge0,
        $$
$$
h_k=O\left(\frac{x^\frac{3}{2}}{k^{\frac{5}{4}}}\right)\,\,\,\,
\hbox{for}\,\,\,\,\alpha=0;\,\,\,\,\,\,
h_k=O\left(\frac{x^\frac{3}{2}}{k^{\frac{3}{4}}}\right)\,\,\,\,
\hbox{for}\,\,\,\,\alpha=1,
        $$
$$
4\beta=2\alpha +1,\,\,\,\,\,\,2N=2k+1+\alpha,
        $$
where the notation $\,O(u)\,$ means "of the order of $\,\,u\,$
as $u\rightarrow 0$".

\subsection{Convergence of the series involving Laguerre polynomials} 

{\it Lemma 1.} The series 
\begin{equation}
s \sum_{k=1}^{+\infty}(-1)^k\frac{1}{k}L_{k-1}^{(1)}(s)
e^{ik\phi}                                      
 \label{c2_3.5}                          
\end{equation}       
converges in any domain $\,\phi<\pi\,$, $0\le s\le s_1$ and converges uniformly in any
rectangle $|\phi|\le \pi -\delta\,\,$,   $\,\,s_0\le s\le s_1\,\,\,$ 
($\,\pi >\delta>0,\,$ $s_0>0$). \\

\noindent
{\it Lemma 2.} For $-\pi < \phi <\pi,\,$  $s>0$ the following formula holds
\begin{equation}
s\sum_{k=1}^{+\infty}(-1)^k\frac{1}{k}L_{k-1}^{(1)}(s)
e^{ik\phi}=1-\exp\left(\frac{1}{2}s\right)\cdot
\exp\left(i\frac{1}{2}s\tan\left(\frac{1}{2}\phi\right)\right).
\label{c2_3.6}                     
\end{equation}    
 
\noindent                                                    
{\it Proof of Lemma 1.} It follows from \eqref{c2_3.4}   that the series \eqref{c2_3.5}   converges, if the series
\begin{equation}
 \sum_{k=1}^{+\infty}(-1)^k\frac{1}{k^\frac{3}{4}}
  \cos \left( 2x\sqrt {k+1}-
\frac{3}{4}\pi \right) e^{ik\phi}\,\,\,\,(i^2=-1)   
\label{c2_3.7}            
\end{equation}   
 converges.
We will now proceed to show that the series\eqref{c2_3.7} converges in domain $\,\,|\phi|<\pi,\,$ $x>0$.
Take
$$ 
  a_k=\frac{1}{k^\frac{3}{4}}\cos\left(2x\sqrt k-\frac{3}{4}\pi\right)
,\,\,\,\,\,\,
b_k=(-1)^k\,e^{ik\phi},\,\,\,\,\,\,B_N=\sum_{k=1}^{N}b_k,
	$$
then	
$$	
a_k=O\left(\frac{1}{k^{\frac{3}{4}}}\right),\,\,\,\,\,\,a_k-a_{k-1}=O\left(\frac{1}{k^{\frac{5}{4}}}\right),   
       $$
$$
|B_N|\le \frac{2}{1+\cos\phi},
\,\,\,\,\,\,S_N=\sum_{k=1}^{N}a_kb_k,
        $$
where $\,\,S_N\,\,$ is the partial sum of series \eqref{c2_3.7}.
Abel's transformation leads to \cite[]{whitwat}
$$
   S_N=a_NB_N+\sum_{k=1}^{N-1}(a_k-a_{k+1})B_k.            
	$$
Now, it becomes obvious that series \eqref{c2_3.7}  converges, and it converges
uniformly in any rectangle $\,|\phi|\le \pi -\delta\,,$ $x_0\le x\le x_1\,$ ($\,\delta>0\,$ $x_0>0$). \\

\noindent
{\it Proof of Lemma 2.}
Integrating both parts of the series \eqref{c2_3.3} along the line segment
$$
w=\rho_1 \exp(i(\pi+\phi)),\,\,\,\,\,0\le\rho_1\le\rho<1,\,\,\,\,\,
\phi={\rm const},\,\,\,\,\,-\pi <\phi <\pi,
        $$
with the use of substitution $\,w-1=1/z\,,$ we obtain
\begin{equation}
1-e^se^{-sz_{*}}=s\sum_{k=1}^{+\infty}(-1)^k\frac{1}{k}L_{k-1}^{(1)}(s)
e^{ik\phi}\rho^k,               
\label{c2_3.8}                          
\end{equation}    
where $\,z_{*}=1/(1+\rho e^{i\phi})\,$.
\vspace{3mm}

Since the series \eqref{c2_3.7} converges, there exists the limit of the
right-side series \eqref{c2_3.5} as $\,\rho\,$ tends to 1, and,
consequently, (by Abel's theorem on continuity up to the circle of convergence \cite[]{whitwat} formula \eqref{c2_3.6} holds.

\chapter{Plane Gravity Waves in Two-Layered Liquid}

{\small{In chapter 3, equations of the liquid-liquid interface are sought in a parametric form, which maps the trace of the interface in a vertical plane onto a half of a circle. The mapping allows us to express the solution with the use of a countable set of functions. Using the results of chapters 1, the governing equations are rewritten in a particular curvilinear coordinate system, and the preliminary solution in the form of infinite series is obtain with the use of integral operators derived in chapter 2.}}

\section{Gravity waves: governing equations in curvilinear coordinates}  

Mathematical formulation of the problem outlined in Chapter 1 is presented below.
Let at $t<0$ the liquid be at rest and the interface be  
a horizontal plane. We consider the problem assuming 
that at $t=0$ a body of water is disturbed, so  
the  waves start to propagate away from the initially disturbed body of water. 
At $t>0$ the water is acted on by no external force other than gravity.
\vspace{3mm}

Let the curve $\,\,\Gamma\,\,$ (in figure 1) be the trace of the interface 
 in the  plane $\,\,(x,y)\,\,$,  the vertical $x$-axis be oriented upward and the $y$-axis be horizontal; 
 $\,x=f<0,\,$ $y=0$ be the coordinates of the pole $\,O_1\,$ of the polar coordinate system in the $\,(x,y)\,$ plane,
$\,\theta\,$ be the polar angle measured from the positive $x$-axis
in the counterclockwise direction, $\,t\,$ be the time.  
The equilibrium position of the interface is horizontal plane $x=0$. 
\vspace{3mm}

Equations of the interface are sought in parametric form
 \begin{equation} 
  x=W(\theta,t),\, y=(W-f)\,\tan\,\theta,\,    
-\pi/2 < \theta < \pi/2,                            
\label{c3_1.1} 
\end{equation}	
where $\,\,W(\theta,t)\,\,$ is an unknown function that must be found while solving the problem. 
Formally, equations \eqref{c3_1.1} 
for each specified function
$\,\,W\,\,$ describe a family of curves depending on $\,f,\,$ $\,t\,$
 being considered a constant. 
The  value  of $\,\,f\,\,$ determines the horizontal scale of the problem.
Indeed, for any specific function $W(\theta, t)$ and any fixed value of $t$, equations \eqref{c3_1.1} describe a family of curves depending on parameter $f$. According to the equations, at each value of $\theta$ the horizontal coordinate $y$ increases when $|f|$ increases, 
while vertical coordinate $x$ remains unchanged, so the curve 
\eqref{c3_1.1} ``stretches" along the horizontal axis.
\vspace{3mm}

In the $\,\,(x,y)\,\,$ plane, curvilinear coordinates $(\sigma, \theta)$ are defined by the relations
 \begin{equation} 
  x=\sigma+W(\theta,t),\, y=(\sigma+W-f)\,\tan\,\theta,\,    
-\pi/2 < \theta < \pi/2                      
\label{c3_1.2}
\end{equation}		
so the equation of the interface takes the form $\sigma =0$ (the liquid of density $\gamma_2$ occupies the half-space $\sigma <0$). 
\begin{figure}
\centering
	\resizebox{0.7\textwidth}{!}
			{\includegraphics{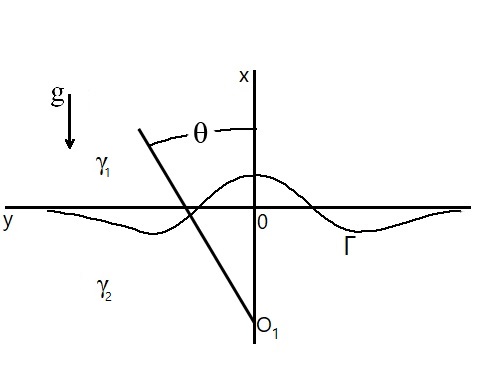}}
	\caption{
	Coordinate systems and sketch of the liquid-liquid interface. 
	}
	\label{c3_qu-.1-.10}
\end{figure} 

For any function $F(\sigma ,\theta ,t)$ the following notations are introduced for one-sided limits  ($\sigma<0$ for negative side  and 
$\sigma>0$ for positive side of the interface):
$$
F_{-}=\lim_{\sigma\to -0}F(\sigma ,\theta ,t),\,\,\,\,\,\,
F_{+}=\lim_{\sigma\to +0}F(\sigma ,\theta ,t).
	$$
	
In chapter 1, subsection 2.1, coordinates $x,\,y$ introduced  as implicit functions  of curvilinear coordinates $u,\,v$. In this section, the  coordinates $x,\,y$ are given as explicit functions of 
$\theta,\,\sigma$. Below, in section 3, the quantities $D_{ij}$ \eqref{c1_2.3} of  chapter 1, subsection 2,  are expressed as functions of $\theta,\,\sigma$.
\vspace{3mm}
	
 In terms of variables $(\sigma, \theta)$, the exact  equations   
\eqref{c1_2.9}, \eqref{c1_2.10}, \eqref{c1_2.11} of chapter 1, section 2,  governing time  evolution of the interface, the doublet density $g(\theta,t)$ and velocity potential $\Phi(\sigma,\theta,t)$ become respectively the forms 
 \begin{equation} 
\pd{W}{t} \left(1- 2\pd{W}{\theta}\hat D_1 \tan\theta \right)=
\hat D_2\,\pd{\hat\Phi}{\theta_-}+
\hat D_3\,\pd{\hat\Phi}{\sigma_-}
\label{c3_1.3}                                       
 \end{equation}    
where
$$
\hat D_1=-\frac{\cos^2\theta}{W-f},\,\,\,\,\,\,
\hat D_2=-\frac{\sin\theta\cos\theta}{W-f} 
-\pd{W}{\theta}\,\frac{\cos^2\theta}{(W-f)^2}
	$$
$$
 \hat D_3=1+
2\frac{\sin\theta\cos\theta}{W-f}\pd{W}{\theta} +
\frac{\cos^2\theta}{(W-f)^2}
\,\left(\pd{W}{\theta} \right)^2,
	$$
 \begin{equation} 
\pd{g}{t}+\pd{g}{\theta}\hpd{U}{t}+
\frac{\hat D_2}{\hat D_3}\hpd{V}{t}\pd{g}{\theta}+
\frac{1}{2}\frac{\hat D_1^2}{\hat D_3}\pd{g}{\theta}
\left(\pd{\hat\Phi}{\theta}_++\pd{\hat\Phi}{\theta}_-\right)+
	$$                     
$$
\gamma
\left(\frac{1}{2}\,\hat q^2_-+W+\hpd{\Phi}{t}_-\right)=0,                                     
\,\,\,\,\,\,\gamma=\frac{\gamma_2}{\gamma_1}-1 
\label{c3_1.4}
\end{equation}             
where $\gamma_1$ 	and $\gamma_2$ are the densities of the liquids.

In equations  \eqref{c3_1.3} and  \eqref{c3_1.4} all terms  are calculated at the point  \linebreak $ Q_1(\sigma_1=0,\,\theta=\theta_1)$ on the interface. 		 
         \begin{equation} 
\Phi(\sigma,\theta,t)=-\frac{D_1}{|D_1|}
\frac{1}{2\pi}\int\limits_{-\pi/2}^{\pi/2}
g(\theta_1,t)A(\sigma,\theta,\theta_1,t)
\left.\frac{d\theta_1}{S}\right|_{\sigma_1=0},  
\label{c3_1.5}           
\end{equation}
$$
A=(\sigma +W-W_1)(f-W_1)+
        $$
$$
\pd{W_1}{\theta_1}\,
[(\sigma+W-f)(\tan\,\theta-\tan\theta_1)-(W_1-f)\,\tan\,\theta_1]
\cos^2\theta_1
	$$	                
$$
S=(\sigma-\sigma_1+W-W_1)^2\cos^2\theta_1+
        $$
$$
[(\sigma-f+W)\tan\theta\cdot\cos\theta_1-
(\sigma_1-f+W_1)\sin\theta_1]^2, 
        $$
$$
W=W(\theta,t),\,\,\,\,\,\,W_1=W(\theta_1,t),\,\,\,\,\,\,
\frac{D_1}{|D_1|}=-1,
        $$
 $S=\cos^2\theta_1R^2$, $R^2$ is the squared distance between points $Q(\sigma, \theta)$ and $ Q_1(\sigma_1,\,\theta_1)$. The subscript 
 $\sigma_1=0$ in \eqref{c3_1.5} denotes that integrand is calculated at  
 $\sigma_1=0$.
       
Equations \eqref{c3_1.1} - \eqref{c3_1.5} are supplemented by boundary conditions  
  \begin{equation}
|W(\theta ,t)|<C(t)\cos^2\theta,\,\,\,\,\, \lim \limits_{\cos \theta \to 0}\frac {\partial W}{\partial \theta}=0,\,\,\,\,\,
|g(\theta,t)|<C(t)                                    
  \label{c3_1.6}                 
  \end{equation}
and initial conditions $W=W(\theta ,0),\,\,g=g(\theta,0)$.
The conditions at infinity assure that the total energy initially supplied to the water by a source of disturbances of finite power remains finite at any moment of time. 
\vspace{3mm}

The problem on gravitational waves is formulated mathematically in terms of nonlinear integro-differential equations 
in two unknown functions $W(\theta,t)$ and $g(\theta,t)$.
Note, that the change of variables \eqref{c3_1.2} is a nonlinear transformation of the equations written  in Cartesian coordinates.
The equations remain valid in time as long as each ray
$\,\theta=\hbox{const}\,$
intersects the free surface at not more than one point.
\vspace{3mm}

The last remark reflects the problem which inevitably arises in
connection with any theoretical work, devoted to the free surface waves, even when Cartesian coordinates are used: how can be checked whether or not the free surface may be described by a function of one spatial coordinate (and time) as the surface evolves with time? The problem has been cleared up only for linear waves. 
In the present book, for nonlinear waves, the situation is similar:
the validity of the leading-order equations (and their solutions) on the positive semiaxis of time is ensured  by Theorems  proved below in Chapter 3. 
That is why it is important, that the leading-order equations are solved exactly.
\vspace{3mm}

The equations  \eqref{c3_1.3} - \eqref{c3_1.5} governing the flow are written in non-dimensional variables. Since the problem has no characteristic linear size, 
the dimensional unit of length, $\,\,L_*,\,\,$ is a free parameter.
But for applications in Chapters 5 and 6, the value of $L_*$,  
as well as the value of $|f|L_*$, will be obtained  from instrumental data.
The dimensional unit of time, $\,\,T_*,\,\,$ is defined
by the relation $\,\,T_*^2g=L_*\,$, where $\,\,g\,\,$ is the acceleration of free fall. The non-dimensional acceleration of free fall is equal to unity.
All parameters, variables and equations are made non-dimensional by 
the quantities $\,L_*,\,T_*,\,P_* $ and 
the density of water $\gamma_*=1000\,$ $\hbox{kg/m}^3$.
\vspace{3mm}

Physical reasons imply that the initial conditions can not be assigned arbitrarily. 
The volume of the "upper" liquid of density $\gamma_1$ detrained from the region $x>0$ into the region $x<0$ displaces the equal volume of the "lower" liquid into the region $x>0$. The volume of the liquid transferred across the equilibrium plane $x=0$ is required to be finite. 
This means that the improper integral 
\begin{equation}
\int\limits_{-\pi/2}^{\pi/2}x(\theta,t)dy(\theta,t)= 
\int\limits_{-\pi/2}^{\pi/2}W\left(\tan\theta\pd{W}{\theta}+
(W-f)\,\frac{1}{\cos^2\theta}\right)\,d\theta   
\label{c3_1.7}
\end{equation}
must be equal to zero.
\vspace{3mm}

To remove the singularities, we accept that 
\begin{equation}
W(\theta,t)=c\cdot\cos^2\theta\cdot V(\theta,t),                                                                 
\label{c3_1.8}
\end{equation}
where $V(\theta,t)$ is a function bounded in any rectangle 
$-\pi/2\le\theta\le\pi/2$, $0\le t\le T$, $c$ is a constant determined by normalization of some sort.
Substituting \eqref{c3_1.8} into \eqref{c3_1.7} and integrating by parts 
we obtain 
$$
\int\limits_{-\pi/2}^{\pi/2}\left[\frac{1}{4}\,c^2\, V^2\cdot(\cos(2\theta)+1) -f\,c V \right]d\theta=0      
	$$

$$
\frac{c}{f}\int\limits_{-\pi/2}^{\pi/2}\frac{1}{4}\, V^2\cdot\cos^2\theta\,d\theta	
-\int\limits_{-\pi/2}^{\pi/2} V\cdot\cos^2\theta\,d\theta	=0
	$$	
In the Chapter 4, we will consider small values of $\mu=c/f$.
\vspace{3mm}

At $\mu\to 0$, we have	
\begin{equation}
\int\limits_{-\pi/2}^{\pi/2} V\cdot\cos^2\theta\,d\theta	=0
\label{c3_1.9}                                         
\end{equation}
Assume that  
$$
V(\theta,0)=\alpha_0(0)+\sum_{k=1}^{+\infty}\alpha_k(0)\cos (2k\theta),\,\,\,\,\,\,                            
\sum_{k=0}^{+\infty}\alpha^2_k(0)<+\infty.
        $$
It follows from \eqref{c3_1.9} that $2a_0(0)+a_1(0)=0$

\section{ Derivation of the governing equations}  
                                                                    
From \eqref{c1_2.1} of chapter 1, section 2, we obtain 	
$$ 
du=\pd{U}{x} dx+\pd{U}{y} dy+\pd{U}{t} dt 
	$$
$$ 
dv=\pd{V}{x} dx+\pd{V}{y} dy+\pd{V}{t} dt 
	$$
$$
D=\pd{U}{x}\pd{V}{y}-\pd{U}{y}\pd{V}{x}\ne 0 
	$$
The inverted system reads  
$$
d\hat x=\frac{1}{\hat D}\left[\hpd{V}{y}du-\hpd{U}{y}dv+
\left(\hpd{U}{y}\, \hpd{V}{t}-\hpd{U}{t}\, \hpd{V}{y}\right)dt\right]
	$$	
	$$
d\hat y=\frac{1}{\hat D}\left[-\hpd{V}{x}du+\hpd{U}{x}dv+
\left(\hpd{U}{x}\, \hpd{V}{t}-\hpd{U}{t}\, \hpd{V}{x}\right)dt\right]
	$$	
which gives
$$
\hpd{V}{y}=\hat D\,\pd{\hat x}{u},\,\,\,\,\,\,
\hpd{V}{x}=-\hat D\,\pd{\hat y}{u},\,\,\,\,\,\,
\hpd{V}{t}=
-\left(\pd{\hat x}{t}\,\pd{\hat y}{u}+
\pd{\hat y}{t}\,\pd{\hat x}{u}\right)
	$$
$$
\hpd{U}{y}=-\hat D\,\pd{\hat x}{v},\,\,\,\,\,\,
\hpd{U}{x}=\hat D\,\pd{\hat y}{v},\,\,\,\,\,\,
\pd{U}{t}=\hat D\left(\hpd {x}{t}\pd{y}{v}+\hpd{y}{t}\hpd{x}{v}\right)
	$$	
$$
\frac{1}{\hat D}=\pd{\hat x}{u}\,\pd{\hat y}{v}-
\pd{\hat y}{u}\,\pd{\hat x}{v}
	$$
	
Now we use relations \eqref{c3_1.2} to calculate derivatives	 and obtain	
$$
\frac{1}{\hat D}=\pd{W}{\theta}\tan\theta-
\left(\pd{W}{\theta}\tan\theta+(\sigma+W-f)\,\frac{1}{\cos^2\theta}
\right)\cdot 1=-\frac{\sigma+W-f}{\cos^2\theta}
	$$		
$$
\hat D=-\frac{\cos^2\theta}{\sigma+W-f}
	$$
Proceeding exactly on the same lines as for $\hat D$, we get from the inverted system	
$$
\hat D_{11}=\frac{\cos^2\theta}{(\sigma+W(\theta,t)-f)^2}
	$$
$$
\hat D_{12}=-\frac{\sin\theta\cos\theta}{\sigma+W(\theta,t)-f}
-\frac{\cos^2\theta}{(\sigma+W(\theta,t)-f)^2}\pd{W}{\theta}
	$$	
$$
 \hat D_{22}=1+
2\frac{\sin\theta\cos\theta}{\sigma+W-f}\pd{W}{\theta} +
\frac{\cos^2\theta}{(\sigma-f+W)^2}
\,\left(\pd{W}{\theta} \right)^2
	$$ 
\begin{equation}
\hpd{V}{x}\,(x-x_1)=-\hat D_1(\sigma+W-W_1)
\left(\pd{W_1}{\theta_1}\tan\theta_1+
\frac{W_1-f}{\cos^2\theta_1}\right)
\label{c3_2.1}
\end{equation}
$$
\hpd{V}{y}\,(y-y_1)=\hat D_1\,\pd{W_1}{\theta_1}
[(\sigma+W-f)\,\tan\,\theta-(W_1-f)\,\tan\,\theta_1]
	$$
$$
\hpd{V}{t}=\pd{W}{t}+2\frac{\sin\theta\cos\theta}{\sigma-f+W}                    
\pd{W}{\theta}
	$$		
$$
\hpd{U}{t}=\frac{\cos\theta(\sin\theta+\cos\theta)}{\sigma-f+W}\,
\pd{W}{t}
	$$
	
Substituting just obtained expressions (at $\sigma=0$) in	
the  equations  \eqref{c1_2.9},  \eqref{c1_2.10}, and \eqref{c1_2.11}  of chapter 1 gives equation \eqref{c3_1.3} and equation \eqref{c3_1.4} governing evolution of the interface and dublets distribution respectively, and \eqref{c3_1.5}.

Alternative way to prove \eqref{c3_1.5} is shown in the next lines:
$$
\Phi=-\frac{1}{2\pi}\int\limits_{\Gamma} g(\theta_1)\pd{}{n}\ln R\,dl=
-\frac{1}{2\pi}\int\limits_{\Gamma} g(\theta_1)
\frac{1}{R^2}\,\frac{1}{2}\,\pd{R^2}{n}\,dl,
	$$
$$
R^2=(x(\theta,\sigma)-x(\theta_1,\sigma_1))^2+
(y(\theta,\sigma)-y(\theta_1,\sigma_1))^2,
	$$	
$$
dl=\sqrt{(dx)^2+(dy)^2}=\sqrt{\left(\pd{\hat x}{\theta_1}\right)^2+\left(\pd{\hat y}{\theta_1}\right)^2}\,d\theta_1=\frac{\sqrt{\hat D_{22}}}{|\hat D_1|} d\theta_1.
	$$
$$
\sqrt{D_{22}}\,\frac{1}{2}\,\pd{R^2}{n}=\pd{V}{x}(x(\theta,\sigma)-x(\theta_1,\sigma_1))+\pd{V}{y}(y(\theta,\sigma)-y(\theta_1.\sigma_1))
	$$	
From equations \eqref{c3_2.1} we obtain	
$$
\frac{1}{2}\,\pd{R^2}{n}=\frac{D_1}{\cos^2\theta_1\sqrt{D_{22}}}\,A,
	$$
$$
\frac{1}{R^2}\,\frac{1}{2}\,\pd{R^2}{n}\,dl=\frac{1}{S}\,\frac{D_1}{|D_1|}A,\,\,\,\,\,\,S=R^2\cos^2\theta_1,
	$$
$$
\frac{1}{R^2}\, \frac{D_1}{\cos^2\theta_1\sqrt{D_{22}}}A\,
\frac{\sqrt{\hat D_{22}}}{|\hat D_1|} du=
\frac{1}{S}\,\frac{D_1}{|D_1|}A.
	$$	
	
\section{Series expansion of  velocity potential}   

To solve equations \eqref{c3_1.3} and \eqref{c3_1.4}, a usefull 
 procedure for finding one-sided limits of velocity potential  $\Phi$ is proposed below, bearing in mind that the integral 
\eqref{c3_1.5} must be evaluated at $\sigma_1=0$, and one-sided limits are to be found at $\sigma\to 0$ from positive (negative) side of the interface.
\vspace{3mm}

Consider the fraction $\frac{1}{S}$ involved in \eqref{c3_1.5} 
$$
S=(\sigma-\sigma_1+W-W_1)^2\cos^2\theta_1+
        $$
$$
[(\sigma-f+W)\tan\theta\cdot\cos\theta_1-
(\sigma_1-f+W_1)\sin\theta_1]^2, 
        $$
Let $S_0$ denote the value of $S$ at $W=W_1=0$:
 $$
S_0=(\sigma-\sigma_1)^2\cos^2\theta_1+
        $$
$$
[(\sigma-f)\tan\theta\cdot\cos\theta_1-
(\sigma_1-f)\sin\theta_1]^2, 
        $$
Define        
$$
\rho=\frac{S-S_0}{S_ 0}
	$$
 At $\sigma\ne 0$ the distance between points $Q_1$ on the interface and $Q$ outside the interface is positive, and consequently $S>0$ and $S_0>0$ irrespective of the values of other variables, but $S=0,\,S_0=0$ at $\sigma_1=0$, $\sigma=0$, and $\theta=\theta_1$ simultaneously.  
Nevertheless,  the ratio $\rho$ is bounded,  if  
$|W-W_1|<k|\sin\theta-\sin\theta_1|$ where $k$ is a constant.  
\vspace{3mm}
 
 At $\sigma=\sigma_1=0$, after some algebra  and trigonometry  we find 
$$		
\cos^2\theta(S-S_0)=(W-W_1)^2\cos^2\theta_1\cos^2\theta+
	$$		
$$
[(W-W_1)\sin\theta\cos\theta_1+(W_1-f)\sin(\theta-\theta_1)]^2
	$$
$$
\cos^2\theta S_0=f^2\sin^2(\theta-\theta_1)
	$$	
and boundedness of $\rho$ becomes obvious.
The difference $ S-S_0$ is a  polynomial of degree  $2$ in two variables $W$ and $W_1$:
$$
 S-S_0=aW+bW_1+V(W,W_1)
 	$$
$$
a=\pd{S_0}{\sigma},\,\,\,\,\,\,b=\pd{S_0}{\sigma_1}
	$$
$$	
V(W,W_1)=\frac{\cos^2\theta_1}{\cos^2\theta}W^2
-2\frac{\cos\theta_1}{\cos\theta}\,\cos(\theta_1-\theta)WW_1+
\frac{1}{\cos^2\theta}W^2_1, 
	$$
where
$$
\frac{\cos ^2\theta_1}{\cos^2\theta}=
\frac{\partial^2S_0}{\partial\sigma^2},\,\,\text{and so on}.
	$$
$$
\frac{S_0}{S}=\frac{1}{1+\rho},\,\,\,\,\,\,
\frac{1}{S}=\frac{1}{S_0}\,\frac{1}{1+\rho},\,\,\,\,\,\,
\frac{1}{S}=\frac{1}{S_0}\,\sum\limits_{n+0}^{+\infty}(-1)^n\rho^n
	$$	
Now the velocity potential $\Phi$ may be rewritten as	
$$ 
\Phi(\sigma,\theta,t)=\frac{1}{2\pi}\int\limits_{-\pi/2}^{\pi/2}
g(\theta_1,t)A\frac{1}{1+\rho}
\left.\frac{d\theta_1}{S_0}\right|_{\sigma_1=0}  
	$$
\begin{equation}
F=g(\theta_1,t)A\,\frac{1}{1+\rho},\,\,\,\,\,\,
\Phi(\sigma,\theta,t)=\hat H(F)
\label{c3_3.1}
\end{equation}
$$
A=(\sigma -\sigma_1)f-W_1(\sigma -\sigma_1+W-W_1)+f(W-W_1)
	$$        
$$
(\sigma-f+W)\,\pd {W_1}{\theta_1}\,\,
(\tan\theta\cdot\cos^2\theta_1-
\sin\theta_1\cdot\cos\theta_1)
        $$     
Linear operator $\hat H(F)$ is defined by formula \eqref{c2_2.9} of  Chapter 2 .
\vspace{3mm}
  
Since the ratio
$$
\frac{S_0}{S}=\frac{1}{1+\rho}
	$$
is a bounded function in two variables $W$ and $W_1$, it can be expanded  in the  series
$$
\frac{S_0}{S}=1+\sum\limits_{n=1}^{+\infty}V_n,\,\,\,\,\,\,
V_n=\sum\limits_{i+j=n}\frac{W^iW_1^j}{i!j!}\frac{\partial^n S_0}{\partial\sigma^i\,\partial\sigma_1^j},
	$$
convergent within a small circle $W^2+W^2_1<\delta^2$.

\chapter{Nonlinear Theory of Specific Wave Packets} 

{\small{Assuming that the horizontal dimensions of the initially disturbed body of  water are much larger than the magnitude of the surface displacement in the wave origin, a small parameter is introduced, and leading-order equations are obtained. A numerable set of exact solutions of leading-order  equations (specific wave packets) are obtained in closed form. It is shown that any wave group emanating from the origin is a non-linear blend of finite or infinite number of the specific wave packets.}} 

\section{ Leading-order equations}   

From now on we assume that the waves are generated by 
an initial disturbance to the two-layered liquid, and the horizontal dimensions of the initially disturbed water body are much larger than the magnitude of the interface displacement in the wave origin. 
If  $c$ denotes the maximum of the interface displacement  
in the wave origin, the above assumption means that the ratio 
$\,\mu =c/f$ is a small quantity.
\vspace{3mm}

We seek  series expansions of the form  
$$
W(\theta,t)=c [W_0(\theta,t)+\mu W_1(\theta,t)+\mu^2 W_2(\theta,t)+...]
	$$ 
$$
g(\theta,t)=c\sum\limits_{k=0}^{+\infty}
\mu^kg_k(\theta,t)
	$$
$$
\Phi(\theta,\,\sigma,t)=c\sum\limits_{k=0}^{+\infty}
\mu^k\Phi_k(\theta,\,\sigma,t)
	$$
 By expanding all terms of the  equations 
 governing time  evolution of the interface, the doublet density $g(\theta,t)$ and velocity potential $\Phi(\sigma,\theta,t)$ (equations  \eqref{c3_1.3}, \eqref{c3_1.4},  and \eqref{c3_1.5}  of Chapter 3) in powers of $\mu$, and equating coefficients of like powers of $\mu$, we obtain the leading-order equations in unknown functions  $W_0(\theta,t),\,\,g_0(\theta,t)$: 
\begin{equation}
\pd{W_0}{t}=\pd{\Phi_0}{\sigma_-}+
\frac{\sin\theta\cdot\cos\theta}{f}\,\pd{\Phi_0}{\theta_-}                                           \label{c4_1.1}
\end{equation}
\begin{equation}
\pd{g_0}{t}+\gamma\left(\pd{\Phi_0}{t}_-+W_0\right)=0,
\,\,\,\,\,\,\gamma=\frac{\gamma_2}{\gamma_1}-1  
\label{c4_1.2}                           
 \end{equation}
\begin{equation}
\left.\Phi_0=\left.\frac{1}{|f\sigma|}\,\hat H(f\sigma g_0)\right|_{\sigma_1=0}=
\frac{f\sigma}{|f\sigma|}\,\hat H(g_0)\right|_{\sigma_1=0} 
 \label{c4_1.3}                                     
\end{equation}
\vspace{3mm}

Integral operator $H(F)\equiv \hat H(F)$ and its  eigenfunctions and eigenvalues  are  defined in Chapter 2 and specified here as
\begin{equation}
H(F(\theta))=|\alpha(1-\alpha)|\int\limits_{-\pi/2}^{\pi/2}
\frac{F(\theta_1)\,d\theta_1}{a\cos^2\theta_1+b\sin^2\theta_1-c\sin(2\theta_1)} 
\label{c4_1.4}
\end{equation}
$$
\alpha=\frac{f}{f-\sigma},\,\,\,\,\,\,a=(1-\alpha)^2+\tan^2\theta,\,\,\,\,\,\,
b=\alpha^2,\,\,\,\,\,\,c=\alpha\,\,\tan\theta
	$$
(in curvilinear coordinates $(\sigma, \theta)$  equation of the interface takes the form $\sigma =0$).
\vspace{3mm}

The eigenfunctions and eigenvalues of the linear operator  $\hat H(F)$ are found to be 
\begin{equation}
2\hat H(c_1\cos (2n\theta)+c_2\sin (2n\theta))=
 \mu_n(c_1\cos (2n\beta)+c_2\sin (2n\beta)),      
\label{c4_1.5}
\end{equation}
$$
\cos (2\beta)=\frac {2\alpha\cos^2\theta -1}{z},\,\,\,\,\,\,
\sin (2 \beta)=\frac {\alpha \sin (2\theta)}{z},
        $$
$$
z=\sqrt {1-4\alpha(1-\alpha)\cos^2\theta},
        $$
$$
 \mu_n=\frac {1}{z^n}\,\,\,\,\hbox{for}\,\,\,\, \alpha (1-\alpha)<0;\\
\,\,\,\,\,\,\hat \mu_n=z^n\,\,\,\,\hbox{for}\,\,\,\,
 \alpha (1-\alpha)>0;
        $$              
In the limit case $\sigma\to 0$  we have
$$
\alpha\to 1,\,\,\,\,\,\,z\to 1,\,\,\,\,\,\,
\cos (2\beta)\to \cos (2 \theta),\,\,\,\,\,\,
\sin (2\beta)\to \sin (2 \theta)
	$$ 
        Equations \eqref{c4_1.1} - \eqref{c4_1.3} are supplemented by boundary conditions    
  \begin{equation}
|W(\theta ,t)|<C(t)\cos^2\theta,\,\,\,\,\, \lim \limits_{\cos \theta \to 0}\frac {\partial W}{\partial \theta}=0,\,\,\,\,\,
|g(\theta,t)|<C(t)                                    
  \label{c4_1.6}        
  \end{equation}
and initial conditions $W=W(\theta ,0),\,\,g=g(\theta,0)$. 
Although the liquid is infinite in extent, boundary conditions \eqref{c4_1.6} ensure that the total energy supplied to the liquid by a source of disturbances is finite. 

\section{ Solution to the leading-order equations}  

Solution to the equations  \eqref{c4_1.1} - \eqref{c4_1.3}
 satisfying  boundary conditions \eqref{c4_1.6}
is sought by expanding unknown functions in trigonometric series: 
\begin{equation}
W_0(\theta,t)=\cos^2\theta\sum_{k=0}^{+\infty}
[a_k(t)\cos (2k\theta)+b_k(t)\sin (2k\theta)],   
\label{c4_2.1}                     
\end{equation}
                                                 
$$
g_0(\theta,t)=\sum_{k=0}^{+\infty}
[\rho_k(t)\cos (2k\theta)+e_k(t)\sin (2k\theta)]. 
        $$
assuming that series \eqref{c4_2.1} converge uniformly at $t=0$.
The constants $a_n(0)$, $b_n(0)$ determine the initial displacement to the interface,  
the constants $\rho_n(0)$, $e_n(0)$ determine the initial velocity field. 

\subsection{Countable system of ordinary differential equations for time-dependent coefficients of double series solution } 

With the use of equations  \eqref{c4_1.3} and \eqref{c4_1.5}, we obtain
$$
\Phi_0=\frac{1}{2}\,\frac{f\sigma}{|f\sigma|}
\sum_{k=0}^{+\infty}\mu_k[\rho_k(t)\cos(2k\beta)+
e_k(t)\sin (2k\beta)]        
	$$
The derivatives of the velocity potential $\Phi$ with respect to $\theta,\,\sigma,\, t$ can be obtained as follows.
At $f<0,\,\,\sigma<0$ we have
$$
\Phi_0=\frac{1}{2}
\sum_{k=0}^{+\infty}\frac{1}{z^k}[\rho_k(t)\cos(2k\beta)+
e_k(t)\sin (2k\beta)] ,      
	$$	
and at $\sigma=-0$	
$$
\Phi_0|_-=\frac{1}{2}\,
\sum_{k=0}^{+\infty}[\rho_k(t)\cos(2k\theta)+
e_k(t)\sin (2k\theta)]=\frac{1}{2}\,g_0,        
	$$		
$$
\pd{\Phi_0}{t}_-=\frac{1}{2}\sum_{k=0}^{+\infty}
[\rho'_k(t)\cos (2k\theta)+e'_k(t)\sin (2k\theta)]=
\frac{1}{2}\,\pd{g_0}{t}
	$$
By the way, 	equation \eqref{c4_1.2}	becomes
\begin{equation}
\pd{g_0}{t}+\varepsilon_*W_0=0,\,\,\,\,\,\,
\varepsilon_*=\frac{\gamma_2-\gamma_1}{\gamma_2+\gamma_1}.
\label{c4_2.2}
\end{equation}
$$
\sin\theta\cos\theta\,\pd{\Phi_0}{\theta}_-=
\frac{1}{2}\sum_{k=0}^{+\infty}
\cos(2k\theta)[(k-1)\rho_{k-1}-(k+1)\rho_{k+1}]+
	$$		
$$
\sin(2k\theta)[(k-1)e_{k-1}-(k+1)e_{k+1}]
	$$
Now we write		
$$
\left.\pd{\Phi_0}{\sigma}_-=\frac{1}{f}\,\pd{\Phi_0}{\alpha}\right|_{\alpha=1}
	$$	
It follows from \eqref{c4_1.5} for operator $\hat H$ that
$$
\pd{}{\alpha}\left(\frac{\cos(2k\beta)}{z^k} \right)_{\sigma=-0}=                                             
-k\cos(2k\theta)-k\cos(2(k+1)\theta),   
	$$	
$$
\pd{}{\alpha}\left(\frac{\sin(2k\beta)}{z^k} \right)_{\sigma=-0}=                                             -k\sin(2k\theta)-k\sin(2(k+1)\theta),     
	$$
and so	
$$	
\pd{\Phi_0}{\sigma}_-=\frac{1}{2f}\,
\sum_{k=0}^{+\infty}[-k\rho_k(t)(\cos(2k\theta)+\cos(2(k+1)\theta))   
	$$
$$
-e_kk(\sin(2k\theta)+\sin(2(k+1)\theta)]    
	$$			
Inserting  the expressions for derivatives in the leading-order equations  
\eqref{c4_1.1}, \eqref{c4_1.2}, and  \eqref{c4_1.3}, 
 we obtain the following initial value problem for time-dependent coefficients of the series \eqref{c4_2.1}:
\begin {equation}
\frac{d\nu_k}{dt}=\varepsilon_*\left(\frac{1}{2}z_{k-1}+z_k+
\frac{1}{2}z_{k+1}\right),\,\,\,\,\,\,k=1,\,2,\,\dots ,
\label{c4_2.3}                          
\end{equation}        
\begin {equation}
\frac{dz_k}{dt}=-\frac{k}{|f|}\nu_k,\,\,\,\,\,\,
\varepsilon_*=(\gamma_2-\gamma_1)/(\gamma_2+\gamma_1), 
\label{c4_2.3a}                          
\end{equation}    
\begin {equation}
a_k(t)=z_k(t),\,\,\,\,\,\,\rho_k(t)=\nu_k(t),\,\,\,\,\hbox{if}\,\,\,\,
z_k(0)=a_k(0)\,\,\,\,\,\,\,,\rho_k(0)=\nu_k(0);
\label{c4_2.4}                         
\end{equation}         
\begin {equation}
b_k(t)=z_k(t),\,\,\,\,\,\,e_k(t)=\nu_k(t),\,\,\,\,\hbox{if}\,\,\,\,
z_k(0)=b_k(0)\,\,\,\,\,\,\,e_k(0)=\nu_k(0)
\label{c4_2.5}                            
\end{equation}         
Thus, coefficients $a_k(t)$ and $\rho_k(t)$ are defined by numerical system of ordinary differential equations \eqref{c4_2.3}
with initial conditions \eqref{c4_2.4}, and the same equations with initial conditions \eqref{c4_2.5}
define coefficients $b_k(t)$ and $e_k(t)$. 
If the liquid is initially still, $\rho_k(0)=0$; if initially the interface is 
in the equilibrium position $x=0$, $a_k(0)=0$. 
 
\subsection{Exact solution to the countable system of differential equations}   

Define new independent variable $\tau$ by the relation 
$$
\tau=t\sigma_*,\,\,\,\,\,\,\sigma^2_*= \frac{\varepsilon_*}{2|f|}.
	$$ 
  Differentiating and using the differential equation \eqref{c4_2.3a} to eliminate 
  $\nu_k$ from \eqref{c4_2.3} , we obtain  the system
\begin{equation}
 \frac{d^2z_k}{d\tau^2}=-k [z_{k-1}+2z_k+z_{k+1}]\,\,\,\,\,k\ge 1      
 \label{c4_2.6}
 \end{equation}
\begin{equation}
\nu_k=-\frac{\varepsilon_*}{2k\sigma_*}\,\frac{dz_k}{d\tau}.
\label{c4_2.6a}
\end{equation}
Solution to the system is given by    		
\begin{equation}
 z_k(\tau)=(-1)^k\int\limits_0^{+\infty}
ue^{-u}L_{k-1}^{(1)}(u)
\left[\varphi(u)\cos(\tau\sqrt {u})+\frac{2\sigma_*}{\varepsilon_*}\, 
\psi(u)\frac{\sin(\tau\sqrt {u})}{\sqrt{u}}\right]\,du,
\label{c4_2.7}
\end{equation}
\begin{equation}
\nu_k=(-1)^k\int\limits_0^{+\infty}\frac{1}{k}
ue^{-u}L_{k-1}^{(1)}(u)
\left[\frac{\varepsilon_*}{2\sigma_*}\varphi(u)\sqrt {u}\sin(\tau\sqrt {u})-\psi(u)\cos(\tau\sqrt {u}) \right]\,du. 
\label{c4_2.7a}
\end{equation}	
where $\varphi(u)$ and $\psi(u)$ are arbitrary functions, $L_k^{(1)}(u)$ are Laguerre polynomials (see  Chapter 2).

This can be verified by substituting \eqref{c4_2.7}  into equation  
\eqref{c4_2.3} (the recursion relation and the orthogonality integral  for Laguerre polynomials should be taken into account).

At $\tau=0$ we have 	
$$
 z_k(0)=(-1)^k\int\limits_0^{+\infty}
 ue^{-u}L_{k-1}^{(1)}(u)\varphi(u)\,du,
	$$
$$
\nu_k(0)=(-1)^{k_1}\int\limits_0^{+\infty}
\frac{1}{k} ue^{-u}L_{k-1}^{(1)}(u)\psi(u)\,du.
	$$	
Initial conditions \eqref{c4_2.4}, as well as \eqref{c4_2.5},	are satisfied with
$$
\varphi(u)=\sum_{j=1}^{+\infty}(-1)^jz_j(0)
L_{j-1}^{(1)}(u)\frac{1}{j},
	$$
$$
\psi(u)=\sum_{j=1}^{+\infty}(-1)^j\nu_j(0)
L_{j-1}^{(1)}(u).
	$$	     
If $z_k(0)$ and $\nu_k(0)$ tend to 0 fast enough, series \eqref{c4_2.6} can be integrated term by term  to give
\begin{equation}
z_k(t)=(-1)^k\sum_{j=1}^{+\infty}(-1)^j\frac{1}{j}z_j(0)l_{kj}(\tau)+\frac{2\sigma_*}{\varepsilon_*}
(-1)^k\sum_{j=1}^{+\infty}(-1)^j\nu_j(0)h_{kj}(\tau),
\label{c4_2.8}                             
\end{equation}  
$$
l_{kj}(\tau)=2\int\limits_0^{+\infty}
x^3e^{-x^2}L_{k-1}^{(1)}(x^2)L_{j-1}^{(1)}(x^2)\cos(\tau x)\,dx, 
        $$
$$
l_{kk}(0)=k,\,\,\,\,\,\,l_{kj}(0)=0,\,\,\,\,\hbox{for}\,\,\,\,k\ne j.
        $$
$$
h_{kj}(\tau)=2\int\limits_0^{+\infty}
x^2e^{-x^2}L_{k-1}^{(1)}(x^2)L_{j-1}^{(1)}(x^2)\sin(\tau x)\,dx, 
        $$ 
        
\subsection{Double series solution to the leading-order equations}
Substituting \eqref{c4_2.8}  into the equations\eqref{c4_2.1} - \eqref{c4_2.4} we obtain, after some algebra, the exact solution of the leading-order equations in the form of functional double series 
\begin{equation} 
x=cW_0(\theta,t),\\\ y=(x-f)\tan\theta, \\\-\pi/2< \theta < \pi/2                
\label{c4_2.9}                                
\end{equation}
\begin{equation}
W_0(\theta,t)= \sum_{n=1}^{+\infty} (-1)^{n-1}\frac{1}{2n}
[a_n(0)C_n(\tau,\theta)+b_n(0)S_n(\tau,\theta)]+                 
\label{c4_2.10}                      
\end{equation}               
$$
\frac{1}{\sqrt{2|f|}}\sum_{n=1}^{+\infty} (-1)^n
[\rho_n(0)H_n(\tau,\theta)+e_n(0)G_n(\tau,\theta)] 
      $$
\begin{equation}
\pd{g_0}{t}=-\varepsilon_*W_0(\theta,t),\,\,\,\,\,\,t=\tau\,{\sqrt{2|f|}} 
\label{c4_2.11}
\end{equation}
where
\begin{equation}
C_n(\tau,\theta)=-2\cos^2\theta\cdot\sum_{k=1}^{+\infty}
(-1)^kl_{kn}(\tau)\cos(2k\theta)           
\label{c4_2.12}                            
\end{equation}
\begin{equation}
S_n(\tau,\theta)=-2\cos^2\theta\cdot\sum_{k=1}^{+\infty}
(-1)^kl_{kn}(\tau)\sin(2k\theta)                
\label{c4_2.13}                              
\end{equation}
\begin{equation}
H_n(\tau,\theta)=-2\cos^2\theta\cdot\sum_{k=1}^{+\infty}(-1)^k
h_{kn}(\tau)\cos(2k\theta)        
\label{c4_2.14}                            
\end{equation}        
\begin{equation}
G_n(\tau,\theta)=-2\cos^2\theta\cdot\sum_{k=1}^{+\infty}(-1)^k
h_{kn}(\tau)\sin(2k\theta)          
\label{c4_2.15}                               
\end{equation}       
\begin{equation}
l_{kn}(\tau)=2\int\limits_0^{+\infty}
x^3e^{-x^2}L_{k-1}^{(1)}(x^2)L_{n-1}^{(1)}(x^2)\cos(\tau x)\,dx                                 \label{c4_2.16}                            
\end{equation}                     
$$
h_{kn}(\tau)=2\int\limits_0^{+\infty}
x^2e^{-x^2}L_{k-1}^{(1)}(x^2)L_{n-1}^{(1)}(x^2)\sin(\tau x)\,dx   
        $$ 
$$
k=1,2,\dots,\,\,\,\,n=1,2,\dots,\,\,\,\,\,\,
l_{kn}=l_{nk}
	$$           
$$
l_{kk}(0)=k,\,\,\,\,\,\,l_{kn}(0)=0,\,\,\,\,\hbox{for}\,\,\,\,k\ne n,\,\,\,\,\,\,\frac{dh_{kn}}{d\tau}=l_{kn}
	$$
The right hand side of  \eqref{c4_2.16}
 involves the Laguerre polynomials,
$\,\,L_k^{(1)}(u),\,\,$ defined by the recursion relation 
\begin{equation}
 L_0^{(1)}=1,\,\,\,L_1^{(1)}(u)=2-u,
\label{c4_2.17}                                   
\end{equation}  
$$
\hbox{for}\, k\ge 2\,\,\,\,\,\,kL_k^{(1)}(u)=(2k-u)L_{k-1}^{(1)}(u)-kL_{k-2}^{(1)}(u)                      
	$$ 	    
Though the function $W_0(\theta,t)$ is a linear combination of functions 
\eqref{c4_2.12} - \eqref{c4_2.15},  the waves \eqref{c4_2.9} are still nonlinear. 
In the wave theory, the principle of linear superposition 
states, in particular, that when two waves overlap, 
the actual displacement of any point of the free surface, at any time, 
is obtained by adding two displacements. The waves \eqref{c4_2.9} does not obey the principle (the implicit form of the free surface  
is  $x=cW_0(\hbox{arctangent}(y/(x-f)),t)$). 
        
\section{ Summing up of the series involved in the leading-order solution. \\ Three theorems }   
Three theorems are formulated and proved in this section. 
\vspace{3mm}

\noindent
{\bf Theorem 1.}   
In any rectangle $|\theta|\le\pi/2-\delta$,
 $ 0\le \tau \le T,\,\,$  ($0<\delta<\pi/2,\, T>0 $
may  be assigned arbitrarily) series \eqref{c4_2.12} through \eqref{c4_2.15} converge uniformly with regard to $ \theta $ and $ \tau $.
\bigskip

\noindent
{\bf Theorem 2.} On the interval $\,|\theta|<\pi/2\,$  the sums of the series are given by
\begin{equation}
C_n(\tau,\theta)=\int\limits_0^{+\infty}
x^3e^{-x^2/2}L_{n-1}^{(1)}(x^2)\cos\phi         
\cos(\tau x)\,dx              
 \label{c4_3.1}                                   
\end{equation}         
\begin{equation}
S_n(\tau,\theta)=\int\limits_0^{+\infty}
x^3e^{-x^2/2}L_{n-1}^{(1)}(x^2)\sin\phi           
\cos(\tau x)\,dx            
 \label{c4_3.2}                                    
\end{equation}          
\begin{equation}
H_n(\tau,\theta)=\int\limits_0^{+\infty}
x^2e^{-x^2/2}L_{n-1}^{(1)}(x^2)\cos\phi                 
\sin(\tau x)\,dx            
\label{c4_3.3}                  
\end{equation}         
\begin{equation}
G_n(\tau,\theta)=\int\limits_0^{+\infty}
x^2e^{-x^2/2}L_{n-1}^{(1)}(x^2)\sin\phi                 
\sin(\tau x)\,dx          
\label{c4_3.4}                        
\end{equation}           
where $\phi=1/2\,x^2\tan\theta$.
\bigskip

\noindent
{\bf Theorem 3.} The following expansions hold:  
\begin{equation}
C_n(\tau,\theta)=\sum_{k=1}^{+\infty}
2^k\,k!\,b_{n,k}(\tau)(\cos\theta)^{k+1}\cos((k+1)\theta)     
\label{c4_3.5}                            
\end{equation}                        
$$
S_n(\tau,\theta)=\sum_{k=1}^{+\infty}2^k\,k!\,
b_{n,k}(\tau)(\cos\theta)^{k+1}\sin((k+1)\theta)     
	$$
The coefficients $b_{n,k}(\tau)$ are defined by the following power series:  
\begin{equation}
F_{n,1}\equiv x^2L_{n-1}^{(1)}(x^2) \cos(\tau x) = 
\sum_{k=1}^{+\infty}b_{n,k}(\tau)(x^2)^k    
\label{c4_3.6}                           
\end{equation}        
In any rectangle $|\theta|\le\pi/2$, $ 0\le \tau \le T,\,\,$ 
($T>0$ may  be assigned arbitrarily) 
series \eqref{c4_3.6} converges uniformly with regard to $ \theta $ and $ \tau $. 

Similar expansions hold for $H_n(\tau,\theta)$ and $G_n(\tau,\theta)$.

\subsection{ Proof of Theorem 1}  
When in the series  \eqref{c4_2.12} and \eqref{c4_2.13} the products of trigonometric
functions are decomposed into sums, the rearranged series take the form
\begin{equation} 
\hat C_m(\tau,\theta)=-\frac{1}{2}\sum_{k=0}^{+\infty}
(-1)^k\hat l_{km}(\tau)\cos(2k\theta)         
\label{c4_3.7}                             
\end{equation}       
\begin{equation}
\hat S_m(\tau,\theta)=-\frac{1}{2}\sum_{k=0}^{+\infty}
(-1)^k\hat l_{km}(\tau)\sin(2k\theta)     
\label{c4_3.8}                                
\end{equation}       
$$
\hat l_{km}=-l_{k+1,m}+2l_{km}-l_{k-1,m},\,\,\,\,\,\,
\hat l_{0,m}=-I_{1,m}
        $$
or,  by  \eqref{c4_2.16} and \eqref{c4_2.17},                                             
\begin{equation}
\hbox{for}\,\,\,k\ge 1\,\,\,\hat l_{km}(\tau)=\frac{2}{k}\int\limits_0^{+\infty}
x^5L_{k-1}^{(1)}B(x)\,dx      
\label{c4_3.9}                            
\end{equation}
\begin{equation}
\hat l_{0,m}=-l_{1,m}(\tau)=-2\int\limits_0^{+\infty}
x^3B(x)\,dx        
\label{c4_3.9a}
\end{equation}
where       
$$
B(x)=\exp{(-x^2)}L_{m-1}^{(1)}(x^2)\cos(\tau x)
	$$
It follows from \eqref{c4_2.16}  and asymptotic formula for Laguerre polynomials (Chapter 2)  that \linebreak $l_{nm}=O(1/n^{1/4})$ 
 $\hbox{for}\,\,{0\le{\tau}}\le{T},\,\,$ where $\,\,T\,\,$ and 
$\,\,m\,\,$ are fixed. 

The difference of the $n$-th partial sums of series \eqref{c4_2.12} and 
\eqref{c4_3.7} tends to zero as $n$ tends to $\,\,+\infty,\,\,$ since 
$\,\,l_{nm}\to 0\,\,$
as $\,\,n\to{+\infty}.\,\,$ This means, that either both series converge
to the same sum or both series diverge.  
 
It follows from the asymptotic formula  for Laguerre polynomials and \eqref{c4_3.9}  that $\hat l_{nm}=O(1/n^{5/4})$. Now we conclude that 
series \eqref{c4_3.7} converges absolutely and uniformly throughout the region
$\,\,{0\le{\tau}}\le{T},\,\,|\theta|\le \pi/2\, $ 
and series \eqref{c4_2.12} converges uniformly in the same region.

The same result for series \eqref{c4_2.13} - \eqref{c4_2.15}  can be proved in a similar manner.  This completes the proof of Theorem 1.

\subsection{ Proof of Theorem 2}  
By the Lemma 2 of Chapter 2, subsection 3.2, 
\begin{equation}
x^2\sum_{k=1}^{+\infty}(-1)^k\frac{1}{k}L_{k-1}^{(1)}(x^2)\cos(2k\theta)=1-\exp\left(\frac{1}{2}\,x^2\right)\cos\phi,                 
\label{c4_3.10}                      
\end{equation}     	
$$	
\phi=\frac{1}{2}\,x^2\tan \theta  
	$$
It follows from \eqref{c4_3.7}, \eqref{c4_3.9}, and \eqref{c4_3.9a} that
$$
\hat C_m(\tau,\theta)=-l_{1m}(\tau)+\frac{1}{2}\sum_{k=1}^{+\infty}(-1)^k\hat l_{km}(\tau)\cos(2k\theta), 
	$$	
According to Theorem 1, he series \eqref{c4_3.7} converges uniformly within the square $\,\,{0\le{\tau}}\le{T},\,\,|\theta|\le \pi/2$.
So we can write 	
$$
\frac{1}{2}\sum_{k=1}^{+\infty}(-1)^k\hat l_{km}(\tau)\cos(2k\theta)  =
	$$
$$
-\int\limits_0^{+\infty}x^3\exp{(-x^2)}L_{m-1}^{(1)}(x^2)
\cos(\tau x)\left(x^2\sum_{k=1}^{+\infty}(-1)^k\frac{1}{k} L_{k-1}^{(1)}(x^2)\cos(2k\theta)\right)\,dx.    
	$$
With  the use of lemma 2 we obtain	
$$
-\int\limits_0^{+\infty}x^3\exp{(-x^2)}L_{m-1}^{(1)}(x^2)
\cos(\tau x)\left[1-\exp\left(\frac{1}{2}\,x^2\right)\cos\phi\right]dx.
	$$		
Consequently	
$$	
\hat C_m(\tau,\theta)=	
\int\limits_0^{+\infty}x^3\exp\left(-\frac{1}{2}\,x^2\right)L_{m-1}^{(1)}(x^2)\cos\phi\cos(\tau x)\,dx=C_m(\tau,\theta).
	$$		
Equalities \eqref{c4_3.2} - \eqref{c4_3.4} can be proved in the similar way.

In the proof of Theorm 1, it was found that the right-hand parts in 
\eqref{c4_3.1} - \eqref{c4_3.4} drop as $n^{-1/4}$. 

 Series \eqref{c4_2.9} converges absolutely, if values of $a_n,\,\,$ $b_n,\,\,$ 
$\rho_n,\,\,$ and $e_n$ drops, for example, as $1/n$.

Series \eqref{c4_2.8} converges absolutely and uniformly with regard to 
$ \theta $ and $ \tau $, if the initial free surface shape $W(\theta,0)$ and initial doublet distribution $\nu(\theta,0)$ are smooth functions and, consequently, can be expanded in a convergent Fourier series.

\subsection{ Proof of Theorem 3}  
 Write 
$$
C_m=\int\limits_0^{+\infty}F_{m,1}\pd {V_1}{x}\,dx,\,\,\,\,\,\,
\pd {V_1}{x}= xe^{-x^2/2} \cos\phi    
	$$
The function $F_{m,1}$ is defined by \eqref{c4_3.6}.  

In any rectangle 
$0\le x\le a$, $0\le \tau\le T$ series \eqref{c4_3.6} 
converges uniformly with regard to $x$ and $\tau$.

Integration by parts gives
$$
C_m=\cos\theta\cdot \int\limits_0^{+\infty}F_{m,2}\pd {V_2}{x}\,dx
	$$
where 
$$
xF_{m,2}=\pd {F_{m,1}}{x},\,\,\,\,\,\, 
F_{m,2}=2\sum_{k=1}^{+\infty}kb_{m,k}(\tau)(x^2)^{k-1}
	$$
$$
\pd {V_2}{x}= xe^{-x^2/2}\left[\sin\theta \cos\phi+   
\cos\theta \sin\phi\right]     
	$$
$$
V_2=-\cos\theta e^{-x^2/2}
\left[\sin(2\theta) \cos\phi+    
\cos(2\theta) \sin\phi\right] ,\,\,\,\,\,\,    
\phi=1/2\,x^2\tan\theta
	$$

Performing the integration by parts again and again we come to the first 
equality \eqref{c4_3.5}   
The second equality \eqref{c4_3.5}  can be proved in a similar manner.

It is not difficult to find that at any fixed value of $m$, coefficients 
$2^kk!b_{m,k}$ go to zero fast enough as $k\rightarrow +\infty$ 
to ensure uniform convergence of series \eqref{c4_3.6}. 

For $m=1$ we have 
$$
F_{1,1}=x^2\cos(\tau x)
	$$
$$
2^kk!b_{1,k}= 
(-1)^{k-1}\frac{2k}{(2k-3)!!}(\tau^2)^{k-1}\,\,\,\,\,(k=1,\,2,\,\cdots )
	$$
$$
(2k-1)!!=1\cdot 3\cdot 5\cdot\dots (2k-1),\,\,\,\,\,\,(-1)!!=0
	$$
If $m=2$, then
$$
F_{2,1}=x^2(2-x^2)\cos(\tau x)
	$$
$$
2^kk!b_{2,k}=
(-1)^{k-1}\frac{4k}{(2k-1)!!}(\tau^2)^{k-1}+
(-1)^{k-1}\frac{4k(k-1)}{(2k-3)!!}(\tau^2)^{k-2}
	$$

Expression \eqref{c4_3.5}  shows that $W_0(\theta, t)$ satisfy boundary condition \eqref{c4_1.6}.

\section{ Specific wave packets}  
If in \eqref{c4_3.5}  coefficient $b_{m,1}$ does not equal zero, $W_0(\theta,t)$ drops as  $1/y^2$ when $y$  increases without bound. But the departure of the interface from its equilibrium position may drop faster: for any given natural number $n$ and suitable initial conditions 
the departure drops as $1/y^{n+1}$ (and even faster). 
Indeed, there exist such value of $\lambda_k$ that
$$                          
u^n=\sum_{k=0}^n\lambda_kL_k^{(1)}(u)
	$$
so at initial conditions in \eqref{c4_2.8} corresponding  to $\lambda_k$ we obtain 
$$
W_0(\tau,\theta)=I_{2n+1},
	$$
$$ 
 I_{2n+1}(\tau,\theta)=\int\limits_0^{+\infty}
x^{2n+1}e^{-x^2/2}\cos\phi\cos(\tau x)\,dx   
	$$
In this case, expansion \eqref{c4_3.6}) should be replaced by
$$
F(x)\equiv x^{2n}\cos(\tau x) = 
\sum_{k=n}^{+\infty}b_k(\tau)(x^2)^k  
        $$
and, consequently, 
$$
 I_{2n+1}(\tau,\theta)=\sum_{j=0}^{+\infty}(-1)^j\,2^{n+j}\cdot (n+j)!\,
\frac{\tau^{2j}}{(2j)!}
(\cos\theta)^{n+1+j}\cos((n+1+j)\theta)
	$$
$$
 I_{2n+1}(0,\theta)=2^n\cdot n!\,(\cos\theta)^{n+1}\cos((n+1)\theta)
	$$
The integral is normalized by   $2^n\cdot n!$:
$$
I_{2n+1}(\tau,\theta)=2^n\cdot n!\,\hat I_{2n+1}(\tau,\theta),\,\,\,\,\,\,
\hat I_{2n+1}(0,0)=1 
	$$
$$
\hat I_{2n+1}(0,\theta)=(\cos\theta)^{n+1}\cos((n+1)\theta)
	$$
Functions $C_m(\tau,\theta)$ are linear combinations of integrals 
$I_{2n+1}(\tau,\theta)$; for instance, $C_1(\tau,\theta)=I_3(\tau,\theta),\,$ 
$C_2=2I_3-I_5,\,$ $2C_3=2I_7-6I_5+6I_3$. 

Similar results for $S_m(\tau,\theta)\,$, $H_m(\tau,\theta)\,$, and  $G_m(\tau,\theta)$ 
can be obtained in the similar way. 

We define specific wave packets by equations 
\begin{equation} 
x=c_nI_{2n+1},\,\,\,\, y=(x-f)\tan\theta 
\label{c4_ 4.1}             
\end{equation}  
$$
I_{2n+1}(\tau,\theta)=\int\limits_0^{+\infty}
x^{2n+1}e^{-x^2/2}\cos\left(\frac{1}{2}x^2\tan\theta\right)\cos(\tau x)\,dx;   
	$$
$2n+1$ is referred to as the packet number.

Equations of other three sets of wave packets will be obtained from \eqref{c4_2.12}, \eqref{c4_2.13} and \eqref{c4_2.14} if the product of trigonometric functions in the integrands are replaced by one of the following expressions
$$
\cos\phi\sin(\tau x)/x,\,\,\,\,\,\,
\sin\phi\cos(\tau x),
	$$
$$
\sin\phi\sin(\tau x)/x,\,\,\,\,\,\,\phi=\frac{1}{2}x^2\tan\theta
	$$
\vspace{3mm}

By the three Theorems, any wave group on the 
interface is a nonlinear mixture of finite or infinite (it depends on initial conditions) set of the specific wave packets of different numbers,  and evolution of each packet in the mixture is not influenced by evolution of the others. 
 
%
%
%
%
%
%
%
%
%
%
%
%

\chapter{Deep-Water Free Surface Gravity Waves}
 
{\small{In this chapter, an original ``law of similarity'' for the wave packets is obtained. The analytical solution and the ``law of similarity''  are used to test the theory of packets against experiments performed in a water tank by Feir (1967)  and  Yuen \& Lake (1982).}}

\section{Time evolution of specific wave packets on the free surface} 

The concept of ``free surface'' is
used when the upper of the two liquids separated by the interface 
is a fictitious liquid of zero density ($\gamma_1=0$). 
It is assumed that the flow is  irrotational, the pressure on the free surface is constant, and the Bernoulli's equation holds inside the water. The equilibrium position of the free surface is horizontal plane $x=0$.  At $\varepsilon_*=1$ all equations and analytical results obtained in Chapter 4  remain valid for the free surface water waves.
\vspace{3mm}

Figures 1 - 5 illustrate evolution of wave profiles specified by equations (at $f=-10$)
\begin{equation}
x=0.4\,\hat I_{23}(\tau,\theta)+0.075\,\hat I_5(\tau,\theta),\,\,\,\,\, |y|=(x-f)|\tan\theta|                                  
\label{c5_1.1}               
  \end{equation}	
\begin{equation}
x=0.4\,\hat I_{23}(\tau,\theta),\,\,\,\,\,|y|=(x-f)|\tan\theta|   
 \label{c5_1.2}                
  \end{equation}	
	
\begin{equation}
x=0.075\,\hat I_5(\tau,\theta),\,\,\,\,\,|y|=(x-f)|\tan\theta| 
 \label{c5_1.3}             
  \end{equation}	  
  At each value of $\tau$, the profiles are drawn in the same scale. 
For each of the profiles, initially, at $\tau=0$, the crest of the wave 
of maximum height is situated on the ray $\theta=0$, i.e., on the $x$-axis; the vertical coordinates of the crests are: $x=0.4$ for 
\eqref{c5_1.2}, $x=0.075$ for \eqref{c5_1.3}, and, consequently, $x=0.475$  for \eqref{c5_1.1}.
The wave group \eqref{c5_1.1} gradually disintegrates into its components \eqref{c5_1.2} and \eqref{c5_1.3}, and each of the components doesn't affect the evolution of the another one. 
\vspace{3mm}

\begin{figure}[ht]
\centering
	\resizebox{0.8\textwidth}{!}
		{\includegraphics{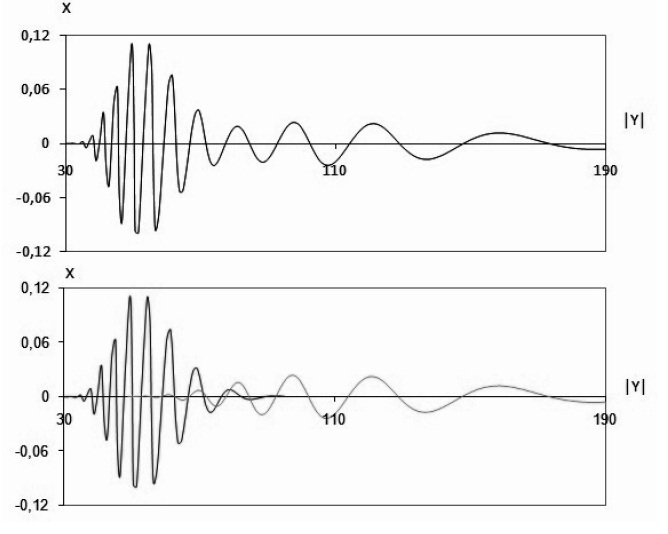}}  
	\caption{
	Instantaneous profiles of the waves \eqref{c5_1.1} (upper panel), 
\eqref{c5_1.2} (bottom panel, the curve on the left), and\eqref{c5_1.3} (bottom panel, the curve on the right) at $\tau=25$.  
	}
\end{figure} 

We note in passing that, in case of plane waves, the term `wave group' 
means a system of waves in which the amplitude dies away on either side of the wave of maximum height; the term `wave' means a
wave-group's section which consists of one crest and the adjacent trough and is singled out by three successive zeros of the group;
the level difference between a crest and the following or the preceding trough  is referred to as the wave-height (or height of the wave); 
the term `wave packet' means a wave group in which 
the amplitude dies away very quickly with distance from the wave of maximum height; figures 2 and 3 show, that the wave-height of the wave of maximum height of the wave group \eqref{c5_1.2} is about 0.21 at 
$\tau=25$ and 0.18 at $\tau=50$.
System of harmonic waves $x=\cos(k y-\omega t)$ on the free surface infinite in extent is not a wave group.
\vspace{3mm}

The wave group \eqref{c5_1.1} is a nonlinear combination of wave packets \eqref{c5_1.2} and \eqref{c5_1.3}. 
The figures show that the wave packets \eqref{c5_1.2} and \eqref{c5_1.3}  propagate at different speeds: the horizontal distance between the crests of the waves of maximum 
height is approximately 70 at $\tau=25,\,$ 115 at $\tau=50,\,$ 260 at $\tau=100,\,$440 at $\tau=200,\,$ and 690 at $\tau=300$.
The greater the number of a packet, 
the shorter the wavelength of the packet's carrier wave component and  
the slower the packet travels. 
The wavelength and velocity of simple harmonic waves on deep water 
are related in a similar way.
\vspace{3mm}

\begin{figure}[ht]
\centering
	\resizebox{0.8\textwidth}{!}
		{\includegraphics{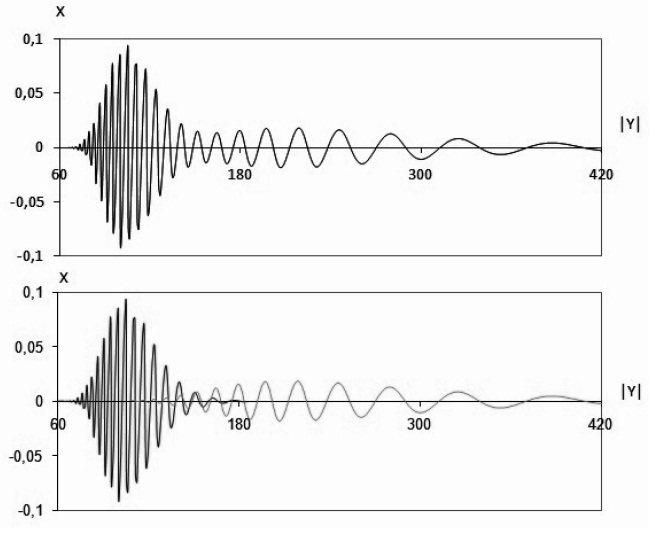}} 
	\caption{
  Same as Figure 1, but $\tau=50$.
	}
\end{figure} 

At any particular moment of time, the system of waves contains only one wave of maximum height (WMH) on semi-axis $y>0$  
(the situation with two waves of equal maximum height can be ignored), so the zeros of WMH constitute a 'natural frame of reference' for other zeros. 
The wave of maximum height is singled out by three zeros and consists 
of a crest and the trough following or preceding the crest.
Let $\theta_r(\tau)$ and $\theta_f(\tau)$  denote two of the three zeros 
of the WMH which correspond to minimum and maximum of the three 
$|\tan\theta(\tau)|$ respectively. By the zero $\theta_f(\tau)$ we define the front of WMH at the instant $\tau$, by $\theta_r(\tau)$ the rear of the wave is defined.  
\vspace{3mm}
 
Zeros of the wave packet
\begin{equation} 
x=c_nI_{2n+1},\,\,\,\, y=(x-f)\tan\theta 
\label{c5_1.4}                           
\end{equation}  
$$
I_{2n+1}(\tau,\theta)=\int\limits_0^{+\infty}
x^{2n+1}e^{-x^2/2}\cos\left(\frac{1}{2}x^2\tan\theta\right)\cos(\tau x)\,dx;      
	$$
 are defined by the equation $I_{2n+1}(\tau;\theta)=0$ and 
(at fixed value of $\tau$) are situated in the numbered rays 
$\theta=\theta_k(\tau)$ ($k$ is the number of a zero). 
For brevity we will use the term `zero $\theta(\tau)$' to denote a zero 
of a wave.
\vspace{3mm}

\begin{figure}[ht]
\centering
	\resizebox{0.8\textwidth}{!}
		{\includegraphics{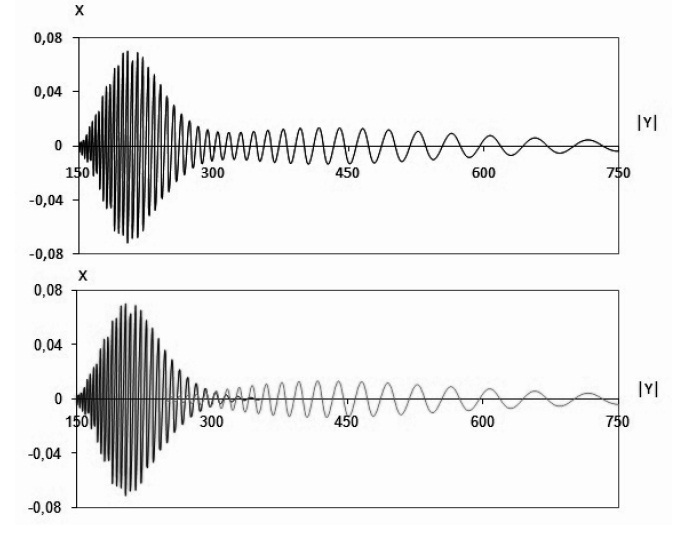}} 
	\caption{
  Same as Figure 1, but $\tau=100$.
	}
\end{figure} 

Zeros $\theta_k(\tau)$ of the wave packet \eqref{c5_1.4} are independent of $f$ and $c_n$. This leads to the following 
{\bf Assertion}: At any given value of $\tau$

i) the vertical coordinates of the crests and the troughs (and, consequently, 
the waveheights) are independent of $f$;

ii) the ratio of the distances $y_{k+1}(\tau)-y_k(\tau)$ and
$y_k(\tau)-y_{k-1}(\tau)$ between any successive zeros of the waves is independent of $f$.
\vspace{3mm}

In the course of time, the number of zeros of the waves  
increases, so the existence of the functions $\theta_k(\tau)$ on the whole semi-axis $\tau>0$ is not assured. 
Horizontal velocity of a zero of the wave is $d y_k/d t=u_k(\tau)\sqrt{0.5|f|}$, where $u_k(\tau)=d\tan\theta_k(\tau)/d\tau$.
The definition does not guarantee the existence of
differentiable (and even continuous) functions $\tan\theta_k(\tau)$, 
but the average rate
$$
u_k=[\tan\theta(\tau+\Delta\tau)-\tan\theta(\tau)]/\Delta\tau
	$$
exists, and this gives us a useful piece of information about the speed 
of waves as we will see in section 2. 
\vspace{3mm}

\begin{figure}[ht]
\centering
	\resizebox{0.8\textwidth}{!}
		{\includegraphics{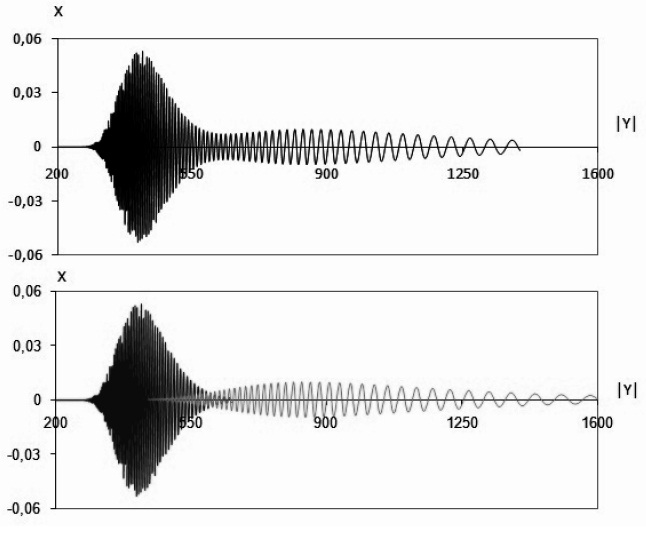}}
	\caption{
  Same as Figure 1, but $\tau=200$.
	}
\end{figure} 

Let $L_*$ be the dimensional unit of length, then $T_*=\sqrt{L_*/g}\,$ is the dimensional unit of time.
At the instant $t_*$, the dimensional coordinate of the zero $\theta=\theta(\tau)$ is 
$$
y_*(t_*)=|f|L_*\tan\theta(\tau), \,\,\,\,\,\,
t_*=\tau\sqrt{2|f|}\cdot T_*
	$$ 	
and, consequently, the ratio 
\begin{equation}
\lambda(\tau)=\frac{y_*(t_*)}{gt_*^2}=
\frac{1}{2}\,\,\frac{\tan\theta(\tau)}{\tau^2}      
\label{c5_1.5}                          
\end{equation}	
depends on $\tau$ only. 
\vspace{3mm}

 Given a fixed value of $\tau$, $\tan\theta(\tau)$  can be calculated from equations of a wave packet, and corresponding value of $\lambda(\tau)$ can be obtained from \eqref{c5_1.5}. 
Calculations show that $\tan\theta$ and $\lambda$ are monotone functions of $\tau$, so 
$\tau$ and $\tan\theta$ can be calculated for given value of $\lambda$.
\vspace{3mm}
 
\begin{figure}[ht]
\centering
	\resizebox{0.8\textwidth}{!}
		{\includegraphics{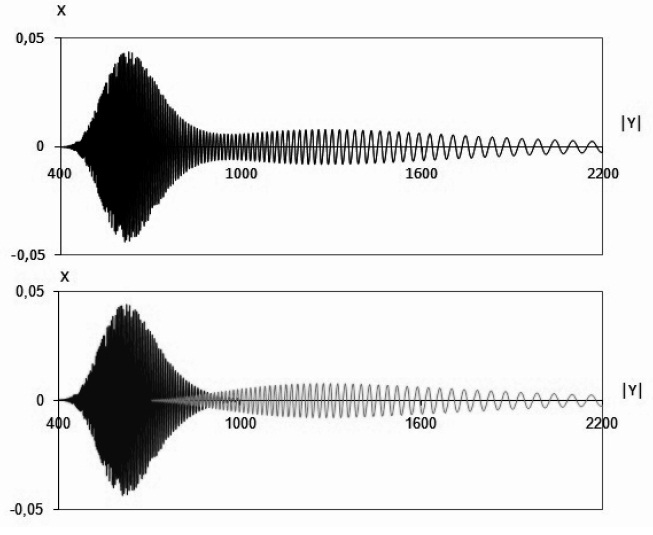}}
	\caption{
  Same as Figure 2, but $\tau=300$.
	}
\end{figure} 

The figures show that the initial free surface displacement turns into two 
wave groups which run away from $x$-axis in opposite directions. From the 
figure we notice that with time the waves leave a region about 
the vertical plane $y=0$. As there is no flow across the plane, 
the fluid motion would be unaffected if the plane was replaced by 
rigid barrier. From now on we will consider wave groups propagating 
in a fixed direction, say, along the semiaxis $y>0$.

\section{Estimation of dimensional parameters of a packet}

The front of the wave of maximum height of packet \eqref{c5_1.2} travels a distance of $y_*$ from the origin $y=0$ in a time $t_*$. Below for specific values of $y_*$, $\,t_*$ and  initial height of the packet, $H_*$, the following dimensional parameters of the packet are estimated:
\begin{description}
\item
$L_*\,\,$ - the dimensional unit of length; 
\item
$h_*(t_*)$ - the maximum height of the packet at the instant $t_*$; 
\item
$V_*\,\,$ - the average speed of the packet during the travel time $t_*$; 
\item
$v_*\,\,$ - the speed of the WMH at the instant $t_*$; 
\item
$l_*\,\,$ - the length of WMH at the instant $t_*$; 
\item
$l_{**}\,\,$ - the length of the packet at the instant $t_*$.
\end{description}
Algorithm for calculating values of the parameters may be summarized as follows: 
\vspace{3mm}

Calculate the value of $y_*/t_*^2$ and then find the front of the wave 
of maximum height of packet \eqref{c5_1.2} such that $\tau$ and $\tan\theta_f(\tau)$ 
satisfy equation \eqref{c5_1.5}; also find $\tan\theta_r(\tau)$, non-dimensional height $h(\tau)$ of the WMH, 
and $\tan\theta_f(\tau+\Delta\tau)$ at small $\Delta\tau$. 
\vspace{3mm}

From equations $H_*=c\,L_*$ and $|f|L_*|\tan\theta_f(\tau)|=y_*\,\,$ find dimensional estimates for vertical $L_*$ and horizontal 
$a=|f|L_*$ parameters of length ($c$ is the non-dimensional height of the packet at $t=0$). The height and the length of the WMH at the instant $t_*$ are estimated as $h_*(t_*)=h(\tau)L_*$ and 
$\,\,l_*=a[|\tan\theta_f(\tau)|-|\tan\theta_r(\tau)|]$ respectively. 
The speeds are estimated as  
$$
V_*=y_*/t_*,\,\,\,\,\,\,
\,v_*=\sqrt{0.5ag}\,[|\tan\theta_f(\tau+\Delta\tau)|-|\tan\theta_f\tau)|]/ \Delta\tau.
	$$
Technique for estimating the length of the packet is shown below in example 1. 
\vspace{3mm}

The estimates of packet parameters are independent of $c$ (in 
\eqref{c5_1.2} $c=0.4$) and the estimate of $a$ is independent of $H_*$.\\
\vspace{5mm}

{\bf Examples}
 \vspace{3mm}

{\it Example 1.} At $y_*=8.53 m$, $t_*=27.4\,$s, $H_*=0.75\,$cm, 
$g=9,8\,\,\hbox{m/s}^2,\,\,$  $2y_*/(gt_*^2)=0.0023$ and from equations (8.2), we obtain $\tau=91$,
$$
\tan\theta_f(91)=19.150,\,\,\,\,\tan\theta_r(91)=18.610,\,\,\,\,\tan\theta_f(91.1)=19.192 
	$$
$$
\,\,\,\,h(91)=0.075,\,\,\,\,L_*=1.86 \,\,\hbox{cm},\,\,\,\,a=|f|L_*=0.445\,\,\hbox{m}
	$$
$$
h_*=1.4 \,\,\hbox{mm},\,\,\,\,l_*=0.24\,\,\hbox{m},\,\,\,\,
V_*=0.311\,\,\hbox{m/s},\,\,\,\,v_*=0.620\,\,\hbox{m/s}
	$$
An estimate of instantaneous length of a wave packet based on a record of the packet depends on sensitivity of the wave amplitude gauge used for the record.
\vspace{3mm}

At the instant $t_*$ the body of the packet is bounded by 
$y_{min}=|f|L_* \tan\theta_{min}$ and  $y_{max}=|f|L_* \tan\theta_{max}$ where 
$\tan\theta_{min}$ and $\tan\theta_{max}$ are to be found from equations \eqref{c5_1.2} at obtained value of $\tau$. The body of a packet includes waves with amplitudes greater than a preassigned level which depends on the gauge sensitivity. 
\vspace{3mm}

Inequality $|x|>0.01$ corresponds to waves with amplitude greater than 0.2 mm. 
At the instant $t_*$ the waves are located in the interval $14.3\,a<y<25.9\,a$ 
(at $\tau=91\,\,$ $\tan\theta_{min}=14.3,\,$ $\,\tan\theta_{max}=25.9$). 
The length of the interval is estimated as 
$l_{**}=(25.9-14.3)\cdot 0.45\,\,\hbox{m}=5.2\,\,$m. 
\vspace{3mm}

In the interval $15\,a<y<25\,a$ the inequality $|x|>0.02$ holds. 
At the instant $t_*$ the length of the interval is estimated as 
$l_{**}=10\cdot a=4.5\,\,$m. 
\vspace{3mm}

At the instant $t_*$ in the interval $15.5\,a<y<22.3\,a$ 
inequality $|x|>0.03$ is satisfied, so amplitudes of the waves in the interval 
are greater than 0.6 mm. The length of the interval is estimated as 
$l_{**}=6.8\cdot a=3.0\,\,$m. 

Depending on the preassigned amplitude level, one of the three values may be 
taken as the estimate of the packet length at the instant $t_*$.
\vspace{3mm}

The above estimates of the average speed of the packet 
and speed of the wave of maximum height 
equal to estimates of group and phase velocities respectively based on 
linear theory for waves of infinitesimal amplitude: 
at phase velocity of 0.62 m/s of sinusoidal carrier wave of a linear packet,  
the carrier frequency is $\omega=15.8\,\,\hbox{s}^{-1}$,  
the wavenumber is 
 $k=\omega^2/g= 25.47\,\hbox{m}^{-1}$, and
the group velocity is $v=0.5\omega/k=0.31\,\,$ m/s. 
\vspace{5mm}

{\it Example 2.} At $y_*=8.53\,\,$m,  $t_*=24\,\,$s, $H_*=0.5\,$cm, 
$g=9.8\,\,\hbox{m/s}^2$, 
 $2y_*/(gt_*^2)=0.0030$, we obtain 
 $$
 \tau=70,
 $$
$$
\tan\theta_f(70)=14.765,\,\,\,\,
\tan\theta_r(70)=14.227,
\,\,\,\,\tan\theta_f(70.1)=14.807
	$$
$$
h(70)=0.084,\,\,\,\,L_*=1.25 \,\,\hbox{cm},\,\,\,\,a=|f|L_*=0.578\,\,\hbox{m}
	$$
$$
h_*=1.05 \,\,\hbox{mm},\,\,\,\,\,\,\,\,l_*=0.31\,\,\hbox{m},\,\,\,\,
V_*=0.355\,\,\hbox{m/s},\,\,\,\,v_*=0.707\,\,\hbox{m/s}
	$$
\vspace{3mm}

{\it Example 3.} At $y_*=8.53\,\,$m,  $t_*=28.8\,\,$s, $H_*=0.7\,$cm, 
$g=9.8\,\,\hbox{m/s}^2$,  $2y_*/(gt_*^2)=0.0021$, we obtain  
$$
\tau=100,
$$
$$
\tan\theta_f(100)=20.990,\,\,\,\,\tan\theta_r(100)=20.450,\,\,\,\,\tan\theta_f(100.1)=21.032
	$$
$$
h(100)=0.072,\,\,\,\,L_*=1.75 \,\,\hbox{cm},\,\,\,\,a=|f|L_*=0.406\,\,\hbox{m}
	$$
$$
h_*=1.3 \,\,\hbox{mm},\,\,\,\,\,\,\,\,l_*=0.22\,\,\hbox{m},\,\,\,\,
V_*=0.296\,\,\hbox{m/s},\,\,\,\,v_*=0.592\,\,\hbox{m/s}
	$$
The section $170<y<250$ in figure 4 (bottom panel, the curve on the left) of 
the packet includes only waves with amplitudes greater than 0.2 mm. The length 
of the packet is estimated as $(25-17)\cdot a=3.2\,$ m.                                                      
\vspace{3mm}

The examples show that the average speed of the packet of finite amplitude during the travel time $t_*$ is a half of the speed of the WMH at the instant $t_*$; exactly the same relationship exists between group and phase velocities in linear theory of deep water waves of infinitesimal amplitude; the average speed of the packet \eqref{c5_1.2} and the speed of the WMH are equal, respectively, to the group and phase velocities of a linear wave packet provided that sinusoidal carrier wave of the linear packet and the wave of maximum height of nonlinear packet \eqref{c5_1.2} have the same wavelength. 
\vspace{3mm}

The reason for the relationship between the four speeds 
is that 
 the leading-order term $\Phi_0(x,y,t)$ of velocity
potential and the function $W(y,t)$ satisfy equations
\begin{equation}
\pd {W(y,t)}{t}=\pd {\Phi_0}{n_-},\,\,\,\,\,\,\,\,\,\,\, \pd {\Phi_0}{t_-}+W=0
\label{c5_2.1}
\end{equation}
where subscript ``n'' is used for the derivative in normal direction to the free surface.
Equations (2.1) are the  boundary conditions on the free surface: 
continuity of the normal velocity and the pressure continuity. 

 If the boundary conditions are shifted from the evolving free surface to the equilibrium position of the surface, 
the normal derivative reads $\partial \Phi_0/\partial x|_{x=0}$,
 the equations \eqref{c5_2.1} reduce to equations of classical linear theory of infinitesimal waves.  

\section{ Testing of the theory against experiments performed in water tanks}    

We use some results from wave pulse experiments presented in \cite[]{feir} and \cite[]{yuen}
bearing in mind that the aim of Feir is to study the effects of finite amplitude of a wave group on the group shape and the frequency distribution over the group,
while the main purpose of the work by Yuen \& Lake is to 
decide whether the predictions based on asymptotics of solution of the Schr\''odinger equation correspond with experimental results. 
Below the term ``pulse'' is reserved for experimental wave groups 
 to distinguish them  from the wave packets \eqref{c5_1.1} - \eqref{c5_1.3}.
\vspace{3mm}

\begin{figure}[ht]
\centering
	\resizebox{0.7\textwidth}{!}
		{\includegraphics {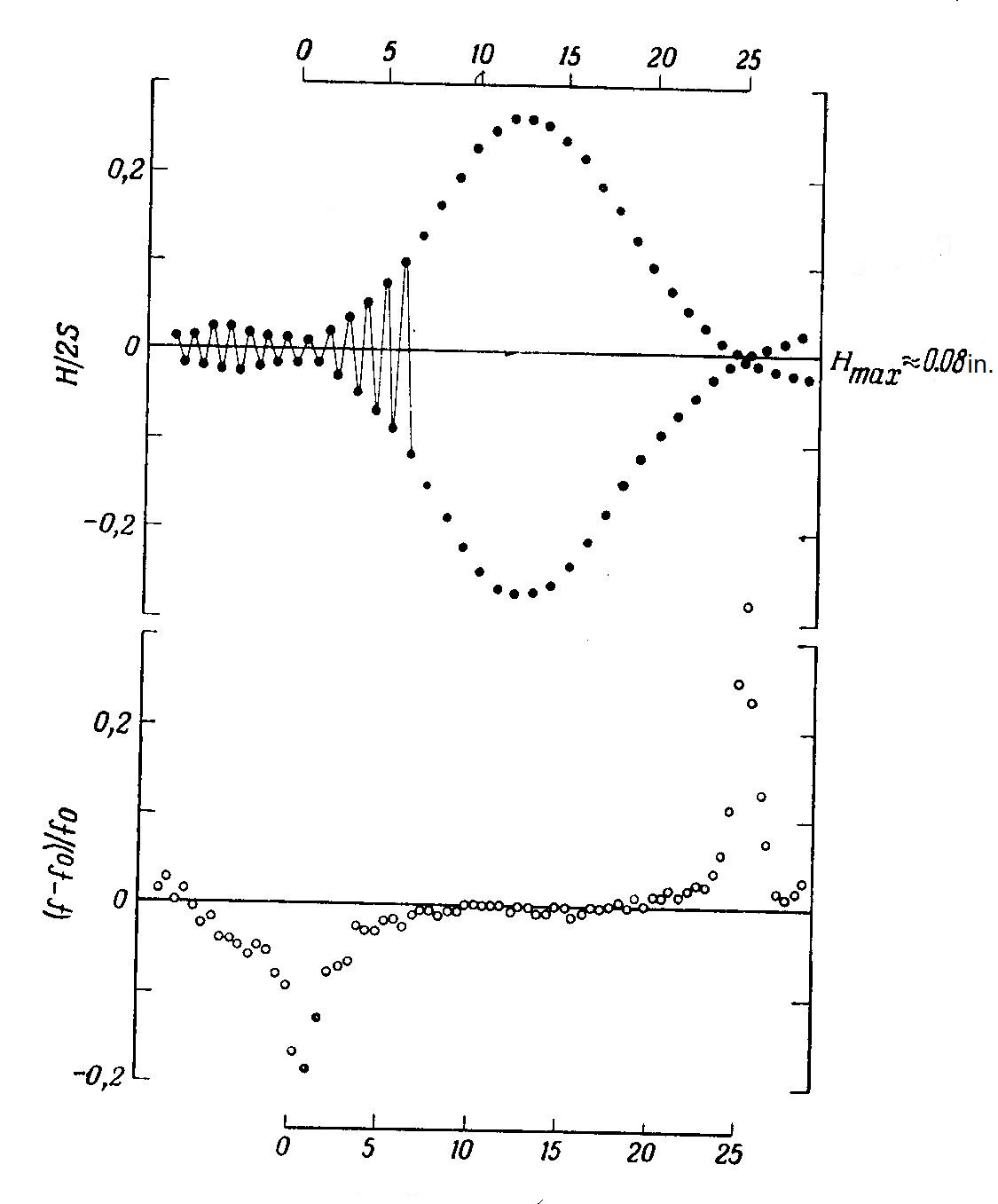}} 
	\caption{
   Pulse data in Feir's experiment. Distance from wavemaker, 28 ft. Wavemaker  frequency, $f_0= 2.50c/s$; stroke S=0.075 in.; pulse period 10.00 s (reproduced from (Feir 1967)).
	}
\end{figure} 

In \cite[]{feir}, a number of records of wave pulses are shown generated in a water tank equipped with a wavemaker. 
The wavemaker was oscillating at a carrier frequency of 2.5 Hz with amplitude varying slowly from minimum to some maximum and back to minimum; the range of the maximum amplitude of the wavemaker stroke was from 0.19 cm for the wave pulses of small amplitude to 1.5 cm for the pulses of large amplitude; in the records the range of maximum water surface displacement is from 1 mm to 10 mm. 
\vspace{3mm}

\begin{figure}
\centering
	\resizebox{0.9\textwidth}{!}
		{\includegraphics {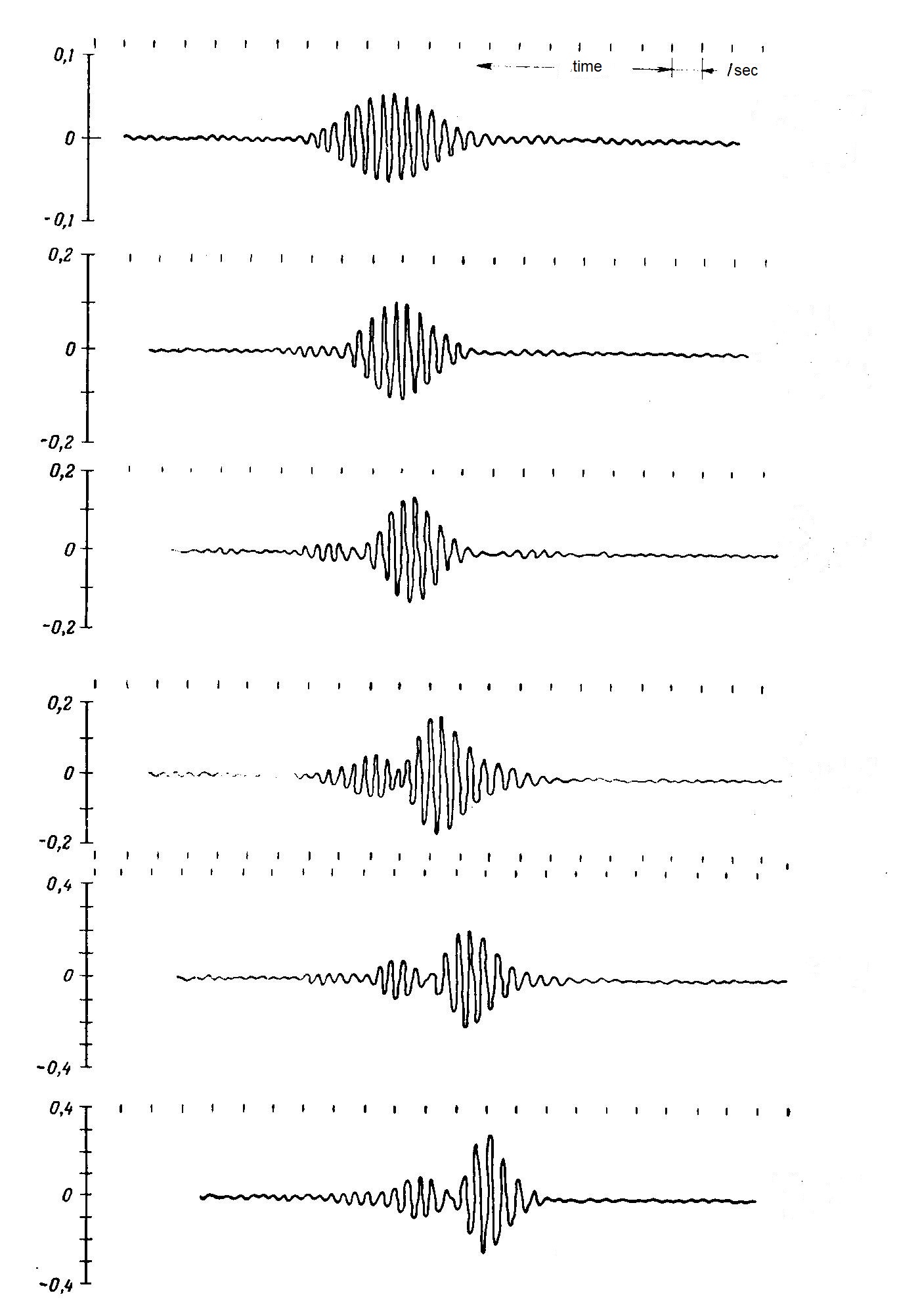}} 
	\caption{
   Wave pulse records at 28 ft distance from the wavemaker (reproduced from (Feir 1967), to be compared with analytical solutions in figures 1-5).
	}
\end{figure} 

When comparing the pulse with the packet, one should keep in mind that in  figures 6 and 7, each wave pulse was  recorded during some time interval at  a fixed point of the water tank, so the leading waves 
of longer length  are shown in the left side of the figures,
while in figures 1 - 5 the packet profiles are shown on some space interval at a fixed point in  time, so the leading waves of longer length are  in the right side of the figures.
 In (Feir 1967), time intervals between neighboring crests and troughs of the pulse were measured, and inverse of the intervals  were named as frequencies.
\vspace{3mm}

The record of the wave pulse shown in fig. 6 was made 
at the distance  8.53 m from the wavemaker. 
The upper part of  figure 6 shows the group of 25 individual waves which passes the gauge for 10 s. This means  that the mean period of the waves is 0.4 s which equals to the wavemaker period. In the record, maximum water surface displacement is about 1 mm. Periods of the individual waves were measured, and the results are shown in the lower part of the figure 6: in the figure, periods of 18 waves in the middle of the wave pulse are close to 0.4 s; periods of the waves at the ends of the pulse differ from the wavemaker period up to 20\%. According to (Feir 1967), such frequency distribution is typical for pulses of small amplitude. (The maximum water surface displacement of 6 mm at the distance of 8.53 m from the wavemaker is considered to be large.) 
 The waves in the middle of the pulse move slower than 
the leading waves and faster than the trailing waves. 
\vspace{3mm}

Profile of packet \eqref{c5_1.2} at $\tau=100$ (figure 4, bottom panel, the curve on the left) is similar to the pulse profile shown in figure 6: in figure 4, bottom panel, $160<y<300$, we can clearly see 25 individual waves. 
  In [Feir, 1967], the length of the wave pulse shown in figure 6, the travel time $t_*$ from the wavemaker to the gauge and the maximum amplitude $H_*$ of the pulse are not shown. Nevertheless, the first two quantities can be estimated as follows: by the linear theory, a wave packet with carrier frequency of 2.5 Hz travels at the group velocity of $0.31\,\,\hbox{m/s}$. This suggests that the length of the wave pulse is $0.31\,\,\hbox{m/s}\cdot 10\,\,\hbox{s}=3.1\,\,\hbox{m}$, and the travel time from the wavemaker to the gauge is about 27.5 s. The ratio of the distance  $D=8.5$ m to the length of the pulse is about three.
These values of the travel time and length of the experimental wave pulse are in good agreement with the estimates obtained in the three examples above. 
\vspace{3mm}

 In figure 7, records of six pulses are shown corresponding to six  values of  wavemaker stroke. The three records shown at the bottom  of the figure  correspond to amplitude of the water surface displacement of 5 - 6 mm.
Feir comments on the figure 7: ``... as the pulse moves farther down the tank, it separates into two distinct groups. The frequency in this groups varies erratically, but the trend seems to be that the frequency of the leading group is lower than the frequency of the trailing group.''
\vspace{3mm}

This experimental result presents the essential argument for the nonlinear theory of wave packets.
The wave group shown in figure 1 with time disintegrates in two wave packets (figure 4);  the leading packet move faster than the trailing one;  the distance between zeros of the leading packet is larger than the distance in the trailing packet.
In figures 2 and 3, we can clearly see the pattern of the packets overlapping, which demonstrate irregularity in the distance between zeros of the mix of two packets.
\vspace{3mm}

Figure 8 is a reproduction of Figure 1 from  (Yuen \& Lake 1982). 
Evolution of three wave envelopes is shown in cases $A,\,$ $B\,\,$ and $C$ corresponding to three initial profiles (given at $x=0$) symmetric about their peaks: ($A$) envelope soliton, ($B$) a  hyperbolic secant envelope, ($C$) a sine envelope of the same amplitude as in the case $A$; the amplitude scale of the hyperbolic secant envelope is reduced by a factor of 2.5 compared to the case $A$; in the three cases the carrier frequency of the wavemaker was 
$2\,$ Hz; the experiments were performed in a water tank 3 feet deep.  The wave envelopes propagate along the tank from the left to the right. 
\vspace{3mm}

During the time interval of the observation, in case $A$, the shape of the envelope in the figure seems to be almost unchanged with time; in cases $B$ and $C$, an initial wave group gradually turns into a structure which consists of a relatively `tall' wave packet following a `low' one and long chains of waves of small amplitude in the front and in the rear of the system of packets.

\begin{figure}
\centering
	\resizebox{\textwidth}{!}
		{\includegraphics{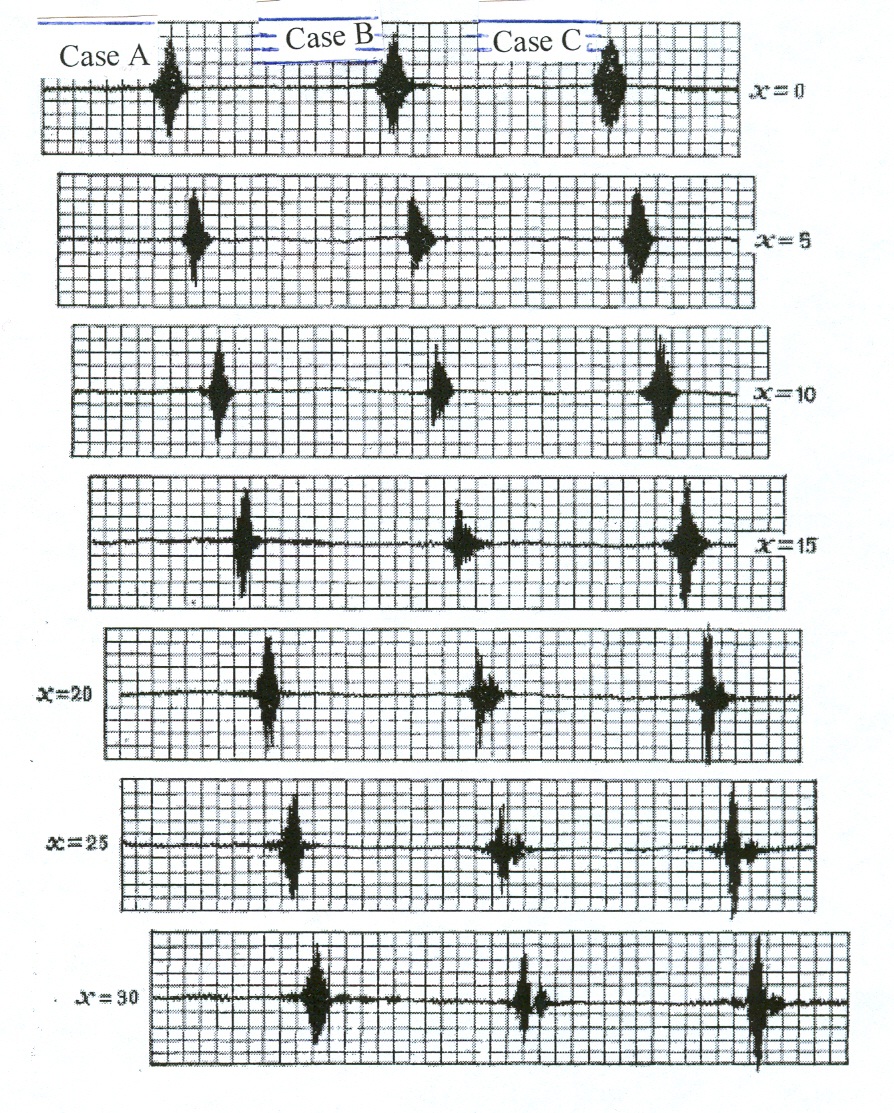}} 
	\caption{
  Evolution of three wave envelope pulses (reproduced from
   (Yuen \& Lake 1982), to be compared with analytical solutions in figure 9.).
	}
\end{figure} 

In (Yuen \& Lake 1982), experimental measurements are not presented, 
actual dimensions of the envelopes as well as the ratio of vertical 
to horizontal scales and the individual wave profiles are not shown. 

We can obtain an idea about the horizontal and vertical scales only 
for the initial envelope soliton (figure 8, case $A$, $x=0$).

The shape of the envelope soliton is given by 
$x=a\,{\rm sech}\,u,\,$ \linebreak$u=\sqrt{2}k^2a\,s$, $ka\approx 0.14$; 
$s=0$ is the vertical axis of symmetry of the shape, 
$a$ is the height of the envelope;  maximum slope of the tangent 
to the envelope is reached at $u=0.882$; coordinates of the point of tangency are $s=32.14a,\,\,\,x=0.7a$; at the level of the point, i.e., on the level 
of 0.7 of the envelop's height, the width of the envelope is nearly 64 times as great as the height.

But in figure 8, case A, at $x=0$, the envelope's width on the level 
0.7 of it's height is approximately of 0.2 of the height, so the ratio of 
the vertical scale to the horizontal scale is about 250.
\vspace{3mm}

Figure 9 shows the same profiles as the figures 1 - 5, but the horizontal scales are reduced significantly.
In figure 9, column $A_1$, the envelopes seem to be of the same height and the same width.
The resemblance between columns $B$ (figure 8) and $B_1$ (figure 9) is doubtless. 
\vspace{3mm}

At $\tau=25$ (figure 2, bottom panel, the curve on the left) the height of the packet \eqref{c5_1.2} equals 0.1136, the packet's width at the level 0.7 of its height is equal to 6.6, 60 times as great as the height (as for envelope soliton with $ka=0.144$). The length of the packet increases with time, while the height decreases (figures 2 - 6, bottom panels, the curves on the left; formation of a chain of waves of small amplitude in the front of the packet is demonstrated in the top panels).  
\vspace{3mm}

\begin{figure}
\centering
	\resizebox{0.9\textwidth}{!}
		{\includegraphics{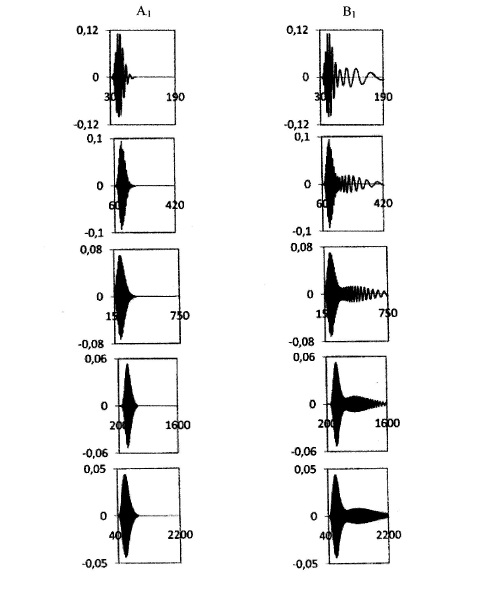}} 
	\caption{
  Same as figures 1 - 5 (top to bottom) at $\tau=25,\,$ 
$\tau=50,\,$ $\tau=100,\,$ $\tau=200,\,$ $\tau=300\,$ respectively, 
but on reduced horizontal scale.  Column $A_1$: profiles of packet \eqref{c5_1.2},            
column $B_1$: profiles of packet \eqref{c5_1.1}  .
	}
\end{figure} 

One can estimate some parameters of a pulse  observed in [Yuen \& Lake,1982]: 
wavemaker frequency  $f_0=4\pi s^{-1}$, wave length $l=0.39$m, 
  each wave pulse travelled the distance $D=9.14$m, 
 the distance is 23 times as great as the wave length,  
   group velocity $v=0.39 m$, the travel time $t=D/v=23.4 s$.
In (Yuen \& Lake 1982), the pulse duration and its length, and the ratio of the distance
to the pulse length are not specified. It is reasonable to assume that this ratio is about 3, since the estimates for the pulse of Yuen \& Lake are close to the corresponding pulse parameters in (Feir 1967).
For such value of the ratio and slow variation in the pulse parameters, the use of asymptotic statements of the weakly nonlinear wave theory seems to be unpersuasive.
\vspace{3mm}

The resemblance of the experimental pulse profiles in (Feir 1967) to the theoretical profiles in Figs. 1-5, and
the experimental envelope profiles in (Yuen \& Lake 1982) to the theoretical profiles in Fig. 9, as well as the agreement between the theoretical and experimental estimates of certain packet parameters allow us to conclude that
the profiles reproduced in Figs. 1-5 correspond (for given values of the parameter $a$) to time intervals
comparable with the pulse observation time in the water tank.




\chapter{Specific Packets and Theory of Long Waves Traveling Across an Ocean}  

\section{Specific packet $H_1$ as a mathematical model for long waves}   

The  free surface long waves are modeled by the specific wave packet 
\begin{equation}
x(t)=c\,\frac{1}{\sqrt{2|f|}}H_1(\theta,\tau),\,\,\,\, y=(x-f)\tan\theta, \,\,\,\,\,\,-\pi/2< \theta < \pi/2          
\label{c6_1.1}
\end{equation}
$$
H_1(\theta,\tau)=\int\limits_0^{+\infty}
x^2e^{-x^2/2}\cos\left(\frac{1}{2}x^2\tan\theta\right)\sin(\tau x)\,dx,
\, \,\,\,\,\,t=\tau\,{\sqrt{2|f|}}  .   
	$$
By the moment $t=0$ the free surface has not yet been displaced 
from its mean level (the horizontal plane $x=0$), but the velocity field has already become different from zero: 
$$
 H_1(\theta, 0)=0,\,\,\,\,\,\,\,\,
\left.\pd{H_1}{\tau}\right|_{\tau=0}=
2\cos^2\theta\cdot\cos(2\theta)
	$$
	This means that the water at rest initially is set in motion  by a sudden change in the velocity field. 
\vspace{3mm}

Figure 1 displays profiles of the packet \eqref{c6_1.1} at 
$\tau=1,\,5,\,10,\,15,\,50,\,100.$
The wave packet \eqref{c6_1.1} travels faster then any other specific packet and with time leaves behind the other packets. 
The model is adopted as a rough mathematical model for 
propagation of long waves through an open sea, and is not 
intended for detailed quantitative description of the tsunamis referred to below.
\vspace{3mm}

All equations are written in non-dimensional variables.
Since the problem has no characteristic linear size, 
the dimensional unit of length, $\,\,L_*,\,\,$ is a free parameter.
But for applications in the section 6, the value of $L_*$,    
as well as the value of $|f|L_*$, will be obtained  from instrumental data.
The dimensional unit of time, $\,\,T_*,\,\,$ is defined
by the relation $\,\,T_*^2g=L_*\,$, where $\,\,g\,\,$ is the acceleration
of free fall. The non-dimensional acceleration of free fall is equal to unity.
All parameters, variables and equations are made non-dimensional by 
the quantities $\,L_*,\,T_*,\,P_* $ and 
the density of water $\gamma_*=1000\,$ $\hbox{kg/m}^3$

\begin{figure}[ht]
	\resizebox{\textwidth}{!}
		{\includegraphics{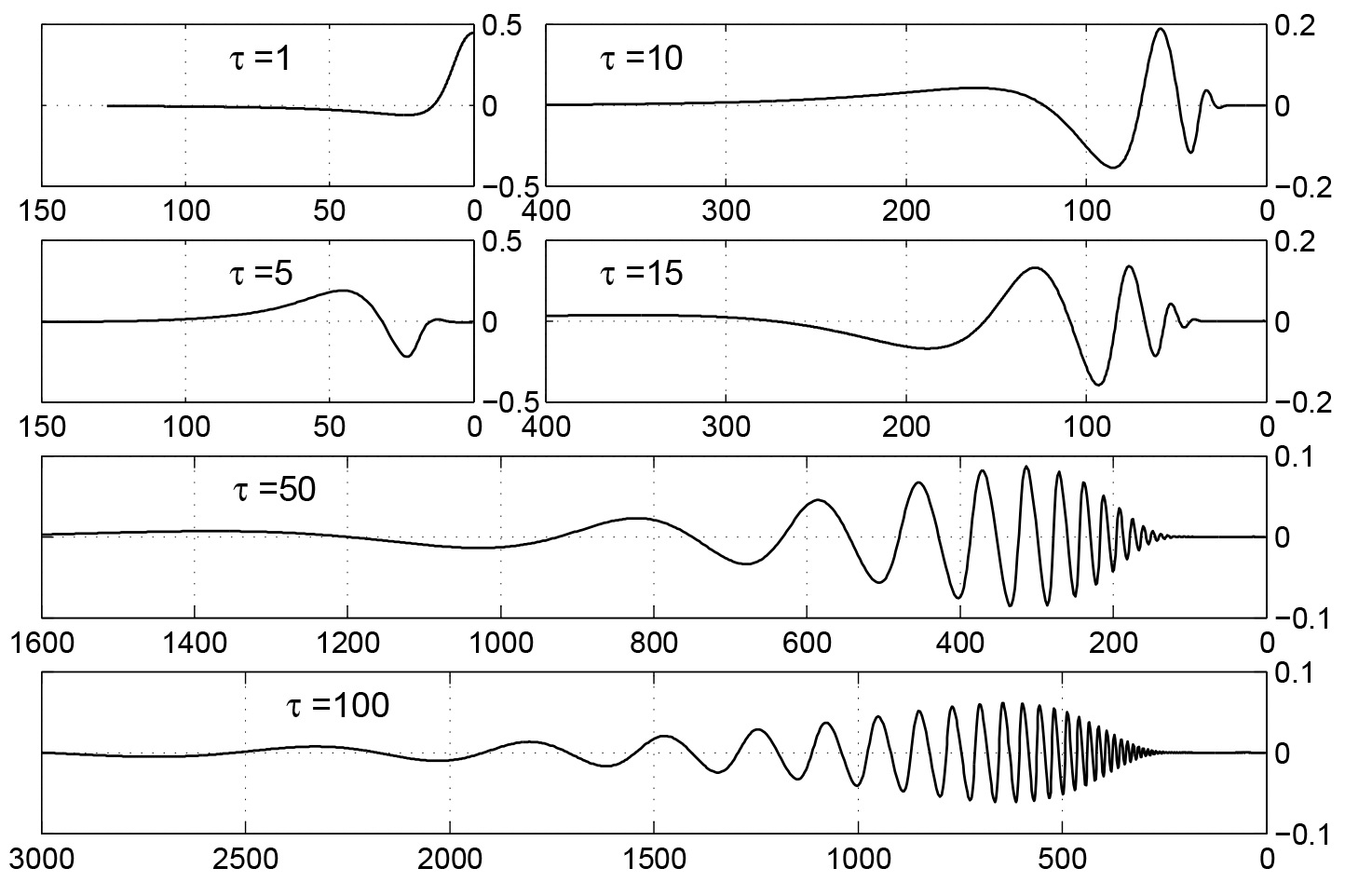}} 
	\caption{
Profiles of the packet \eqref{c6_1.1} at different values of $\tau$ shown in the figure near the corresponding curves, $c=0.5,\,\,\,f=-10$.  	
	}
	\label{c6_qu-.1-.10}
\end{figure} 

\section{Tsunami measurements in an open ocean}

To be assured that the horizontal dimensions of the wave origin are much larger than the magnitude of the free surface displacement in the origin, data on long waves traveling across the Pacific Ocean
generated by the  Kuril 01/2007,  Kuril 11/2006, and Peruvian 08/2007 earthquakes are used to test the theoretical model \eqref{c6_1.1}.
\vspace{3mm}

The data is obtained from tsunami records  at locations of the deep-ocean bottom-pressure recorders (DART buoys, DART - Deep-ocean Assessment and Reporting of Tsunami).    
Bottom recorders are deployed in the Pacific Ocean at a depth of 3000 - 5000 meters. Waves arriving at a deep-ocean buoy is a mixture of tidal, seismic, and gravity waves which come from the tsunami origin itself. Records of the buoys and their locations can be found on the USA National Data Buoy Center public website 
 (http://www.ndbc.noaa.gov/dart.shtml).
\vspace{3mm}

Figure 2 shows the  Kuril 01/2007 tsunami records de-tided 
using low-pass Buttherworth filter with 150 min cut-off; the residuals are 
non-tidal components, i.e., the mixture of seismic signal  (of shorter period) and of long ocean waves generated by the earthquakes. 
\begin{figure}[ht]
	\resizebox{\textwidth}{!}
		{\includegraphics{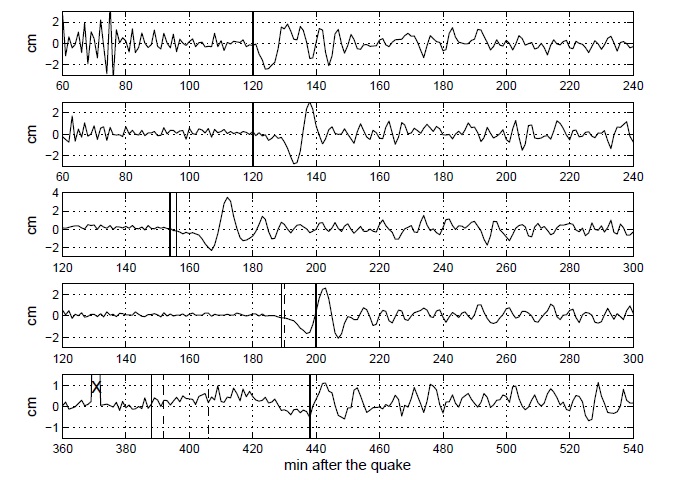} }  
	\caption{
De-tided records of the Kuril 01/2007 tsunami at DART buoys (top to bottom) 	21413, 21414, 46413, 46408, 46419. Vertical lines mark the arrival of the front of the WMH (thick solid line) and its estimate given in subsection 8.
 Cross marks a trigger pulse (signals send by an operator).
	}
\end{figure} 

Data extracted from the records at the DARTs  
are summarized in Tables 1 - 3, where the travel time 
$t_*\,\hbox{min}$ of the wave of maximum height (WMH),
the maximum wave-height, $H_*\,\hbox{cm}$, and
the horizontal distance between the gauge sensors and the centre (epicentre of the earthquake) of 
the wave origin, $y_*\,\hbox{km}$, are shown for each of the DARTs. 
Figure 2 and Table 1 show the de-tided DART records and data  for tsunami triggered by the earthquake near Kuril Islands in January 2007. 

\begin{center}
TABLE 1. The 2007 Kuril tsunami: WMH as recorded at the DART buoys.

Data 1.
 \begin{footnotesize}
\begin{tabular}{|p{15mm} |p{10mm} |p{10mm} |p{10mm} | p{15mm} | } 
\hline $\hbox{Buoy}$ & $t_*\,\,\hbox{min.}$ & $H_*\,\,\hbox{cm} $ & 
  $y_*\,\, \hbox{km}$ & $y_*/t_*^2$ \\
\hline   21413    &     120    &    4.0    &    1762  &  0.122361   \\
           21414    &     120    &    5.7    &    1804  &  0.125278   \\
           46413    &     154    &    5.7    &    2253  &  0.094999   \\
           46408    &     200    &    4.5    &    2660  &  0.066500   \\
           46419    &     438    &    1.6    &    5470  &  0.028513   \\
\hline 
\end{tabular}\\
\end{footnotesize}
\end{center}

\begin{center}
TABLE 2. The 2006 Kuril tsunami: WMH as recorded at the DART buoys. 

Data 2.
\begin{footnotesize}
\begin{tabular}{|p{15mm} |p{10mm} |p{10mm} |p{10mm} | p{15mm} | } 
\hline $\hbox{Buoy}$ & $t_*\,\,\hbox{min.}$ & $H_*\,\,\hbox{cm} $ & 
  $y_*\,\, \hbox{km}$ & $y_*/t_*^2$ \\
\hline     46413    &     155    &    8.5    &    2331  &  0.097024   \\
           46408    &     238    &    7.5    &    2735  &  0.048284   \\
           46402    &     270    &    10.0   &    3127  &  0.042894   \\
           46403    &     365    &    7.5    &    3581  &  0.026879   \\
\hline 
\end{tabular}\\
\end{footnotesize}
\end{center} 
\begin{center}
TABLE 3. The Peruvian tsunami: WMH as recorded at the DART buoys. 

Data 3.
\begin{footnotesize}
\begin{tabular}{|p{5mm} |p{15mm} |p{10mm} |p{10mm} |p{10mm} | p{15mm} | } 
\hline $\hbox{i}$ & $\hbox{Buoy}$ & $t_*\,\,\hbox{min.}$ & $H_*\,\,\hbox{cm} $ & 
  $y_*\,\, \hbox{km}$ & $y_*/t_*^2$ \\
\hline     1  & 32401    &     49     &    7.0    &    713   &  0.296960   \\                                                     
           2  & 32411    &     198    &    1.7    &    2561  &  0.065325   \\
           3  & 51406    &     444    &    3.9    &    5320  &  0.026986   \\
           4  & 46412    &     624    &    2.0    &    6921  &  0.017775   \\
\hline 
\end{tabular}\\
\end{footnotesize}
\end{center}

\section{Specific packet $H_1$: computed characteristics of the wave of maximum height }                          

It is shown in Chapter 5, formula (1.5) , that for any zero 
$\theta=\theta(\tau)$  of a wave of  specific  packet the ratio 
$$
\lambda(\tau)=\frac{y_*(t_*)}{gt_*^2}=
\frac{\tan\theta(\tau)}{2\tau^2}              
	$$
depends only on $\tau$.
When the distance is measured in kilometers and time in minutes, it is convenient to rewrite the ratio  as
\begin{equation}
\lambda_*(\tau)=\frac{y_*(t_*)}{t_*^2}\,\hbox{km/min$^2$}=
17.64\,\frac{\tan\theta(\tau)}{\tau^2}\,\hbox{km/min$^2$}.             
\label{c6_3.1}
\end{equation}                                                   
 Given a fixed value of $\tau$, $\tan\theta(\tau)$  can be calculated from equations \eqref{c6_1.1} of the wave packet, and corresponding value of $\lambda_*(\tau)$ can be obtained from \eqref{c6_3.1}. 

 \begin{figure}[ht]
	\resizebox{\textwidth}{!}
		{\includegraphics{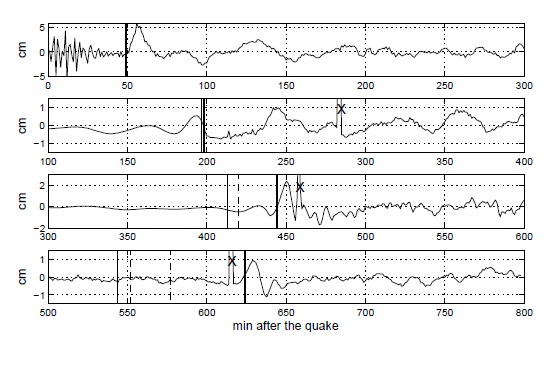} }  
	\caption{
De-tided records of the Kuril 11/2006 tsunami at DART buoys (top to bottom) 	46413, 46408, 46102, 46403.  Vertical lines mark the arrival of the front of the WMH (thick solid line) and its estimate given in subsection 8.
 Cross marks a trigger pulse (signals send by an operator).
	}
\end{figure} 
 
Table 4 shows computed characteristics of the wave of WMH: 
$\tan\theta_f,\,$ $\lambda_*(\tau),\,$ and  \linebreak
$\Delta(\tau)=H(\tau)\sqrt{2|f|}\,\,$ ($\,\,cH(\tau)$ is the maximum wave height) obtained from \eqref{c6_1.1} and \eqref{c6_3.1} at $c=1$ $f=-10$).
By the zero $\theta_f(\tau)$ we define the front of the wave of maximum height.
\newpage

\begin{center}
TABLE 4. Computed characteristics of the wave of maximum height.
\end{center}
\begin{footnotesize}

\begin{center}
\begin{tabular}{|p{10mm}  |p{10mm}  |p{10mm}  |p{10mm}  |p{10mm}  | p{10mm} | p{10mm}| } 
\hline $\tau$      & 25      & 30      & 35      & 40      & 45      & 50     \\ 
 $\tan\theta_f  $    & 19.032  & 19.818  & 23.687  & 27.587  & 31.508  & 32.531 \\                           
 $\lambda_*$         & 0.5371  & 0.3884  & 0.3411  & 0.3041  & 0.2744  & 0.2295  \\
 $\Delta(\tau) $         & 0.6326  & 0.5927  & 0.5653  & 0.5380  & 0.5126  & 0.4911  \\
$\lambda_*\cdot\tau$ & 13.429  & 11.653  & 11.938  & 12.166  & 12.351  & 11.477 \\
 $u $              &         & 0.1572  & 0.7738  & 0.7800  & 0.7842  & 0.2046  \\
\hline
\end{tabular}\\
\end{center}

\begin{center}
\begin{tabular}{|p{10mm}  |p{10mm}  |p{10mm}  |p{10mm}  |p{10mm}  | p{10mm} | p{10mm}| }
\hline $\tau$      & 55      & 60      & 65     & 70     & 75     & 80     \\ 
 $\tan\theta_f  $    & 36.396  & 40.268  & 44.159 & 48.057 & 49.114 & 52.974 \\                           
 $\lambda_*$         & 0.2122  & 0.1973  & 0.1844 & 0.1730 & 0.1540 & 0.1460 \\
 $\Delta(\tau) $         & 0.4738  & 0.4577  & 0.4420 & 0.4267 & 0.4155 & 0.4035 \\
$\lambda_*\cdot\tau$ & 11.673  & 11.839  & 11.984 & 12.110 & 11.552 & 11.681 \\
 $u $              & 0.7730  & 0.7744  & 0.7782 & 0.7796 & 0.2114 & 0.7720  \\
\hline
\end{tabular}\\
\end{center}

\begin{center}
\begin{tabular}{|p{10mm}  |p{10mm}  |p{10mm}  |p{10mm}  |p{10mm}  | p{10mm} | p{10mm}| }
\hline $\tau$      & 90     & 100    & 110    & 120    & 130    & 140    \\
 $\tan\theta_f  $    & 58.000 & 65.695 & 70.734 & 78.421 & 83.470 & 88.559 \\                           
 $\lambda_*$         & 0.1263 & 0.1159 & 0.1031 & 0.0960 & 0.0871 & 0.0797 \\
 $\Delta(\tau)  $        & 0.3840 & 0.3661 & 0.3507 & 0.3371 & 0.3255 & 0.3152 \\
$\lambda_*\cdot\tau$ & 11.368 & 11.589 & 11.343 & 11.528 & 11.326 & 11.158 \\
 $u $              & 0.5026 & 0.7695 & 0.5039 & 0.7687 & 0.5049 & 0.7606  \\
\hline
\end{tabular}\\
\end{center}

\begin{center}
\begin{tabular}{|p{10mm}  |p{10mm}  |p{10mm}  |p{10mm}  |p{10mm}  | p{10mm} | p{10mm}| }
\hline $\tau$      & 150    & 160     & 170     & 180     & 200     & 220    \\
 $\tan\theta_f  $    & 96.203 & 101.294 & 108.935 & 116.607 & 126.767 & 142.072  \\                           
 $\lambda_*$         & 0.0754 & 0.0698  & 0.0665  & 0.0635  & 0.0559  & 0.0518   \\
 $\Delta(\tau) $         & 0.3050 & 0.2960  & 0.2884  & 0.2804  & 0.2676  & 0.2554   \\
$\lambda_*\cdot\tau$ & 11.313 & 11.168  & 11.304  & 11.427  & 11.181  & 11.391   \\
 $u $              & 0.7644 & 0.5091  & 0.7641  & 0.7672  & 0.5081  & 0.7652  \\
\hline
\end{tabular}\\
\end{center}

\begin{center}
\begin{tabular}{|p{10mm}  |p{10mm}  |p{10mm}  |p{10mm}  |p{10mm}  | p{10mm} | p{10mm}| }
\hline $\tau$      & 240     & 260     & 280     & 300     & 310     & 320   \\
 $\tan\theta_f  $    & 154.802 & 167.532 & 180.265 & 192.996 & 198.071 & 205.727 \\ 
 $\lambda_*$         & 0.0474  & 0.0437  & 0.0406  & 0.0378  & 0.0364  & 0.0354  \\
 $\Delta(\tau) $         & 0.2446  & 0.2367  & 0.2281  & 0.2204  & 0.2175  & 0.2141  \\
$\lambda_*\cdot\tau$ & 11.378  & 11.368  & 11.362  & 11.348  & 11.271  & 11.341  \\
 $u $              & 0.6365  & 0.6366  & 0.6366  & 0.6365  & 0.5075  & 0.7656  \\
\hline
\end{tabular}\\
\end{center}

\begin{center}
\begin{tabular}{|p{10mm}  |p{10mm}  |p{10mm}  |p{10mm}  |p{10mm}  | p{10mm} | p{10mm}| }
\hline $\tau$      & 330     & 340     & 360     & 370     & 380     & 390   \\
 $\tan\theta_f  $    & 210.802 & 218.458 & 231.189 & 236.267 & 243.920 & 249.002 \\ 
 $\lambda_*$         & 0.0341  & 0.0333  & 0.0315  & 0.0304  & 0.0298  & 0.0289  \\
 $\Delta(\tau) $         & 0.2108  & 0.2077  & 0.2019  & 0.1991  & 0.1965  & 0.1950  \\
$\lambda_*\cdot\tau$ & 11.268  & 11.334  & 11.328  & 11.264  & 11.323  & 11.262  \\
 $u $              & 0.5075  & 0.7656  & 0.6365  & 0.5078  & 0.7653  & 0.5082  \\
\hline
\end{tabular}\\
\end{center}

\begin{center}
\begin{tabular}{|p{10mm}  |p{10mm}  |p{10mm}  |p{10mm}  |p{10mm}  | p{10mm} | p{10mm}| }
\hline $\tau$      & 400     & 410     & 420     & 440     & 460     & 470      \\
 $\tan\theta_f  $    & 254.094 & 261.734 & 269.387 & 279.559 & 292.291 & 299.931  \\                           
 $\lambda_*$         & 0.0280  & 0.0275  & 0.0269  & 0.0255  & 0.0244  & 0.0239   \\
 $\Delta(\tau) $         & 0.1926  & 0.1902  & 0.1879  & 0.1836  & 0.1796  & 0.1776   \\                                                                               
$\lambda_*\cdot\tau$ & 11.205  & 11.261  & 11.314  & 11.208  & 11.209  & 11.257   \\
 $u $              & 0.5092  & 0.7640  & 0.7653  & 0.5086  & 0.6366  & 0.7640  \\
\hline
\end{tabular}\\
\end{center}
\begin{center}
\begin{tabular}{|p{10mm}  |p{10mm}  |p{10mm}  |p{10mm}  |p{10mm}  | p{10mm} | p{10mm}| }
\hline $\tau$      & 480     & 490     & 500     & 520     & 540     & 550   \\
 $\tan\theta_f  $    & 305.023 & 312.663 & 317.762 & 330.494 & 343.227 & 350.867 \\                           
 $\lambda_*$         & 0.0233  & 0.0230  & 0.0224  & 0.0216  & 0.0208  & 0.0205  \\
 $\Delta(\tau) $         & 0.1758  & 0.1744  & 0.1729  & 0.1706  & 0.1674  & 0.1659  \\
$\lambda_*\cdot\tau$ & 11.209  & 11.256  & 11.211  & 11.211  & 11.212  & 11.253  \\
 $u $              & 0.5092  & 0.7640  & 0.5099  & 0.6366  & 0.6366  & 0.7640  \\
\hline
\end{tabular}\\
\end{center}

\begin{center}
\begin{tabular}{|p{10mm}  |p{10mm}  |p{10mm}  |p{10mm}  |p{10mm}  | p{10mm} | p{10mm}| }
\hline $\tau$      & 560     & 580     & 600     & 620     & 630     & 640   \\
 $\tan\theta_f  $    & 355.959 & 368.691 & 381.424 & 394.155 & 401.796 & 406.888 \\                           
 $\lambda_*$         & 0.0200  & 0.0193  & 0.0187  & 0.0181  & 0.0178  & 0.0175  \\
 $\Delta(\tau)$         & 0.1644  & 0.1616  & 0.1588  & 0.1563  & 0.1550  & 0.1538  \\
$\lambda_*\cdot\tau$ & 11.213  & 11.213  & 11.214  & 11.214  & 11.250  & 11.215  \\
 $u $              & 0.5092  & 0.6366  & 0.6366  & 0.6365  & 0.7641  & 0.5092  \\
\hline
\end{tabular}\\
\end{center}
\end{footnotesize}

In Table 4, values of the ratios $u=\Delta\tan\theta_f/\Delta\tau$ are given for each two neighboring columns (for instance, for the columns $\tau=90$ and $\tau=100$, we find $u=(65.695-58.000)/(100-90)=0.7695$).
During a time interval $t$ min, the front of the WMH travels a distance 
$y(t)$ km at the average speed  
\begin{equation}
V=\frac{y(t)}{t}=
\lambda_*(\tau)\cdot \tau \sqrt{2|f|}\cdot \frac{T_*}{60}\,\hbox{km/min}.        
\label{c6_3.2}
\end{equation}                                   
The value of the average speed during time interval 
$\Delta \tau=\tau_{n+1}-\tau_n$ equals
\begin{equation}
v=u\,\sqrt{|f|L_*g/2}\,\hbox{m/s},\,\,\,\,\,\,              
u=[\tan\theta_f(\tau_{n+1})-\tan\theta_f(\tau_n)]/\Delta\tau,    \label{c6_3.3}
\end{equation}                     
where $\tau_n$ and $u$ are given in Table 4. 
The values of $\lambda_*\tau$ and $u$ seems to suggest 
that the average speed of the front of WMH is nearly constant: 
in the interval $75\le\tau\le 640$, $\lambda\tau$ ranges from 11.158 to 11.680. 
Calculations show that the the length of the wave  of maximum height equals to
\begin{equation}
l =L_*|f|(\tan\,\theta_f-\tan\,\theta_r)\approx 5L_*|f| \,\,\,\,\,\,\hbox{for}\,\,\,\,\, 30<\tau<640.              
\label{c6_3.4}
\end{equation} 	

\section{Theoretical characteristics of the WMH at locations of the 
the deep-ocean bottom-pressure recorders (DART buoys)}                     

Table 4 shows that $\tan\theta_f$ and $\lambda_*$ are monotone functions of $\tau$, so for given value of $\lambda_*$ the values of $\tau$ and $\tan\theta_f$ can be calculated using equations \eqref{c6_1.1} or estimated using Table 4. 
For each DART location, setting $\lambda_*=y_*/t_*^2$ (see Tables 1 - 3) and using Table 4, we obtain the results shown in Tables 5 - 7.
\begin{center}
TABLE 5. 
For Data 1 (Kuril 2007): theoretical characteristics of the WMH at locations of DART buoys.  

\begin{footnotesize}
\begin{tabular}{|p{3mm}|p{15mm} |p{15mm} |p{15mm} |p{15mm} | p{15mm}| } 
\hline $i$&$\hbox{Dart}$ & $\lambda $ & $\tau $   & $\tan\theta_f(\tau)$ & $\Delta$ \\ 
\hline 1 & 21413     & 0.122361 & 93.787  &  60.914  & 0.374984 \\
         2  &21414     & 0.125278 & 90.983  &  58.756  & 0.378995 \\
         3  &46413     & 0.094999 & 121.125 &  78.987  & 0.335795 \\
         4  &46408     & 0.066500 & 170.000 &  108.935 & 0.288400  \\
         5  &46419     & 0.028513 & 394.300 &  251.192 & 0.193968 \\
\hline
\end{tabular}\\
\end{footnotesize}
\end{center}

Values of $\tau,\,\,$ $\tan\theta_f(\tau),\,\,$ $\Delta(\tau)$ in the first line of Table 5  are obtained with the use of $\lambda_*= 0.122361$ (in the first line of Table 1) and Table 4 as follows. 

We see from Table 4 that 
$0.1159<\lambda_*=0.122361<0.1263,\,\,$ $90<\tau<100,\,\,$ 
$0.5800<\tan\theta_f(\tau)<65.695,\,\, $ $0.3661<\Delta(\tau)<0.3840$.

Method of linear interpolation gives $\tau=93.787,\,\,$ 
$\tan\theta_f(\tau)=60.914,\,\,$ $\Delta(\tau)=0.374984$. 
The rest lines in Table 5 are obtained in the same way. 

Tables 6  and 7 are similar to Table 5 and obtained in similar way.
\newpage

\begin{center}
TABLE 6. 
For Data 2 (Kuril 2006):  theoretical characteristics 
of the WMH at locations of the DART buoys. 

\begin{footnotesize} 
\begin{tabular}{|p{3mm}|p{15mm} |p{15mm} |p{15mm} |p{15mm} | p{15mm}| } 
\hline $i$&$\hbox{Dart}$ & $\lambda $ & $\tau $   & $\tan\theta_f(\tau)$ & $\Delta$ \\ 
\hline  1& 46413     & 0.097024 & 118.558  &  78.411  & 0.339047 \\
          2& 46408     & 0.048284 & 235.888  &  154.791 & 0.246741 \\
          3& 46402     & 0.042894 & 265.038  &  180.234 & 0.234534 \\
          4& 46403     & 0.026879 & 420.300  &  279.545 & 0.187836 \\
\hline
\end{tabular}\\
\end{footnotesize}
\end{center}
\begin{center}
TABLE 7. For Data 3 (Peruvian): theoretical characteristics 
of the WMH  at locations of the DART buoys. 

\begin{footnotesize}

\begin{tabular}{|p{3mm}|p{15mm} |p{15mm} |p{15mm} |p{15mm} | p{15mm}| } 
\hline $i$&$\hbox{Dart}$ & $\lambda_* $ & $\tau $   & $\tan\theta_f(\tau)$ & $\Delta$ \\ 
\hline 1& 32401     & 0.296960 & 41.206   &  28.533  & 0.531882  \\
         2& 32411     & 0.065325 & 173.917  &  111.939 & 0.285267  \\
         3& 51406     & 0.026986 & 418.567  &  268.290 & 0.188230  \\
         4& 46412     & 0.017775 & 630.833  &  402.220 & 0.154900  \\
\hline
\end{tabular}\\
\end{footnotesize}
\end{center}
\bigskip

\section{ Estimators of parameters of  specific packet $H_1$}  

For the front of the wave of maximum height (as for any zero) the following formulas hold 
$$
y(t_*)=-f\tan\theta_f(\tau)\cdot L_*>0,\,\,\,\,\,\,
t_*=\tau \sqrt{2|f|L_*/g}.
	$$

If at each locality $i$ data for the waves {\eqref{c6_1.1}} were obtained from the records exactly, the functions 
$$
S(a)=\sum_{i=1}^n\left(y_{*i}(t_{*i})-a\cdot \tan\theta_f(\tau_i)\right)^2,
	$$
and
$$
F(a)= \sum_{i=1}^n \left(60t_{*i}-\tau_i \sqrt{\frac{2a}{g}}\right)^2
	$$
would be equal to zero at $\,a=|f_*|$, where $|f_*|=|f|L_*$
(n=5 for Data 1).  

But using equations \eqref{c6_1.1} and values 
of $y_{*i},\,$ $t_{*i},\,$ $H_{*i}\,$ (obtained with errors) we estimate the value of $a$ by minimizing $S(a)$ or $F(a)$. 

Minimum value of $S(a)$ occurs at
\begin{equation}
a=|f_{*1}|=\frac{ \sum_i {y_{*i}\tan\theta_{1i}}  }
{\sum_i {\tan^2\theta_{1i}}}\,\,\hbox{km},       
\label{c6_5.1}
\end{equation}      

while minimum value of $F(a)$ is reached at
\begin{equation}
a=|f_{*2}|=\left(\frac{\sum_i {t_{*i}\tau_i}}{\sum_i {\tau^2_i}}\cdot 60\right)^2 4.9\,\,\hbox{m}.                                      
\label{c6_5.2}
\end{equation}
In the Table 4, the quantity $\Delta(\tau)=H(\tau)\sqrt{2|f|}$ is independent 
of $f$ ($cH(\tau)$ is the height of the WMH). 
If the data in Tables 1 - 3 were measured exactly at each locality $i$, the function 
$$
D(s)=\sum_i\left(c\Delta(\tau_i)\frac{1}{\sqrt{2|f|}}\,L_* -H_{*i}\right)^2=
\sum_i\left(\Delta(\tau_i)s-H_{*i}\right)^2,
\,\,\,\,\,\,
s=\frac{1}{\sqrt{2|f|}}\,L_*c=\frac{1}{\sqrt{2a}}L_*^{3/2}c
	$$
would be equal to zero at true value of $s$.

At actual values of $H_{*i}$ the value of $s$ is estimated by minimizing  
$D(s)$, which leads to the estimator 
\begin{equation}
s^2=\frac{L_*^3c^2}{2a}=\eta^2,\,\,\,\,\,\,\eta=\frac{B}{A},\,\,\,\,\,\,  A=\sum_i \Delta^2(\tau_i),\,\,\,\,\,\,
B=\sum_i \Delta_iH_{*i},                     
\label{c6_5.3}
\end{equation}
where $\eta$ is independent of $c$ and $f$, $a$ is estimated by
\eqref{c6_5.1}  or \eqref{c6_5.2}.

For Data 1, Tables 2 and 5, estimators \eqref{c6_5.1}, \eqref{c6_5.2}, and \eqref{c6_5.3}  give 
$$
A=0.5178,\,\,\,\,\,\,B=7.1823,\,\,\,\,\,\,\frac{B}{A}=13.9
	$$
\begin{equation}
a_1=23.413\,\,\,\hbox{km},\,\,\,\,\,\,a_2=23140\,\,\,\hbox{m},\,\,\,\,\,\,s=\eta=13.9\,\text{cm}.
\label{c6_5.4}
\end{equation}

Tables 2 and 6 (for Data 2) lead to the values 
$$
A=0.2261,\,\,\,\,\,\,B=8.4865,\,\,\,\,\,\,\frac{B}{A}=37.5
	$$
\begin{equation}
a_1=15.424\,\,\,\hbox{km},\,\,\,\,\,\,a_2=15785\,\,\,\hbox{m},\,\,\,\,\,\,s=\eta=37.5\,\text{cm}.
\label{c6_5.5}
\end{equation}
	
For Data 3, Tables 1 and 5, estimators \eqref{c6_5.1}, \eqref{c6_5.2}, and \eqref{c6_5.3}  give 
\begin{equation}
a=a_1=18300\,\,\hbox{m},\,\,\,\,\,\,
a=a_2=18278\,\,\hbox{m},\,\,\,\,\,\,s=\eta=12\,\,\hbox{cm} 
\label{c6_5.6}
\end{equation}

\section{ Estimation of the wave origin parameters}      

\subsection{Effective length of wave origin}  

Here  the water surface above the disturbed body of water is referred to as the wave origin. 
By model \eqref{c6_1.1}, at the moment $t=0$ the free surface has not yet been displaced 
from its mean level (the horizontal plane $x=0$), but the velocity field has already become different from zero. 
\vspace{3mm}

Calculations show that during some time interval, say $0\le\tau\le\tau_1$, 
a water hill in the form of a rounded solitary elevation symmetric with respect to
the vertical $x$-axis is appearing on the water surface (Fig. 2).
For the model \eqref{c6_1.1}, the height of the hill increases and reaches its maximum at $\tau=\tau_1=0.842$. 
On the interval $0\le\tau\le\tau_1$ there is only one zero 
$\theta(\tau)$ in the water surface (at $y>0$), and the zero is almost 
immovable. Only at $\tau>\tau_1$ the heap of water begins to spread out and to turn into a wave group, which runs away from the wave origin. The time interval $0\le\tau\le\tau_1$ is referred 
to as the interval of formation of the wave origin. 
\vspace{3mm}

The quantity $a=|f|L_*$ is the characteristic horizontal scale of the wave origin, so we determine the effective length, $L_{ef}$, of the origin by $L_{ef}=2ka$, the non-dimensional effective length $l_{ef}=2k|f|$. The body of water in the region $|y|<ka$ 
may be referred to as effective wave origin. 
The value of $k$ may be assigned to meet different conditions; for instance, the magnitude of 
the sea surface displacement on the boundary of the effective wave origin does not exceed a given value. 
\vspace{3mm}

It is found from \eqref{c6_1.1}, that the  free surface displacement in the  wave originating area at
 $\tau=\tau_1=0.842$  is given by 
$$
x=\frac{1}{\sqrt{2|f|}}\,H_1(\theta,\tau_1).
	$$
The quantity $b=\sqrt{2|f|}\,x$ is independent of $f$. The values of the quantity corresponding to different values of $k=\tan\theta$ are shown in Table 8.

\begin{center} 
TABLE 8. Values of $h$ corresponding to some values of $k$.

\begin{footnotesize}
\begin{tabular}{|p{10mm} |p{11mm} |p{11mm} |p{11mm} |p{11mm}| p{11mm}| p{11mm}| p{11mm}|} 
\hline
$k$    & 0.0     &2.0         & 2.1        & 2.2        & 2.3        & 2.4    & 2.5     \\ 
$b$    &1.036  & -0.1342  &-0.1361  &-0.1364  & -0.1355  & -0.1336  &
-0.1311   \\
\hline
\end{tabular}\\
\begin{tabular}{|p{10mm} |p{11mm} |p{11mm} |p{11mm} |p{11mm}| p{11mm}| p{11mm}| p{11mm}|} 
\hline 
$k$    & 2.6        & 2.7         & 2.8        & 2.9         & 3.0         & 3.5     & 4.0     \\ 
$b$    &-0.1280  & -0.1247  &-0.1211  &-0.1174  & -0.1136  & -0.1098  &-0.1060 \\
\hline
\end{tabular}\\
\begin{tabular}{|p{10mm} |p{11mm} |p{11mm} |p{11mm} |p{11mm}| p{11mm}| p{11mm}| p{11mm}|} 
\hline 
$k$    & 4.5        & 5.0         & 5.5        & 6.0         & 6.5         & 7.0     & 8.0     \\ 
$b$    &-0.0666  & -0.0563  &-0.0480  &-0.0413  & -0.0358  & -0.0314  &-0.0245 \\
\hline
\end{tabular}\\
\end{footnotesize}
\end{center}

For the model {\eqref{c6_1.1}}, $x<0$ for $k> 1.3$. 

By \eqref{c6_3.4}, the dimensional length of the wave of maximum height equals to $5a$,  where $5$ is the portion of $y$-axis which corresponds to the dimensional length of the wave.
Effective length of the wave origin equals 
\begin{equation}
L_{ef}=l_{ef}L_*=2k|f|L_*=2ka
\label{c6_6.1}
\end{equation}	
At $k=2.5$, the dimensional wave origin effective length is 
$$
L_{ef\,1}=5a=5\cdot 23.413 \,\text{km}=117\,\text{km}.
	$$
$$
L_{ef\,2}=5a=5\cdot 18.300 \,\text{km}=91\,\text{km}.
	$$
At any value of $k$, the ratio $L_{ef\,1}/L_{ef\,2}=1.28.$	
Here and below the subscripts 1 and 2 correspond to the Data 1 (the 2007 Kuril tsunami) and  Data 2 (the 2006 Kuril tsunami) respectively.

\subsection{ Water elevation in  the wave origin}  

From Table 8, we find $\max \Delta(\tau)=\Delta(0.842)=1.036\,$ 
(the maximum is reached at $\theta=0$). This means that 
the maximum of the water elevation, $h$, in the wave origin is given as
\begin{equation}
h=1,036cL_*\frac{1}{\sqrt{2|f|}}=1.036\eta \,\,\text{cm}.             
\label{c6_6.2}   
\end{equation}
where values of $\eta$ are given by \eqref{c6_5.4} and \eqref{c6_5.5} .
Formula \eqref{c6_6.2}  lead to the following estimates:  
\begin{equation}
 \eta_1=13.9\,\,\,\text{cm}, \,\,\,h_1=14.4\,\,\hbox{cm};
\,\,\,\,\,\,\eta_2=37.5\,\,\,\text{cm},\,\,\, h_2=38.8\,\,\hbox{cm},                          
\label{c6_6.3}  
\end{equation}
From Table 8 we obtain the non-dimensional level difference between maximum and minimum surface displacement in the wave origin 
at  $\tau=0.842$:
\begin{equation}
x_{max}(0.842)-x_{min}0.842)=\Delta(\tau_1)=1.036+0.1364=1.1724,
\label{c6_6.4}
\end{equation} 
Now for dimensional level differences in the two wave origins we write	
$$
X_{max}(0.842)-X_{min}(0.842)=1.1724\cdot 13.9\,\text{cm}=
16.3\,\text{cm}\,\,\,\,\text{for Data 1}
	$$
$$
X_{max}(0.842)-X_{min}(0.842)=1.1724\cdot 37.5\,\text{cm}=
43.9\text{cm}\,\,\,\,\text{for Data 2}
	$$	
At $k=2.5$ the free surface displacement, $d$, on the boundary of the origin equals
$$
d_1=b\cdot 13.9=0.1311\cdot 13.9\,\text{cm}  =1.8\,\text{cm}.
	$$	
$$
d_2=0.1311\cdot 37.5\,\text{cm}  =4.9\,\text{cm},
	$$	
$$
d_2=1.8\,\text{cm},\,\,\,\,\,\,L_{ef\,2}=201\,\text{km}\,\,\,
\text{at} \,\,\,k=5.5,\,\,\,117/201=0.58
	$$.

\subsection{ Duration of wave origin formation.}   

Duration $t$ of the wave origin formation is estimated 
by ($g=9.8\,\,\hbox{m/s}^2$, $a$ is measured in meters)
$$
t^*=\tau_1\sqrt{2|f|}T_*=0,842\sqrt{2a/g}\,\,\,\text{s},                                 
	$$
which gives  $t=58\,\hbox{s}$ for Data 1, and $t=47\,\hbox{s}$ for Data 2. 

\subsection{ Estimation of the waves energy}   

It is supposed above, initially still water is set in motion at $t=0$
by an impulsive force. Kinetic energy supplied to the water by the force can be estimated by integral (involving velocity potential 
and its normal derivative) evaluated over the sea surface \cite[]{miln}.
The energy supplied to the vertical layer of the water 
between two parallel planes at the distance $a=|f|L_*$ apart is given by   
\begin{equation}
E = \frac{1}{8}\pi |f| 
c^2\cdot\gamma gL_*^4\,\,\,\hbox{J}=\frac{1}{4}\,\pi\gamma g a^2\eta^2 \,\,\,\hbox{J}.             
\label{c6_6.5}       
\end{equation}    
With the values of $a$ and $\eta$ obtained in sections 4, we find the energy estimates as 
\begin{equation}
E_1= 8.03\cdot 10^7\,\,\text{J},    
\,\,\,\,\,\,\,E_2= 2.58\cdot 10^8 \,\,\text{J},\,\,\,\,\,\,E_2=3.2 E_1     
\label{c6_6.6}
\end{equation}                        
The subscripts 1 and 2 correspond to the Data 1 and Data 2 respectively. 

\section{ Theoretical estimation of waves parameters} 

\subsection{Estimation of length of the wave of maximum height}

By \eqref{c6_3.4} the WMH have the length  $l\approx5a$
$$
l_1\approx 117\,\,\,\,\hbox{km},\,\,\,\,\,\,
l_2\approx 76\,\,\,\,\hbox{km}
	$$ 
obtained from Data 1 and 2 respectively. 
The smaller value of $a$, the smaller effective size $l_{ef}=2ka$ of the wave origin and wave length, the higher dominant frequency of the WMH. 

\subsection{ Speed of the wave of maximum height.} 

For the 2007 Kuril tsunami, using data of Table 2, we find that actual values of average speed of the front of WMH (during time interval $0-t_*$) arrived at the buoys locations are 
$$
V_{*1}=\frac{1762\cdot 60}{120}=881\,\hbox{km/h},\,\,\,\,\,\,
V_{*2}=902\,\hbox{km/h},
	$$
$$
V_{*3}=878\,\hbox{km/h},\,\,\,\,\,\,V_{*4}=798\,\hbox{km/h},\,\,\,\,\,\,
V_{*5}=749\,\hbox{km/h}
	$$
with arithmetic mean $838\,\text{km/h}$.
We introduce the notation $a\text{(m)}=20$ which means that the number 20 is non-dimensional while $a$ is measured in meters: $a=20$ m.
The travel time $t_*$ is estimated as 
\begin{equation}
t^*=\frac{\tau}{60}\cdot\sqrt{\frac{2a\text{(m)}}{9.8}}\,\,\hbox{min}.                         
 \label{c6_7.1}                                 
\end{equation}                           
By \eqref{c6_3.2} and \eqref{c6_7.1}, speed of the front of WMH of the packet \eqref{c6_1.1} is
$$
V\text{km/min}=\frac{y_*}{t_*}=\lambda_* t_*=\lambda_* \tau\, \text{km/min}^2 \cdot\frac{1}{60}\,
\sqrt{\frac{2a\text{(m)}}{9.8}}\,\,\hbox{min}.
	$$
Substituting $l=5a \,\,\,\,\lambda_* \tau\approx 11.3$ (from Table 1) we obtain the relation between speed of the front of WMH and the length of the wave of maximum height as
\begin{equation}	
V^*=\frac{11.3}{60}\cdot \frac{1}{\sqrt{24.5}}\sqrt{l(m)}=
0.1902\sqrt{l(m)}\,\,\,\text{km/min}  
\label{c6_7.2}                               
\end{equation}  	
\begin{equation}
V^*=\frac{11.3}{60}\cdot \frac{50}{3}
\,\sqrt{\frac{l\text{(m)}}{24.5}}\,\,\,\text{m/s}=
0.634\sqrt{l\text{(m)}}\,\,\,\,\,\hbox{m/s}  
\label{c6_7.2a}
\end{equation}
\begin{equation}
V_*=6.34\sqrt{10}\, \sqrt{l\text{(km)}}\,\,\text{km/min}=20.03\sqrt{l(km)}\,\,\text{m/s}  
\label{c6_7.2b}
\end{equation}
At $l=117\,\text{km}$ formula \eqref{c6_7.2b} gives 
$$
V^*_1=20.03\,\sqrt{177}\,\,\,\hbox{m/s}=
271.82\,\,\,\hbox{m/s}=978\,\,\,\hbox{km/h},
	$$
and
$$
V^*_2=20.03\,\sqrt{76}\,\,\,\hbox{m/s}=
174.62\,\,\,\hbox{m/s}=629\,\,\,\hbox{km/h},
	$$
for Data 1 and Data 2 respectively.	
Instantaneous speed of the wave of maximum height  is estimated as 
$$
v_*=uT_*=u\sqrt{2a\,g}=u\sqrt{17.6\,a\text{(m)}}\,\,\text{m/s}.
	$$
where $a$ is measured in meters, $g=9.8\,\text{m/s}^2$.
Substituting $5a=l$ and taking $u=(0.50+0.76)/2$  (from Table 1) we obtain
\begin{equation}
v_*=1.19\,\sqrt{l\text{(m)}}\,\,\,\text{m/s}. 
\label{c6_7.3} 
\end{equation}         
The ratio of speed of the front of WMH to the instantaneous speed of the wave of maximum height is estimated as
$$
\frac{V_*}{v_*}=\frac{0.634}{1.19}=0.53
	$$
for any length of WMH.
\vspace{3mm}

Formula \eqref{c6_7.2a} leads to relationship between the speed of the front of WMH and length of the wave:
\begin{equation}
\frac{2\pi}{9.8l(m)}{V^*}^2=0.256, 
\label{c6_7.4}
\end{equation}
It is surprising that the non-dimensional ratio \eqref{c6_7.4} is obtained from bottom-less model \eqref{c6_1.1} (''deep water'' in terms of linear theory).
Echoing the linear theory, we define parameter $d(m)$ by equation 
\begin{equation}
\tanh\frac{2\pi d(m)}{l(m)}=0.256,\,\,\,\,\,\,
\frac{2\pi d(m)}{l(m)}=0.288
\label{c6_7.5}
\end{equation}
From  \eqref{c6_7.5} we find for Data 1
$$	
\frac{2\pi d}{177}=0,26,\,\,\,\,\,\,d=7.328  \,\,\,\hbox{km},
	$$
 and	
$$
\frac{2\pi d}{76}=0,26,\,\,\,\,\,\,d=3.146  \,\,\,\hbox{km}
	$$		
for Data 2.
Note that water depth in Northwest Pacific basin is about 6 -7 km 
while in the Northeast  basin is about 3 - 4 km (see location of the darts mentioned in Tables 1 and 2).	

\subsection{Evolution of the wave heights in the packet $H_1$}

The height of the wave of number $k$ in the packet \eqref{c6_1.1} is estimated as 
\begin{equation}
H^*=|x_{max}-x_{min}|\cdot c\frac{\Delta_k \cdot L_*}{\sqrt{2|f|}}=
c\Delta_k\sqrt{\frac{L_*^3}{2a}}=\Delta_k(\tau) \eta.       \label{c6_7.6}                      
\end{equation}   	
For the wave of MH the value of $\Delta_k(\tau)=\Delta(\tau)$ is shown in Table 4.

With time in the packet a central part is developing which contains waves of nearly equal heights. One can see the central part in the interval $250<y<400$ at $\tau=50$, and in the interval $500<y<700$ at $\tau=100$ (fig. 1). The length of the central part increases with time. In the central part, $\Delta_k(\tau)\approx \Delta(\tau)$.
It was found above from data 1 that $\eta_1=13.9$ cm and $\eta_2=37.5$ cm from data 2. 
  By \eqref{c6_6.4}, the maximum wave heights 
are estimated as $H_1^*=13.9 \Delta(\tau_*)$ cm for data of 2007 and 
$H_2^*=37.5 \Delta(\tau_{**})$ cm for the data of 2006.
 If $\tau_*=\tau_{**}$, the ratio of the maximum wave heights $H_2^*/H_1^*=2.71$, so the ratio of wave heights in the central parts of the tsunami waves approximately equals 2.71. 

The ratio is equal 2.71, if the wave heights are taken at the same value of $\tau$. Under this condition, from  \eqref{c6_7.1},\eqref{c6_3.1} and \eqref{c6_3.4}, we get 
\begin{equation}
\frac{t_{*2}}{t_{*1}}=\sqrt{\frac{a_2}{a_1}}=\sqrt{\frac{l_2}{l_1}}
=\sqrt{\frac{y_{*2}}{y_{*1}}},\,\,\,\,\, \frac{y_{*2}}{l_2}=\frac{y_{*1}}{l_1} .
\label{c6_7.7}                             
\end{equation}                                        	
where subscripts 1 and 2 correspond to Data 1 (the 2007 Kuril tsunami) and Data 2 (the 2006 Kuril tsunami) respectively. 

When the conditions \eqref{c6_7.7} are  satisfied approximately,  the ratio of the maximum wave heights 
${H_2^*} \ {H_1^*}$ of the Kuril tsunamis may be greater or less than 2.71.

It seems in \cite[]{rabin} the calculations were performed  at proper values of time and at proper numbers of waves (of crests). 

	\subsection{Comparison the theoretical estimates with results of numerical study}    

 Below the term `tsunami source' means an initially disturbed body of  the Earth, `wave source' means a disturbed body of water.
The water surface above the wave source is referred to as the `wave
origin'. 

In the available literature, as in \cite[]{rabin}, a  wave source model include the assumption that a body of sea water is pushed up by a piston which represents  the instantaneous rise of the sea bottom.  
According to the ``piston'' assumption, initially the sea water is still, but the sea surface is displaced from its mean level so that  the initial water elevation in the wave origin is approximately the same as the unknown submarine crustal displacement. 

In this situation, all what can be done is to constrain the wave origin and to decide the shape of initial water elevation in the origin. 

Usually a set of assumed distributions of the initial water elevation in the wave origin are  examined in order to produce tsunami waveforms  similar to the actual tsunami records,  at least, at some locations.

 According to the packet theory, and in contrast with most numerical studies, it is supposed that  
 the ``quake'' is instantaneous event, after which the wave origin begins to form on the free water surface, and the formation takes some time. All wave estimates depend on horizontal dimensions of the wave origin.
The horizontal dimensions of the wave origin obtained here are adequate to the wave source constructed in \cite[]{rabin} where a numerical study of the two tsunamis triggered by the earthquakes of the 2006 and 2007 near the Kuril Islands is presented.  
  
 As to comparison of quantitative parameters, one can read in  \cite[]{rabin}: 
 \begin{enumerate}
\item	 
``The wave energy of the 2007 Kuril tsunami is reduced
compared to that of 2006 Kuril tsunamis'' (p.115). 
\item
``at remote sites ...the ratio of 2006/2007 far-field wave heights is typically around 3:1'' (p.115). 
\item
The  maximums  of sea surface  amplitudes (from mean sea level to peak)  in the wave origins equal  $\approx 1.9$ m (for 2006 tsunami) and $\approx 2.6$ m (for 2007 tsunami); the ratio of the displacements is $1.9/2.6=0.73$ (p.112).   
\item
``..the 2007 tsunami had higher dominant frequency'' (p.115). 
\end{enumerate}

Four following theoretical estimates correspond to the above assertions. The estimates are obtained from DART records, not from assumed water elevation:
\begin{enumerate}
\item
Theoretical estimates  of the  energy supplied to the water in both cases relate as $E_2=3.2 E_1$, where the subscripts 1 and 2 correspond to the Data 1 and Data 2, (subsection 6.4). 
\item
It follows from the model \eqref{c6_1.1}, that at remote sites, the ratio  of the maximum wave heights ${H_2^*}/{H_1^*}$ may be about 2.71 (subsection 7.3).
\item
the  ratio of the theoretical sea surface displacements  in the wave origin equals 
$X_{max 2}/X_{max 1}=38.8/14.3=2.71$   (subsection 6.2)
\item
Wave lengths obtained from Data 1 and Data 2 equal 117 km and 76 km respectively. Higher frequency corresponds to the wave of shorter length.
\end{enumerate}

\section{Theoretical forecast based on the wave records} 

In this section, the model \eqref{c6_1.1} is tested against instrumental data obtained from DART records. The test is independent of numerical modeling.

In Table 1, the DARTs are arranged from top to bottom according the tsunami arrival time. Consider the situation when the first two lines in Table 1 are known, but the records of the next three buoys are not yet obtained. 

Bellow, starting from the first two lines of Table 1, a line of forecasts of the WMH arrival time and amplitude at the next three buoys is produced corresponding to the timeline of the DART records.

For each of the next three buoys $\tan\theta_f=y_*/a$ is calculated, which is then used to  locate the values of $\tau$ and $\Delta$ between two appropriate consecutive values from Table 4. 

Then, for each buoy, the travel time of the WMH and its height at the locations of the buoys are estimated by  \eqref{c6_7.1}  and \eqref{c6_7.4}. 

\subsection{ The forecast based on Data 1 (2007 Kuril tsunami)}   

\subsubsection{The forecast based on the first two DART records.}

The first two lines of Table 1,
the first two lines of Table 5, 
and formulas \eqref{c6_5.1} and \eqref{c6_5.3} 
give 
$$
a=|f_{*1}|=29.783 \,\,\hbox{km},\,\,A=0,28425,\,\,B=3,066208\,\,\text{cm}
\,\,\eta=12.877\hbox{cm}
	$$
 
For the buoy 3, we obtain 
$$
\tan\theta_f=\frac{y_*}{a}=\frac{2253}{29.783}=75.648.
	$$
From Table 4 we see that $70.734<75.648<76.421$.
This gives intervals for $\tau$ and $\Delta$:
$$
110<\tau<120,\,\,\,\,0.3371<\Delta<0.3607
	$$
Linear interpolation gives the estimates $\tau=116,3921,\,\,\,\,\Delta=0,342007$.
\vspace{3mm}

The travel time of the WMH and its height at the location 
of the buoy 3 are estimated by \eqref{c6_7.1} and \eqref{c6_7.4}
($a$ is taken in meters) as 
$$
t^*=151\,\,\text{min},\,\,\,\,\,\,H^*=4.4\,\,\text{cm}
	$$
{\t For buoys 4 and 5} we get
$$
\tan\theta_f=\frac{y_*}{a}=\frac{2660}{29.783}=89.313,\,\,\,\,\,\,
\tan\theta_f=\frac{y_*}{a}=\frac{5470}{29.783}=183.663
	$$
The forecast for the buoys 4 and 5 is obtain on the same lines as for buoy 3.	

\begin{center}
TABLE 9. For Data 1: the forecast based on two DART records.

\begin{footnotesize}
\begin{tabular}{|p{10mm} |p{10mm} |p{10mm} |p{10mm} | p{10mm} |p{10mm} |}  
\hline $\hbox{Buoy}$ & ${\tan}\theta_f $ & $\tau $   & $\Delta$  
&$t^*\,\,\hbox{min}$ & $H^*\,\,\hbox{cm}$ \\ 
\hline  
         3     & 75.785  & 116.570  &  0.341764 & 151  & 4.4  \\           
         4     & 89.313  & 140.987  &  0.314194 & 183  & 3.4  \\
         5     & 183.663 & 285.338  &  0.223815 & 370  & 2.9  \\
\hline
\end{tabular}\\
\end{footnotesize}
\end{center}

\subsubsection{The forecast based on three DART records.} The following forecast is based on the measurements on DART buoys 21413, 21414, and 46413. 

The first three lines of Table 1, the first three lines of Table 5,  
and formulas   \eqref{c6_5.1} and \eqref{c6_5.3}  give 
$$
a=|f_{*1}|=29.197 \,\,\,\hbox{km},\,\,\,\,\,\,\eta=14.04\,\hbox{cm}.
	$$
\begin{center}
TABLE 10. For Data 1: the forecast based on three DART records.

\begin{footnotesize}
\begin{tabular}{|p{10mm} |p{10mm} |p{10mm} |p{10mm} | p{10mm} |p{10mm} |}  
\hline $\hbox{Buoy}$ & ${\tan}\theta_f $ & $\tau $   & $\Delta$  
&$t^*\,\,\hbox{min}$ & $H^*\,\,\hbox{cm}$ \\ 
\hline  
          4     & 91.116  & 143.305  &  0,311829 & 184 & 4.4  \\
          5     & 183.305 & 291.131  &  0.223815 & 374 & 3.1  \\

\hline
\hline
\end{tabular}\\
\end{footnotesize}
\end{center}
\bigskip

\subsubsection{The forecast based on four DART records.} The following forecast is based on the measurements on DART buoys 21413, 21414, 46413, and 46408. 

The first four lines of Table 1, the first four lines of Table 5,  
and formulas   \eqref{c6_5.1} and \eqref{c6_5.3}  give  
$$
a=|f_{*1}|=26,952 \,\,\,\hbox{km},\,\,\,\,\,\,\eta=10.926\,\hbox{cm}.
	$$
\begin{center}
TABLE 11. For Data 1: the forecast based on four DART records.
\begin{footnotesize}
\begin{tabular}{|p{10mm} |p{10mm} |p{10mm} |p{10mm} | p{10mm} |p{10mm} |}  
\hline $\hbox{Buoy}$ & ${\tan}\theta_f $ & $\tau $   & $\Delta$  
&$t^*\,\,\hbox{min}$ & $H^*\,\,\hbox{cm}$ \\ 
\hline  
          5     & 202.949 & 316.371  &  0,215841 & 390 & 2.1  \\
\hline
\end{tabular}\\
\end{footnotesize}
\end{center}

For Data 1, the results are summarized in Tables 12  where 
for each of the mentioned DART buoy's locations the forecast of travel
time $t_k^*$ (in minutes) of the wave of maximum height and its height $H_k^*$ (in centimeters) are shown; the subscript $k$ shows that the forecast is based on measurements obtained from $k$ buoys. 

The actual travel time $t_*$ and height $H_*$ are repeated from Table 1. 
\begin{center}
TABLE 12.  For Data 1: the forecasts based on the model \eqref{c6_1.1}.
\begin{footnotesize} 
\begin{tabular}{|p{5mm} |p{6mm} |p{6mm}|p{6mm} |p{6mm} | p{6mm}| p{6mm} | p{6mm} | p{6mm} | | } 
\hline $\hbox{Buoy}$  & $t_2^* $ & $t_3^* $ & $t_4^*$ & $t_*$ &$H_2^*$&$H_3^*$ & $H_4^*$ & $H_*$   \\
\hline    
         3   & 151 &  -  &   -  & 154 & 4.4  & -  &  -  &  5.7 \\
         4   & 183 & 184 &  -   & 200 & 3.4  &4.4 &  -  &  4,5 \\      
         5   & 343 & 374 & 390  & 438 & 2.9  &3.1 & 2.4 &  1,6 \\
\hline
\end{tabular}\\
\end{footnotesize}
\end{center}
In figure 3 vertical lines mark the arrival of the front 
of the WMH (thick solid line), its estimate with the first and second buoy records for the next three buoys (thin solid), its estimate with the first three buoy records for the next two buoys (dashed), its estimate with the four buoys for the last one (dash-dot). Cross marks a trigger pulse (signals send by an operator).

\subsection{  The forecast based on Data 2 (2006 Kuril tsunami)}   

The Data 2 obtained from the records are shown in Table 2. The results of forecasting are presented in Table 13.
\begin{center}
TABLE 13. For Data 2: the forecasts based on the model \eqref{c6_1.1}.
\begin{footnotesize}
\begin{tabular}{|p{10mm} |p{8mm} |p{8mm} |p{8mm} | p{8mm}| p{8mm} | p{8mm} | p{8mm} |
p{8mm} | p{8mm} | p{8mm} | } 
\hline $\hbox{Buoy}$ & $t_1^* $ & $t_2^* $   & $t_3^*$ & $t_*$ &$H_1^*$ & $H_2^*$ & $H_3^*$ & $H_*$  \\
\hline 
         2     & 187 & -   &  -   & 238 & 7.8 & -   & -   & 7.5 \\ 
         3     & 214 & 257 &  -   & 270 & 7.3 & 6.6 & -   & 10.0 \\
         4     & 243 & 295 &  307 & 365 & 6.9 & 6.2 & 6.8 & 7.5 \\
\hline
\end{tabular}\\
\end{footnotesize}
\end{center}
\bigskip

\subsection{ The forecast based on Data 3 (2007 Peruvian tsunami)}  

\begin{center}
TABLE 14. For Data 3: the forecasts based on the model \eqref{c6_1.1}.
\begin{footnotesize}
\begin{tabular}{|p{10mm} |p{8mm} |p{8mm} |p{8mm} | p{8mm}| p{8mm} | p{8mm} | p{8mm} | p{8mm} | } 
\hline $\hbox{Buoy}$ & $t_1^* $ & $t_2^* $   & $t_3^*$ & $t_*$ &$H_1^*$ & $H_2^*$ & $H_3^*$ & $H_*$  \\
\hline 
         32411     & 192 & -   &  -   & 198 & 3.9 & -   & -   & 1.7 \\
         51406     & 396 & 411 &  -   & 444 & 2.8 & 2.3 & -   & 3.9 \\
         46412     & 518 & 540 &  576 & 624 & 2.4 & 2.0 & 2.1 & 2.0 \\
\hline
\end{tabular}\\
\end{footnotesize}
\end{center}
\bigskip

The forecast results show earlier arrivals of the WMH ($t^*<t_*$). 
For Data 1 the percentage errors equal $(1-192/198)\cdot 100\%=3.0\%$ 
for buoy 32411, $(1-411/444)\cdot 100\%=7.4\%$ for buoy 51406, and 
$(1-576/624)\cdot 100\%=7.8\%$ for buoy 46412. 

The estimates $H_i^*$ of the maximum wave heights at the buoys exhibit no regularity: at the buoys, some of the estimates are close to the actual values  values $H_*$, and some  of the deviations  $|H_i^*-H_*|$ is from $0$ to $H_*$. Though the packet \eqref{c6_1.1} is a poor model for  application to actual long waves in an open ocean,
  the theoretical estimates $t_k^*,\,H_k^*$ are adequate to corresponding instrumental values $t_*,\,H_*$.

\chapter{Nonlinear Gravity Waves Generated By Variable Pressure Acting On The Free Surface}

{\small{In this chapter, an initial-value problem is solved analytically in a case
when the water initially at rest is forced to move by external pressure 
 force of limited power distributed over a large area in the free surface, but is otherwise arbitrary.  The solution is exists
 at all times and adequately describes the forced waves up to the breaking point.}}
   
\section { Problem outline  and basic equations } 	

\begin{figure}
 \centering
	\resizebox{0.8\textwidth}{!}
		{\includegraphics{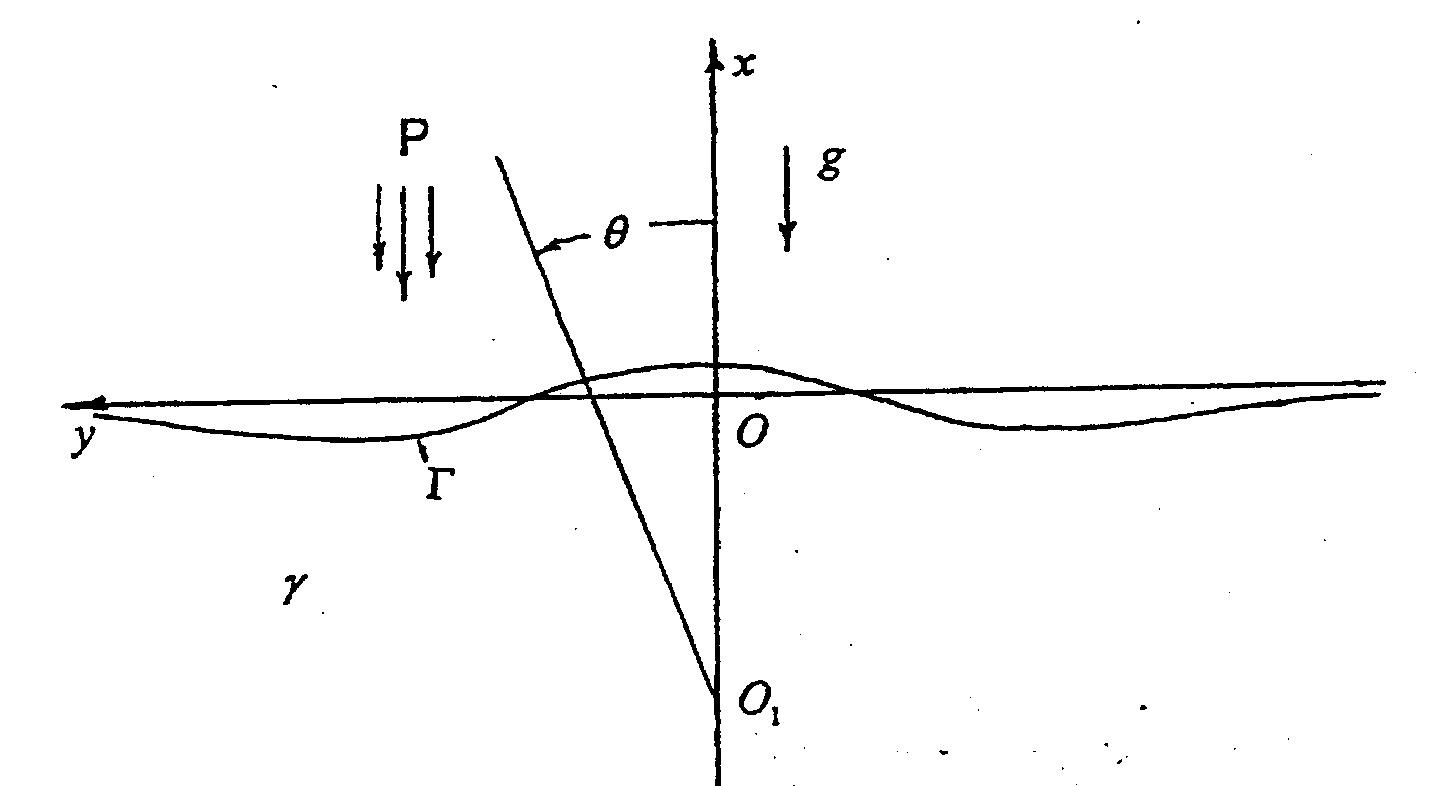}} 
	\caption{
Flow diagram
	}
	\label{c7_fg1}
\end{figure}

Motion of ideal heavy liquid of constant density $\gamma$
is considered below assuming that the liquid moves 
two-dimensionally parallel to a vertical plane, say $x,\,y$-plane, with 
$x$-axis oriented upward and $y$-axis in horizontal direction (Figure \ref{c7_fg1}).
The liquid is bounded from above by a free surface $S$ 
infinite in horizontal directions and fills the half-space below the surface. 
It is convenient to think of the liquid as being contained between two vertical planes parallel to that of $(x,y)$.
\vspace{3mm}

Let the curve $\Gamma$ in Fig. \ref{c7_fg1} be the trace of the free
surface $S$ in the $(x,y)$ plane,
$\,x=f\,<\,0,\,$ $y=0$ be the coordinates of the pole $\,O_1\,$ of the polar coordinate system in the $\,(x,y)\,$ plane,
$\,\theta\,$ be the polar angle measured from the positive $x$-axis
in the counterclockwise direction, $\,t\,$ be the time.
When the external pressure, 
$\,P=P_0\,$,  is constant and the water is still, the free surface is a horizontal plane $\,\,x=0\,\,$. Then the liquid is set in motion by a variable ``gauge'' pressure, $\,P_*\,$.   
\vspace{3mm}

Below, all equations are written in  dimensionless variables.
Since the problem has no characteristic linear size, the dimensional
unit of length, $\,\,L,\,\,$ may be chosen arbitrarily, the dimensional
unit of time, $\,\,T,\,\,$ is defined
by the relation $\,\,T^2g=L\,$, where $\,\,g\,\,$ is the acceleration
of free fall. The dimensionless acceleration of free fall is equal to unity.
Let $\,\,\gamma_0\,\,$ be some characteristic density and
$\,\,P_0\,\ne\,0\,\,$ be some characteristic pressure (for example, the
constant ``atmospheric'' pressure on the free surface in static equilibrium).
All parameters, variables and equations are made dimensionless by the
quantities $\,L,\,T,\,\gamma_2,\,P_0 $ (if $\gamma_2$ is the density of the liquid, then $\gamma=1$; non-dimensional external pressure acting on the free surface is $P=1+P^*$). 
\vspace{3mm}

Following Chapter 3,
equations of the free surface are sought in the parametric form
$$
x=W(\theta,t),\ \  y=(W-f)\,\tan\,\theta,\ \ \ 
-\pi/2 < \theta < \pi/2,                  
       $$
       
In the $\,\,(x,y)\,\,$ plane, curvilinear coordinates $(\sigma, \theta)$
are defined by the relations
$$
  x=\sigma+W(\theta,t),\, y=(\sigma+W-f)\,\tan\,\theta,\,    
-\pi/2 < \theta < \pi/2                       
	$$ 	
so the equation of the interface takes the form $\sigma =0$ (the liquid of density $\gamma_2$ occupies the half-space $\sigma <0$). 

Equations  for the free surface evolution and  for velocity potential $\Phi$ are rewritten from Chapter 3:
\begin{equation} 
\pd{W}{t}+2\frac{\sin\theta\cos\theta}{W-f}\,\pd{W}{\theta}+\hat D_2\,\pd{\hat\Phi}{\theta_-}+
\hat D_3\,\pd{\hat\Phi}{\sigma_-}=0,
\label{c7_1.1}                                        
 \end{equation}     
where
$$
\hat D_1=-\frac{\cos^2\theta}{W-f},\,\,\,\,\,\,
\hat D_2=-\frac{\sin\theta\cos\theta}{W-f} 
-\pd{W}{\theta}\,\frac{\cos^2\theta_1}{(W-f)^2}
	$$
$$
 \hat D_3=1+
2\frac{\sin\theta\cos\theta}{W-f}\pd{W}{\theta} +
\frac{\cos^2\theta}{(W-f)^2}
\,\left(\pd{W}{\theta} \right)^2,
	$$
         \begin{equation} 
\Phi(\sigma,\theta,t)=-\frac{D_1}{|D_1|}
\frac{1}{2\pi}\int\limits_{-\pi/2}^{\pi/2}
g(\theta_1,t)A(\sigma,\sigma_1,\theta,\theta_1,t)
\left.\frac{d\theta_1}{S}\right|_{\sigma_1=0},  
\label{c7_1.2}           
\end{equation}
$$
A=(\sigma -\sigma_1+W-W_1)(f-W_1)+
        $$
$$
(\sigma-f+W)\,\pd {W_1}{\theta_1}\,\,
(\tan\theta\cdot\cos^2\theta_1-
\sin\theta_1\cdot\cos\theta_1)
        $$     
$$
S=(\sigma-\sigma_1+W-W_1)^2\cos^2\theta_1+
        $$
$$
[(\sigma-f+W)\tan\theta\cdot\cos\theta_1-
(\sigma_1-f+W_1)\sin\theta_1]^2, 
        $$
$$
W=W(\theta,t),\,\,\,\,\,\,W_1=W(\theta_1,t),\,\,\,\,\,\,
\frac{D_1}{|D_1|}=-1
        $$
 The subscript $\sigma_1=0$ in \eqref{c7_1.2} denotes that integrand is calculated at  $\sigma_1=0$.      
In the region $\sigma<0$, the pressure $P$ is related to velocity potential $\Phi$ by Bernoulli's equation. 
As $\sigma\rightarrow -0$, the Bernoulli's equation gives 
$$
P_{-}(\theta,t)-1=-\left[\pd {\Phi}{t_{-}}-
\frac{1}{2}D_2\left(\pd {\Phi}{\sigma_{-}}\right)^2+
\frac{1}{2}\left(D\pd {\Phi}{\theta_{-}}\right)^2+W(\theta,t)\right],                                                            
      $$
so the nonlinear condition of the pressure continuity across the free surface may be written as 
\begin{equation}
-\left[\pd {\Phi}{t_{-}}-
\frac{1}{2}D_2\left(\pd {\Phi}{\sigma_{-}}\right)^2+
\frac{1}{2}\left(D\pd {\Phi}{\theta_{-}}\right)^2+W(\theta,t)\right]=\,P^*(\theta,t;f),           
\label{c7_1.3}                        
\end{equation}
where $\,\,P^*(\theta,t;f)\,\,$ is supposed to be a given function. 
At infinity along the free surface the following boundary conditions are imposed:
\begin{equation}
|W(\theta ,t)|<C(t)\cos^2\theta, \,\,\,\lim \limits_{\cos \theta \to 0}
\pd {W}{\theta}=0,\,\,\,\,\,|g(\theta, t)|<C(t),             
\label{c7_1.4}                               
\end{equation}
where $C(t)$ is a positive quantity independent of $\,\,\theta\,\,$.
It follows from \eqref{c7_1.4} that the liquid is at rest at infinity. 
Initial conditions are taken in the form
\begin{equation}
t=0:\,\, \, \, \,g(\theta,0)=0,\,\,\,\,\,W(\theta,0)=0.
\label{c7_1.5}                            
\end{equation}
The conditions mean that the water initially at rest is forced to move by external pressure $P_*$ ( $P_*\equiv 0$ at $t<0$)

\subsection{ Theoretical model for the external pressure} 
It is supposed that the gauge pressure, $\,\,P^*\,\,$,
on the free surface (between two vertical planes of flow at unit distance apart) is due to ``compressor'' of finite power. This condition of finite power is formulated mathematically as 
$$
\left|\int \limits_\Gamma P^*dl\right|=\left|\int \limits_\Gamma P^*
\sqrt{D_2}\,\frac{W(\theta,t)-f}{\cos^2\theta}\,d\theta\right|\le b<+\infty,
          $$
where $\,ds=1\,\cdot dl\,$ is area element of the free surface (1 is the distance 
between two vertical planes which confine the liquid), $\,P^* ds\,$ is 
the pressure force acting on the surface element, $b$ is a constant independent of $f$.
From this condition, it follows that the pressure $\,\,P^*\,\,$ must vary
inversely as $\,f\,$, and directly as the product of 
$\,\,\cos^2\theta\,$ and a function which 
remains bounded as $\,\,|f|\to+\infty\,$:
\begin{equation}
P^*(\theta,t;f)=\frac{a\gamma}{f}\cos^2\theta\,F(\theta,t;f),  
\label{c7_1.6}                    
\end{equation}
The factor $\,\,|a\,\gamma/f|\,\,$ may
be thought of as an amplitude of the pressure $\,\,P^*\,$.
\vspace{3mm}

It is convenient to take that the function $F$ is given in the form 
of Fourier series: 
\begin{equation}
F(\theta,t)=\sum_{k=1}^{+\infty}\alpha_k(t)\cos(2k\theta)+
\beta_k(t)\sin(2k\theta),\,\,\,\,\,\,
\sum_{k=1}^{+\infty}\sqrt{{\alpha_k}^2+{\beta_k}^2}<+\infty,                                                       
\label{c7_1.7}                          
\end{equation}
where functions $\,\,\alpha_k(t),\,\,\beta_k(t)\,\,$ are not specified
but are required to be functions that are of interest in applications; 
for example, the functions are almost-periodical of the form 
\begin{equation}
\alpha_k(t)=\sum_jA_{kj}\cos(\Omega_{kj}t)+B_{kj}\sin(\Omega_{kj}t).
\label{c7_1.8}                              
\end{equation}                      
Increasing the value of $\,\,|f|\,\,$ does lessen the pressure
$\,\,P^*\,\,$ on the free surface by spreading the finite
pressure's power  over a region of larger horizontal size. If,
for instance, $F(\theta,t)=\alpha_k(t)\cos(2k\theta)$ and the part
of the free surface is considered where $|\tan\theta|>2$, then the
external pressure amplitude in that part of the surface is less
than $1/5$ of the pressure amplitude in the entire free surface,
and, as a consequence of the relation between the amplitude and
the rate of work done by the pressure force, the power supplied to the
liquid across that part of the free surface is about $5\%$ of the power
supplied across the entire free surface. That is why (for the
example) we can adopt that the interval $|y|<2|f|$ in the free
surface is a zone of variable pressure, and the size $2|f|$ of the
interval may be considered as the characteristic horizontal scale
of the problem.   
\vspace{3mm}

The magnitude of the free surface displacement varies jointly as
 the quantity $ a/f $, so the ratio of the magnitude to the typical
horizontal size of the disturbed region in the surface is
approximately proportional to the
quantity $\,\,\varepsilon=a/f^2$.

\section{Waves emitted from  an extended zone of variable pressure}          
                                                
Extended zone of variable pressure
corresponds to small values of $\,\,|\varepsilon|$.

\subsection{ Leading-order equations}      
 
For $\,\,|\varepsilon|<<1,\,\,$ the solution of the initial-boundary
value problem \linebreak 
\eqref{c7_1.1} - \eqref{c7_1.5} is sought in the form of series in powers of 
$\,\,\varepsilon $:
$$
W(\theta,t)=\frac{a}{f} [W_0(\theta,t)+\varepsilon W_1(\theta,t)+\varepsilon^2 W_2(\theta,t)+...]
	$$ 
$$
g(\theta,t)=\frac{a}{f}\sum\limits_{k=0^{+\infty}}
\varepsilon^kg_k(\theta,t)
	$$
$$
\Phi(\theta,\,\sigma,t)=\frac{a}{f}\sum\limits_{k=0^{+\infty}}
\varepsilon^k\Phi_k(\theta,\,\sigma,t)
	$$ 

Using the technique of Chapter 4 
we obtain the leading-order equations in unknown functions  $W_0(\theta,t),\,\,g_0(\theta,t)$: 
\begin{equation}
\pd{W_0}{t}=\pd{\Phi_0}{\sigma_-}+
\frac{\sin\theta\cdot\cos\theta}{f}\,\pd{\Phi_0}{\theta_-}                                                         
\label{c7_2.1}                             
\end{equation}
\begin{equation}
\Phi_0=\left.\hat H(g_0)\right|_{\sigma_1=0}
\label{c7_2.2}                         
\end{equation}
\begin{equation}
-\left(\pd{\Phi_0}{t}_-+\frac{\sin\theta\cos\theta}{f}\,
\pd{\Phi_0}{\theta}_- \right)=\cos^2\theta\,F(\theta,t).
\label{c7_2.3}                        
\end{equation}
$$
F(\theta,t)=\sum_{k=1}^{+\infty}\alpha_k(t)\cos(2k\theta)+
\beta_k(t)\sin(2k\theta).
	$$   
Linear integral operator $\hat H(u)$ is defined by formula \eqref{c2_2.9} of Chapter 2.  Eigenfunctions and eigenvalues of the operator are also obtained in that chapter.      

\subsection{Numerical system of ordinary differential equations for time-dependent coefficients of double series solution}
Solution to the equations  \eqref{c7_2.1} - \eqref{c7_2.3} 
 satisfying  boundary conditions \eqref{c7_1.4} and initial conditions 
\eqref{c7_1.5} is sought in the form of trigonometric series: 
\begin{equation}
W_0(\theta,t)=\cos^2\theta\sum_{k=0}^{+\infty}
[a_k(t)\cos (2k\theta)+b_k(t)\sin (2k\theta)],   
	$$                                                
$$
g_0(\theta,t)=\sum_{k=0}^{+\infty}
[c_k(t)\cos (2k\theta)+e_k(t)\sin (2k\theta)]. 
\label{c7_2.4}                        
\end{equation}       
Technique for solving the leading-order equations is similar to that employed in Chapter 4. 
Inserting series \eqref{c7_2.4} in the equations  
\eqref{c7_2.1} - \eqref{c7_2.3}, and  making use of  eigenfunctions and eigenvalues  of operator $\hat H(u)$,  
we obtain the following initial value problem for coefficients of the series \eqref{c7_2.4}: 
\begin{equation}
a_k^{'}=-\frac{k}{|f|}c_k,  \,\,\,\,\,\,    
c_k^{'}=\frac{1}{2}a_{k-1}+a_k+\frac{1}{2}a_{k+1}+
\frac{1}{2}\alpha_{k-1}+\alpha_k+\frac{1}{2}\alpha_{k+1} ,
\label{c7_2.5}                 
\end{equation}       
$$
a_k(0)=c_k(0)=0\, \, \, \left(k=0,1,2...\right),\, \, \,
\alpha_{-1}=\alpha_0=a_{-1}=a_0(t)=0.
        $$
If in \eqref{c7_2.5} we replace $\,\,a_k,\,\,c_k,\,\,\alpha_k\,\,$ by 
$\,\, b_k,\,\,e_k,\,\,\beta_k $ respectively, we obtain initial value problem  
for $\,\, b_k(t),\,\,e_k(t) $.
Now the problem \eqref{c7_2.5}  is reduced to 
$$
\frac{d^2}{dt^2}a_k+\frac{k}{2|f|}(a_{k-1}+2a_k+a_{k+1})=
-\frac{k}{2|f|}(\alpha_{k-1}+2\alpha_k+\alpha_{k+1}),                               
	$$	
$$
k\ge 1,\,\,\,\,\,\,a_k(0)=a_k'(0)=0.
	$$
$$    
\tau=\sigma_*t,\,\,\,\,\,\,\sigma_*^2=\frac{1}{2|f_0|},\,\,\,\,\,\,\frac{d^2}{dt^2}=\sigma_*^2\,\frac{d^2}{d\tau^2}
	$$				
\begin{equation}
\frac{d^2}{d\tau^2}a_k+k(a_{k-1}+2a_k+a_{k+1})=
k(\alpha_{k-1}+2\alpha_k+\alpha_{k+1})\,\,\,\,\,\,(n\ge 1)
\label{c7_2.6}         
\end{equation}  	
Having solved the problem \eqref{c7_2.6},  we can find $\,\,c_k(t) $ from  \eqref{c7_2.5}.
     
\subsection{Exact solution to the numerical system of ordinary differential equations}
The leading-order ``response''  $W_0(\theta,t)$ of the free surface to
the gauge pressure depends  on the factor $F(\theta,t;f)$ linearly.
For the linearity, we first confine our attention to a particular
case of the gauge pressure when for a certain $\,m\ge 1 $
\begin{equation}
    F(\theta,t)=\alpha_m(t)\cos(2m\theta),\,\,\,\,\,\,m\ge 1
\label{c7_2.7}                           
\end{equation} 
$$
\alpha_m(t)=A_m\cos(\Omega\,t-\phi_m), \,\,\, \,\,\,
A_m>0 ,\,\,\,\,\,\,\alpha_k(t)\equiv0\,\,\,\text{for}\,\,\,k\ne m                                             
    $$

It can be verified by direct substitution that exact solution 
to the problem \eqref{c7_2.6}, \eqref{c7_2.7}   is given by 
\begin{equation}
a_0(t)\equiv 0, \,\,\,\,\,\, a_k(t)=a_{km}(\tau),\,\,\,\,\,\,k\ge 1,
\,\,\,\,\,\,m\ge 1
\label{c7_2.8}                             
\end{equation} 
$$
a_{km}(\tau)=(-1)^mA_m\left[C_{km}(\tau;\lambda)\cos(\lambda\tau-\phi_m)+
S_{km}(\tau;\lambda)\sin(\lambda\tau-\phi_m) \right],    
          $$
$$
t=\sigma_{*}\tau ,\,\,\,\, \, \lambda=\sigma_{*}\Omega,\,\,\,\,\,\,
\sigma_{*}=\sqrt{2|f|},
          $$
where
$$
C_{k0}=(-1)^{k+1}2\int\limits_0^{+\infty}x^3e^{-x^2}L_{k-1}^{(1)}(x^2)\Delta_1\,dx,\,\,\,\,\,\,k\ge 1,  
	$$     
$$
S_{k0}=(-1)^{k+1}2\int\limits_0^{+\infty}x^3e^{-x^2}L_{k-1}^{(1)}(x^2)\Delta_2\,dx,   
	$$
$$
m\ge 1\, \, \,C_{km}=(-1)^k\frac{2}{m}\int\limits_0^{+\infty}
x^5e^{-x^2}L_{k-1}^{(1)}(x^2)L_{m-1}^{(1)}(x^2)\Delta_1\,dx,
	$$   %
$$ 
S_{km}=(-1)^k\frac{2}{m}\int\limits_0^{+\infty}x^4e^{-x^2}L_{k-1}^{(1)}
(x^2)L_{m-1}^{(1)}(x^2)\Delta_2\,dx,
	$$
$$
\Delta_1=\frac{\cos(\lambda\tau)-\cos(x\tau)}{\lambda^2-x^2},\,\,\,\,
\Delta_2=\frac{x\sin(\lambda\tau)-\lambda\sin(x\tau)}{\lambda^2-x^2}.
        $$ 
The right hand side of the integrals involve Laguerre polynomials,
$\,\,L_k^{(1)}(u)\,\,$, defined by recursion relations
\begin{equation}
L_0^{(1)}(u)=1,\,\,\,L_1^{(1)}(u)=2-u,\,\,\,\,
\label{c7_2.9}           
\end{equation}           
$$
k\ge 1\,\,\,\,\,\,
(k+1)L_{k+1}^{(1)}(u)-(2k+2-u)L_k^{(1)}(u)+(k+1)L_{k-1}^{(1)}(u)=0.                                                        
        $$
By the relations 
$$
C_{km}(\tau;\lambda)=(-1)^k\left[H_{km}(\tau;\lambda)\cos(\lambda\tau)+
G_{km}(\tau;\lambda)\sin(\lambda\tau)\right],
        $$
$$
S_{km}(\tau;\lambda)=(-1)^k\left[H_{km}(\tau;\lambda)\sin(\lambda\tau)-
G_{km}(\tau;\lambda)\cos(\lambda\tau)\right]  
        $$
we define functions $H_{km}(\tau; \lambda)$ and $G_{km}(\tau; \lambda)$.
Decomposition of the fraction 
$$
\frac{2\lambda}{\lambda^2-x^2}=\frac{1}{\lambda+x}+\frac{1}{\lambda-x},
	$$
gives 
$$
H_{k0}(\tau;\lambda)=-2\int\limits_0^{+\infty}x^2e^{-x^2}L_{k-1}^{(1)}(x^2)f_1\,dx,  
        $$

$$
G_{k0}(\tau;\lambda)=-2\int\limits_0^{+\infty}x^2e^{-x^2}L_{k-1}^{(1)}(x^2)f_2\,dx,         
        $$
\begin{equation}
H_{km}(\tau;\lambda)=\frac{2}{m}\int\limits_0^{+\infty}x^4e^{-{x}^2}L_{k-1}^{(1)}(x^2)L_{m-1}^{(1)}(x^2)f_1dx,  
 \label{c7_2.10}                              
\end{equation} 
\begin{equation}
G_{km}(\tau;\lambda)=\frac{2}{m}\int\limits_0^{+\infty}x^4e^{-{x}^2}
L_{k-1}^{(1)}(x^2)L_{m-1}^{(1)}(x^2)f_2dx   
\label{c7_2.11}                               
\end{equation} 
$$
(k\ge 1,\,\,\, m\ge1),
        $$
\begin{equation}
f_1=-\frac{1}{\lambda+x}\sin^2\left(\frac{\lambda+x}{2}\tau\right)+\frac{1}{\lambda-x}\sin^2\left(\frac{\lambda-x}{2}\tau\right),                                                     
\label{c7_2.12}                           
\end{equation}
\begin{equation}
 f_2=\frac{1}{2}\frac{1}{\lambda+x}\sin\left((\lambda+x)\tau\right)-
\frac{1}{2}\frac{1}{\lambda-x} \sin\left((\lambda-x)\tau\right).                                                    
\label{c7_2.13}                              
\end{equation}
In terms of functions $H_{km}$ \eqref{c7_2.10} and  $G_{km}$ \eqref{c7_2.11} solution of the problem \eqref{c7_2.6}  (for the particular case of the gauge pressure \eqref{c7_2.7}) reads 
\begin{equation}
a_{k}(\tau)=a_{km}=(-1)^{k+m}A_m\left[H_{km}(\tau;\lambda)\cos(\lambda\tau-\phi_m)-\right.
	$$
$$	
\left.G_{km}(\tau;\lambda)\sin(\lambda\tau-\phi_m) \right].
\label{c7_2.14},\,\,\,\,\,\,k\ge 1         
\end{equation} 

\subsection{ Double series solution to the leading-order equations}
For the pressure law \eqref{c7_2.7},  
using \eqref{c7_2.14}, equation \eqref{c7_2.4} 
is rewritten in the form of functional double series 
$$
W_0(\theta,t)=\cos^2\theta\sum_{k=1}^{+\infty}\cos(2k\theta)
\left((-1)^k\sum_{m=1}^{+\infty}(-1)^ma_{km}(\tau)\right)=
	$$
$$
\cos^2\theta\sum_{k=0}^{+\infty}\cos(2k\theta)
(-1)^k\sum_{m=1}^{+\infty}(-1)^mA_m\left[H_{km}(\tau;\lambda)
\cos(\lambda\tau-\phi_{m})-\right.
	$$
$$	
\left.G_{km}(\tau;\lambda)\sin(\lambda\tau-\phi_{m}) \right]
	$$
$$
t=\sigma_{*}\tau ,\,\,\,\, \, \lambda=\sigma_{*}\Omega,
\,\,\,\,\,\,\sigma_{*}=\sqrt{2|f|},
          $$	
After interchanging the order of summation in the double sum, we obtain
$$
W_0=\sum_{m=1}^{+\infty}(-1)^mA_m\cos(\lambda\tau-\phi_{m})
\left(\cos^2\theta\sum_{k=1}^{+\infty}
\cos(2k\theta)H_{km}(\tau;\lambda)\right)  
	$$	
$$
-\sum_{m=1}^{+\infty}(-1)^mA_m\sin(\lambda\tau-\phi_{m})
\left(\cos^2\theta\sum_{k=1}^{+\infty}
\cos(2k\theta)G_{km}(\tau;\lambda)\right),  
	$$	
or  
\begin{equation} 
W_0(\theta,t)=\sum_{m=1}^{+\infty}(-1)^mA_m[R_m\cos(\tau\lambda-\phi_m)
+Q_m\sin(\tau\lambda-\phi_m)], 
\label{c7_2.15}                                
\end{equation}
where
\begin{equation} 
R_m=R_m(\tau,\theta,\lambda)=\cos^2\theta \sum_{k=1}^{+\infty}(-1)^k
H_{km}(\tau;\lambda)\cos(2k\theta),              
\label{c7_2.16}          
\end{equation}
\begin{equation} 
Q_m=Q_m(\tau,\theta,\lambda)=\cos^2\theta \sum_{k=1}^{+\infty}(-1)^k
G_{km}(\tau;\lambda)\cos(2k\theta).          
\label{c7_2.17}         
\end{equation}

\section{ Forced free surface waves:  summing up of the series involved in the leading-order solution. Three theorems}   
                  
\subsection{ Statement  of basic theorems}  

{\bf Theorem 1.}  For any finite time-interval ($ 0\le \tau \le T,\,\, T>0 $ may  be assigned arbitrarily) series \eqref{c7_2.16} 
 and \eqref{c7_2.17}   converge uniformly with regard to 
 $ \theta $ and $ \tau $.\\

{\bf Theorem 2.} On the interval $\,|\theta|<\pi/2\,$ the following
formulas hold
$$
  R_0(\tau,\theta;\lambda)=\frac{1}{2}\int\limits_0^{+\infty}x^4
\exp\left(-\frac {1}{2}x^2\right)
f_1\cos\left(\frac {1}{2}x^2\tan\theta \right)\,dx,
	$$
$$
  Q_0(\tau,\theta;\lambda)=\frac{1}{2}\int\limits_0^{+\infty}x^4
\exp\left(-\frac {1}{2}x^2\right)
f_2\cos\left(\frac {1}{2}x^2\tan\theta \right)\,dx,
	$$
for $m\ge 1$
\begin{equation}
  R_m(\tau,\theta;\lambda)=-\frac{1}{2m}\int\limits_0^{+\infty}x^4
\exp\left(-\frac {1}{2}x^2\right)
L_{m-1}^{(1)}(x^2)f_1\cos\left(\frac {1}{2}x^2\tan\theta \right)\,dx,                                           
\label{c7_3.1}                             
\end{equation}	
\begin{equation}
  Q_m(\tau,\theta;\lambda)=-\frac{1}{2m}\int\limits_0^{+\infty}x^4
\exp\left(-\frac {1}{2}x^2\right)
L_{m-1}^{(1)}(x^2)f_2\cos \left(\frac {1}{2}x^2\tan\theta \right)\,dx,                                               
\label{c7_3.2}                            
\end{equation}		
where functions $\,f_1\,$ and $\,f_2\,$ are specified by \eqref{c7_2.12}
  and \eqref{c7_2.13} \\

{\bf Theorem 3.} Corresponding to any preassigned value of 
 $\,\tau\,$ there exists a quantity $\,c(\tau)\,$ such that
\begin{equation}
|\tan^2\theta \cdot R_m(\tau,\theta)|<c(\tau),\,\,\,
|\tan^2\theta \cdot Q_m(\tau,\theta)|<c(\tau).    
\label{c7_3.3}                               
\end{equation}

\subsection{Proof  of basic theorems }	
	
     The proof of Theorems 1 and 2 essentially employs the asymptotic formula for Laguerre polynomials 
\begin{equation}
 x^3e^{-x^2}L_k^{(1)}(x^2)=\frac{1}{\sqrt \pi}\root 4\of{k+1}
 e^{-\frac{x^2}{2}}l_k,
\label{c7_3.4}                
\end{equation}
	$$
 l_k=x^\frac{3}{2}\cos\left( 2x\sqrt{k+1}-\frac{3}{4}\pi \right)+r_k,
$$	
	$$
    r_k=O\left(\frac{x^\frac{1}{2}}{\sqrt k}\right)+
    O\left(\frac{x^\frac{9}{2}}{\sqrt k}\right)
    +O\left(\frac{x^7}{k^\frac{3}{4}}\right),\,\,\,\,\,x\ge0, 
     $$
where the notation $\,O(u)\,$ means ``of the order of $\,\,u\,$''.
The following two lemmas proved in Chapter 2 will be also used:\\

 {\it Lemma 1.} The series
\begin{equation}
 \sum_{k=1}^{+\infty}(-1)^k\frac{1}{k}L_{k-1}^{(1)}(\lambda^2)
e^{i2k\theta}  
\label{c7_3.5}             
\end{equation}
converges at any point $\,\theta\,$ and converges uniformly in any
segment $\,|\theta|\le \pi /2 -\delta\,,$ $\,\delta>0\,$.\\

 {\it Lemma 2.} For $-\pi/2 < \theta <\pi/2$ the following formula holds
\begin{equation}
s\sum_{k=1}^{+\infty}(-1)^k\frac{1}{k}L_{k-1}^{(1)}(s)
e^{i2k\theta}=1-\exp\left(\frac{1}{2}s\right)\cdot
\exp\left(i\frac{1}{2}s\tan\theta\right).    
\label{c7_3.6}             
\end{equation}
\vspace{5mm}

{\it  Proof of Theorem 1}\\

\noindent
When in the series \eqref{c7_2.16}, \eqref{c7_2.17}  
the products of trigonometric
functions are decomposed into sums, the rearranged series take the form
\begin{equation}
\hat{R}_m(\tau,\theta)=\frac{1}{4}\sum_{k=0}^{+\infty}(-1)^{k+1}h_{km}  \cos(2k\theta),        
  \label{c7_3.7}       
\end{equation}
\begin{equation}
  \hat{Q}_m(\tau,\theta)=\frac{1}{4}\sum_{k=0}^{+\infty}(-1)^{k+1}g_{km}
  \cos(2k\theta),        
  \label{c7_3.8}               
\end{equation}
$$             
  h_{km}=H_{k-1,m}-2H_{k,m}+H_{k+1,m},\,\,\,\,\,\,
  g_{km}=G_{k-1,m}-2G_{k,m}+G_{k+1,m},  
            $$
$$
  H_{-1,m}=H_{0,m}=0,\,\,\,\,\,\,G_{-1,m}=G_{0,m}=0,\,\,\,\,\,\,
  h_{0,m}=H_{1,m},\,\,\,\,\,\,   g_{0,m}=G_{1,m}
       $$
With the use of equations \eqref{c7_2.10}, 
\eqref{c7_2.11} 
and  the recursion relations \eqref{c7_2.9} among Lagueerre polynomials, 
we obtain      (with $k\ge 1,\, m\ge1$):
\begin{equation}
    h_{km}(\tau;\lambda)=-\frac{2}{mk}\int\limits_0^{+\infty}x^6e^{-{x}^2}
    L_{k-1}^{(1)}(x^2)L_{m-1}^{(1)}(x^2)f_1dx,     
  \label{c7_3.9}                
\end{equation}
\begin{equation}
    g_{km}(\tau;\lambda)=-\frac{2}{mk}\int\limits_0^{+\infty}x^6e^{-{x}^2}
    L_{k-1}^{(1)}(x^2)L_{m-1}^{(1)}(x^2)f_2dx,\,\,  
 \label{c7_3.10}                     
\end{equation}
It follows from \eqref{c7_2.10}, \eqref{c7_2.11},  and \eqref{c7_3.4} that
$$
     H_{km}(\tau;\lambda)=O\left(\frac{1}{k^\frac{1}{4}}\right),\,\,\,\,\,\,
     G_{km}(\tau;\lambda)=O\left(\frac{1}{k^\frac{1}{4}}\right)
     $$
$ \hbox{for}\,\,{0\le{\tau}}\le{T},\,\,$ where $\,\,T\,\,$ and $\,\,m\,\,$ are
fixed. The difference of the $n$-th partial sums of series 
\eqref{c7_2.16} and \eqref{c7_3.9}
tends to zero as $n$ tends to $\,\,+\infty,\,\,$ since $\,\,H_{km}\to 0\,\,$
as $\,\,k\to{+\infty}.\,\,$ This means, that either both series converge
to the same sum or both series diverge ($\,\,\hbox{for}\,\,{0\le{\tau}}\le{T}
\,$).
Making use of relations \eqref{c7_3.4} and \eqref{c7_3.7},  we conclude that \linebreak 
$\,\,h_{km}(\tau;\lambda)=O\left(1/{k^\frac{5}{4}}\right),\,\,$ and,
consequently,
series \eqref{c7_3.7} converges absolutely and uniformly throughout the region $\,\,{0\le{\tau}}\le{T},\,\,|\theta|\le \pi/2$, series \eqref{c7_2.16} converges uniformly.

The same result for series \eqref{c7_2.17} can be proved in a similar
manner.  This completes the proof of theorem 1.
\vspace{5mm} 

{\it Proof of Theorem 2} \\

\noindent
By \eqref{c7_3.8}
$$
h_{0m}=H_{1,m}=
\frac{2}{m}\int\limits_0^{+\infty}x^4e^{-{x}^2}(x^2)L_{m-1}^{(1)}(x^2)f_1dx,  
	$$
By \eqref{c7_3.7}
$$
R_m(\tau,\theta)=\hat{R}_m(\tau,\theta)=-\frac{1}{2m}\int\limits_0^{+\infty}x^4e^{-{x}^2}(x^2)L_{m-1}^{(1)}(x^2)f_1dx+  
	$$
$$
\frac{1}{4}\sum_{k=1}^{+\infty}(-1)^{k+1}h_{km}  \cos(2k\theta) 
	$$
The last series converges absolutely and uniformly and may be integrated term by term. With the use of \eqref{c7_3.9} and lemma 2 we find
$$
\frac{1}{4}\sum_{k=1}^{+\infty}(-1)^{k+1}h_{km}  \cos(2k\theta) =
	$$	
$$
\frac{1}{2m}\int\limits_0^{+\infty}x^4e^{-{x}^2}L_{m-1}^{(1)}(x^2)f_1
\left(\sum_{k=1}^{+\infty}(-1)^{k}\frac{1}{k}x^2\cos(2k\theta)
L_{k-1}^{(1)}(x^2)\right)\,dx=
	$$	
$$
\frac{1}{2m}\int\limits_0^{+\infty}x^4e^{-{x}^2}L_{m-1}^{(1)}(x^2)f_1\left(1-\exp\left(\frac{1}{2}x^2\right)\cdot
\cos\left(\frac{1}{2}x^2\tan\theta\right) \right)\,dx.
	$$
Consequently
$$
\hat{R}_m(\tau,\theta)=-\frac{1}{2m}\int\limits_0^{+\infty}x^4
\exp\left(-\frac {1}{2}x^2\right)
L_{m-1}^{(1)}(x^2)f_1\cos\left(\frac {1}{2}x^2\tan\theta \right)dx
	$$				
The equality \eqref{c7_3.2} can be proved in a similar way.\\

{\it Proof of Theorem 3}\\

\noindent
We can rewrite \eqref{c7_3.1} as follows
$$
  R_m=\int\limits_0^{+\infty}F(x)x\cos \left(\frac {1}{2}x^2\tan\theta \right)\,dx,
  	$$
$$
  \hbox{where}\,\,\,\,\,  F(x)=-\frac{2}{m}x^3\exp \left(-\frac {1}{2}x^2 \right)L_{m-1}^{(1)}(x^2)f_1.
	$$
It follows from \eqref{c7_2.12} that $\,f_1(-x)=-f_1(x)\,$, the function
$\,f_1\,$ possesses derivatives of any order at any point
$\,x\ge0\,$. Thus, the function $\,F(x)\,,$ too, possesses all
derivatives and has a zero of the 4th order at $\,x=0\,$.
Integration by parts gives
$$
\tan\theta R_m=-\int\limits_0^{+\infty}F'(x)\sin \left(\frac {1}{2}x^2
\tan\theta \right)\,dx.
$$
The right-side integral converges absolutely. Next two
applications of the integration by parts lead to the first inequality
\eqref{c7_3.3}.
The second inequality \eqref{c7_3.3} can be proved in a similar way.
Theorem 3 shows, that   the function $W_0$ (\eqref{c7_2.15})  satisfy the boundary conditions \eqref{c7_1.4}
\vspace{3mm}

Similar three theorems can be proved for the case of pressure law
$$
    F(\theta,t)=\beta_m(t)\sin(2m\theta),
	$$
$$
\beta_m(t)=B_m\cos(\Omega\,t-\phi_m), \,\,\, \,\,\,
B_m>0 ,\,\,\,\,\,\,\beta_k(t)\equiv0\,\,\,\text{for}\,\,\,k\ne m.  
    $$
In particular,    
$$
\cos^2\theta \sum_{k=1}^{+\infty}(-1)^k
H_{km}(\tau;\lambda)\sin(2k\theta)=
	$$
$$	
-\frac{1}{2m}\int\limits_0^{+\infty}x^4
\exp\left(-\frac {1}{2}x^2\right)
L_{m-1}^{(1)}(x^2)f_1\sin\left(\frac {1}{2}x^2\tan\theta \right)\,dx,            
	$$
$$
\cos^2\theta \sum_{k=1}^{+\infty}(-1)^k
G_{km}(\tau;\lambda)\sin(2k\theta)=
	$$
$$
-\frac{1}{2m}\int\limits_0^{+\infty}x^4
\exp\left(-\frac {1}{2}x^2\right)
L_{m-1}^{(1)}(x^2)f_2\sin\left(\frac {1}{2}x^2\tan\theta \right)\,dx,            
	$$	
\vspace{3mm}

Under what conditions the first term of the expansion gives a reasonable approximation to solution of full nonlinear problem? 
\vspace{3mm}

The situation is typical for nonlinear mechanics. 
The usual technical work with expansions leads to a sequence of systems
of linear non-homogeneous equations. Every non-homogeneous system of the sequence involves terms corresponding to applied ``forces'' or ``sources''. The corresponding homogeneous equations are identical. It is impossible to answer the question unless the exact solution of the leading-order equations is studied for arbitrarily varying external pressure.
\vspace{3mm}

Now we can evaluate the importance of Theorems 1, 2, and 3.

It can be shown that under conditions \eqref{c7_1.4}  the ``force'' terms have the form $\cos^2\theta\cdot F(\theta, t)$ similar to \eqref{c7_1.6}.
Having expanded the function $F(\theta, t)$ in a trigonometric series similar to \eqref{c7_1.7}, we can find the solution of corresponding linear system with the use of the theorem 2.	
\vspace{3mm}

Theorems 1, 2, and 3 show that the ``input'' and the
``output'' of the integro-differential equations (i.e., the applied ``forces'' and the solution) are related by a bounded operator. This means that the solution of leading-order equations  give a reasonable approximation to the solution of the full nonlinear problem for small values of the parameter  $\varepsilon=a/f^2$.
The proof of this assertion is beyond the scope of the book. 
Similar assertion for Lamb's problem on vortex bifurcation was proved
in \cite[]{mind1984}.   
          
\subsection{Extension of the results for the pressure with continuous spectrum}        

The results can be extended to include the case of external pressure 
of continuous frequency spectrum. In this case equation  \eqref{c7_1.8} is replaced by 
$$
 \alpha_k(t)=\int\limits_0^{+\infty}[A_k(\lambda)\cos(\lambda\tau)+
 B_k(\lambda)\sin(\lambda\tau)]d\lambda=\alpha_k^*(\tau),
	$$
where the integral is supposed to be  absolutely convergent.

Solution to the problem \eqref{c7_2.6} takes the form 
\begin{equation}
a_k(t)=(-1)^k\sum_{m=0}^{+\infty}(-1)^ma_{km}(\tau),  
\label{c7_3.11}                      
\end{equation} 
$$
a_{km}(\tau)=\int\limits_{0}^{+\infty} A_m(\lambda)
\left[H_{km}(\tau;\lambda)\cos(\lambda\tau)+
G_{km}(\tau;\lambda)\sin(\lambda\tau) \right]\,d\lambda +
          $$
$$
\int\limits_{0}^{+\infty} B_m(\lambda)\left[H_{km}(\tau;\lambda)\sin(\lambda\tau)-
G_{km}(\tau;\lambda)\cos(\lambda\tau) \right]\,d\lambda.  
$$	
It is easy to check that 
$$
\pd {f_1}{\tau}=-\sin(x\tau)\cos(\lambda\tau),\,\,\,\,\,\,
\pd {f_2}{\tau}=-\sin(x\tau)\sin(\lambda\tau),
	$$
and, by \eqref{c7_2.10}  and  \eqref{c7_2.11}, 
\begin{equation}
H_{km}(\tau;\lambda)=\frac{1}{m}\int\limits_0^{\tau}l'_{km}(\tau_1)
\cos(\lambda\tau_1)\,d\tau_1,
\label{c7_3.12}                   
\end{equation}	                                            
$$
G_{km}(\tau;\lambda)=\frac{1}{m}\int\limits_0^{\tau}l'_{km}(\tau_1)\sin(\lambda\tau_1)\,d\tau_1,
 \,\,\,\,\,\,l'_{km}(\tau)=dl_{km}/d\tau
	$$
$$
l_{km}(\tau)=2\int\limits_0^{+\infty}
x^3e^{-x^2}L_{k-1}^{(1)}(x^2)L_{m-1}^{(1)}(x^2)\cos(\tau x)\,dx,                                                  
        $$
Inserting \eqref{c7_3.12}  in \eqref{c7_3.11}  and performing the integration with respect to $\,\lambda\,$, we obtain   
$$
a_{km}(\tau)=\frac{1}{m}\int\limits_0^{\tau}l'_{km}(\tau -\tau_1) 
\alpha_m^{*}(\tau_1)\,d\tau_1,
	$$
which gives (accompanied by \eqref{c7_3.11} ) solution to the problem \eqref{c7_2.6}, no matter 
whether the functions $\alpha_k(t)$ have Fourier transform or not.  
If 
$$
P^{*}=0\,\,\,\,\,\hbox{for}\,\,\,\, \,t<0,
	$$
$$
\hbox{and}\,\,\,\,
P^{*}=\frac{a\gamma}f\cos^2\theta\sum_{k=1}^{+\infty}A_k\cos(2k\theta)\,\,\,\,
(A_k=\hbox{const})\,\,\hbox{for}\,\,t>0, 
	$$
i.e., at $t=0$ ``nonuniform'' external pressure is applied to the free surface suddenly, then for transient process formula
\eqref{c7_3.11}  with 
$$
a_{km}=A_m\cdot\frac{1}{m}[l_{km}(\tau)-l_{km}(0)]
	$$
holds. In this case 
$\lim\limits_{t\to {+\infty}}a_k(t)=-A_k$, 
and the stationary free surface shape is 
$$
 x=-\frac {a}{f}\cos^2\theta\sum_{k=1}^{+\infty}A_k\cos(2k\theta),\,\,\,\,
 y=(x-f)\tan\theta.    
	$$
This result may be obtained directly from equations \eqref{c7_2.6}.

\section{ Steady-state free surface waves}  

  In this section, we are interested in what happens when the time,
$\,\,t\,\,$, goes to  $\ +\infty $.
We will prove the existence of the limits
\begin{equation}
R_m^*(\theta;\lambda)=\lim\limits_{\tau \to +\infty}
R_m(\tau, \theta;\lambda),\,\,\,\,\,\,
Q_m^*(\theta;\lambda)=\lim\limits_{\tau \to +\infty}Q_m(\tau,
\theta;\lambda),   
\label{c7_4.1}                            
\end{equation} 
and, consequently, the existence of the steady-state solution of
the leading-order equations.
\vspace{3mm}

The equations of the steady-state free surface wave are
\begin{equation}
x=\frac{a}{f}W_0^*(\theta,\tau),\,\,\,\,\,\, y=(x-f)\tan\theta,
 \label{c7_4.2}                    
\end{equation}
$$
W_0^*(\theta,\tau)=(-1)^mA_m[R_m^*(\theta;\lambda)
\cos(\tau\lambda-\phi_m)
+Q_m^*(\theta;\lambda)\sin(\tau\lambda-\phi_m)].  
        $$
{\it Lemma 3. } 
If $G(x)$ is a differentiable function defined in the semi-axes $x\ge 0$ 
such that $\int\limits_0^{+\infty}|G(x)|\,dx<+\infty$, 
$f_1$ and $f_2$  are determined by \eqref{c7_2.12}  and 
\eqref{c7_2.13} , then 
\begin{equation}
\hbox{(i)}\,\,\,\,\,\,\,\lim\limits_{\tau \to +\infty}
     \int\limits_0^{+\infty}G(x)f_1\,dx=I_1+I_2+I_3,     
 \label{c7_4.3}                
\end{equation}   
$$ 
I_1=-\frac{1}{2}\int\limits_0^{+\infty}\frac{G(x)}{x+\lambda}dx,\,\,\,\,
I_3=-\frac{1}{2}\int\limits_{2\lambda}^{+\infty}\frac{G(x)}{x-\lambda}dx,                                                          
	$$          
$$
I_2=-\frac{1}{2}\int\limits_0^{\lambda}
\frac{G(\lambda +s)-G(\lambda -s)}{s}ds,
	$$
\begin{equation}
\hbox{(ii)}\,\,\,\,\,\,\,\lim\limits_{\tau \to +\infty}
\int\limits_0^{+\infty}G(x)f_2\,dx=-\frac{1}{4}\pi G(\lambda).  
  \label{c7_4.4}             
\end{equation}      
Define $r(\tau)$ by 
\begin{equation}
\int\limits_0^{+\infty}G(x)f_1\,dx=I_0+I_1+I_3+r(\tau),          \label{c7_4.5}                  
\end{equation}    
where
\begin{equation}
I_0=-\int\limits_0^{2\lambda}\frac{G(x)}{x-\lambda}\sin^2\left(\frac{\lambda
-x}{2}\tau \right)\,dx,            
  \label{c7_4.6}                
\end{equation}         
 $I_1$, $I_3$ are determined in \eqref{c7_4.3}. 
\vspace{3mm}

According to the well-known property of trigonometric integrals
$$
\int\limits_a^bF(x)\cos(\tau x)dx \rightarrow 0
\,\,\,\,\,\hbox{for}\,\,\,\,\,  \tau \rightarrow +\infty, 
$$
$ r(\tau)\rightarrow 0 $ as $\tau \rightarrow +\infty $. 
Integral $I_0$ is decomposed into sum of two integrals
corresponding to two segments, $[0,\lambda]$ and $[\lambda , 2\lambda]$.
With the use of substitution $x=\lambda-s$ for the first segment
and $x=\lambda+s$ for the second, we get
$$
I_0=\int\limits_0^{\lambda}\frac{G(\lambda+s)-G(\lambda-s)}{s}
\sin^2\left(\frac{s\tau}{2} \right)ds.
$$
Since $2\sin^2\left(s\tau/2 \right)=1-\cos(s\tau) $, it is obvious 
that $I_0 \rightarrow I_2$ as $\tau\rightarrow+\infty$. 
\vspace{3mm}

To prove the formula \eqref{c7_4.4}  the equality  
$$
2\int\limits_0^{+\infty}\frac{\sin u)}{u}du=\pi
        $$
is used.
$$
\int\limits_0^{+\infty}G(x)f_2\,dx=
\frac{1}{2}\int\limits_0^{+\infty}G(x)\frac{1}{\lambda+x}\sin\left((\lambda+x)\tau\right)\,dx-
	$$
$$	
\frac{1}{2}\int\limits_0^{+\infty}G(x)\frac{1}{\lambda-x} \sin\left((\lambda-x)\tau\right)\,dx                                               
	$$
$$
\lim\limits_{\tau\to +\infty}\int\limits_0^{+\infty}G(x)\frac{1}{\lambda+x}\sin\left((\lambda+x)\tau\right)\,dx=0
	$$	
$$	
-\frac{1}{2}\int\limits_0^{+\infty}G(x)\frac{1}{\lambda-x} \sin\left((\lambda-x)\tau\right)\,dx =
-\frac{1}{2}\int\limits_{0}^{+\infty}G\left(\frac{z}{\tau}+\lambda\right)\frac{1}{z} \sin z\,dz\to -\frac{\pi}{4}G(\lambda)                                              
	$$	
The proof is similar to that for equalities \eqref{c7_4.3}.
\vspace{3mm}

Application of Lemma 3 to $  R_m(\tau,\theta;\lambda)$    and $  Q_m(\tau,\theta;\lambda)$ (see formulas \eqref{c7_3.4}) and \eqref{c7_3.5}) 
  leads to

{\bf Theorem 4.}  On the interval $\,|\theta|<\pi/2\,$

(i) There exist the limits \eqref{c7_4.1};

(ii) The limits are
\begin{equation}
R_{m}^*(\theta;\lambda)=I_1+I_2+I_3,  
  \label{c7_4.7}                 
\end{equation}     
$$
I_1=\int\limits_0^{+\infty}\frac{G(x)}{x+\lambda}dx,\,\,\,\,
I_2=\int\limits_0^{\lambda}\frac{G(\lambda +s)-G(\lambda -s)}{s}ds,	
\,\,\,\,\,\,I_3=\int\limits_{2\lambda}^{+\infty}\frac{G(x)}{x-\lambda}dx,   	
	$$
\begin{equation}
Q_{m}^*(\theta;\lambda)=\pi G(\lambda), 
  \label{c7_4.8}               
\end{equation}    
where
\begin{equation}
G(x)=v_m(x)\cos\left(\frac{1}{2}x^2\tan\theta\right),\,\,\,\,\,\,
v_m(x)=\frac{1}{4m}x^4L_{m-1}^{(1)}(x^2)e^{-x^2/2}. 
  \label{c7_4.9}            
\end{equation}    

{\bf Theorem 5.} For large values of $|\tan\theta|$ the following
asymptotic formula holds
$$
R_{m}^*(\theta;\lambda)=-\pi v_m(\lambda)\sin\left(\frac{1}{2}\lambda^2
|\tan\theta|\right) +O\left(\frac {1}{|\tan\,\theta|}\right)
        $$

\bigskip
{\it Proof of Theorem 5.} \\

\noindent
Integrals $I_1$ and $I_3$ in \eqref{c7_4.7}
vanish as $|\tan\theta|\to\infty$, so at large values of
$|\tan\theta|$ asymptotic expression holds $R_{m}^*\approx I_2$.
By the equality
\begin{equation}
v_m(\lambda+s)=v_m(\lambda)+sp_m(s,\lambda)      
  \label{c7_4.10}            
\end{equation}
we define the differentiable function $p_m(s,\lambda)$. With the
use of \eqref{c7_4.10} we rewrite \eqref{c7_4.7} in the form
$$
R_{m}^*(\theta;\lambda)=v_m(\lambda)\int\limits_0^{\lambda}\frac
{\cos\big(\frac12(\lambda+s)^2\tan\theta\bigr)-\cos\bigl(
\frac12(\lambda-s)^2\tan\theta\bigr)}{s}ds +O\left(\frac{1}{\tan\theta}\right).   
        $$
Applying the well-known trigonometric identities we obtain
$$
\int\limits_0^{\lambda}\frac
{\cos\big(\frac12(\lambda+s)^2\tan\theta\bigr)-\cos\bigl(
\frac12(\lambda-s)^2\tan\theta\bigr)}{s}ds=
	$$
$$
-\sin\left(\frac12 \lambda^2
|\tan\theta|\right)I_4-\cos\left(\frac12 \lambda^2|\tan\theta|\right)I_5,
        $$
where
$$
I_4=2\int\limits_0^{\lambda}\cos\left(\frac12 s^2|\tan\theta|\right)
\frac{\sin(\lambda s|\tan\theta|)}{s}ds,
	$$
$$
I_5=2\int\limits_0^{\lambda}\frac{\sin\left(\frac12 s^2|\tan\theta|\right)}{s^2}\sin(\lambda s |\tan\theta|)ds.
        $$
By Lemma 5 of Chapter 2 we conclude that 
$$
\lim\limits_{|\tan\theta|\to+\infty}I_4=2\lim\limits_{|\tan\theta|\to+\infty}
 \int\limits_0^{\lambda ^2
|\tan\theta|}\cos\left(\frac{u^2}{2\lambda^2 |\tan\theta|}\right)
\frac{\sin u}{u}du=\pi,\,\,\,\,\,\,
\lim\limits_{|\tan\theta|\to+\infty}I_5=0,
        $$
Now one can easily complete the proof of theorem 5.

\section{ Hydrodynamic interpretation of the theorems.} 

\begin{figure}[ht]
	\resizebox{\textwidth}{!}
		{\includegraphics{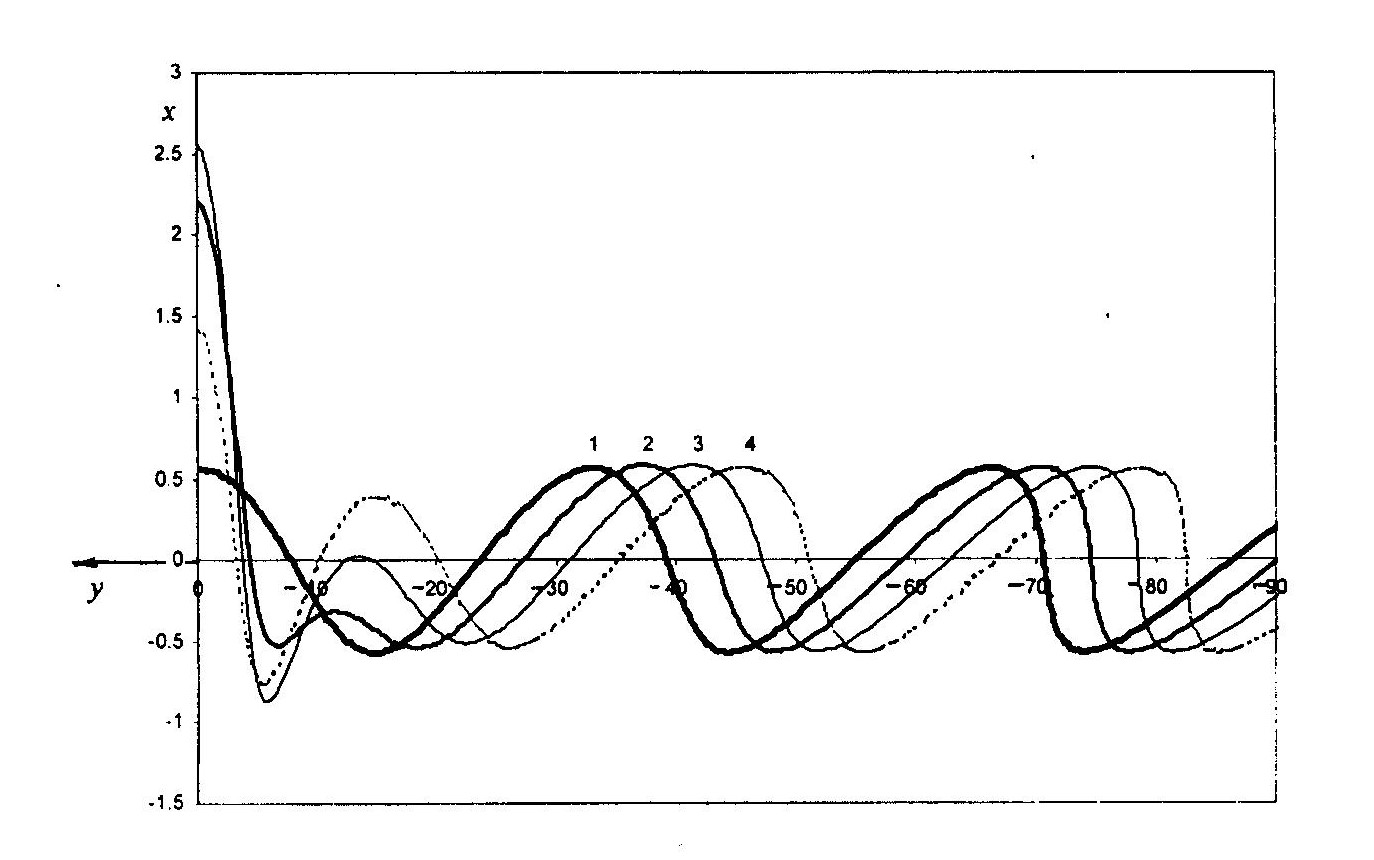}} 
	\caption{
 Successive profiles of the steady-state free surface wave \eqref{c7_4.2} at four instants, a quater of the period apart: 1 - for $\lambda\tau=0,\,$ 
2 - for $\lambda\tau=\pi/2\,$ 3 - for $\lambda\tau=\pi,\,$ 
4 - for $\lambda\tau=3\pi/4\,$; $\lambda=2,\,$ $m=3,\,$ $f=-10,\,$ $aA_3=-10,\,$ $|y<90|$.  
	}
	\label{c7_fg2}
\end{figure}

\begin{figure}[ht]
	\resizebox{\textwidth}{!}
		{\includegraphics{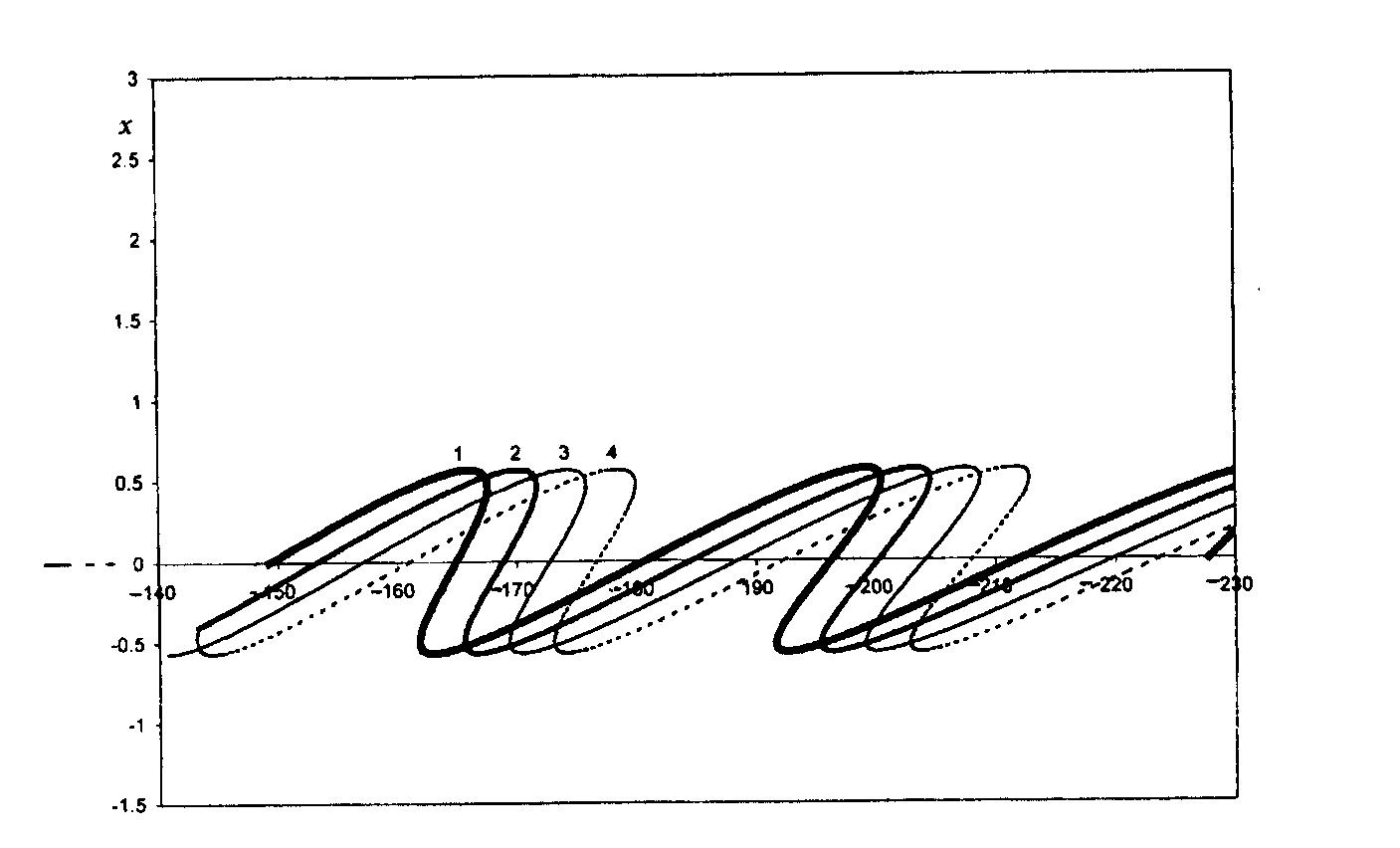}} 
	\caption{
 Same as Fig. \ref{c7_fg2}, but for $140<|y|<230$.
	}
	\label{c7_fg3}
\end{figure}

\begin{figure}[ht]
	\resizebox{\textwidth}{!}
		{\includegraphics{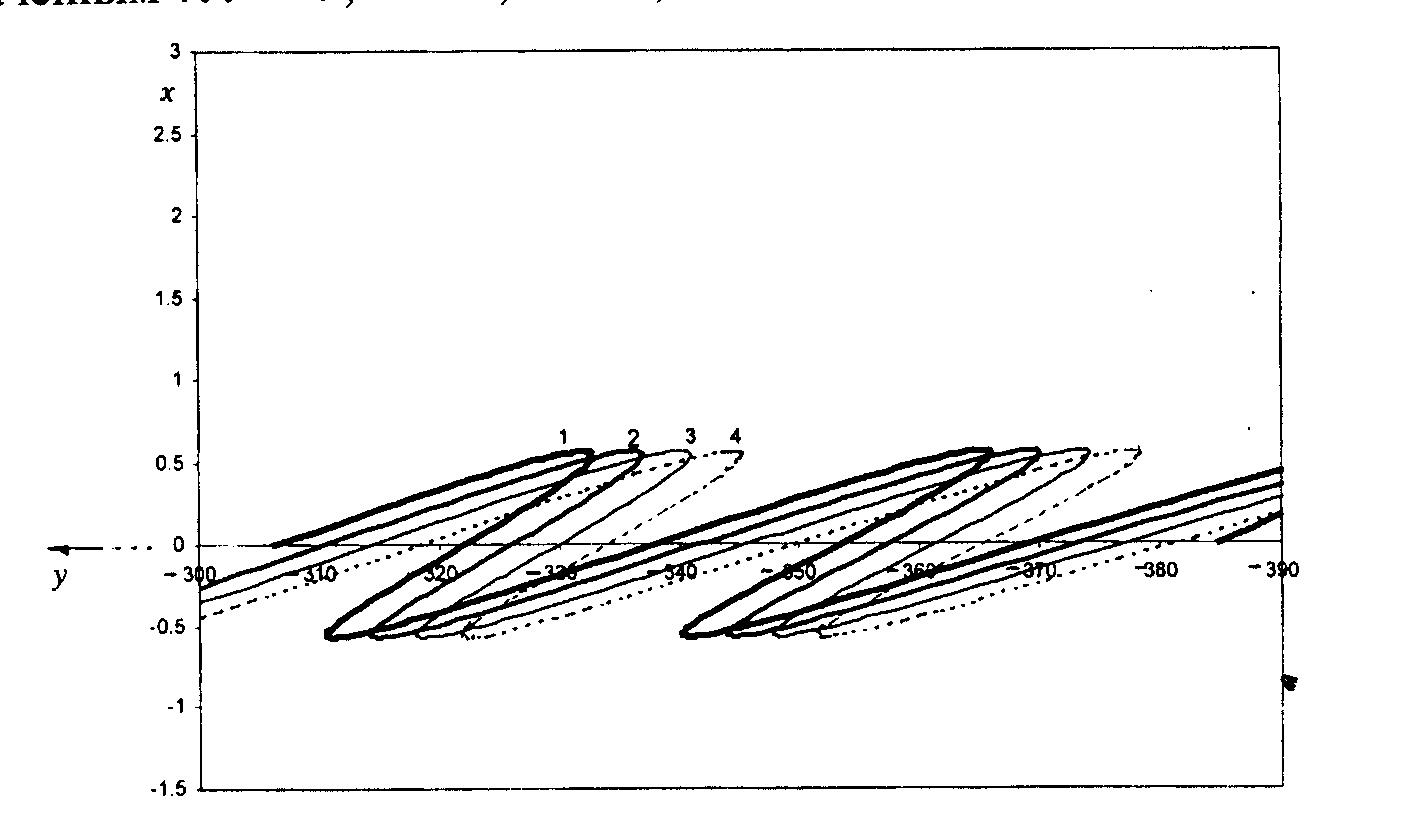}} 
	\caption{
Same as Fig. \ref{c7_fg2}, but for $300<|y|<390$.
	}
	\label{c7_fg4}
\end{figure}

We consider the case of $L_{m-1}^{(1)}(\lambda^2)\ne 0$ unless indicated otherwise.
Figures \ref{c7_fg2} - \ref{c7_fg4} offer comprehensive view of the theoretical steady-state wave \eqref{c7_4.2}. The wave is symmetrical about the vertical plane $y=0$. Calculations
were performed using the relations \eqref{c7_4.3} - \eqref{c7_4.6}
 with $m=3$, $f=-10$, $-aA_3=10$, $\lambda =2$. The three figures
correspond to the three regions:
$|y|<9|f|$, $14|f|<|y|<23|f|$, and $30|f|<|y|<39|f|$ of the same
size, respectively. Each of the figures shows four successive
profiles of the free surface at four instants, a quarter of the
period apart: $\lambda \tau =0, \pi/2, \pi, 3\pi/2$,
respectively.

For large values of $|\tan\theta|$ (at points far from the zone of variable pressure) we obtain the following simple approximation to the equations of the theoretical steady-state wave:
$$
x=l_m(\lambda )\sin\left(\lambda \tau-\phi _m
-\frac{1}{2}\lambda^2|\tan\theta|\right),
	$$
$$
y=(x-f)\tan\theta,\ \ \ \
l_m(\lambda)=(-1)^m\frac{a}{f}A_m\pi v_m(\lambda)
        $$
or
\begin{equation}
x=l_m(\lambda )\sin\left(\lambda \tau-\phi _m
-\Omega^2\frac{|y|}{1+\frac{x}{|f|}}\right).    
  \label{c7_5.1}                   
\end{equation}      
These equations follow as an immediate consequence of \eqref{c7_4.2} and Theorems 4 and 5.

We conclude from \eqref{c7_5.1} that (for large values of $|\tan \theta|$) a point of the free surface with given elevation $x=x_0={\rm const}$, $y=y(t)$ moves horizontally with velocity
$$
\frac{dy}{dt}=\left(1+\frac{x_0}{|f|}\right)\frac{1}{\Omega}\frac{y(t)}{|y(t)|},
        $$
so the points of higher elevation travel faster.
Since the crests of the wave move faster than the troughs, the horizontal
distance between a crest and the following  trough increases when the wave runs away from the origin. 
 This phenomenon is shown in figures \ref{c7_fg2} - \ref{c7_fg4}.

Theorem 3 shows that $R_m\rightarrow 0$ and
$Q_m\rightarrow 0$ as $|\theta|\rightarrow \pi /2$, $t$ being
considered as constant. On the other hand, by theorem 4,
$R_m\rightarrow R_m^*$ and $Q_m\rightarrow Q_m^*$ as $t\rightarrow +\infty$, treating $\theta$ as a constant. Thus, for sufficiently large values of $t$ there exists a region $|\theta|<|\theta_1|$ where the advancing free surface wave is close to the theoretical steady-state one, and there exists a region $|\theta|>|\theta_2|>|\theta_1|$ where the free surface displacement from its equilibrium position is very small. In the course of time, the size of the first region increases gradually. Calculations show that the profiles of the free surface become close to that shown in figures \ref{c7_fg2} - \ref{c7_fg4}:
in the region $|y|<90$ at $\lambda \tau =28\pi$,
in the region $|y|<230$ at $\lambda \tau =74\pi$,
in the region $|y|<390$ at $\lambda \tau =120\pi$, respectively.

Fig. \ref{c7_fg2} shows the steepening of the wave front up to time when  breaking point ocurred  and a short time after  passing the breaking point. 

There is a similarity between Figure \ref{c7_fg2} and Figures 10.10.12 obtained by numerical calculations and shown in \cite[]{stoker}.

Fig \ref{c7_fg3} and \ref{c7_fg4} show ``overhanging'' of the wave crest beyond the point where the breaking begins. 

There is no doubt that the theoretical free surface wave is unstable.
That is why only a finite set of billows was
observed in experimental studies \cite[]{caul}, \cite[]{chomaz}.

\section{ Energy absorption functional }  

     The average power of the pressure force, $\,\,N(t)\,\,$, is
determined by
\begin{equation}
N(t)=\frac{1}{t}\int\limits_0^t\left[-\int\limits_\Gamma{P^{*}q_n}\,dl\right]dt,                
 \label{c7_6.1}                 
\end{equation}
where $\,\,1\,dl\,\,$ is an area element in the free surface, $\,\,q_n\,\,$
is the velocity of the surface in the direction of the normal. 
The power is averaged over time-interval $\,\,(0,t) $.
\vspace{3mm}

Using \eqref{c7_1.7}  and \eqref{c7_1.8}, we find (up to small terms of order of $\varepsilon$)
\begin{equation}
P^*\cdot dl=-a\gamma \sum_{k=0}^{+\infty}[\alpha_k(t)\cos(2k\theta)+\beta_k(t)\sin(2k\theta)]\,d\theta,
 \label{c7_6.2}                    
\end{equation}                                             
\begin{equation}
q_n=\frac{a}{f}\pd {W_0}{t}=\frac{a}{f}\cos^2\theta 
[a'_k(t)\cos(2k\theta)+b'_k(t)\sin(2k\theta)].
 \label{c7_6.3}                      
\end{equation}
The last expression follows from the kinematic condition which means that a liquid particle in the free surface can have no velocity relative to the surface in the direction of the normal.
\vspace{3mm}

Substituting \eqref{c7_6.2} and \eqref{c7_6.3} into \eqref{c7_6.1} and performing the integration, we get
$$
N(t)=-\frac{a^2\gamma\pi}{8|f|}\sum_{k=0}^{+\infty}
\frac{1}{t}\int\limits_{0}^{t}
\left[\alpha_k(t_1)P'_k(t_1)+\beta_k(t_1) R'_k(t_1)\right]\,dt_1,
	$$

$$
P_k(t_1)= a_{k-1}(t_1)+2a_k(t_1)+a_{k+1}(t_1),\,\,\,\,\,\, 
R_k(t_1)= b_{k-1}(t_1)+2b_k(t_1)+b_{k+1}(t_1).
	$$
Integration by parts leads to
$$
N(t)=-\frac{a^2\gamma\pi}{8|f|}\sum_{k=0}^{+\infty}\frac{1}{t}
\left[\alpha_k(t)P_k(t)+\beta_k(t) R_k(t)\right]- 
\frac{a^2\gamma\pi}{8|f|}\sum_{k=0}^{+\infty}I_k,
	$$
$$
I_k=\frac{1}{t}\int\limits_{0}^{t}
\left[\alpha'_k(t_1)P_k(t_1)+\beta'_k(t_1) R_k(t_1)\right]\,dt_1,
	$$ 
$$
I_0=\frac{1}{t}\int\limits_{0}^{t}\alpha'_0(t_1)P_0(t_1)\,dt_1.
	$$
The first summand tends to zero as $t$ increases without bound, so
$$
N^{*}=\lim_{t\to {+\infty}}N(t)=-\frac{a^2\gamma\pi}{8|f|} 
\lim_{t\to {+\infty}}\sum_{k=0}^{+\infty}I_k.
	$$
At
$$
\alpha_k(t)=A_k\cos(\Omega t)+B_k\sin(\Omega t),\,\,\,\,\,\,
\beta_k(t)= A^{*}_k\cos(\Omega t)+B^{*}_k\sin(\Omega t) 
	$$
(all functions are of frequency $\Omega ;\,$ $k=0,\,1,\,2,\dots$), 
$\sigma_{*}=\sqrt {2|f|}$,  $t=\sigma_{*}\tau$, 
$\lambda =\sigma_{*}\Omega$,  $\lambda\tau=\Omega t$, 
it follows from \eqref{c7_2.7}  - \eqref{c7_2.9}  that
$$
k\ge 1\,\,\,\, a_{k-1}(t)+2a_k(t)+a_{k+1}(t)=
	$$
$$
\cos(\lambda\tau)\cdot 
\sum_{j=1}^{+\infty}(-1)^{k+j}
\left(A_jh_{kj}-B_jg_{kj}\right)+ 
	$$
$$
\sin(\lambda\tau)\cdot
\sum_{j=1}^{+\infty}(-1)^{k+j}
\left(A_jg_{kj}+B_jh_{kj}\right),
	$$
where
$$
h_{kj}(\tau;\lambda)=-H_{k-1,j}(\tau;\lambda)+2H_{kj}(\tau;\lambda)-
H_{k+1,j}(\tau;\lambda),
	$$
$$
g_{kj}(\tau;\lambda)=-G_{k-1,j}(\tau;\lambda)+2G_{kj}(\tau;\lambda)-
G_{k+1,j}(\tau;\lambda).
	$$
Functions $H_{kj}(\tau;\lambda)$ and	 $G_{kj}(\tau;\lambda)$ are defined by \eqref{c7_2.10} and \eqref{c7_2.11} respectively.
\vspace{3mm}

By Lemma 3, there exist limits  
$$
H^{*}_{kj}(\lambda)= H_{kj}(+\infty;\lambda),\,\,\,\,
G^{*}_{kj}(\lambda)= G_{kj}(+\infty;\lambda),
	$$

$$
h^{*}_{kj}(\lambda)= h_{kj}(+\infty;\lambda),\,\,\,\,
g^{*}_{kj}(\lambda)= g_{kj}(+\infty;\lambda).
	$$
Using  Lemma 3, we find
\begin{equation}
G^{*}_{kj}(\lambda)=\pi\cdot \frac{1}{j}\lambda^4e^{-\lambda^2}
L^{(1)}_{k-1}(\lambda^2) L^{(1)}_{j-1}(\lambda^2),\
 \label{c7_6.4}               
\end{equation}                                         
\begin{equation}
g^{*}_{kj}(\lambda)=-\pi\cdot \frac{1}{kj}\lambda^6e^{-\lambda^2}
L^{(1)}_{k-1}(\lambda^2) L^{(1)}_{j-1}(\lambda^2).    
 \label{c7_6.5}             
\end{equation}
Straitforward calculations give
$$
M=\lim_{t\to {+\infty}}\sum_{k=0}^{+\infty}
\frac{1}{t}\int\limits_{0}^{t}
\alpha'_k(t_1)P_k(t_1)\,dt_1=
	$$

$$
\frac{\lambda}{2\sigma_{*}}\cdot
\sum_{k,j=1}^{+\infty}(-1)^{k+j}
\left[h^{*}_{kj}(B_kA_j-A_kB_j)+ g^{*}_{kj}(A_kA_j+B_kB_j)\right].
	$$
Exploiting the symmetry of $M$ with respect to indices $k$ and $j$ 
($M$ is invariant under interchange of $k$ and $j$) and equality \eqref{c7_6.5}, we find 
$$
M=\frac{\pi}{2\sigma_{*}}\lambda^7e^{-\lambda^2}\cdot
\sum_{k,j=1}^{+\infty}(-1)^{k+j}
\frac{1}{kj}L^{(1)}_{k-1}(\lambda^2) L^{(1)}_{j-1}(\lambda^2)(A_kA_j+B_kB_j),
	$$
or
$$
M=\frac{\pi}{2\sigma_{*}}\lambda^7e^{-\lambda^2}\left[
\left(\sum_{k=1}^{+\infty}(-1)^k\frac{1}{k}A_k L^{(1)}_{k-1}(\lambda^2)\right)^2+ 
\left(\sum_{k=1}^{+\infty}(-1)^k\frac{1}{k}B_k L^{(1)}_{k-1}(\lambda^2)\right)^2
\right].
	$$
The sum 
$$
\lim_{t\to {+\infty}}\sum_{k=0}^{+\infty}
\frac{1}{t}\int\limits_{0}^{t}
\beta'_k(t_1)P_k(t_1)\,dt_1
	$$
can be calculated in the similar way.
Thus, we find 
\begin{equation}
N^{*}=-\frac{\pi^2\gamma a^2}{16|f|\sqrt {2|f|}} \cdot  V(\lambda),
 \label{c7_6.6}                  
\end{equation}

$$
V(\lambda)= \lambda^7e^{-\lambda^2}\left[
\left(\sum_{k=1}^{+\infty}(-1)^k\frac{1}{k}A_k L^{(1)}_{k-1}
(\lambda^2)\right)^2+
\left(\sum_{k=1}^{+\infty}(-1)^k\frac{1}{k}B_k L^{(1)}_{k-1}
(\lambda^2)\right)^2\right.+
	$$

$$
\left.\left(\sum_{k=1}^{+\infty}(-1)^k\frac{1}{k}A^{*}_k L^{(1)}_{k-1}
(\lambda^2)\right)^2+
\left(\sum_{k=1}^{+\infty}(-1)^k\frac{1}{k}B^{*}_k L^{(1)}_{k-1}
(\lambda^2)\right)^2\right].
	$$
This power is absorbed by the liquid in the steady-state  wave motion.

If  $\alpha_m(t)=A_m\cos(\lambda \tau)+B_m\sin(\lambda\tau)\, $ 
for certain $m\ge 1$, $\alpha_k(t)=0$ for $k\ne m,\,$ 
and $\beta_k(t)=0,\,$, then the "partial" power absorbed by the liquid due to harmonic "input" $\alpha_m(t)$ is  
$$
N_m=\frac{\pi^2\gamma a^2}{16|f|\sqrt {2|f|}}\lambda^7e^{-\lambda^2} 
(L^{(1)}_{m-1}(\lambda^2))^2\frac{1}{m^2}(A^2_m+B^2_m).
	$$
From \eqref{c7_6.6}  we see that the absorbed power due to some "inputs" $\alpha_k(t)\,$ 
is not equal to the sum of the corresponding "partial" powers, but the additivity with respect to the frequency spectrum of the external pressure holds, i.e., 
if, for example,  
$$
\alpha_k(t)=\sum_j[A_{kj}\cos(\Omega_jt)+ B_{kj}\sin(\Omega_jt)],
	$$
then  
\begin{equation}
N=\frac{\pi^2\gamma a^2}{16|f|\sqrt {2|f|}}\sum_j \cdot V(\lambda_j),\,\,\,\,\,\, 
\lambda_j\sqrt {2|f|}=\Omega_j.     
 \label{c7_6.7}                 
\end{equation}	
Formula \eqref{c7_6.7}  shows that the averaged energy absorbed by the liquid per unit time 
is equal to the sum of the absorbed "partial" energies corresponding to the points of frequency spectrum of the external pressure.
\vspace{3mm}

The functional \eqref{c7_6.7} may be referred to as the energy absorption functional.

Let $\,\,0<u_{k1}<u_{k2}<...<u_{kk}\,\,$ be zeros of the Laguerre polynomial
$\,\,L_k^{(1)}(u).\,\,$ If
$$
\alpha_k=\sum_{j=1}^{k-1}A_{kj}\cos(\tau\sqrt{u_{k-1,j}})+
B_{kj}\sin(\tau\sqrt{u_{k-1,j}}),
        $$
then the absorption functional is equal to zero. This means that the
steady-state wave does not transport energy at infinity. Only standing waves have this property.

\section{ Nonlinear standing waves.  }         

     For simplicity, assume  the pressure obeys the law
\begin{equation}
P^*=\frac{a\gamma}f\cos^2\theta\cos(2m\theta)\alpha_m(t).    
 \label{c7_7.1}                  
\end{equation}
$$
\alpha_m(t)=A_m\cos(\lambda_{mj}\tau-\phi_m),\,\,\,\,\,\,
\lambda_{mj}^2=u_{m-1,j},\,\,\,\,\,\,L_{m-1}^{(1)}(u_{m-1,j})=0
	$$          
Existence of the limits  
$$   
 G_{m}^*(\theta)= G_{m}(+\infty,\theta;\lambda_{mj})=0,
 	$$
$$   
 R_{m}^*(\theta)= R_{m}(+\infty,\theta;\lambda_{mj}),
 	$$ 
is assured by theorem 4, so  the nonlinear standing wave on the 
free surface is  described by equations 
\begin{equation} 
x=\frac{a}{f}R^*(\theta)\cdot\alpha_m(t),\,\,\,\,\,\, y=(x-f)\tan\theta,  
 \label{c7_7.2}                
\end{equation}
which means that the free surface oscillates synchronous with the external pressure.    
By the factor theorem, $L_{m-1}^{(1)}(u)=(u-u_{m-1,j})r_{m-2}(u)$, where $r_{m-2}(u)$ is a polynomial of degree $m-2$. 
\vspace{3mm}

It follows from   \eqref{c7_2.10}  and  Lemma 3  that 
$$
   H_{km}^*(\lambda_{mj})= H_{km}(+\infty;\lambda_{mj})= 
-\frac{1}{m}\int\limits_0^{+\infty}u^2e^{-u}L_{k-1}^{(1)}(u)r_{m-2}(u)du.  
	$$	
Making use of the identity $u^2=u(u-u_{m-1,j})+u_{m-1,j}u$ 
 (due to orthogonality of Laguerre polynomials) we obtain 
$$
H_{km}^*=-\delta_{km}-\frac{1}{m}u_{m-1,j}\int\limits_0^{+\infty}
ue^{-u}L_{k-1}^{(1)}(u)r_{m-2}(u)du,     
	$$
where $\delta_{km}$ is the Kronecker simbol, so 
$$
H_{km}^*=0,\,\,\,\,\hbox{for}\,\,k>m;\,\,\,H_{mm}^*=-1,
	$$
and, consequently,  $R_m^*(\theta)$ is a polynomial in 
$\cos(2\theta)$:
\begin{equation} 
R_m=R_m(\tau,\theta,\lambda)=\cos^2\theta \sum_{k=1}^m(-1)^k
H^*_{km}\cos(2k\theta),               
\label{c7_7.3}                 
\end{equation}
The nodes of the standing wave \eqref{c7_7.1} are determined by the equation $R^*(\theta)=0 $. This wave has a finite number of nodes in the free surface infinite in extent. In contrast with it, a classical standing wave has an infinite set of nodes.
Similarly, when 
$$
P^*=\frac{a\gamma}f\cos^2\theta\sin(2m\theta)\beta_m(t),\,\,\,\,\,
 \beta_m(t)=A_m\cos(\lambda_{mj}\tau)+B_m\sin(\lambda_{mj}\tau) 
	$$
we obtain 
$$
W_0(\theta,t)=W_0^{**}(\theta)\beta_m(t),\,\,\,\,\,\,
W_0^{**}(\theta)=\cos^2\theta\sum_{k=1}^{m}(-1)^{k+m}H_{km}^*(\lambda_{mj})
\sin(2k\theta).
        $$	
Parameters $H_{km}^{*}$ of the standing wave \eqref{c7_7.1} may be found directly from equations \eqref{c7_2.6}. For the pressure \eqref{c7_7.1}  the equations have a particular solution of the type
$$
 a_k(t)= (-1)^{k+m}H_{km}^{*}(\lambda_{mj})\alpha_m(t)\, ,\,\,\,a_k=0 \,\,
 \hbox{for}\,\,k>m.   
	$$
Inserting  $ a_k(t)$ into equations \eqref{c7_2.6} leads to the following results:
\begin{equation}
 H_{mm}^{*}(\lambda_{mj})=-1,\,\,\,H_{m-1,m}^*=\frac{1}{m}u_{m-1,j},   
\label{c7_7.4}                 
\end{equation}   
$$
  H_{m-2,m}^{*}=\left(2-\frac1{m-1}u_{m-1,j}\right)\frac{1}{m}u_{m-1,j},
	$$                                                      
$$
kH_{k-1,m}^*-(2k-u_{m-1,j})H_{km}^*+kH_{k+1,m}^*(\lambda_{mj})=0,
	$$
$$
k=1,2,...m-2;\,\,\,H_{0,m}^*=0\,\,\,(m>1).
	$$
{\it Example 1.}   Let the external pressure law be
$$
 P^{*}=\frac{a \gamma}{f}\cos^2\theta[\alpha(t)\cos(4\,\theta)+\beta(t) \sin(4\,\theta)],\,\,\,\,\,\,(m=2)
	$$
$$
 \alpha(t)=A_1\cos(\Omega t)+A_2\sin(\Omega t),
	$$
$$
 \beta(t)=B_1\cos(\Omega t)+B_2\sin(\Omega t),    
	$$
$ {\Omega}^2 2|f|=u_1,\,$, where $\, u_1=2\,$ is the zero of polynomial 
 $\,L_1^{(1)}(u)=2-u\,$, so $\,\lambda=\sqrt{2}\,$. 
In this case ($m=2$)
$$
  H_{2,2}^{*}(\lambda)=-1,\,\,\,\,\,\,\, H_{1,2}^{*}(\lambda)=1.
	$$
The free surface wave is a nonlinear combination of two standing waves:
$$
 x=-2\cdot \frac{a}{f}\cos^3\theta[\alpha(t)\cos(3\theta)+\beta(t)\sin(3\theta)], 
\,\,\,\,\,\, y=(x-f)\tan\theta.
	$$
Although the wave does not transfer energy to infinity, it has not (in general) 
fixed nodes: the nodes oscillate along the free surface. The amplitude of 
the oscillations depends on the phase difference between the pressure's components. 
If in particular  
$\,\alpha(t)=c\beta(t)\,$, i.e. $\,\alpha(t)\,$ and $\,\beta(t)\,$ 
are in phase or have a phase difference of $\pi$, then the sum 
$$
     cos^3\theta[\alpha(t)\cos(3\theta)+\beta(t)\sin(3\theta)]  
	$$
reduces to 
$$
  \cos^3\theta[c\cos(3\theta)+\sin(3\theta)]\beta(t),
	$$
and the nodes of the wave immovable in the rays $ \theta=\theta^{*}=\hbox{const}$,  
$(x=0, \,y=-f\tan\theta^{*}).$
\vspace{5mm}

{\it Example 2.} Let the external pressure law be
$$
 P^{*}=\frac{a \gamma}{f}\cos^2\theta[\alpha(t)\cos(6\,\theta)+
\beta(t)\sin(6\,\theta)],\,\,\,\,\,\,(m=3)
	$$
$$
 \alpha(t)=A_1\sin(\Omega_1 t)+B_1\cos(\Omega_1 t)+
 A_2\sin(\Omega_2 t)+B_2\cos(\Omega_2 t),
	$$
$$
 \beta(t)=A_1^{*}\sin(\Omega_1 t)+B_1^{*}\cos(\Omega_1 t)+
 A_2^{*}\sin(\Omega_2 t)+B_2^{*}\cos(\Omega_2 t),
	$$
$$
 \Omega_1^{2}=\frac{u_1}{2|f|},\,\,\, \Omega_2^{2}==\frac{u_2}{2|f|},\,\,\,
 u_1=3-\sqrt{3},\,\,\, u_2=3+\sqrt{3},  
	$$
where $\, u_1,\,\,$  $\,u_2\,$ are zeros of polynomial $\,L_2^{(1)}(u)=1/2u^2-3u+3.\,$ 
\vspace{3mm}

\begin{figure}[ht]
	\resizebox{\textwidth}{!}
		{\includegraphics{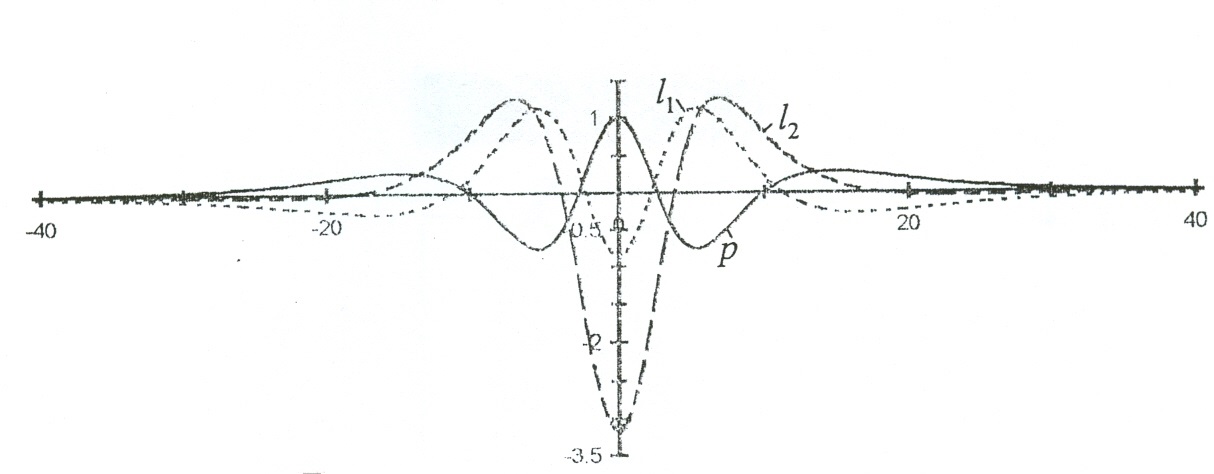}} 
	\caption{
To example 1,\,\,\,
even profiles of pressure $P=\cos^2\theta\cos(4\theta$  (solid curve) and   standing waves       
$l_1:\,\,\,x=V_1(\theta),\,\,\,l_2:\,\,\,x=V_2(\theta),\,\,\,
 y=(x-f)\tan\theta$ (dashed lines).
	}
	\label{c7_fg5}
	
\end{figure}[ht] 
 \begin{figure}
	\resizebox{\textwidth}{!}
		{\includegraphics{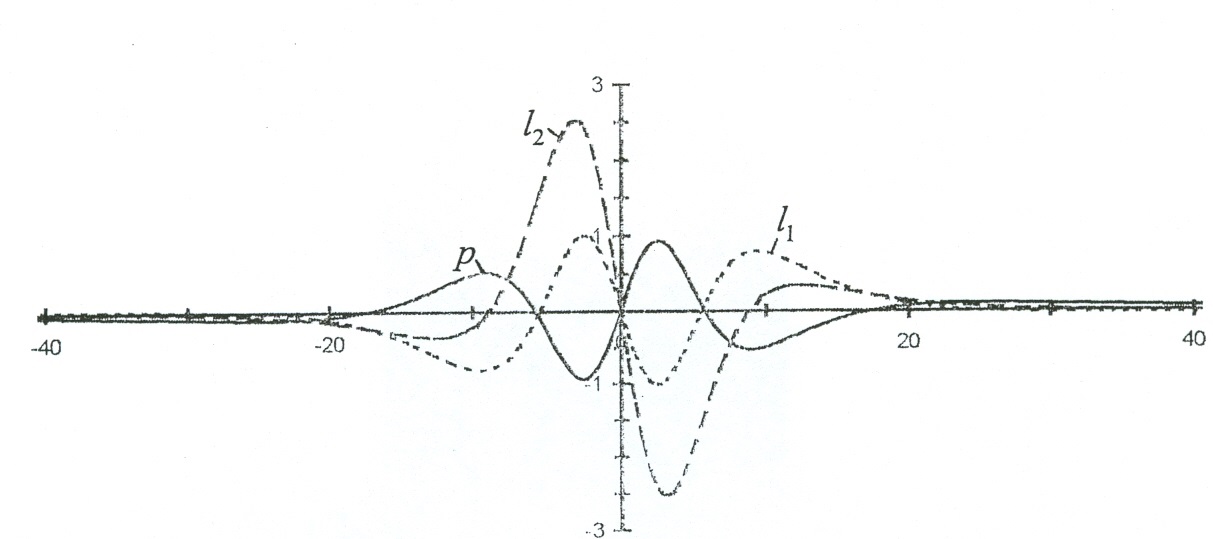}} 
	\caption{
To example 2,\,\,\,
Odd profiles of pressure $P=\cos^2\theta\sin(4\theta)$  (solid curve) and  standing waves
$l_1:\,\,\,x=V*_1(\theta),\,\,\,l_2:\,\,\,x=V*_2(\theta),\,\,\,
 y=(x-f)\tan\theta$ (dashed lines).
	}
	\label{c7_fg6}
\end{figure} 

 \begin{figure}[ht]
 \centering
	\resizebox{0.8\textwidth}{!}
		{\includegraphics{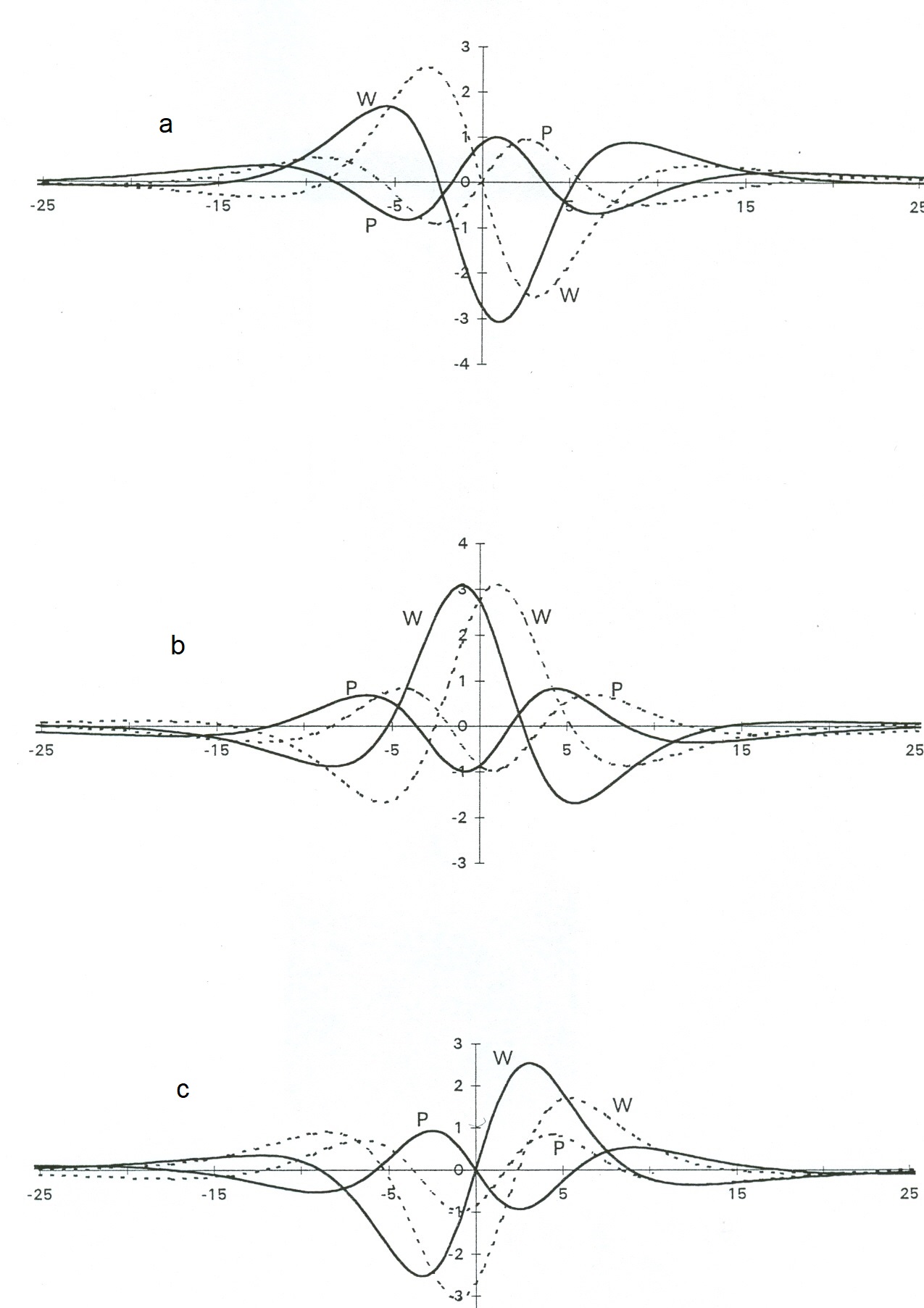}} 
	\caption{
 Profiles of the external pressure $P$ \eqref{c7_7.5} and 
free surface wave $W$ \eqref{c7_7.6} 
at instants $t_k=(2k+1)\delta t,\,\,\delta t=\pi/(6\Omega_2)$ for  
$k=0,\,2,\,4$  (solid lines) and $k=1,\,3,\,5$ (dashed lines); a: $\,\,k=0,\,\,1$; b: $\,\,k=2,\,\,3$;  c: $\,\,k=4,\,\,5$.
	}
	\label{c7_fg7}
\end{figure} 

The pressure generates a combination of four standing waves corresponding to 
$\,\lambda_1=\sqrt{u_1}\,$ and $\, \lambda_2=\sqrt{u_2}\,$. Formulas \eqref{c7_7.4}  give (with $\,(m=3)\,$)
$$
H_{3,3}^{*}(\lambda_1)=H_{3,3}^{*}(\lambda_2)=-1,\,\,\,\,\,
H_{2,3}^{*}(\lambda_1)=1-\sqrt{3}/3,     
	$$
$$
H_{2,3}^*(\lambda_2)=1+\sqrt{3}/3,\,\,\,\,\,
H_{1,3}^{*}(\lambda_1)=-H_{1,3}^{*}(\lambda_2)=\sqrt{3}/3.
	$$
The free surface equations are found to be 
$$
  x=\frac{a}{f}W_0(\theta,t),\,\,\,\,y=(x-f)\tan\theta,   
	$$
$$
  W_0=[A_1\sin(\Omega_1 t)+B_1\cos(\Omega_1 t)]V_1(\theta)+              
	$$
$$
      [A_2\sin(\Omega_2 t)+B_2\cos(\Omega_2 t)]V_2(\theta)+ 
	$$
$$
      [A_1^{*}\sin(\Omega_1 t)+B_1^{*}\cos(\Omega_1 t)]V_1^{*}(\theta)+  
	$$
$$
      [A_2^{*}\sin(\Omega_2 t)+B_2^{*}\cos(\Omega_2 t)]V_2^{*}(\theta),
	$$
$$
 V_1(\theta)=2\cos^3{\theta}\left[\frac{\sqrt3}{3}\cos(3\theta)-
\cos(5\theta)\right],
	$$
$$
 V_2(\theta)=-2\cos^3{\theta}\left[\frac{\sqrt3}{3}\cos(3\theta)+
\cos(5\theta)\right], 
	$$
$$
V_1^{*}(\theta)=2\cos^3{\theta}\left[\frac{\sqrt3}{3}\sin(3\theta)-
\sin(5\theta)\right],    
	$$
$$
V_2^{*}(\theta)=-2\cos^3{\theta}\left[\frac{\sqrt3}{3}\sin(3\theta)+
\sin(5\theta)\right].    
	$$
Figure \ref{c7_fg5} displays even profiles of the pressure profile and the profiles of two standing waves $l_1$ and $l_2$ corresponding to $\beta(t)\equiv 0,\,$ $B_1=B_2=0$:  
$$
P:\,\,\,\,x=\cos^{2}\theta\cos(6\theta),\,\,\,\,  y=(x-f)\tan\theta ,
\,\,\,\,\,\,f=-10;
	$$
$$
  l_1:\,\,\,\,x=V_1(\theta),\,\,\,\,y=(x-f)\tan\theta;  
	$$
$$
  l_2:\,\,\,\,x=V_2(\theta),\,\,\,\,y=(x-f)\tan\theta. 
	$$
Figure \ref{c7_fg6} represents the odd profiles: 
$$
  P:\,\,\,\,x=\cos^{2}\theta\sin(6\theta),\,\,\,\, y=(x-f)\tan\theta ,
\,\,\,\,\,\,f=-10;
	$$
$$
  l_1:\,\,\,\,x=V_1^{*}(\theta),\,\,\,\, y=(x-f)\tan\theta ;
	$$
$$
  l_2:\,\,\,\,x=V_2^{*}(\theta),\,\,\,\, y=(x-f)\tan\theta .  
	$$
\vspace{5mm}

If, for example, 
$$
B_2=A_2^{*}=1,\,\,A_1=A_1^{*}=A_2=0,\,\,B_1=B_1^{*}=B_2^{*}=0,
	$$
then
\begin{equation}
P^{*}=\frac{a\gamma}{f}\cos^2\theta\cos(6\theta-\Omega_2{t})  \label{c7_7.5}              
\end{equation} 
and combination of the standing waves results in the wave 
\begin{equation}
x=-\frac{a}{f}2\cos^3\theta\left[\frac{\sqrt3}{3}\cos(3\theta-\Omega_2{t})+\cos(5\theta-\Omega_2{t})\right],\,\,\,\,\,\,  
y=(x-f)\tan\theta   
\label{c7_7.6}                   
\end{equation}   

%
%

\chapter{Waves In A Two-Layered Liquid Generated By An Oscillating Cylinder}        

{\small{In this chapter, problem on gravity waves excited in two layered liquid by an oscillating solid cylinder is solved analytically assuming that the ratio of the cylinder radius to the distance to the interface is small. The solution is obtained with  no-flow condition on the  true surface of cylinder (as well as with nonlinear conditions on the liquid-liquid interface) and is valid at all times, from zero (the time of setting initial conditions) to infinity. According to the solution, a chain of waves with ''overhanging''  develops when the  wave  gets away from the oscillating cylinder.
Letting the time go to infinity, the steady state waves are found.}}

\section{Integro-differential equations to the problem }    

Consider two ideal heavy liquids separated by a liquid-liquid interface infinite in horizontal directions. The half-space above the interface is occupied by a homogeneous liquid of density $\gamma_1$, while a homogeneous liquid of density 
$\gamma_2$ fills the half-space below the interface. The two-layered liquid is stably stratified, i.e., $\gamma_1 <\gamma_2 $. 
Diffusion across the interface and surface tension are neglected. 
A long circular rigid cylinder is submerged in the stratified liquid 
at a distance from the interface,  the axis of the cylinder is normal to the $x,\,y$ plane. 
\vspace{3mm} 

Initially, at $t=0$, the liquids and cylinder are at rest and the interface is in its static equilibrium position, a horizontal plane. Then the cylinder starts to move at right angle to 
its axis  so that the motion set up in the liquid is expected to be 
in two dimensions.                                             
The flow diagram is shown in figure 8.1. 
In the figure, $\Gamma_1$ and $\Gamma$ are the sections of the cylinder and of the interface in the $x,y$-plane respectively; 
 the $x$- and $y$-axes are absolute (``Earth-fixed'') axes.
The $x$-axis oriented upward and the $y$-axis in the horizontal direction; the point $Q_1$ is fixed in the cylinder  $x=f(t),\,y=\varphi(t)$ are coordinates of $O_1$; $x_1$- and $y_1$-axes (parallel to the immovable $x$- and $y$-axes, respectively) with the origin $O_1$ move translationally; 
$r,\theta$ are polar coordinates (with the pole $O_1$) in the $x_1,y_1$-plane;  $g$ is acceleration of free fall. 
The horizontal plane $x=0$ is the static (equilibrium) position of the interface. 
\vspace{3mm} 
 
All variables and equations have been made non-dimensional by 
radius of the cylinder, $R_0$,  time, $T_*$,  defined by the relation  $gT_*^2=R_0$, the density $\gamma_2$,  so the equation of the surface of the cylinder equals unity, and  the non-dimensional acceleration of free fall is also unity. 
From now on, it is assumed that distance  between the interface and the cylinder is much greater than radius, i.e.,$|f(t)|>>1$. The shape of the interface is described by equations 
$$
x=W(\theta,t),\,\,\,\,\,\,y=(x-f)\tan\theta+\varphi(t).
	$$ 
In the $x,y$-plane we introduce polar coordinates ($r,\,\,\theta$) and 
 coordinates ($\sigma,\,\,\theta$) defined by the equations 
$$
x=\sigma+W(\theta,t),\,\,\,\,\,\,y-\varphi(t)=(x-f(t))\tan\theta.
        $$
\begin{figure}
\centering
      	\resizebox{0.7\textwidth}{!}
	 	{\includegraphics{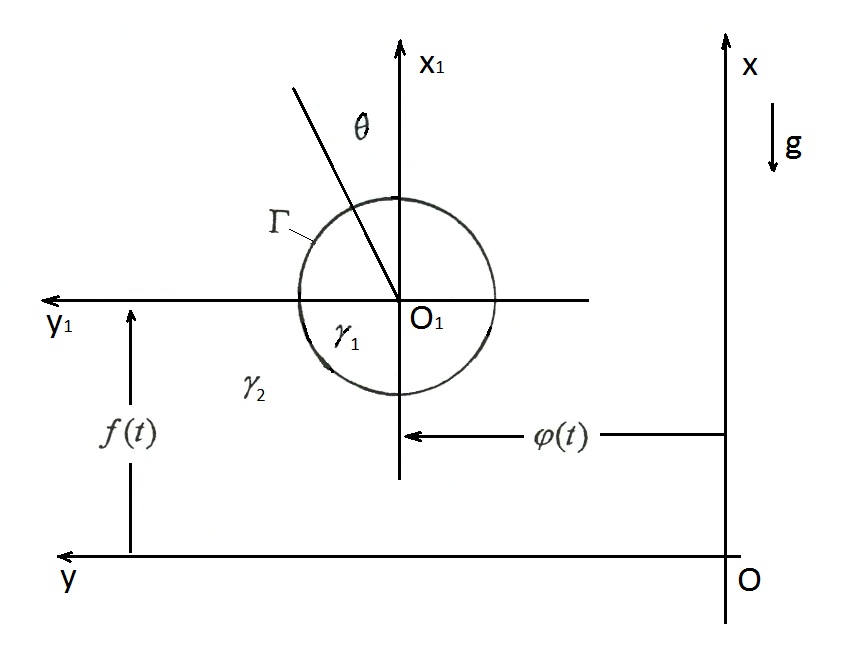}}   
	  \caption{Flow diagram}
	 \end{figure} 	  
In these curvilinear coordinates, equation of the solid surface becomes 
$r =1$ ( $0\le\theta\le 2\pi$ on the solid surface $\Gamma_1$), while the equation of the interface take the form $\sigma =0$ (the region 
$\sigma <0$ is occupied by the lower liquid 
of density  $\gamma_2$; on the interface $-\pi/2\theta<\pi/2$, if $f<0$, and $\pi<\theta<2\pi$, if $f>0$).      
\vspace{3mm}

Any function $F(r,\theta ,t)\,\,(r\ge 1)$ expressed in curvilinear coordinates $\sigma,\,\,\theta$ is notated as $F^{*}(\sigma ,\theta ,t)$, i.e. $F(r,\theta ,t)=F^{*}(\sigma ,\theta ,t)$ where 
$\sigma=r\cos\theta +f-W(\theta ,t).$
Notations are introduced  for one-sided limits 
$$
F_{+}=\lim_{r\to 1+0}F(r,\theta ,t),\,\,\,\,\,\,
F^{*}_{+}=\lim_{\sigma\to +0}F^{*}(\sigma ,\theta ,t),\,\,\,\,\,\,
F^{*}_{-}=\lim_{\sigma\to -0}F^{*}(\sigma ,\theta ,t).
        $$                

The fluid flow engendered by the moving cylinder is expected to posses 
a generalized velocity potential. 
The generalized velocity potential $\Phi$ of the flow is sought in the form of the doublet distribution over the  lines $\Gamma_1$ and $\Gamma$:
\begin{equation}
\Phi(r,\theta,t)=\Phi_1(r,\theta,t)+\Phi_2(r,\theta,t),  
 \label{c8_1.1}           
\end{equation}   
$$   
\Phi_1(r,\theta,t)=-
\frac{1}{4\pi}\int\limits_{\Gamma_1}g_1(\theta_1,t)\pd {}{r_1}
\left.\ln R_1\,d\theta_1\right|_{r_1=1},
	$$
$$
R_1=r^2+r_1^2-2rr_1\cos(\theta_1-\theta),\,\,\,\,\,\,
\Gamma_1:\,\,\,\,0\le \theta_1\le 2\pi,
	$$	
$$
\Phi_2^{*}(\sigma,\theta,t)=\frac{f}{|f|}\,
\frac{1}{4\pi}\int\limits_{\Gamma}\frac{2g_2(\theta_1,t)}{R_2^{*}}
\left.A(\sigma,\sigma_1,\theta,\theta_1,t)\,d\theta_1\right|_{\sigma_1=0},
        $$
$$
A=(\sigma-\sigma_1+W-W_1)(f-W_1)+
        $$
$$
(\sigma+W-f)\pd {W_1}{\theta_1}({\rm tg}\theta\cdot\cos^2\theta_1-
\sin\theta_1\cdot\cos\theta_1),
        $$
$$
R_2^{*}=(\sigma-\sigma_1+W-W_1)^2\cos^2\theta_1+
        $$
$$
[(\sigma+W-f)\tan\theta\cdot\cos\theta_1-(\sigma_1+W_1-f)\sin\theta_1]^2,
        $$
$$
W=W(\theta,t),\,\,\,\,\,\,W_1=W(\theta_1,t),
	$$
$$
\Gamma :\sigma_1=0,\,\,\,\,-\frac{\pi}{2}<\theta_1<\frac{\pi}{2},
\,\,\,\,\hbox{if}\,\,\,\,f<0;
\,\,\,\,\frac{\pi}{2}<\theta_1<\frac{3\pi}{2},
\,\,\,\,\hbox{if}\,\,\,\,f>0.
        $$
        
A point at the cylinder surface and the liquid particle which are in contact have the same velocity normal to the surface. 
This normal-velocity condition at the solid surface expressed as 
\begin{equation}
\pd {\Phi}{r_+}=f'\cos\theta+\varphi'\sin\theta   
 \label{c8_1.2}          
\end{equation}
provides the governing equation for the dipole density $g_1(\theta,t)$.         

The pressure continuity across the liquid-liquid interface is assured by 
equation 
\begin{equation}
\pd {g_2}{t}+\left[D(f'\sin\theta-\varphi'\cos\theta)+\frac{1}{2}
D^2\left(\pd {\Phi^{*}}{\theta_+}+\pd {\Phi^{*}}{\theta_-}\right)\right.-
        $$
$$
\left.\frac{1}{2}D_1\left(\pd {\Phi^{*}}{\sigma_+}+
\pd {\Phi^{*}}{\sigma_-}\right)\right]\pd {g_2}{\theta}+
\varepsilon_1\left[\pd {\Phi^{*}}{t_-}+
(f'\sin\theta-\varphi'\cos\theta)D\pd {\Phi^{*}}{\theta_-}\right.-
        $$
$$
\left.\frac{1}{2}D_2\left(\pd {\Phi^{*}}{\sigma_-}\right)^2+
\frac{1}{2}\left(D\pd {\Phi^{*}}{\theta_-}\right)^2+W(\theta,t)\right]=0,
\label{c8_1.3}         
\end{equation}   
$$
D=\frac{\cos\theta}{W-f},\,\,\,\,\,\,
D_1=\left(\sin\theta+D\cdot\pd {W}{\theta}\right)D,
        $$
$$
D_2=1+(D_1+D\sin\theta)\pd {W}{\theta},\,\,\,\,\,\,
\varepsilon_1 =1-\frac{\gamma_2}{\gamma_1}<0,
         $$
which is also the governing equation for the dipole density $g_2(\theta,t)$. 

Equation governing the interface configuration is
\begin{equation}
\pd {W}{t}=D_2\pd {\Phi}{\sigma_-}-D_1\pd {\Phi}{\theta_-}-
(f'\sin\theta-\varphi'\cos\theta)D\pd {W}{\theta}.   
\label{c8_1.4}           
\end{equation}  
Conditions at infinity are taken in the form
\begin{equation}
|W(\theta ,t)|<a(t)\cos^2\theta, \,\,\,\lim \limits_{\cos \theta \to 0}
\frac {\partial W}{\partial \theta}=0,
\label{c8_1.5}        
\end{equation}        
$$
|g_1(\theta,t)|<a(t),\,\,\,\,\,\,|g_2(\theta,t)|<a(t),
	$$      
where $a(t)$ is independent of $\theta$ and bounded on any interval $0\le t\le T$.  
Conditions imposed on $W(\theta ,t)$ require the volume of the liquid transferred across 
the equilibrium plane $x=0$ to be finite at any given value of $t$. 

Since $g_1(\theta,t)$ is supposed to be a bounded function, 
$$
\frac {1}{r}\pd {\Phi_1}{\theta}\rightarrow 0,\,\,\,\,\,\,
\pd {\Phi_1}{r}\rightarrow 0,\,\,\,\,\hbox{as}\,\,\,\,
r\rightarrow +\infty,
        $$
i.e. the potential $\Phi_1(r,\theta, t)$ does not produce fluid velocity at infinity.

Also, it follows from the conditions \eqref{c8_1.5} that ${\bf grad} \Phi_2(P)\rightarrow 0$ 
as $P$ tends to infinity, so the liquid is at rest at infinity.  

Equations for $u(t)=f'(t)$ and $v(t)=\varphi'(t)$, the velocity components of the centre $O_1$ of the cylinder, and initial conditions are to be appended to equations \eqref{c8_1.1} - \eqref{c8_1.4}. 
Below initial conditions are taken in the form 
$$
t=0\,\,\,\,g_1=g_2=0,\,\,\,\,W=0,\,\,\,\,f=f_0,\,\,\,\,\varphi=u=v=0\,\,\,\,(|f_0|>>1),                                    
        $$
which means that the cylinder starts to move in the fluid initially at rest.

We consider the case when the cylinder is moved with velocity $Q(t)$ through the liquid  along an inclined straight line in the $x,y$-plane:
\begin{equation}
u(t)=Q(t)\cos\alpha,\,\,\,\,\,\,v(t)=Q(t)\sin\alpha,\,\,\,\,\,\,
Q(0)=0,    
\label{c8_1.6}           
\end{equation}  
where $\alpha$ is the angle between the line and the positive $x$-axis.              
When the cylinder is speeding up with constant acceleration $a$, then $Q(t)=at$; 
$Q(t)=a\omega\sin(\omega t)$ when the cylinder oscillates harmonically, and so on.

Equations \eqref{c8_1.1} - \eqref{c8_1.4}  
 constitute the system of nonlinear
integro-differential equations in three unknown functions  
$g_1(\theta,t)$, $g_2(\theta,t)$,$W(\theta,t)$.
The solutions of the system determine 
the velocity field which satisfy the Euler's equations and
the exact boundary conditions on the evolving free surface
 $\sigma=0$.

The range of $\theta$ is from $0$ to $2\pi$ on the surface of the cylinder, where the function $g_1(\theta,t)$ is defined; on the interface the range of $\theta$ is from $-\pi/2$ to $\pi/2$ or from 
$\pi/2$ to $3\pi/2$ depending on the sign of $f(t)$ 
($f(t)\cos\theta <0$ on the interface).

The equations are valid as long as the cylinder moves at a distance from the interface, 
and each ray $\theta ={\rm const}$  
and the interface intersect no more than at one point. 

\section{ The leading-order equations and  double series solution for waves generated by an oscillating cylinder}  

\subsection{Leading-order equations}   

 The problem  will be considered in the case when  the circular cylinder
 oscillates almost periodically along an inclined straight line. For the case we write
\begin{equation}
f(t)=f_0+ah(t)\cos\alpha,\,\,\,\,\varphi (t)=ah(t)\sin\alpha,  
\,\,\,\,\,\,h(0)= \sum\limits_{j=1}^{+\infty}\alpha_j=0
\label{c8_2.1}         
\end{equation}  
$$
h(t)=\sum\limits_{j=1}^{+\infty}[\alpha_j\cos(\omega_jt)+\beta_j\sin(\omega_jt)],
\,\,\,\,\sum\limits_{j=1}^{+\infty}(\alpha_j^2+\beta_j^2)^{1/2}<+\infty,\,\,\,\,|f_0|>>1.
        $$
Equations \eqref{c8_1.1} - \eqref{c8_1.4} as well as the boundary and initial conditions  $f(0)=f_0$, 
$\varphi (0)=ah(0)\sin\alpha$ remain valid; 
$Q(t)=ah'(t)$  $Q(0)=0$ in \eqref{c8_2.1}.  

Assuming that the amplitude of the oscillations is small compared to the distance, $|f_0|$, between the cylinder and the liquid-liquid interface, we solve the problem by expanding unknown functions in powers of  $\varepsilon =a/f_0$:
$$
g_1(\theta,t)=a[g_{1,0}(\theta,t)+\varepsilon g_{1,1}(\theta,t)+
\varepsilon^2 g_{1,2}(\theta,t)+\dots ],
        $$
$$
g_2(\theta,t)=a[g_{2,0}(\theta,t)+\varepsilon g_{2,1}(\theta,t)+
\varepsilon^2 g_{2,2}(\theta,t)+\dots ],        
        $$
$$
W(\theta,t)=a[W_{0}(\theta,t)+\varepsilon W_{1}(\theta,t)+
\varepsilon^2 W_{2}(\theta,t)+\dots ].
        $$
Velocity potential is represented by expansion 
$$
\Phi (r,\theta,t)=a[\Phi_{(0)}(r,\theta,t)+
\varepsilon \Phi_{(1)}(r,\theta,t)+
\varepsilon^2\Phi_{(2)}(r,\theta,t)+\dots ],
        $$
$$
\Phi^* (\sigma,\theta,t)=a[\Phi^*_{(0)}(\sigma,\theta,t)+
\varepsilon \Phi^*_{(1)}(\sigma,\theta,t)+
\varepsilon^2\Phi^*_{(2)}(\sigma,\theta,t)+\dots ].
        $$

Substituting the above series  into \eqref{c8_1.1} - \eqref{c8_1.4}
and equating coefficients of like powers of $\varepsilon$ on both sides, we obtain an infinite sequence of systems of linear integrodifferential equations in unknown terms 
$g_{1,k}(\theta,t)$, $g_{2,k}(\theta,t)$ and $W_k(\theta,t)$.

The leading-order equations are found to be:
\begin{equation}
\Phi_{(0)} (r,\theta,t)= \Phi_{1,0} (r,\theta,t)+ 
\Phi_{2,0} (r,\theta,t),
 \label{c8_2.2}            
\end{equation}
$$
\Phi_{(0)}^{*} (\sigma,\theta,t)= \Phi_{1,0}^{*} (\sigma,\theta,t)+
\Phi_{2,0}^{*} (\sigma,\theta,t).
        $$
$$
\Phi_{1,0}(r,\theta,t)=-\left.\pd {}{r_1}T(g_{1,0})\right|_{r_1=1}\,\,\,\,
\hbox{for}\,\,\,\,r> 1,
        $$
$$
\Phi_{2,0}^{*} (\sigma,\theta,t)=
\left.\frac{\sigma}{|\sigma|}\hat H(g_{2,0})\right|_{\sigma_1=0}
\,\,\,\,\hbox{for}\,\,\,\,\sigma\ne 0,          
        $$
$$
\Phi_{2,0} (r,\theta,t)=\left.\frac{f_0}{|f_0|}\tilde H(g_{2,0})\right|_{\sigma_1=0}, 
	$$
\begin{equation}
\pd {}{t}\,g_{2,0}+\varepsilon_1
\left[\pd {}{t}\,\Phi_{(0)}^{*} (\sigma,\theta,t)+W_0(\theta,t)\right]=0,
\label{c8_2.3}            
\end{equation}
\begin{equation}
\pd {}{t}W_0=\frac{\sin\theta\cdot \cos\theta}{f_0}\,
\pd {}{\theta_+}\,\Phi_{(0)}^{*}+\pd {}{\sigma_+}\,\Phi_{(0)}^{*}, 
\label{c8_2.4}           
\end{equation}
\begin{equation}
\pd {\Phi_{(0)}}{r_{+}}=Q(t)\cos(\theta-\alpha).
\label{c8_2.5}            
\end{equation}

Integral operators $T(f)$, $\hat H(f)$ and $\tilde H(f)$ are defined 
in Chapter 2 (formulas \eqref{c2_1.1}, \eqref{c2_2.9}, \eqref{c2_2.10} respectively);
eigenvalues and eigenfunctions of the operators are given by formulas  \eqref{c2_2.2}, \eqref{c2_2.12}, and \eqref{c2_2.14} of Chapter 2;
parameters appearing in the operators are given by expressions 
$s=r$, $s_1=r_1$, $\alpha =(\sigma -f_0)/(\sigma_1-f_0)$ 
 $2\delta=r/(\sigma_1-f_0)$.

The initial conditions of the form 
$$
t=0,\,\,\,\,\,\,g_{1,0}=g_{2,0}=0,\,\,\,\,\,\,W_0=0,\,\,\,\,\,\,Q(0)=0 
        $$
are appended to the leading-order equations.

\subsection{Double series solution to the leading-order equations}  

Equations \eqref{c8_2.2} - \eqref{c8_2.5} are solved by expanding dependent variables in series of 
eigenfunctions  of the operators involved in the equations:
\begin{equation}
W_0(\theta,t)=\cos^2\theta \sum_{k=1}^{+\infty}a_k(t)\cos(2k\theta-\alpha),
        $$
$$
g_{1,0}(\theta,t)= \sum_{k=1}^{+\infty}\delta_k(t)\cos(k\theta-\alpha),
        $$
$$
g_{2,0}(\theta,t)=\varepsilon_* \sum_{k=0}^{+\infty}v_k(t)\cos(2k\theta-\alpha),
\label{c8_2.6}          
\end{equation}      
$$
\varepsilon_*=\frac{\gamma_2-\gamma_1}{\gamma_2+\gamma_1},\,\,\,\,\,\,
\delta_1(t)=-2Q(t)-\frac{\varepsilon_*}{2|f_0|}v_1(t);
        $$

$$
n\ge 2\,\,\,\,\,\,\delta_n(t)=\varepsilon_*\frac{f_0}{|f_0|}\frac{1}{(-f_0)^n}
\sum_{k=1}^{n}v_k(t)\frac{1}{2^n}C_{k-1}^{n-1},
        $$

$$
C_{k}^{n}=\frac{k!}{n!(k-n)!}.
        $$
      
The functions $z_n(t),\,\,v_n(t)$  $(n=1,\,2,\,\dots )$ are to be found by solving the following initial value problem for infinite set of ordinary differential equations:
\begin{equation}
v_0(0)=0,\,\,\,\,\,\,v_0'(t)=\frac{1}{2}a_1(t)+
\frac{1}{2}\sum_{k=1}^{+\infty}\delta_k'(t)\cdot \frac{1}{(-2f_0)^k}.
	$$
$$ 
a'_n(t)=-\frac{\varepsilon_*}{|f_0|}nv_n(t)+\frac{\varepsilon_*}{|f_0|}nm_n(t)+\frac{1}{f_0^2}h'(t)\delta_{n,1},
        $$
$$
v'_n(t)=\frac{1}{2}(a_{n-1}+2a_n+a_{n+1})+\varepsilon_*\frac{f_0}{|f_0|}m'_n(t)+\frac{1}{f_0}h''(t)\delta_{n,1},
\label{c8_2.7}          
\end{equation}       
$$
\delta_{kj}=1\,\,\,\hbox{for} \,\,k=j,\,\,\,\,\,\
\delta_{kj}=0\,\,\,\hbox{for}\,\,k\ne j,
	$$
$$
m_n(t)=\sum_{j=1}^{+\infty}\beta_{nj}v_j(t),\,\,\,\,\,\,
a_n(0)=v_n(0)=0,\,\,\,\,\,\,n=1,\,2,\,\dots;
        $$
$$
\hbox{for}\,\,1\le j\le n 
\,\,\,\,\, \beta_{nj}=\sum_{k=n}^{+\infty}\frac{1}{f_0^{2k}}\cdot\frac{1}{2^k}C_k^n 
\cdot\frac{1}{2^k}C_{k-1}^{n-1};
	$$
$$
\hbox{for}\,\, j\ge n+1  
\,\,\,\,\,\beta_{nj}=\sum_{k=j}^{+\infty}\frac{1}{f_0^{2k}}\cdot\frac{1}{2^k}C_k^n 
\cdot\frac{1}{2^k}C_{k-1}^{n-1},
	$$
or 
$$
n\beta_{nj}=j\beta_{jn} \,\,\,\, \hbox{for}\,\,\,\,j\le n.
        $$

We express the velocity potential $\Phi_{(0)}$ in terms 
of eigenvalues and eigenfunctions of the operators  $T(f)$, $\hat H(f)$ and $\tilde H(f)$ appearing in the last two equations \eqref{c8_2.2}.

For the neighbourhood $1<r<|f_0|$ of the cylinder we obtain   
$$
\Phi_{(0)}(r,\theta,t)=\frac{1}{2}\sum_{k=1}^{+\infty}\delta_k(t) \cdot 
\frac{1}{r^k}\cdot \cos(k\theta-\alpha)+
\frac{1}{2}\varepsilon_*\frac{f_0}{|f_H|}v_0(t)\cos\alpha +
	$$
$$
\frac{1}{2}\varepsilon_*\frac{f_0}{|f_0|}
\sum_{k=1}^{+\infty}v_k(t)\left[\sum_{j=0}^{+\infty}\left(\frac{-r}{2f_0}\right)^{k+j}
C_{k+j-1}^{k-1}\cos\left((k+j)\theta-\alpha\right)\right],  
	$$
or, taking into account the connection between $\delta_k(t)$ and $v_k(t)$ (see \eqref{c8_2.6}),
\begin{equation}
\left.\Phi_{(0)}\right|_{r=1+0}=
\varepsilon_*\frac{f_0}{2|f_0|}\left[v_0(t)\cos\alpha+
\sum_{n=1}^{+\infty}  
\left(\frac{-1}{f_0}\right)^nl_n(t)\cos(n\theta-\alpha)\right],                                
\label{c8_2.8}          
\end{equation}                                                               
$$
l_n(t)=\sum_{k=1}^n v_k(t)\cdot \frac{1}{2^{n-1}}C_{n-1}^{k-1}
	$$

We approximate the problem \eqref{c8_2.7} by the following initial value problem for infinite set of differential equations:
$$
a'_n(t)=-\frac{\varepsilon_*}{|f_0|}nv_n(t)+
\frac{1}{f_0^2}h'(t)\,\delta_{n,1},
        $$
$$
v'_n(t)=\frac{1}{2}(a_{n-1}+2a_n+a_{n+1})+\frac{1}{f_0}h''(t)\, \delta_{n,1},   
	$$	
$$
a_0(t)\equiv 0,\,\,\,\,\,\,a_n(0)=v_n(0)=0,\,\,\,\,\,\,n=1,\,2,\,\dots .
        $$        		
$$
 \frac{d^2a_n}{dt^2}+\frac{\varepsilon_*}{2|f_0|}n (a_{n-1}+2a_n+a_{n+1})=\frac{1}{f_0^2}
 \left(1-\varepsilon\frac{f_0}{|f_0|}\,n \right) \frac{d^2h}{dt^2}\delta_{n,1}
	$$
 Now new time-variable $\tau$ is defined by
 $$ 
 \tau=\sigma_*t,\,\,\,\,\,\,\sigma_*^2=\frac{\varepsilon_*}{2|f_0|}
 	$$
  and  the last equation  becomes
\begin{equation}
 \frac{d^2}{d\tau^2}   a_1+n\,(a_{0}+2a_1+a_{2})=
\frac{1}{f_0^2}\left(1 -\varepsilon_*\frac{f_0}{|f_0|} \right)
\,\frac{d^2}{d\tau^2}h(\tau)\,\,\,\,\,n\ge 1
\label{c8_2.9}         
\end{equation} 
$$
a_0(t)\equiv 0,\,\,\,\,\,\,a_n(0)=0,\,\,\,\,\,\,n=1,\,2,\,\dots .
        $$
It can be verified by direct substitution that exact solution to the problem \eqref{c8_2.9} is given by 
\begin{equation}
a_n(t)=(-1)^{n+1}\frac{1}{f_0^2}\left(1-\varepsilon_*\frac{f_0}{|f_0|}\right)a_n^{*}(t),
	$$ 
$$
a_n^{*}(t)=\sum_{j}\lambda^2_j[\alpha_jF_{n-1}(\tau,\lambda_j)+
\beta_jS_{n-1}(\tau,\lambda_j)],    
\label{c8_2.10}           
\end{equation}      
$$ 
\frac{\varepsilon_*}{|f_0|}v_n(t)=\frac{1}{nf_0^2}h'(t)\,\delta_{n,1}+
(-1)^n\frac{1}{nf_0^2}\left(1-\varepsilon_*\frac{f_0}{|f_0|}\right)
\frac{da_n^{*}}{dt},
	$$     
$$
\sigma_*=\left(\frac{\varepsilon_*}{2|f_0|}\right)^{1/2},\,\,\,\,\,\,
\tau=\sigma_* t,\,\,\,\,\,\,\lambda_j\sigma_*=\omega_j,\,\,\,\,\,\,
\lambda_i\tau=\omega_jt
        $$
$$
F_n(\tau,\lambda)=\int\limits_0^{+\infty}x^3e^{-x^2}L_n^{(1)}(x^2)
\frac{\cos (\lambda\tau)-\cos (x\tau)}{\lambda^2-x^2}\,dx,
        $$
$$
S_n(\tau,\lambda)=\int\limits_0^{+\infty}x^2e^{-x^2}L_n^{(1)}(x^2)
\frac{x\sin (\lambda\tau)-\lambda\sin (x\tau)}{\lambda^2-x^2}\,dx.
        $$
The integrals for $F_n$ and $S_n$ involve the Laguerre polynomials,
$\,\,L_k^{(1)}(u)\,\,$, defined by the recursion relation 
\begin{equation}
L_0^{(1)}(u)=1,\,\,\,L_1^{(1)}(u)=2-u,\,\,\,\,
	$$          
$$
k\ge 1\,\,\,\,\,\,
(k+1)L_{k+1}^{(1)}(u)-(2k+2-u)L_k^{(1)}(u)+(k+1)L_{k-1}^{(1)}(u)=0.
\label{c8_2.11}        
\end{equation}           

Formulas \eqref{c8_2.10} provide an approximate solution of the problem \eqref{c8_2.7} with a relative error of order $1/f_0^2$ since the terms 
$m_n(t)$ of order $1/f_0^{2n+2}$ have been omitted in \eqref{c8_2.7}.

By relations
$$
F_{n-1}(\tau;\lambda)=H_n(\tau ;\lambda)\cos (\lambda\tau)+
G_n(\tau ;\lambda)\sin (\lambda\tau),
        $$
$$
S_{n-1}(\tau,\lambda=-G_n(\tau, \lambda)\cos (\lambda\tau)+
H_n(\tau,\lambda)\sin (\lambda\tau)
        $$
we define functions $H_n(\tau,\lambda)$ and $G_n(\tau,\lambda)$. 
Decomposition of the fraction 
$$
\frac{2\lambda}{\lambda^2-x^2}=\frac{1}{\lambda+x}+\frac{1}{\lambda-x},
        $$
gives 
\begin{equation}
H_n(\tau,\lambda)=\int\limits_0^{+\infty}x^2e^{-x^2}L_{n-1}^{(1)} (x^2)f_1\,dx,
\label{c8_2.12}         
\end{equation}  
\begin{equation}
G_n(\tau,\lambda)=\int\limits_0^{+\infty}x^2e^{-x^2}L_{n-1}^{(1)} (x^2)f_2\,dx, 
\label{c8_2.13}       
\end{equation}  
\begin{equation}
f_1=-\frac{1}{\lambda+x}\sin^2\left(\frac{x+\lambda}{2}\tau\right)+
\frac{1}{\lambda-x}\sin^2\left(\frac{x-\lambda}{2}\tau\right),
\label{c8_2.14}          
\end{equation}  
\begin{equation}
f_2=\frac{1}{2}\,\frac{1}{x+\lambda}\sin((x+\lambda)\tau)-
\frac{1}{2}\,\frac{1}{x-\lambda}\sin((x-\lambda)\tau),        
\label{c8_2.15}           
\end{equation}   

\begin{equation}
a_n^*=\left[\sum_{j}\lambda^2_jA_j[ H_n(\tau,\lambda_j)\cos(\lambda_j\tau-\alpha)+G_n(\tau,\lambda_j)\sin(\lambda_j\tau-\alpha)]\right]
\label{c8_2.16}          
\end{equation} 
where
$$
a_j=A_j\cos\alpha,\,\,\,\,\,\,\beta_j=A_j\sin\alpha
	$$		
Equations \eqref{c8_2.1}, \eqref{c8_2.10}, \eqref{c8_2.16} lead  to
\begin{equation}
W_0(\theta,t)=4A\cos^2\theta \sum_{k=1}^{+\infty}(-1)^{k+1}
a_n^{*}(t)\cos(2k\theta-\alpha),
\label{c8_2.17}           
\end{equation} 
$$        
4A=\frac{1}{f_0^2}\left(1-\varepsilon_*\frac{f_0}{|f_0|}\right),        
	$$
The interface configuration is described (up to small terms 
of higher order) by the equations
$$
x=aW_0(\theta,t),\,\,\,\,\,\,y=(x-f)\tan\theta +\varphi(t).
	$$	

\section{ Summing up of the double series solution to the leading-order equations}   

Function $W_0(\theta,t)$ depends on $h(t)$ linearly. 
For the linearity, we shall confine our attention to the case when in \eqref{c8_2.1} 
$$
h(t)=\cos (\omega t)-1.                           
        $$

\noindent      
By \eqref{c8_2.10} and \eqref{c8_2.17} in this case (for symplicity $\alpha=0$)
$$
z_n^*(t)=\lambda^2\left[H_n(\tau,\lambda)\cos (\lambda\tau)+
G_n(\tau,\lambda)\sin (\lambda\tau)\right],
        $$
\begin{equation}
W_0(\theta,t)=A\cos\alpha\,\cdot\lambda^2
\left[R_1(\tau,\theta;\lambda)\cos (\lambda\tau)+
Q_1(\tau,\theta;\lambda)\sin (\lambda\tau)\right]+                      
	$$  
$$
A\sin\alpha\,\cdot\lambda^2
\left[R_2(\tau,\theta;\lambda)\cos (\lambda\tau)+
Q_2(\tau,\theta;\lambda)\sin (\lambda\tau)\right],
\label{c8_3.1}            
\end{equation} 
$$
\sigma_* =\left(\frac{\varepsilon_*}{2|f_0|}\right)^{1/2},\,\,\,\,\,\,
\tau=\sigma_* t,\,\,\,\,\,\,\lambda\sigma_* =\omega ,\,\,\,\,\,\,
\lambda\tau=\omega t,
        $$
 where  (with $i^2=-1$)
\begin{equation}
R_1+iR_2=4\cos^2\theta\cdot \sum_{k=1}^{+\infty}(-1)^{k-1}
H_k(\tau ,\lambda)\exp{(i2k\theta)}, 
\label{c8_3.2}         
\end{equation} 
\begin{equation}
Q_1+iQ_2=4\cos^2\theta\cdot\sum_{k=1}^{+\infty}(-1)^{k-1}
G_k(\tau,\lambda)\exp{(i2k\theta)}.  
\label{c8_3.3}        
\end{equation} 
\vspace{5mm}
    
\noindent
Having decomposed the products of trigonometrical functions into sums, we rearrange the series to obtain 
\begin{equation}
\hat R_1+i\hat R_2=\sum_{k=0}^{+\infty}(-1)^k
h_k(\tau,\lambda)\exp{(i2k\theta)}, 
\label{c8_3.4}      
\end{equation}   
\begin{equation}
\hat Q_1+i\hat Q_2=\sum_{k=0}^{+\infty}(-1)^k
g_k(\tau, \lambda)\exp{(i2k\theta)},    
\label{c8_3.5}           
\end{equation}            
$$
h_k=H_{k-1}-2H_k+H_{k+1},\,\,\,\,\,\,H_{-1}=H_0=0,
        $$
$$
g_k=G_{k-1}-2G_k+G_{k+1},\,\,\,\,\,\,G_{-1}=G_0=0.
        $$
\vspace{5mm}

\noindent
By recursion relation for Laguerre polynomials, it follows from 
\eqref{c8_2.12} and \eqref{c8_2.13} that
$$
h_0(\tau,\lambda)=\int\limits_0^{+\infty}x^2e^{-x^2}f_1\,dx,                                                   
\,\,\,\,\,\,\,\,\,  g_0(\tau,\lambda)=\int\limits_0^{+\infty}x^2e^{-x^2}f_2\,dx,  
      $$
\begin{equation}
h_k(\tau,\lambda)=-\frac{1}{k}\int\limits_0^{+\infty}x^4e^{-x^2}
L_{k-1}^{(1)}(x^2)f_1\,dx,   
\label{c8_3.6}        
\end{equation}   
\begin{equation}
g_k(\tau,\lambda)=-\frac{1}{k}\int\limits_0^{+\infty}x^4e^{-x^2}
L_{k-1}^{(1)}(x^2)f_2\,dx.  
\label{c8_3.7}         
\end{equation} 
\newpage

{\bf Theorem 1.} In any rectangle $0\le \tau \le T$, 
$0<\delta\le |\theta|\le\pi/2-\delta$
\begin{description}
\item
(i) series \eqref{c8_3.2} and \eqref{c8_3.3} converge
uniformly with regard to $ \theta $ and $ \tau $,
\item
(ii) series \eqref{c8_3.4} and \eqref{c8_3.5} converge absolutely and 
uniformly with regard to $ \theta $ and $ \tau $,
\item
(iii) series \eqref{c8_3.2} and \eqref{c8_3.4} (respectively, \eqref{c8_3.3} and \eqref{c8_3.5}) converge to the same sum.
\end{description}
\vspace{3mm}

{\bf Theorem 2.} On  the interval $\,|\theta|<\pi/2\,$ 
the following formulas hold
\begin{equation}
R_1+iR_2=\int\limits_0^{+\infty}v(x)\exp{\left(i\frac{1}{2}x^2
\tan\theta\right)}f_1\,dx,   
\label{c8_3.8}         
\end{equation}     
\begin{equation}
Q_1+iQ_2=\int\limits_0^{+\infty}v(x)\exp{\left(i\frac{1}{2}x^2
\tan\theta\right)}f_2\,dx ,  
\label{c8_3.9}       
\end{equation}    
$$
v(x)=x^4e^{-\frac{1}{2}x^2},
        $$
$f_1,\,\,f_2$ are specified by \eqref{c8_2.14} and  \eqref{c8_2.15} respectively.
\bigskip

{\it Proof of Theorem 1.}
 The difference between the $n$-th partial sums of series \eqref{c8_3.2} and \eqref{c8_3.4} is equal to $(-1)^{n-1}[\exp {(i2n\theta)}H_{n+1}+\exp{(i2(n+1)\theta)}H_n]$.

By asymptotic formula \eqref{c2_3.4} of Chapter 2, it follows from \eqref{c8_2.12} that on any interval $0\le \tau \le T$ 
$H_n=O\left(\frac{1}{n^{1/4}}\right)$.      
    
This means, that  series \eqref{c8_3.2} and \eqref{c8_3.4} either converge to the same sum or diverge.

With the use of \eqref{c8_2.12} and \eqref{c8_3.6} we obtain 
$h_n=O\left(\frac{1}{n^{5/4}}\right)$,      
and consequently, series \eqref{c8_3.4} converges absolutely and uniformly, and series \eqref{c8_3.2} converges uniformly to the same sum. 

Convergence of the series \eqref{c8_3.3} and \eqref{c8_3.5} can be proved in a similar way. 
\bigskip

{\it Proof of Theorem 2.}         
By the  Lemmas 1 and 2 of Chapter 2 (subsection 2.2),    
 the series on the left side of the following equation 
 $$
1-x^2\,\sum_{k=1}^{+\infty}(-1)^k\frac{1}{k}L_{k-1}^{(1)}(x^2) \exp{(i2k\theta)}
= \exp{\left(\frac{1}{2}x^2\right)}\exp{\left(i\frac{1}{2}x^2\tan\theta\right)}.
        $$
 converges uniformly in any domain $0\le x\le a$,   
$|\theta|\le \pi /2-\delta$, so the equation may be integrated to give
$$
\int\limits_0^{a}x^4e^{-x^2/2}\exp{\left(i\frac{1}{2}x^2\tan\theta\right)}f_1\,dx=
        $$
$$
\int\limits_0^{a}x^2e^{-x^2}\left[1-x^2\,\sum_{k=1}^{+\infty}(-1)^k 
\frac{1}{k}L_{k-1}^{(1)}(x^2)\exp{(i2k\theta)}\right]f_1\,dx=
        $$
$$
\int\limits_0^{a}x^2e^{-x^2}f_1\,dx-
\sum_{k=1}^{+\infty}(-1)^k\frac{1}{k}\exp{(i2k\theta)}
\int\limits_0^{a}x^4e^{-x^2}L_{k-1}^{(1)}(x^2)f_1\,dx.
        $$
Proceeding to the limit when $\,a \to+\infty\,$ we come to equality \eqref{c8_3.8}.

Equality \eqref{c8_3.9} can be proved in a similar way.

It follows from  Theorem 2 that 
\begin{equation}
W_0(\theta,t)=A\lambda^2
\left[R(\tau,\theta;\lambda,\alpha)\cos (\lambda\tau)+
Q(\tau,\theta;\lambda,\alpha)\sin (\lambda\tau)\right],      
  \label{c8_3.10}         
\end{equation} 
\begin{equation}
R(\tau,\theta;\lambda,\alpha)=\int\limits_0^{+\infty}x^4e^{-\frac{1} {2}x^2} f_1
\cos\left(\frac{1}{2}x^2\tan\theta-\alpha\right)\,dx,                      \label{c8_3.11}      
\end{equation} 
\begin{equation}
Q(\tau,\theta;\lambda,\alpha)=\int\limits_0^{+\infty}x^4e^{-\frac{1} {2}x^2}f_2
\cos\left(\frac{1}{2}x^2\tan\theta-\alpha\right)\,dx,                     
  \label{c8_3.12}        
\end{equation} 
$$
A=\frac{1}{4f_0^2}\left(1-\varepsilon_*\frac{f_0}{|f_0|}\right),
        $$
functions $f_1$ and $f_2$ are specified by \eqref{c8_2.14} and 
\eqref{c8_2.15} respectively.
\begin{figure}
\centering
	\resizebox{0.7\textwidth}{!}
	{\includegraphics{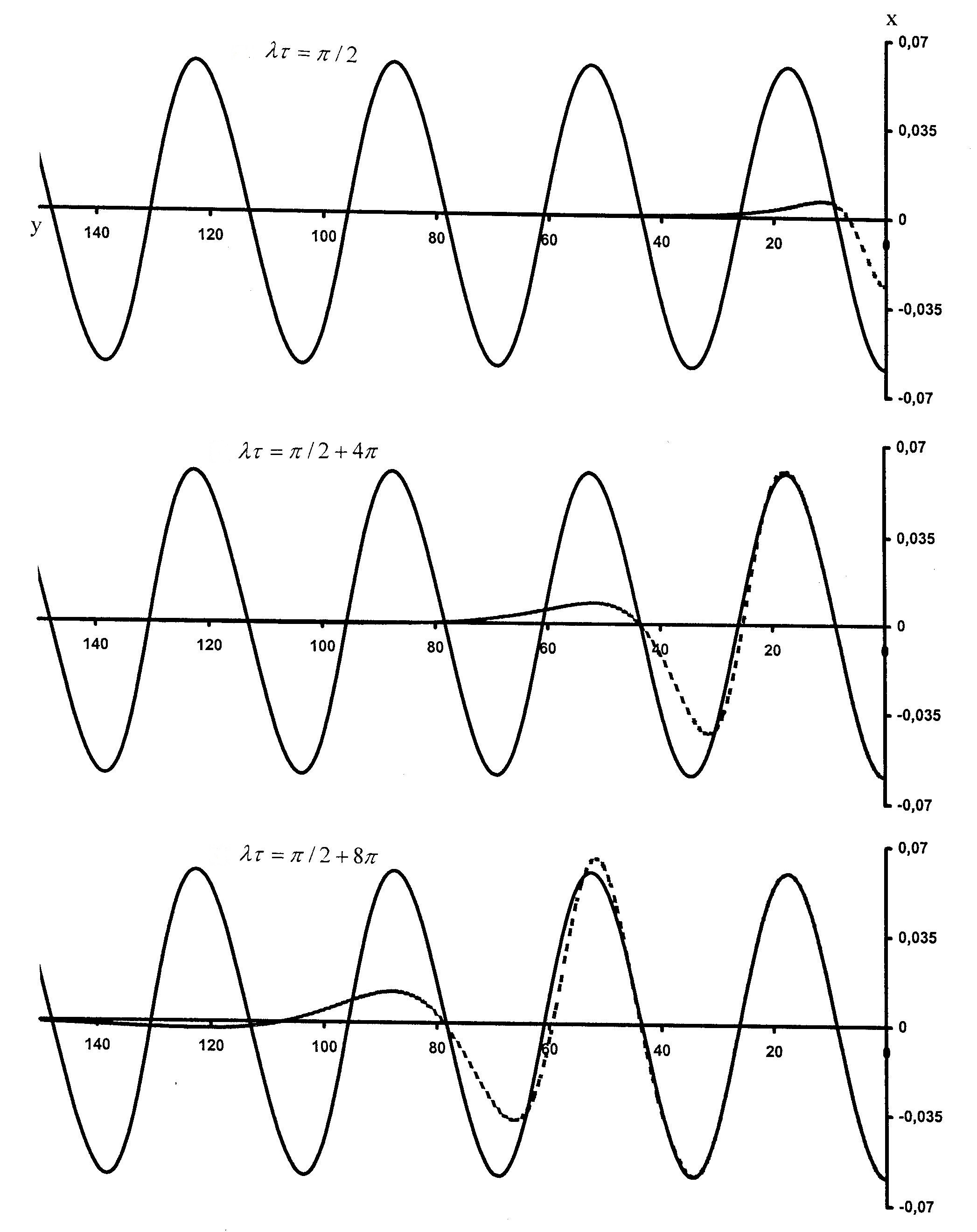}}
	\caption{Solid lines: steady-state wave, dashed line -  transients,
	 $\omega=0.414$ ($\lambda=1.9$) }
	\end{figure} 

For oscillations of general form, setting $\alpha_j=c_j\,\cos\delta_j$ and $\beta_j=c_j\,\sin\delta_j$ in \eqref{c8_2.1} and using \eqref{c8_2.15}, we find 
$$
W_0(\theta,t)=A\sum_j c_j\lambda_j^2
\left[R_j\cos (\lambda_j\tau-\delta_j)+
Q_j\sin (\lambda_j\tau-\delta_j)\right],    
        $$
$$
R_j=R(\tau,\theta;\lambda_j,\alpha),\,\,\,\,\,\,
Q_j=Q(\tau,\theta;\lambda_j,\alpha)
	$$        
$$
\sigma_*=\left(\frac{\varepsilon_*}{2|f_0|}\right)^{1/2},\,\,\,\,\,\,
\tau=\sigma_* t,\,\,\,\,\,\,\lambda_j\sigma_*=\omega_j.
        $$      
Figures 8.2 - 8.4 illustrate development of waves generated 
on the interface by the cylinder which starts to oscillate harmonically with frequencies $\omega=0,414$ ($\lambda=1.9$) and 
$\omega=0,654$ ($\lambda=3$) respectively.

The Figures  represent instantaneous profiles of the evolving interface shape by the dashed lines obtained from equations  \eqref{c8_3.10} - \eqref{c8_3.12}    at $\alpha=0$,
$a=1,\,$ $f_0=-10,\,$ $\varepsilon_*=0.95,\,$ $\sigma_*=0.218$. 
while the solid lines represent the steady-state shape obtained from equation \eqref{c8_4.6}. 
The profiles are symmetrical about the vertical $x$-axis. 

From the figures we see that the transients die out (leaving the final steady-state wave on the interface) 
in the region $0<y<100$ by the time $t=20.5\pi/0.414=125.14$ 
at the frequency $\omega=0.414$ (Fig. 8.2), and 
by the time $t=16.5\pi /0.654=156.04$ 
at the frequency $\omega=0.654$ (Fig. 8.4).  
It is shown below in section 5 that maximum of the energy supplied to the liquid by the  cylinder per unit of time (on average) is reached at resonant frequency (at $\lambda\approx 1.9$, see Fig. 8.7). 
\bigskip

{\bf Theorem 3.} At any fixed value of $\tau$  
functions $R$  \eqref{c8_3.11} and $Q$ \eqref{c8_3.12} satisfy boundary conditions \eqref{c8_1.5}.
\bigskip

{\it Proof of Theorem 3.} Write
$$
R(\tau,\theta;\lambda,\alpha)=\int\limits_0^{+\infty}F_1\,dV,\,\,\,\,\,\,
Q(\tau,\theta;\lambda,\alpha)=\int\limits_0^{+\infty}F_2\,dV
	$$
$$
dV=xe^{-\frac{1} {2}x^2} 
\cos\left(\frac{1}{2}x^2u-\alpha\right)\,dx, 
	$$
$$
F_1=x^3f_1,\,\,\,\,\,\,F_2=x^3f_2,\,\,\,\,\,\,u=\tan\theta
	$$	

It is not difficult to check that	
$$
(1+u^2)V=-\left(e^{-\frac{1} {2}x^2}\left[\cos\left(\frac{1}{2}x^2u-\alpha\right) -u\sin\left(\frac{1}{2}x^2u-\alpha\right)\right] \right)
	$$ 
\begin{figure}
\centering
	\resizebox{0.8\textwidth}{!}
	{\includegraphics{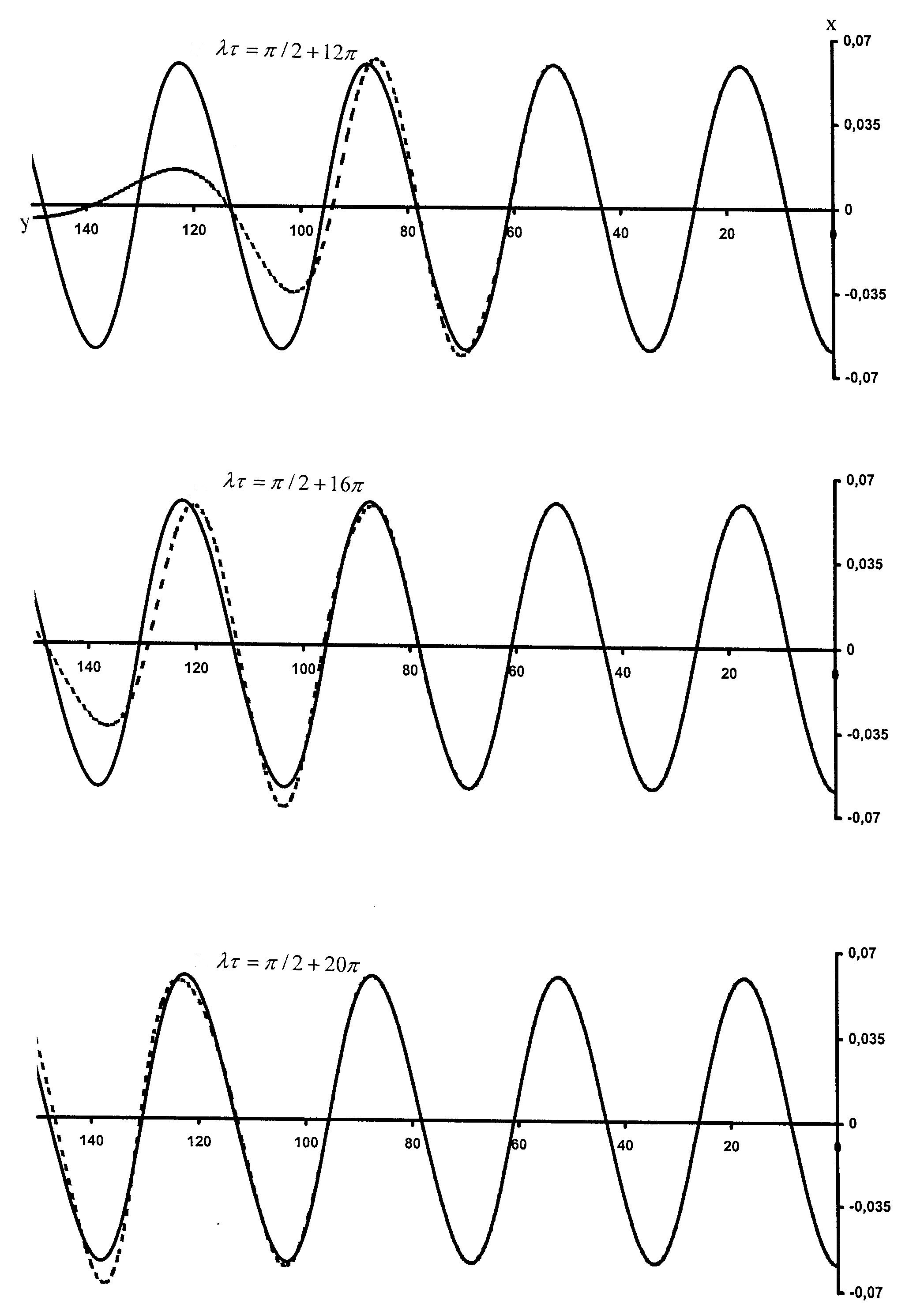}}
	\caption{Solid lines: steady-state wave, dashed line -  transients,
	 $\omega=0.414$ ($\lambda=1.9$) }
	\end{figure} 
\begin{figure}
\centering
	\resizebox{0.8\textwidth}{!}
	{\includegraphics{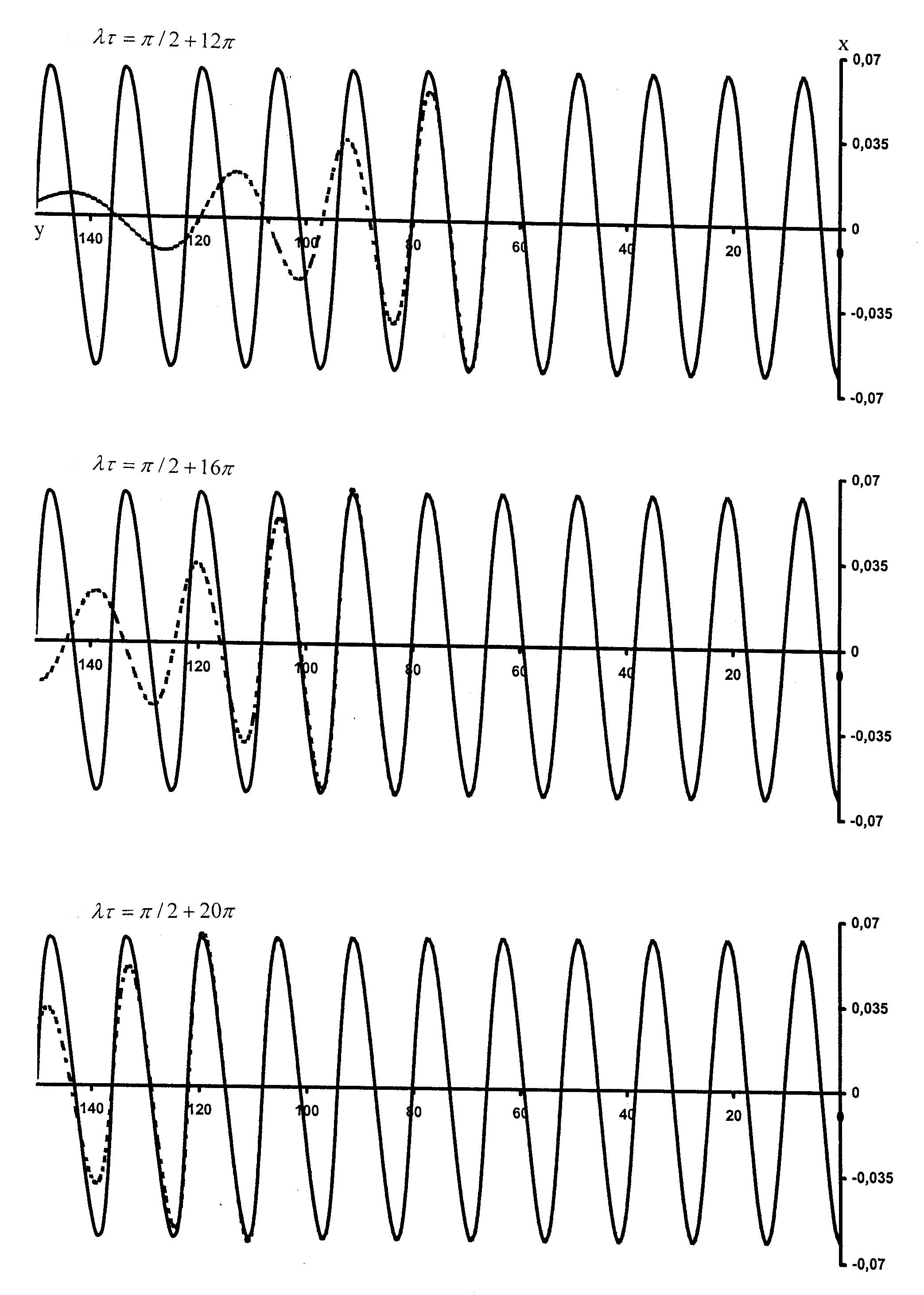}}
	\caption{Same as Figure 8.2, but $\omega=0.654$\,\, ($\lambda=3$)} 
	\end{figure}		
	
Integration by parts gives
 $$
 R=\cos^2\theta\int\limits_0^{+\infty}
e^{-\frac{1} {2}x^2}\left[\cos\left(\frac{1}{2}x^2u-\alpha\right) -u\sin\left(\frac{1}{2}x^2u-\alpha\right)\right]\,dF_1
	$$	
 $$
Q=\cos^2\theta\int\limits_0^{+\infty}
e^{-\frac{1} {2}x^2}\left[\cos\left(\frac{1}{2}x^2u-\alpha\right) -u\sin\left(\frac{1}{2}x^2u-\alpha\right)\right]\,dF_2
	$$
The integrals are bounded on any interval $0\le t\le T$ since  		
the functions $f_1$ and $f_2$ may be expanded in power series of $x-\lambda$ with infinite radius of convergence, and the series converge uniformly throughout the domain $ x\ge 0,\,\,\,\,0\le \tau\le T$.
This remark completes the proof.

Theorem 3 means that the function $W_0(\theta,t)$ \eqref{c8_3.10}  satisfies the boundary conditions \eqref{c8_1.5}.

\section{ Asymptotic behavior of forced waves as the time increases without bound. Steady-state waves.}   
                                                                 
Application of Lemma 3 of Chapter 7 (section 4) leads to   

{\bf Theorem 4.} In the interval $|\theta| <\pi/2$
                            
(i) There exist limits
\begin{equation}
R^{*}(\theta;\lambda,\alpha)=\lim\limits_{\tau \to +\infty}
R(\tau, \theta;\lambda,\alpha),\,\,\,\,\,\,
Q^{*}(\theta;\lambda,\alpha)=\lim\limits_{\tau \to +\infty}Q(\tau,
\theta;\lambda,\alpha);     
  \label{c8_4.1}      
\end{equation}  
(ii) the limits are given by expressions
\begin{equation}
R^{*}(\theta;\lambda,\alpha)=I_1+I_2+I_3, 
  \label{c8_4.2}       
\end{equation}    
\begin{equation}
I_1=\int\limits_0^{+\infty}\frac{G(x)}{x+\lambda}dx,\,\,\,\,
I_3=\int\limits_{2\lambda}^{+\infty}\frac{G(x)}{x-\lambda}dx, \,\,\,\,\,\,
I_2=\int\limits_0^{\lambda}\frac{G(\lambda +s)-G(\lambda -s)}{s}ds,
  \label{c8_4.3}       
\end{equation}  
\begin{equation}
Q^*(\theta;\lambda,\alpha)=-\frac{1}{2}\pi G(\lambda),\,\,\,\,\,\,  
 \label{c8_4.4}         
\end{equation}
\begin{equation}
G(x)=v(x)\cos\left(\frac{1}{2}x^2\tan\theta-\alpha\right),\,\,\,\,\,\,
v(x)=x^4e^{-\frac{1}{2}x^2}  
 \label{c8_4.5}         
\end{equation}

In the case of \eqref{c8_3.1} when the cylinder oscillates harmonically, 
the limit wave is described by the equations
\begin{equation}
x=aW_0^*(\theta,\tau),\,\,\,\,\,\, y=(x-f)\tan\theta +\varphi(t),
	$$
$$
W_0^*(\theta,t)=A\lambda^2[R^*(\theta;\lambda,\alpha)\cos(\lambda\tau)
+Q^*(\theta;\lambda,\alpha)\sin(\lambda\tau)].     
 \label{c8_4.6}         
\end{equation}
\begin{figure}
\centering
      	\resizebox{0.7\textwidth}{!}
	 	{\includegraphics{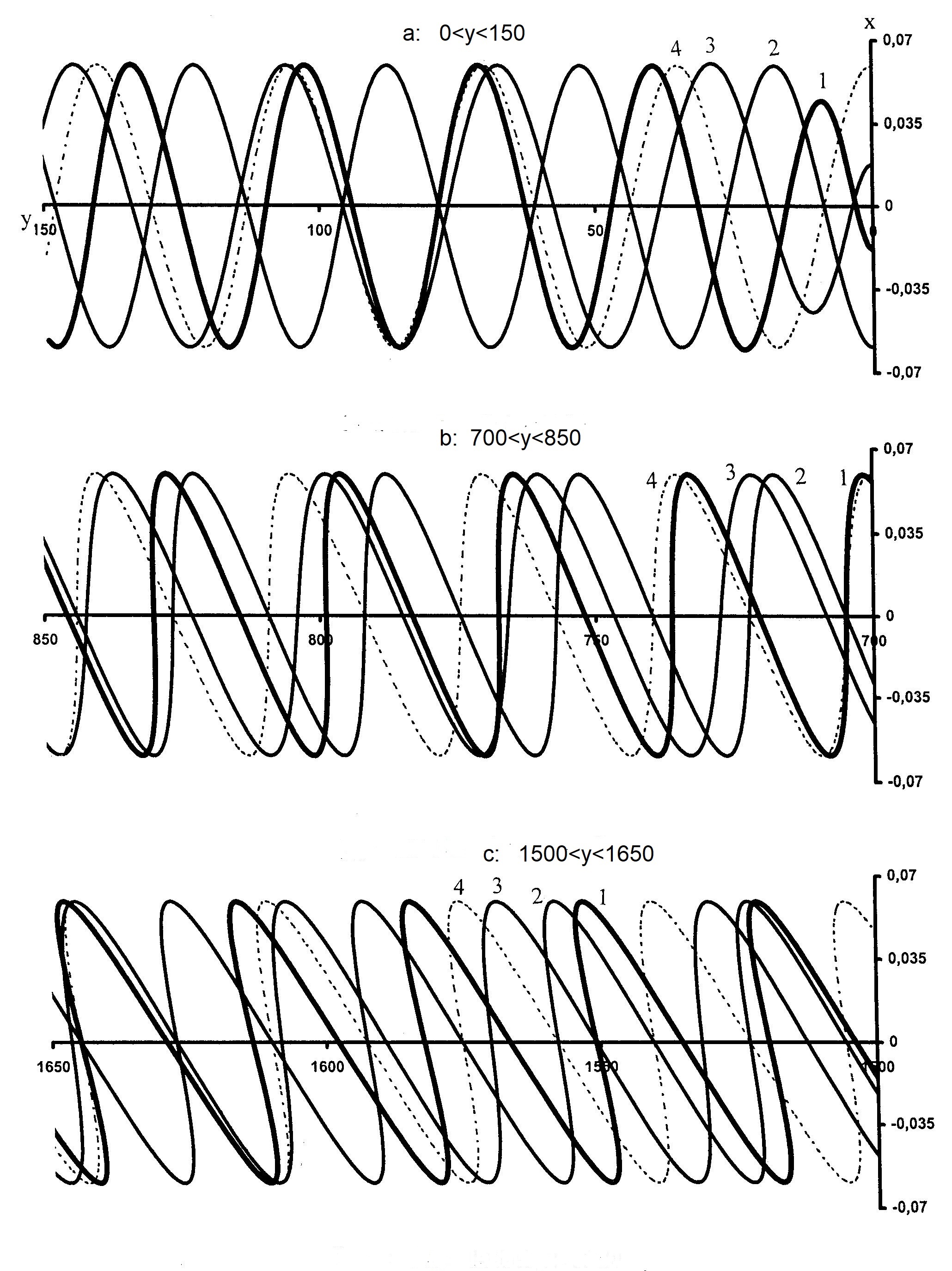}}    
	  \caption{Successive profiles of the limit wave \eqref{c8_4.6},
at four instants, a quarter of the period apart:
1: $\lambda\tau=0,\,\,$ 2: $\lambda\tau=\pi/2,\,\,$ 3: $\lambda\tau=\pi,\,\,$ 4: $\lambda\tau=3\pi/2$, and at different distances from the cylinder ($a=1,\,\,\alpha=0$). }
	 \end{figure} 	
Figure 8.5 shows that  when the wave runs away from the cylinder, the crests of the wave move faster than the troughs, so the horizontal distance between a crest  and the following trough increases;  the wave may steepen sufficiently that overturning occurs. 

Theorem 3 shows that $R(\tau,\theta;\lambda,\alpha)\rightarrow 0$ 
and $Q(\tau,\theta;\lambda,\alpha)\rightarrow 0$ as $|\theta|\rightarrow \pi/2$,  
$t$ being considered as constant. On the other hand, by Theorem 4,
$R\rightarrow R^*$ and $Q\rightarrow Q^*$ as $t\rightarrow+\infty$, 
treating $\theta$ as a constant. Thus, for sufficiently
large values of $t$, there exists a region $|\theta|<|\theta_1|$
where the advancing wave on the interface is close to the steady-state one, and there exists 
a region $|\theta|>|\theta_2|>|\theta_1|$
where the interface displacement from its equilibrium position
is very small. In the course of time, the size of the first region increases
gradually.

\section{The  thrust on the oscillating cylinder.} 

We use the Bernoulli's equation and
 the normal-velocity condition  \eqref{c8_1.2} at the surface of the cylinder, to find the pressure $P_{+}$ on the surface     
$$
P_{+}=\left. P(r,\theta,t)\right|_{r=1+0}=-\gamma_{+}
\left[\pd {\Phi}{t_+}-\frac{1}{2}a^2(h'(t))^2\cos^2(\theta-\alpha)\right. +
        $$
$$
ah'(t)\sin(\theta-\alpha)\pd {\Phi}{\theta_+}+\frac{1}{2}\left(\pd{\Phi}
{\theta_+}\right)^2+\cos\theta+\left.\vphantom{\pd {\Phi}{t_+}}f(t)\right]+
P_0(t),
        $$
where $\gamma_{+}$ is the density of the liquid surrounding the cylinder. 

It follows from \eqref{c8_2.8} that one-sided limit of the velocity potential is 
$$
\Phi_{+}=a\varepsilon_*\frac{f_0}{2|f_0|}\left[v_0(t)\cos\alpha+
\sum_{n=1}^{+\infty}\left(\frac{-1}{f_0}\right)^nl_n(t)\cos(n\theta-\alpha)+O(\varepsilon)\right]
        $$
$$
l_k(t)=\sum_{j=1}^{k}v_j(t)\cdot \frac{1}{2^k}C_k^j,\,\,\,\,\,\,
C_k^j=\sum\limits_{j=1}^k\,\frac{k!}{(k-j)!j!}.
        $$
The pressure at the cylinder may be integrated to give vertical $R_x$ and 
horizontal $R_y$  components of the force (per unit length of the cylinder)
due to liquid-liquid interface:
$$
R_x=-\gamma_+\pi a\varepsilon_*\frac{1}{|f_0|}[\cos\alpha\cdot l'_1(t)+O(\varepsilon)],
        $$
$$
R_y=-\gamma_+\pi a\varepsilon_*\frac{1}{|f_0|}[\sin\alpha\cdot l'_1(t)+
O(\varepsilon)].
        $$
Thus, the wave resistance, $R$, is equal to
\begin{equation}
R=R_x\cos\alpha+R_y\sin\alpha=-\gamma_{+}\pi a\varepsilon_*\frac{1}{|f_0|}
[l'_1(t)+O(\varepsilon)];
\label{c8_5.1}
\end{equation}
The force normal to the velocity of the cylinder (the "lift") is equal to
$$
F=R_x\sin\alpha-R_y\cos\alpha=(a/|f_0|)O(\varepsilon).
        $$

If $h(t)=\cos(\omega t)$, then by \eqref{c8_2.9} 
$$
l_1=\frac{1}{2}v_1(t),\,\,\,\,\,\,
v_1'=\frac{1}{2}(2a_1+a_2)-\frac{\omega^2}{f_0}\cos(\omega\,t).
	$$
By \eqref{c8_2.10}
$$
a_1+\frac{1}{2}\,a_2=\frac{1}{f_0^2}\left(1-\varepsilon_*\frac{f_0}{|f_0|}\right)\left(a_1^*-\frac{1}{2}a_2^*\right)
	$$	
$$  
a_1^*-\frac{1}{2}a_2^*=\left[\sum_{j}\lambda^2_jA_j
[N_j(\tau;\lambda_j) \cos(\lambda_j\tau-\alpha)+
M_j(\tau;\lambda_j)\sin(\lambda_j\tau-\alpha)\right]
	$$	
$$
N_j(\tau;\lambda_j)=[H_1(\tau;\lambda_j)-\frac{1}{2}H_2^*(\tau;\lambda_j)]=
\frac{1}{2}\int\limits_0^{+\infty}x^4e^{-x^2}f_1\,dx,
	$$			
$$
M_j(\tau;\lambda_j)=[G_1(\tau;\lambda_j)-\frac{1}{2}G_2^*(\tau;\lambda_j)]=
\frac{1}{2}\int\limits_0^{+\infty}x^4e^{-x^2}f_2\,dx,
	$$			
		
Let $Q_n(\omega)$ and $Q(\omega)$ denote the power of propulsive force averaged 
over $n$ periods and the limit of the power when $n$ increases without bound. 
For large values of $|f_0|$, using \eqref{c8_2.10} we find 
\begin{equation}
Q_n(\omega)=-\gamma_+\frac{a^2}{2|f_0|^3}\varepsilon_*
\left(1-\varepsilon_*\frac{f_0}{|f_0|}\right)V_n(\omega)<0,
        $$                                                 
$$
V_n(\omega)=-\pi\omega\cdot\frac{1}{nT}\int\limits_{0}^{nT}
\left[a_1^{*}(t)-\frac{1}{2}a_2^{*}(t)\right]\sin(\omega t)\,dt.
 \label{c8_5.2}            
\end{equation}   
Performing the integration with respect to $t$, we get
$$
V_n(\lambda\sigma_*)=\frac{\sigma_*\lambda^3}{2n}
\int\limits_0^{+\infty}x^5e^{-x^2}\frac{1}{(\lambda^2-x^2)^2}
\sin^2\left(2\pi n\,\frac{x}{\lambda}\right)\,dx >0,
	$$
$$
\sigma_*=\left(\frac{\varepsilon_*}{2|f_0|}\right)^{1/2},\,\,\,\,\,\,
\lambda_j\sigma_*=\omega_j.
        $$
Existence of the limits       
$$
\lim_{\tau \to +\infty}N_j(\tau;\lambda_j)(\tau;\lambda)=N_j(\tau;\lambda_j)^*(\lambda),
	$$
$$	
\lim_{\tau \to +\infty}M_j(\tau;\lambda_j)=M_j(\tau;\lambda_j)^*(\lambda)=
-\frac{1}{2}\pi\lambda^2e^{-\lambda^2}L_{n-1}^{(1)}(\lambda^2) 
	$$
is assured by Lemma 3 of Chapter 7.

This result leads to 
$$
\lim_{\tau \to +\infty}\left\{a_1^{*}(t)-\frac{1}{2}a_2^{*}(t)-
\lambda^2[(H_1^*-\frac12H_2^*)\cos(\lambda\tau)+
(G_1^*-\frac{1}{2}G_2^*)\sin(\lambda\tau)]\right\}=0,
	$$
$$
G_1^*-\frac{1}{2}G_2^*=\frac{1}{2}\pi\lambda^5e^{-\lambda^2}
	$$	
and, consequently, by \eqref{c8_5.1}
\begin{equation}
Q(\omega)=-a^2 \nu V(\lambda),\,\,\,\,\,\,
V(\lambda)\equiv \frac{1}{8}\pi^2\sigma_{*}\lambda^7e^{-\lambda^2},  
 \label{c8_5.3}            
\end{equation}   
$$
\nu =\gamma_+\frac{a^2}{2|f_0|^3}\varepsilon_* 
\left(1-\varepsilon_*\frac{f_0}{|f_0|}\right).
	$$
	
According to \eqref{c8_5.2}
, $-Q_n$ is the average energy absorbed 
by the liquid per period  during the first $n$ periods of the cylinder oscillations.

Figure 8.7 displays graphs of $V_n$ for different values of $n$ shown in the figure above the corresponding curves. 
\begin{figure}
\centering
      	\resizebox{0.7\textwidth}{!}
	 	{\includegraphics{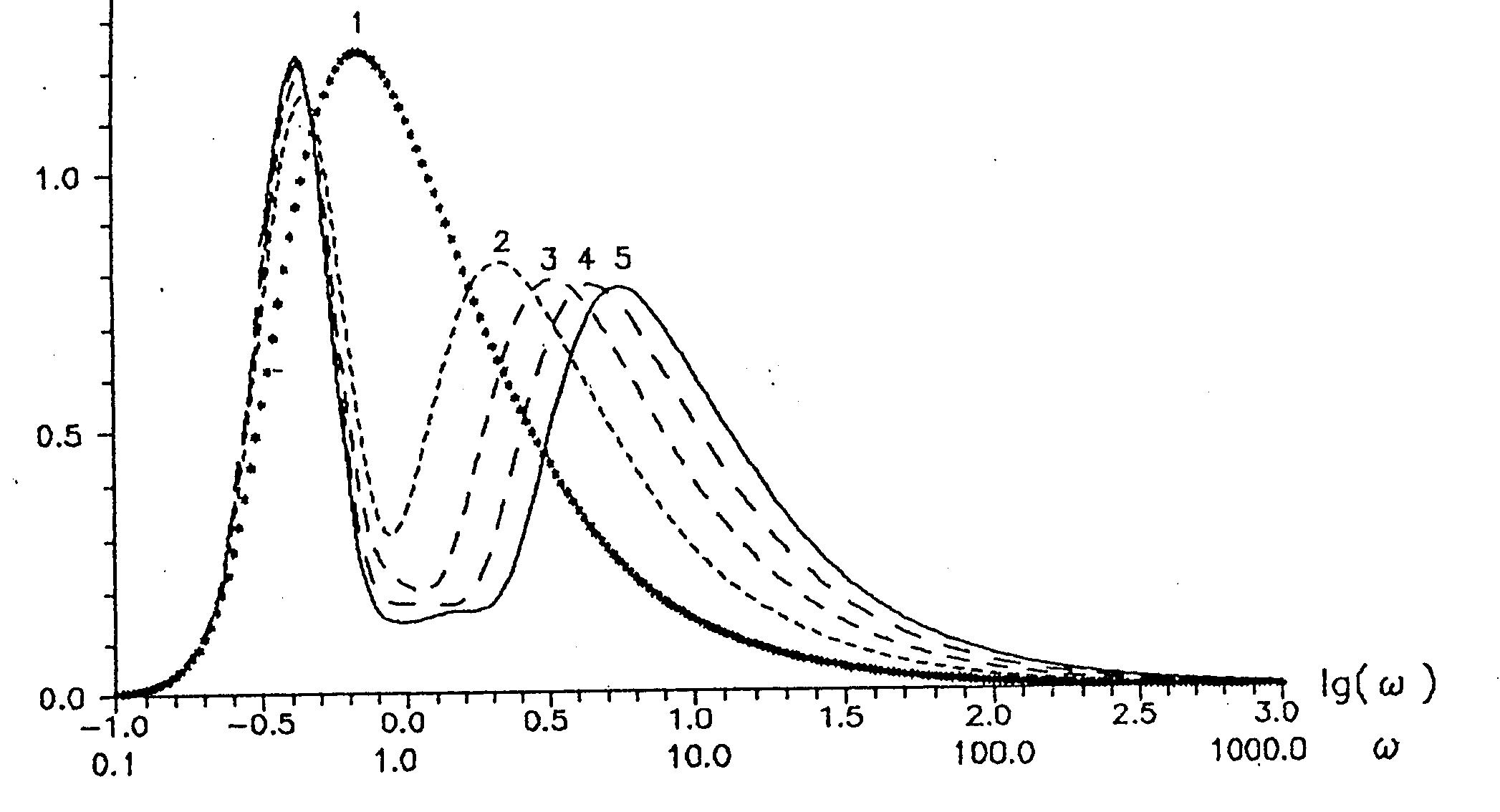}}   
	  \caption{  Graphs of $V_n$ given by \eqref{c8_5.2} 
versus $\,\lg \omega$; 
$n=1 - 5,\,\,$ $\omega$ is the frequency of the cylinder oscillations.}
	 \end{figure} 	  
\begin{figure}
	\centering
      	\resizebox{0.7\textwidth}{!}
	 	{\includegraphics{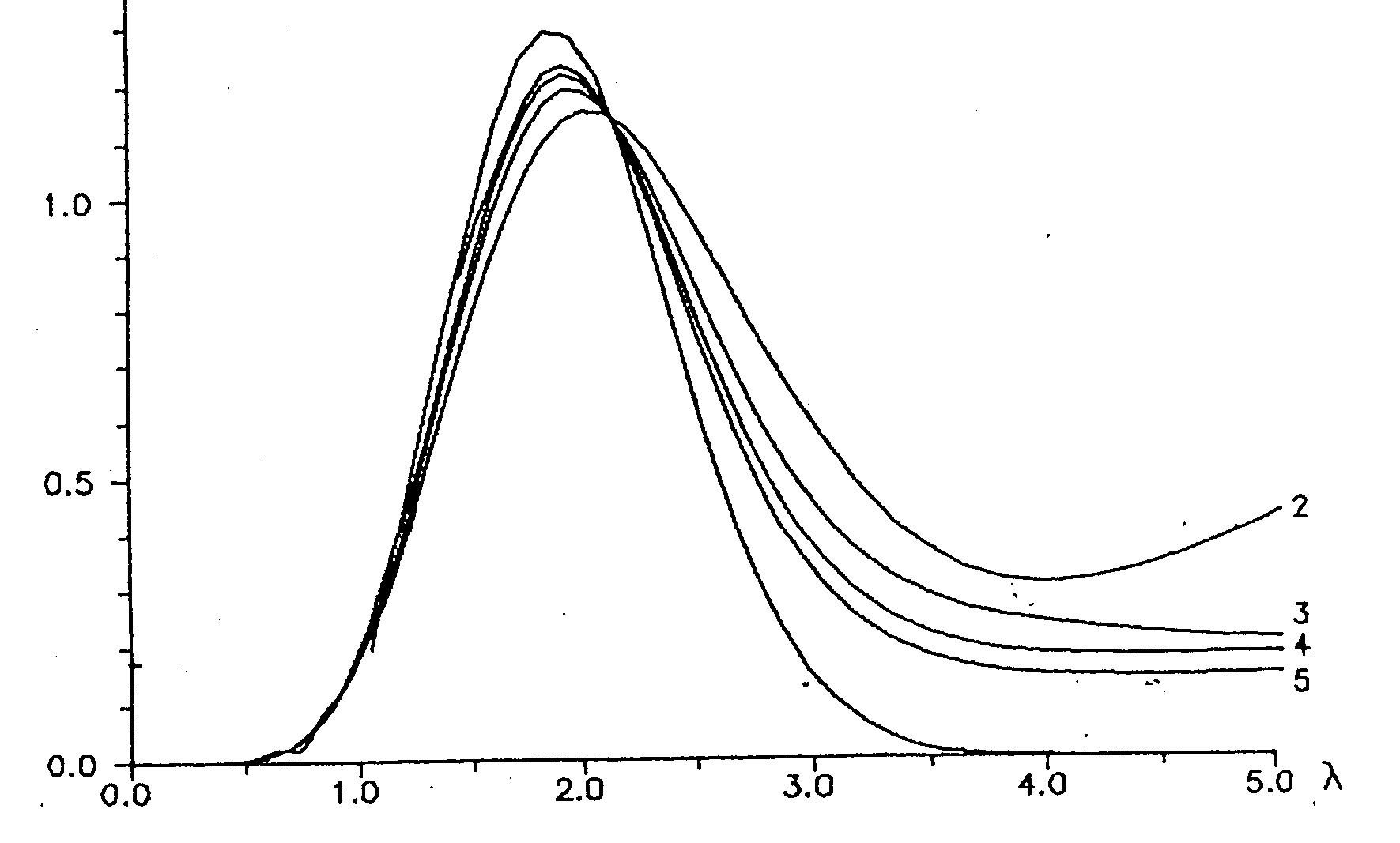}}    
	  \caption{Graphs $V(\lambda)$ given by \eqref{c8_5.3}
and  $V_n(\lambda\sigma_*)$  
versus $\,\lambda$;  $n=1 - 5,\,\,$; $\sigma_*=0,218$.}
	 \end{figure}    
At large values of $n$, $V_n(\lambda\sigma_*)$ is close to 
$V(\lambda)$. 	 	 

In dimensional units the partial averaged energy absorbed by the liquid is
$$
F_*(\Omega)A^2=\frac{1}{2}\pi^2\gamma_+\frac{A^2}{\varepsilon_*^{2}}
\left(1-\varepsilon_*\frac{f_0}{|f_0|}\right)\frac{R_0^5}{g^2}
\Omega^7\exp 
\left(-2\frac{l}{\varepsilon_*}\Omega^2\right).
        $$
where $R_0$ is the radius of the cylinder, $\Omega$ is the angular frequency 
of harmonic mode in \eqref{c8_2.1}, $A$ is the amplitude of the harmonic mode, 
$l=|f_0|R_0>>R_0$ is the distance between the equilibrium position of the interface 
and the cylinder,  $g$ is acceleration due to gravity, 
$\gamma$ is the density of the liquid surrounding the cylinder. 

If the cylinder oscillates below the free surface (at $\varepsilon_*=1$), then  
$$
F_*(\Omega)A^2=\pi^2\gamma A^2\frac{R_0^5}{g^2}\,\Omega^7\exp 
\left(-2\frac{l}{g}\,\Omega^2\right).
        $$
Figure 8.5 displays graphs of $V_n$ for different values of $n$ shown in the figure above the corresponding curves.  
According to \eqref{c8_3.5}, $V_n$ is proportional to the average energy absorbed by the liquid per period  during the first $n$ periods of the cilynder oscillations.

Figure 8.6 shows graphs of $V(\lambda)=V_n (\lambda\sigma_*)$ at 
$\sigma_*=\sqrt {0.95/20}=0.22$.
For $\sigma_*=0.22$, interval $0<\lambda <5$ corresponds to the frequency interval $0<\omega <2.2$. 
In the interval $\lambda >4$, the function $V(\lambda)$ is practically equal to zero. 
The graphs of $V$ and $V_n$ intersect at the point 
$\lambda_0\approx 2,2$  (practically the same for $n=2,\,3,\,4$). From Figure 8 one can see that for $\omega$ fixed in the interval 
$\omega >\omega_0=\lambda_0\sigma_*$, 
$V_n$ first increases and then decreases as $n$ increases. 
This means that the wave resistence is relatively great during the transient process  in the liquid and is small for the steady-state flow; the transient time is relatively long. 
If the frequency is fixed in the interval $\omega <\omega_0$, it is clear from the figures that $V_n$ increases with $n$, i.e., the wave resistance grows with time and reaches its maximum in the steady-state flow; the transient time is relatively short. 
Graph of $V(\lambda)$ represents the energy supplied to the liquid by the  cylinder per unit of time (on average) as function of the frequency 
 $\omega=\lambda\sigma_*$.      
 
 From a certain point of view, the fluid in the problem at hand  is equivalent to a linear oscillator with absorption curve \eqref{c8_5.3} \cite[]{pain}.
 The following result gives an additional 
credence to the similarity (with respect to energy absorption) between the liquid and an oscillator of finite quality  factor. 	 

When the cylinder oscillates  in a general way \eqref{c8_2.1} due to external force, the power of the  force averaged over time interval $t$ is (up to small terms of higher order)
$$
Q_t=-\gamma_+\frac{a^2\varepsilon_*}{|2f_0|^3}
\left(1-\varepsilon^*\frac{f_0}{|f_0|}\right)\frac{1}{t}
\int\limits_0^{t}\left[ a_1^*(t)-\frac{1}{2}a_2^*(t)\right]u_*(t)\,dt.
        $$              
Making use of formulas \eqref{c8_2.1}, \eqref{c8_2.10}, and  \eqref{c8_4.4}, we obtain  (for the steady-state fluid flow)
$$
-Q=-\lim_{t\to +\infty}Q_t=-2\pi\nu\sigma_*\sum_{j}(\alpha_j^2+\beta_j^2)
a^2\lambda_j^3G(\lambda_j)=\nu\sum_jA_j^2V(\lambda_j),
        $$
$$
\lambda_j\sigma_*=\Omega_j,\,\,\,\,\,\,A_j^2=(\alpha_j^2+\beta_j^2)a^2.              
        $$
This result shows that the averaged energy absorbed by the liquid per unit time is equal to the sum of the absorbed ``partial'' energies corresponding to the points  of frequency spectrum of the ``input'' disturbance $ah(t)$, $\nu V(\lambda)$ being  considered as the squared gain-frequency characteristic of an equivalent oscillator. 

%
%
%
%
%

\addcontentsline{toc}{chapter}{Bibliography}


\begin{thebibliography}{99}

\bibitem[Abramovitz\&Stegun(1965)]{abra}
Abramovitz, M. and I.A. Stegun,  Handbook of Mathematical Functions. Dover Publ. Inc, New York, 1965

\bibitem[Caulfield et al(1996)]{caul}
Caulfield C.P., S. Yoshida, W.R. Peltier,
Secondary instability and three-dimensionalization in a laboratory accelerating shear layer with varying density differences,
Dynamics of Atmospheres and Oceans,
Volume 23, Issues 1-4,
1996,
Pages 125-138,
ISSN 0377-0265,
https://doi.org/10.1016/0377-0265(95)00418-1

\bibitem[Chomaz et al(1996)]{chomaz}
Chomaz, J. M., Y. Kadri, P. Bonneton and M. Perrier. Stratified flow over three-dimensional topography. Dynamics of Atmospheres and Oceans, 1996, vol. 23, pp.321-334.

\bibitem[Courant\&Hilbert(1937)]{courant}
Courant, R. and D. Hilbert. Methods of Mathematical Physics, vol. I,II. Interscience Publishers, Inc., New York.

\bibitem[Feir(1967)]{feir}
Feir, J.E., Discussion: Some results from wave pulse experiments, in: A Discussion on Nonlinear Theory of Wave Propagation in Dispersive Systems. Proc. R. Soc. Lond., 1967, vol. 299, pp. 54-58.


\bibitem[Milne-Thomson(1950)]{miln}
Milne-Thomson, L.M. Theoretical Hydrodynamics, Macmillan, New York, 1950

\bibitem[a1(2121)]{aa1}
Mindlin, I.M. Theory of non-stationary vortices in an ideal fluid. Fluid Dynamics, vol. 15,  No 6, 1980, pp. 797-803

\bibitem[Mindlin(1984)]{mind1984}
Mindlin, I.M.  On vorticity-induced waves in a homogeneous
incompressible fluid. PMM, Appl.Math.\& Mech.,
{\bf 48}, 5 (1984), 550-555

\bibitem[a2(2121)]{aa2}
Mindlin, I.M. A new method in nonlinear problems of waves in a heavy stratified liquid excited by a vertically moving body. Fluid Dynamics, vol.26,  No 5, 1991, pp. 763-770

\bibitem[a3(2121)]{aa3}
Mindlin, I.M. Nonlinear waves in a heavy two-layer liquid generated
by an initial disturbance of the horizontal interface. Fluid Dynamics,
{\bf{29}}, 3 (1994), 407-413

\bibitem[a4(2121)]{aa4}
Mindlin, I.M., Nonlinear waves in a heavy two-layer liquid generated by an extended initial disturbance of the horizontal interface: Exact solution. Fluid Dynamics, 1995, vol. 30, no. 6, pp. 943-946.

\bibitem[a5(2121)]{aa5}
Gilman, O.A. and I.M. Mindlin. Analytic investigation of a nonlinear problem of the effect of buoyancy on the evolution of an annular vortex.     
Fluid Dynamics, {\bf{31}}, 1 (1996), 47-56

\bibitem[a6(2121)]{aa6}
Mindlin, I.M. Waves excited by variable pressure on the free surface of a heavy  liquid. Fluid Dynamics,  {\bf{31}},  3 (1996), 418-428. 

\bibitem[Mindlin(1996)]{mind1996}
Mindlin, I.M., Integrodifferentsial’nye uravneniya v dinamike tyazheloi sloistoi zhidkosti (Integro-differential
Equations in Dynamics of a Heavy Layered Liquid), Moscow: Nauka, 1996.

\bibitem[a7(2121)]{aa7}
Gilman, O.A. and I.M. Mindlin. Waves  in a heavy two-layer liquid excited by an oscillating sphere. 
Fluid Dynamics, {\bf{33}},  2 (1998), 252-263

\bibitem[a8(2121)]{aa8}
Gilman, O.A. and I.M. Mindlin. Waves  in a heavy two-layer liquid excited by a cylinder moving at an angle to the horizontal
Fluid Dynamics, {\bf{33}}, 4, (1998), 580-593

\bibitem[a9(2121)]{aa9}
Mindlin, I.M., Nonlinear waves in two-dimensions generated by variable pressure acting on the free surface of
a heavy liquid, J. Appl. Math. Phys. (ZAMP), 2004, vol. 55, pp. 781-799.

\bibitem[a0(2121)]{aa0}
Mindlin, I.M. Water Waves: Theory and Experiments. 
Fluid Dynamics, {\bf{55}}, 4 (2020), 498-510.

\bibitem[Pain(1976)]{pain}
Pain, H.J. The Physics of Vibrations and Waves. Wiley, London (1976) 

\bibitem[Rabinovich et al(2008)]{rabin}
Rabinovich, A. B., Lobkovsky, L. I., Fine, I. V., Thomson, R. E., Ivelskaya, T. N., and Kulikov, E. A.: Near-source observations and modeling of the Kuril Islands tsunamis of 15 November 2006 and 13 January 2007, Adv. Geosci., 14, 105-116, https://doi.org/10.5194/adgeo-14-105-2008, 2008.

\bibitem[Stoker(1953)]{stoker}
Stoker, J.J. Water Waves. Interscience Publishers, Inc.,  New York ,1953

\bibitem[Suetin(1964)]{suetin}
 Suetin, P.K. Classical Orthogonal Polynomials. Nauka, Moscow, 1964 (in Russian)
 
\bibitem[Tyvand\&Miloh(1994)]{tyv}
Tyvand, Peder A.  and Touvia Miloh. Axisymmetric interaction between a vortex ring and a free surface. Physics of Fluids, 6, 224 (1994); doi: 10.1063/1.868070

 
 \bibitem[Whittaker\&Watson(2021)]{whitwat}
 Whittaker, E.T.  and G.N. Watson. A Course of Modern Analysis.
 Cambridge University Press, UK, 1st-5th editions: 1902, 1915, 1920, 1927, 2021.

\bibitem[Yuen\&Lake(1982)]{yuen}
Yuen, H.C. and Lake, B.M., Nonlinear Dynamics of Deep-Water Gravity Waves, Advances in Applied Mechanics,
vol. 22, New York, London: Academic, 1982, p. 67.

\bibitem[Zakharov\&Shabat(1972)]{zakh}
Zakharov, V.E. and Shabat, A.B. (1972) Exact Theory of Two-Dimensional Self-Focusing and One-Dimensional Self-Modulation of Waves in Non-Linear Media. Journal of Experimental and Theoretical Physics, 34, 62.

\end{thebibliography}
\end{document}